\documentclass[aps,prx,reprint,preprintnumbers,superscriptaddress,nofootinbib,longbibliography,floatfix]{revtex4-2}
\pdfoutput=1
\usepackage{rotating}
\usepackage{array}
\usepackage{amsmath}
\usepackage[normalem]{ulem}
\usepackage{slashed}
\usepackage{booktabs}
\usepackage[pdftex,table]{xcolor}
\usepackage{units}
\usepackage{xfrac}
\usepackage{mathtools}
\usepackage{empheq}
\usepackage[]{units}
\usepackage{multirow}
\usepackage{amssymb}
\usepackage{url}
\usepackage{comment}
\usepackage{physics}
\usepackage{color,soul}
\usepackage{bbm}
\usepackage[caption=false]{subfig}
\usepackage{adjustbox}
\usepackage[T1]{fontenc}

\usepackage{hyperref}
\hypersetup{
  colorlinks=true,
  citecolor=blue,
  linkcolor=blue,
  urlcolor=blue
}

\newcommand{\pt}{\mathrm{p_{T}}}
\newcommand{\geant}{\textsc{Geant4}}
\newcommand{\fastjet}{\textsc{FastJet}}

\newcommand{\MGFive}{\textsc{Madgraph5\_aMC@NLO}\ }
\newcommand{\PythiaEight}{\textsc{Pythia8}\ }
\newcommand{\PowhegBox}{\textsc{POWHEG-BoxV2}\ }

\begin{document}

\title{Searching for Anomalies with Foundation Models}

\author{Vinicius Mikuni}
\email{vmikuni@hepl.phys.nagoya-u.ac.jp}
\affiliation{Nagoya University, Kobayashi-Maskawa Institute, Aichi 464-8602, Japan}

\author{Benjamin Nachman}
\email{nachman@stanford.edu}
\affiliation{Department of Particle Physics and Astrophysics, Stanford University, Stanford, CA 94305, USA}
\affiliation{Fundamental Physics Directorate, SLAC National Accelerator Laboratory, Menlo Park, CA 94025, USA}

\begin{abstract}
Foundation models have the potential to expand the discovery reach for anomaly detection searches.  When studying the large OmniLearned foundation model on data from the CMS experiment, unexpected behavior was observed in a mass sideband.  The purpose of this paper is to perform a full analysis, including a complete background estimate, on the phase space selected by the large model.  We find that the background estimation describes the data well in validation regions, but is unable to accurately model the signal region.  We invite further scrutiny of these events and our methods.
\end{abstract}

\maketitle

\vspace{10mm}

\section{Introduction}
\label{sec:intro}

Anomaly detection (AD) is the automated search for new phenomena.  Foundation models (FM) are machine-learned representations that are pre-trained on large datasets and can be used for many downstream tasks.  A number of recent FM proposals have explored their utility for AD~\cite{Mikuni:2024qsr,Mikuni:2025tar,Golling:2024abg,Li:2024htp,Bhimji:2025isp,Hsu:2026sww}.  For example, Ref.~\cite{Mikuni:2024qsr} showed that the OmniLearn FM~\cite{Mikuni:2025tar} can enable full phase-space AD to be sensitive to rare signals.  Ref.~\cite{Bhimji:2025isp} showed how the extended OmniLearned FM can be used to rediscover the top quark without any use of simulations.  The OmniLearned paper introduced three FMs, varying in size; the top quark rediscovery was only demonstrated with the small and medium models.  The large model was not included at the time because of its computational expense.

\begin{figure}
    \centering
    \includegraphics[width=0.9\linewidth]{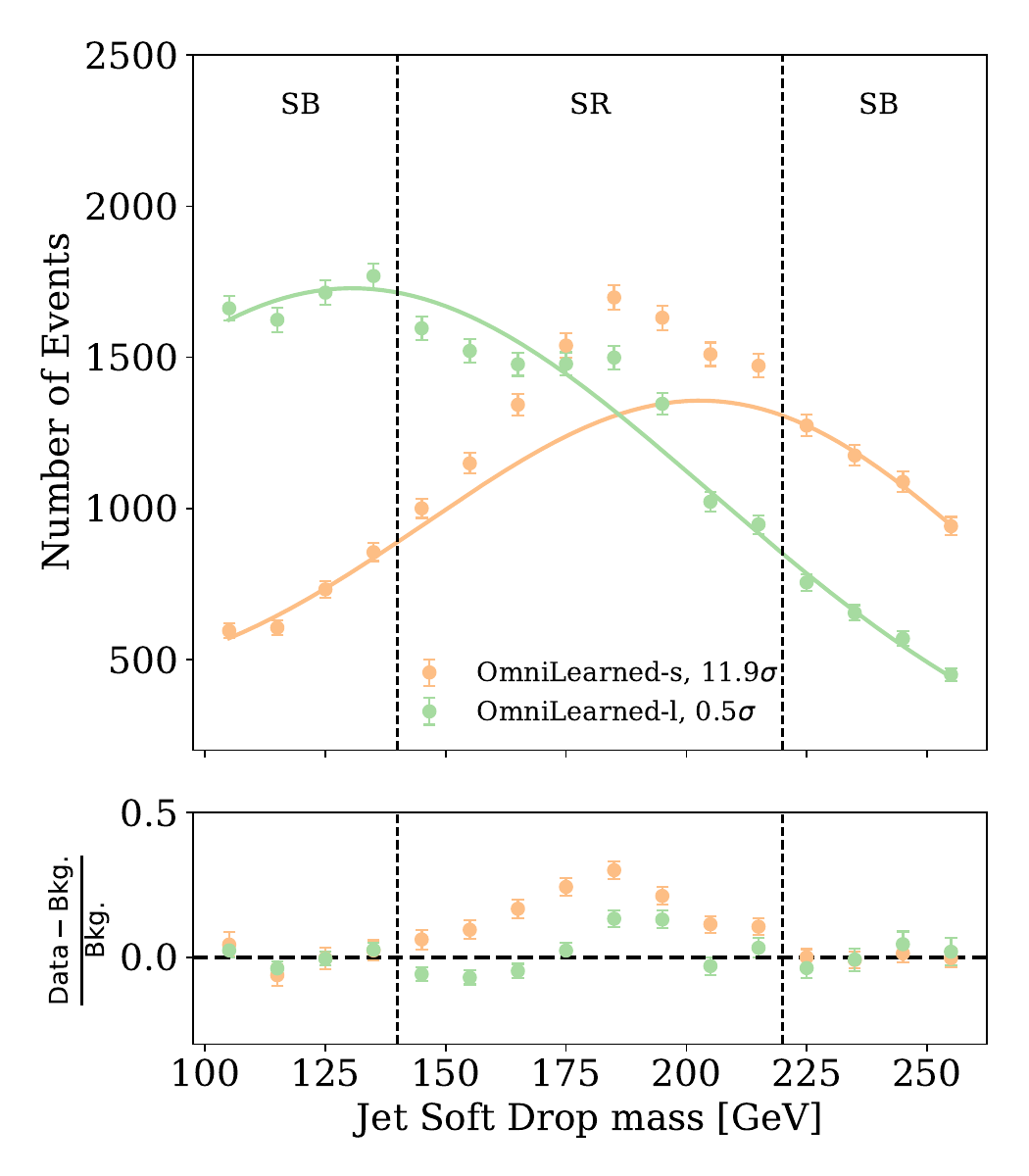}
    \caption{A histogram of the groomed jet mass after selecting the most anomalous jets based on the small or large OmniLearned models.  A parametric fit to the sidebands is also shown.  The fit is excellent for the small model and poor for the large model. }
    \label{fig:omnilearned}
\end{figure}

Since the publication of Ref.~\cite{Bhimji:2025isp}, we have repeated the top quark rediscovery analysis with the large model.  This analysis uses the Aspen Open Jets dataset~\cite{Amram:2024fjg}, a collection of data from the CMS experiment~\cite{Chatrchyan:2008aa}, and focuses on the groomed jet mass~\cite{Larkoski:2014wba} to do resonant anomaly detection.  The results for the small and large models are presented in Fig.~\ref{fig:omnilearned}.  The small model shows a clear excess in the top quark mass window.  The large model has a much weaker top peak in part because the fit is skewed by some unexpected shape in the left sideband.  In order to investigate further, we decided to perform a full analysis with the CMS Open Data directly.  A full analysis requires a rigorous data-driven background estimation with uncertainty quantification and cross-checks using alternative tools.  The purpose of this paper is to present what we found through this study.

The paper is organized as follows. Sec.~\ref{sec:samples} describes the CMS Open Datasets used to perform the full anomaly detection analysis. Sec.~\ref{sec:event} then describes the event selection and anomaly detection score. In  Sec.~\ref{sec:bkg} we discuss the data-driven background estimation method for QCD while Sec.~\ref{sec:unc} describes the sources of systematic uncertainties considered in this work. Sec.~\ref{sec:results} shows the results of the anomaly detection search and our findings. Finally, the paper ends with conclusions and outlook in Sec.~\ref{sec:conclusions}.

\section{Data and Simulated Samples}
\label{sec:samples}

Previously, we used only the measured data to search for anomalies, assuming the background description outside of anomalous events to follow a smoothly falling distribution. In this work, we want to include the simulation of known physics processes while modeling the major background from quantum chromodynamics (QCD) using a data-driven approach.  All simulated and data samples used in this work were processed from the 2016 release of the CMS Open Data. The data consist of proton-proton (pp) collisions with center-of-mass energy 13 TeV and total integrated luminosity of $16.39~\text{fb}^{-1}$\footnote{While the CMS Collaboration collected more than $36~\text{fb}^{-1}$ during the 2016 data-taking period, only this partial dataset is available.}. Our main goal is to investigate particle interactions resulting in pairs of highly energetic jets measured by the detectors. In this setting, the dominant background process is QCD multijet production. Even though the QCD contribution is estimated using a data-driven method, explained in Sec.~\ref{sec:bkg}, we process QCD simulations to validate the background estimation strategy. These samples were generated at leading order (LO) accuracy with \MGFive~\cite{Alwall:2014hca} using up to four partons in the matrix element calculation and interfaced with \PythiaEight~\cite{Sjostrand:2014zea} for showering and hadronization. Our second main process corresponds to the pair production of top quarks (t$\bar{\mathrm{t}}$). Samples are simulated at Next-to-Leading order (NLO) using \PowhegBox~\cite{Alioli:2010xd} interfaced to the NNPDF3.1 NNLO pdf set~\cite{Ball_2017}. Samples are generated in the 5-flavor scheme, where b-quarks are considered massless. The top quark mass is set to 172.5~GeV. Single top production processes are also generated at NLO using the same simulation setting as t$\bar{\mathrm{t}}$. The production of a vector boson with additional jets (W+jets and Z+jets) are simulated at LO with \MGFive, including decays to all flavors of quarks and up to three (W+jets) and four (Z+jets) additional partons at the matrix element level.
Additional smaller background samples are also considered such as the production of top quark pairs with an additional W or Z boson (t$\bar{\mathrm{t}}$V) and diboson production (VV) such as WW, ZZ, and WZ. These processes are simulated using \MGFive for the matrix element calculation, which is interfaced with \PythiaEight for showering and hadronization. Although expected to be negligible, we also consider single Higgs Boson production processes from gluon fusion, vector boson fusion, associated production with vector bosons, and in associated production with top quarks. All these processes are generated at NLO using the \PowhegBox generator and interfaced with \PythiaEight for showering and hadronization. 
All simulated samples include the effects of additional simultaneous pp collisions (pileup) using simulated minimum bias pp collisions that are superimposed on the primary simulated sample. Simulations of the interaction of particles with the detector material are performed by \geant~\cite{GEANT4:2002zbu}.

\section{Event Selection}
\label{sec:event}

We investigate the processes identified as anomalous by our search closely following the workflow previously used in~\cite{Bhimji:2025isp}. Our anomaly score is composed of the sum of multiple classes present during the pretraining phase of \textsc{OmniLearned}. In particular, \textsc{OmniLearned} is trained using 200 different class labels, ranging from specific nodes for known physics processes, such as the jets produced by the decays of top quarks, together with additional generic classes using different combinations of initial partons. The anomaly score is then defined by the sum of output class predictions consisting of particle decays to either 2-, 3-, or 4-prong jets, divided by the sum of all classes associated to QCD jets. The anomaly score is  calculated using the information of all visible particles clustered inside jets. In the CMS Open data, particles interacting with the detector material are reconstructed using the particle-flow (PF) algorithm~\cite{CMS:2017yfk}, combining the information of multiple detectors to improve the reconstruction quality. Jets are  clustered using PF candidates with the \fastjet~\cite{Cacciari:2011ma} package. For this work, we use high transverse momentum ($\pt$) and large-radius jets with distance parameter of 0.8~\cite{CMS-PAS-JME-16-003} and clustered using the anti-$k_{T}$ algorithm~\cite{Cacciari:2008gp}. Trigger paths that require a minimum threshold on jet $\pt$, mass, and the scalar $\pt$ sum of all jets are used to select interesting events. Additional selections are applied to recorded events to ensure high trigger efficiency for both data and simulations. Events are required to have at least two large-radius jets with minimum jet $\pt$ of 450 GeV, absolute pseudo-rapidity $|\eta| < 2.5$, and passing the tight jet identification threshold. Additionally, we require the soft drop mass with parameters $z_{cut} = 0.1$ and $\beta = 0$~\cite{Larkoski:2014wba} of the reconstructed jet to be above 55 GeV for all jets. These selections  ensure that the trigger paths used in this work are fully efficient, verified by measuring the trigger efficiency using particle collisions with an additional isolated muon and passing a single muon trigger. Events containing isolated muons ($\pt$ > 30 GeV, $|\eta| < 2.5$) or electrons ($\pt$ > 15 GeV, $|\eta| < 2.5$) are rejected.

\section{Background Estimation}
\label{sec:bkg}

\begin{figure*}[ht]
    \centering
        \includegraphics[width=.78\textwidth]{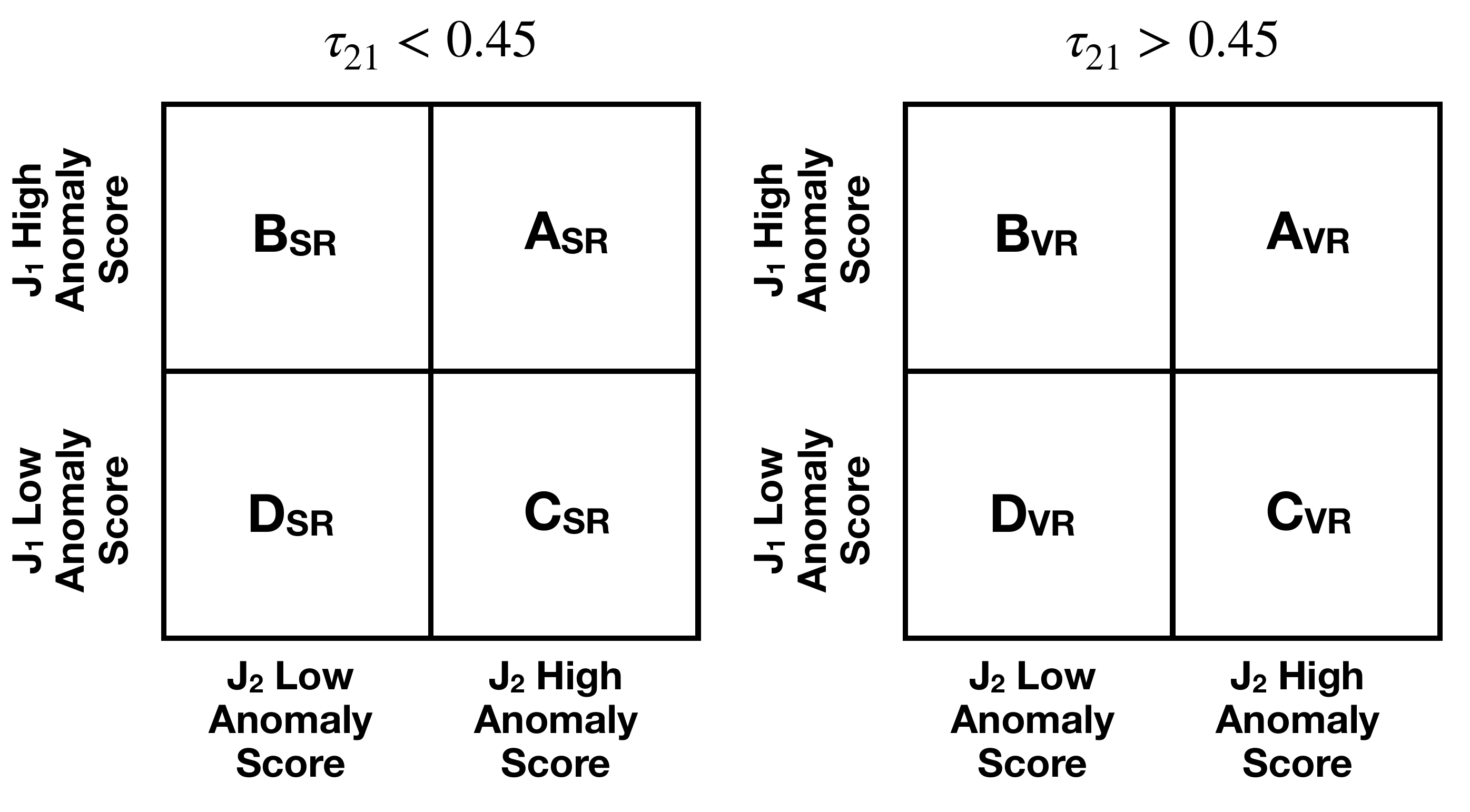}
    \caption{Visual representation of the eight different regions used simultaneously to fit and constrain the contribution of different backgrounds.}
    \label{fig:regions}
\end{figure*}

After basic event selection, the dominant background process is the QCD multijet production. Other physics processes besides QCD are estimated using data-driven simulation-based predictions, including the second most relevant physics process, top quark pair production. Since
we define regions of interest based on the \textsc{OmniLearned} anomaly score, we need to estimate scale factors to correct the simulation efficiencies compared to observed data. A separate validation region is used to determine the simulation efficiencies by splitting the data based on the $\tau_{21}$ subjettiness~\cite{Thaler:2010tr} observable. The region where both jets have $\tau_{21} < 0.45$ is used to select events consistent with jets more likely to contain 2 separate subjets as opposed to a single one. This region is later used to constrain the backgrounds produced by the decays of W and Z bosons. The complementary region, where at least one jet has $\tau_{21} > 0.45$, is then dominated by top quark pair production after reducing the dominant background process from QCD multijet production. The scale factor is then determined as an additional normalization factor that scales all simulated processes by the same amount, common between the two subjettiness regions. During the fit to the data, both subjettiness regions are used simultaneously to determine the scale factor. 

While simulations for QCD production are available, the regions of the phase space  selected by the anomaly score may lead to unreliable predictions. For this reason, we instead use a data-driven approach to determine the contribution of the QCD background in the region of interest. We use the so called ``ABCD'' method, where two independent observables are used to define four regions based on a pass or fail criteria for each observable. The two observables used in this work are the anomaly scores for each of the jets.  
We select the thresholds to create the regions for the ABCD prediction by fixing the data efficiency in the main measurement region (A) to a fixed value. Regions B and C are then defined by collision events where one jet passes and one fails the selection criteria. Finally, region D is where both jets fail the selections.  The estimation of the number of QCD jets in region A, N$_{A}$, is then defined by:
\begin{equation}
    \text{N}_{A} = \frac{\text{N}_{B}\text{N}_{C}}{\text{N}_{D}}.
    \label{eq:abcd}
\end{equation}

We select the data efficiency threshold to define the anomalous region A to be $0.2\%$, resulting in roughly 1000 particle collisions in the most anomalous region. This choice allows us to remove the majority of the QCD background while retaining enough interesting events produced by other physics processes. It is important to note that when performing a targeted search, the selection threshold is chosen to maximize the sensitivity of the process of interest. Since we are performing anomaly detection, our main goal is to remove the dominant background rather than optimizing for a specific physics process. The combination of four regions for the ABCD determination in the two subjettiness regions results in a total of eight independent regions used simultaneously during data extraction to constrain different background processes, as shown in Fig.~\ref{fig:regions}.

\section{Uncertainties}
\label{sec:unc}

A number of experimental and theoretical uncertainties are included in this work as nuisance parameters during signal extraction. Experimental uncertainties include the jet energy scale and resolution, determined from in-situ measurements of the momentum imbalance using dijet, Z+jet, photon+jet, and multijet events and are propagated to the simulations together with their uncertainties~\cite{Khachatryan:2016kdb}. Additional 2\% uncertainties for both the jet mass scale and resolution are considered independently from the jet energy scale and resolution, as recommended~\cite{CMS-DP-2023-044}. Additional experimental uncertainties include corrections from differences in pileup levels between data and simulated samples, the L1 pre-firing correction~\cite{CMS:2020cmk}, and the uncertainty of the total integrated luminosity, estimated to be 2.5\%. Theoretical uncertainties are also included for the different simulated processes used in this work. Independent variations of the renormalization and factorization scales by factors of two are used to determine the uncertainty from the approximations used in the matrix element calculations. The envelope that describes the largest impacts of these variations is used as a single nuisance parameter per physics process. Uncertainties in the parton shower simulation are included by changing the scale at which $\alpha_s$ is evaluated in the shower by factors of two and a half. This is done independently for initial and final state radiation, creating two independent nuisance parameters for each physics process. An additional uncertainty related to the mismatch of the top quark $\pt$ spectrum is included for t$\bar{\mathrm{t}}$ events and estimated based on predictions from NNLO Monte Carlo generators. All theory uncertainties are taken as decorrelated between physics process while the experimental uncertainties are considered correlated. 
Next, we have the uncertainties from the QCD background prediction and scale factor determination from the selection applied based on the anomaly detection score. The ABCD prediction is implemented separately for each bin of the target distribution. To account for the contribution and uncertainty of other physics processes during the ABCD estimation we include all regions necessary for the ABCD calculation in the final fit, i.e the yield normalization for QCD in regions B, C, and D are free parameters and determined simultaneously during the fit while the prediction of the QCD background in region A follows Eq.~\ref{eq:abcd}. The scale factor determination for top quarks is carried out by performing the fit simultaneously between different subjettiness regions. The constant scale factor corrects the t$\bar{\mathrm{t}}$ normalization in region A and is correlated between the two subjettiness regions. Since the contributions from physics processes other than top quark production are comparatively small, deriving an independent scale factor for each individual process is not feasible with the same approach. We therefore assume a common scale factor value shared across all relevant processes in the fit. To account for potential differences in the scale factor determination between physics processes, a conservative 50\% normalization uncertainty is assigned on top of the scale factor correction, applied separately for three groups of processes defined by their final-state signatures: single boson production (W+Jets and Z+Jets), diboson production (WW, ZZ, and WZ), and single top quark production. A simplified Barlow–Beeston approach~\cite{Barlow:1993dm}, with a single nuisance parameter per bin of the distribution, is used to account for the statistical uncertainties due to the limited size of the simulated background samples. The dominant systematic uncertainties are related to the parameter estimation of the ABCD prediction for the QCD background and the limited simulation size in the region with high anomaly score.

\begin{figure}[ht]
    \centering
        \includegraphics[width=.46\textwidth]{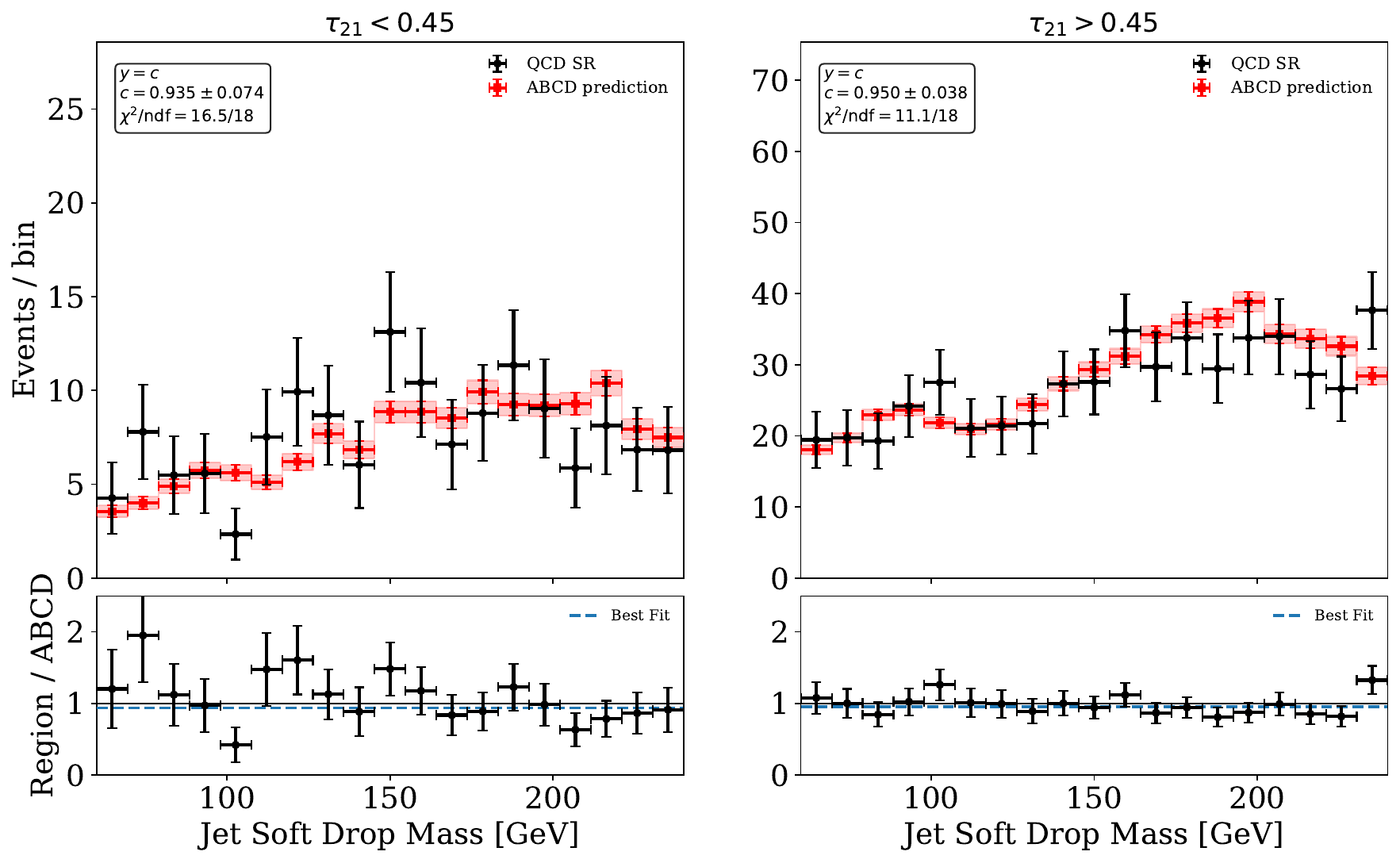}
    \caption{Compatibility test between the ABCD prediction using QCD simulated events and the soft drop mass distribution of QCD events in the region where both jets pass the $0.2\%$ data efficiency selection. A constant fit is performed to determine the compatibility between the prediction and the actual distribution, shown in blue. }
    \label{fig:abcd}
\end{figure}

\section{Results}
\label{sec:results}

We first show the results obtained by the proposed analysis strategy when the small \textsc{OmniLearned} model is used to create the anomaly score. Results are obtained by performing the binned maximum likelihood fit of the leading jet soft drop mass distribution, selected as the jet with highest $\pt$ in the dijet system. We consider the pair production of top quarks as the signal of interest whose overall normalization is determined by the fit. The signal extraction is based on the profile likelihood ratio test statistic~\cite{CMS-NOTE-2011-005}, combining the eight regions shown in Fig.~\ref{fig:regions}. To perform the fit, we use the \textsc{COMBINE}~\cite{CMS:2024onh} statistical package, which is based on the \textsc{ROOFIT}~\cite{Verkerke:2003ir} and \textsc{ROOSTATS}\cite{Moneta:2010pm} software packages.

\begin{figure}[ht]
    \centering
        \includegraphics[width=.23\textwidth]{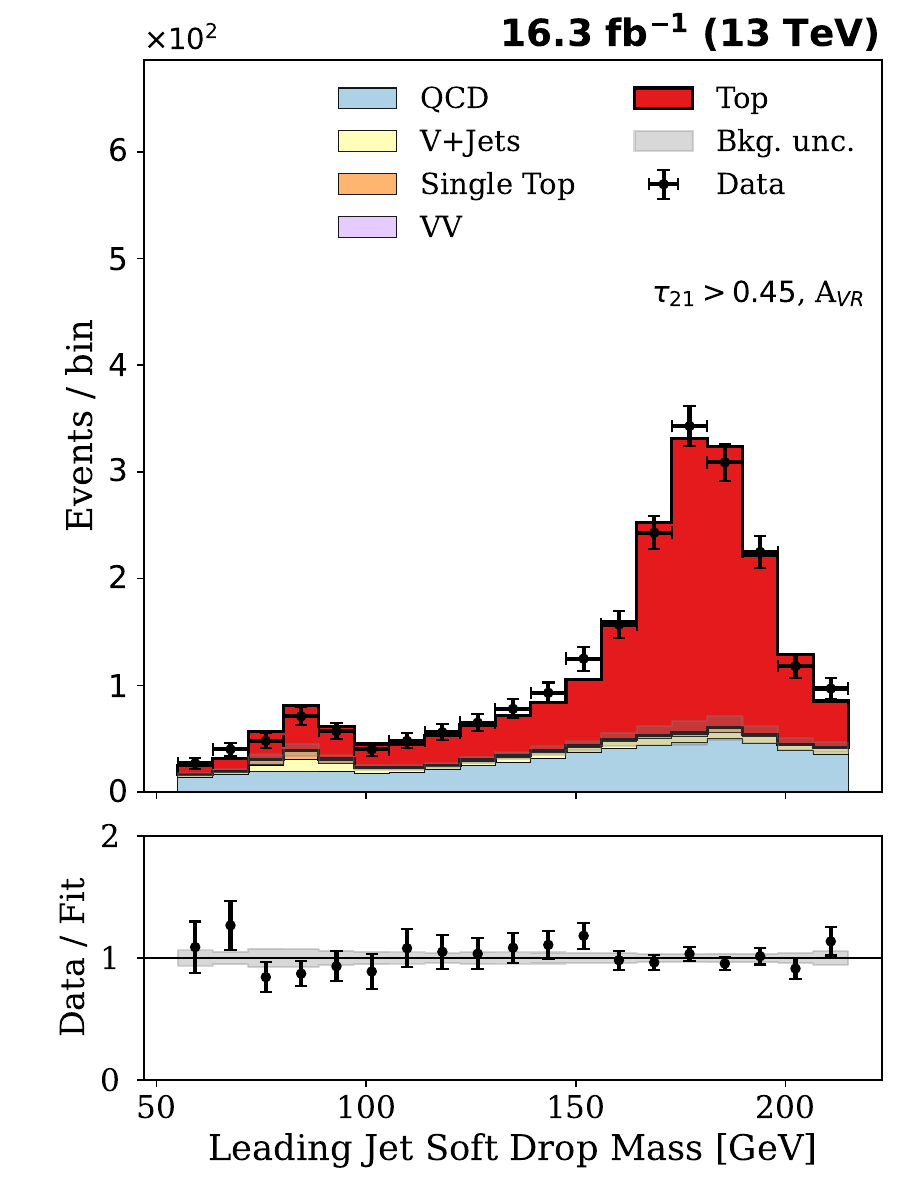}
        \includegraphics[width=.23\textwidth]{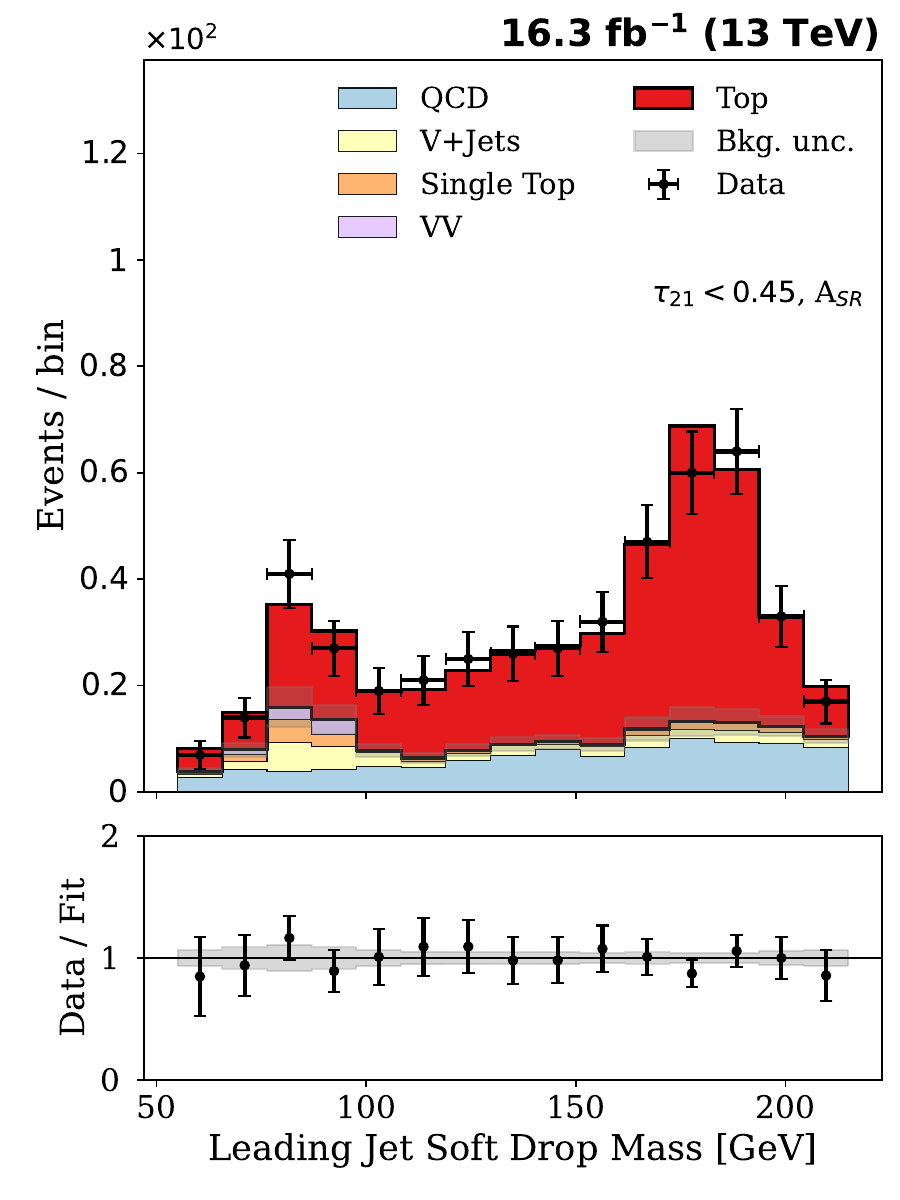}
    \caption{Leading jet soft drop mass where both jets are considered anomalous based on the \textsc{OmniLearned} small model score. The region where both jets have low $\tau_{21}$ values is shown on the right while the region where at least one jet fails the $\tau_{21}$ selection is shown on the left. Shaded regions represent the total background uncertainty.}
    \label{fig:small_results}
\end{figure}

We first validate the ABCD background estimation method by using simulations of QCD multijet production. In Fig.~\ref{fig:abcd}, we show the comparison between the ABCD-predicted QCD multijet background and the observed distribution obtained from QCD simulations in the signal region of interest. We evaluate the compatibility of the ABCD assumption by fitting the ratio between the predicted and calculated values with a constant function and assessing the $\chi^2$ of the fit.

\begin{figure}[ht]
    \centering
        \includegraphics[width=.48\textwidth]{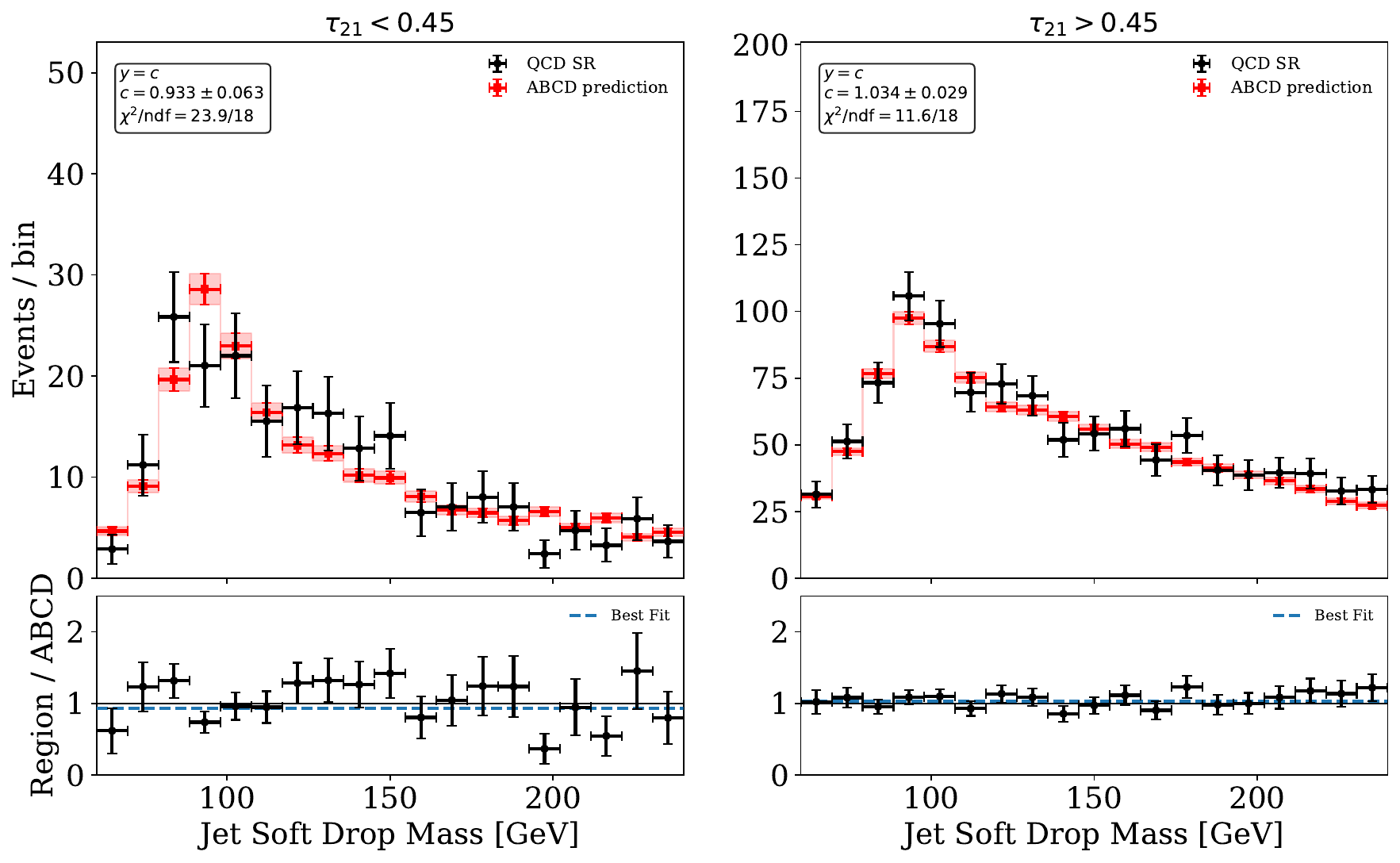}
    \caption{Compatibility test between the ABCD prediction using QCD simulated events and the soft drop mass distribution of QCD events in the region where both jets pass the $0.2\%$ data efficiency selection based on the large \textsc{OmniLearned} model. A constant fit is performed to determine the compatibility between the prediction and the actual distribution, shown in blue. }
    \label{fig:abcd_l}
\end{figure}

In both subjettiness regions we observe a good agreement between the expected and predicted QCD multijet background, compatible with unit within statistical uncertainties. Next, we proceed with the fit to extract the signal contribution using all eight combined regions, with results shown in Fig.~\ref{fig:small_results}. In App.~\ref{app:abcd} we show the fit results for the additional six regions relevant for the ABCD determination of the QCD background.

\begin{figure}[ht]
    \centering
        \includegraphics[width=.23\textwidth]{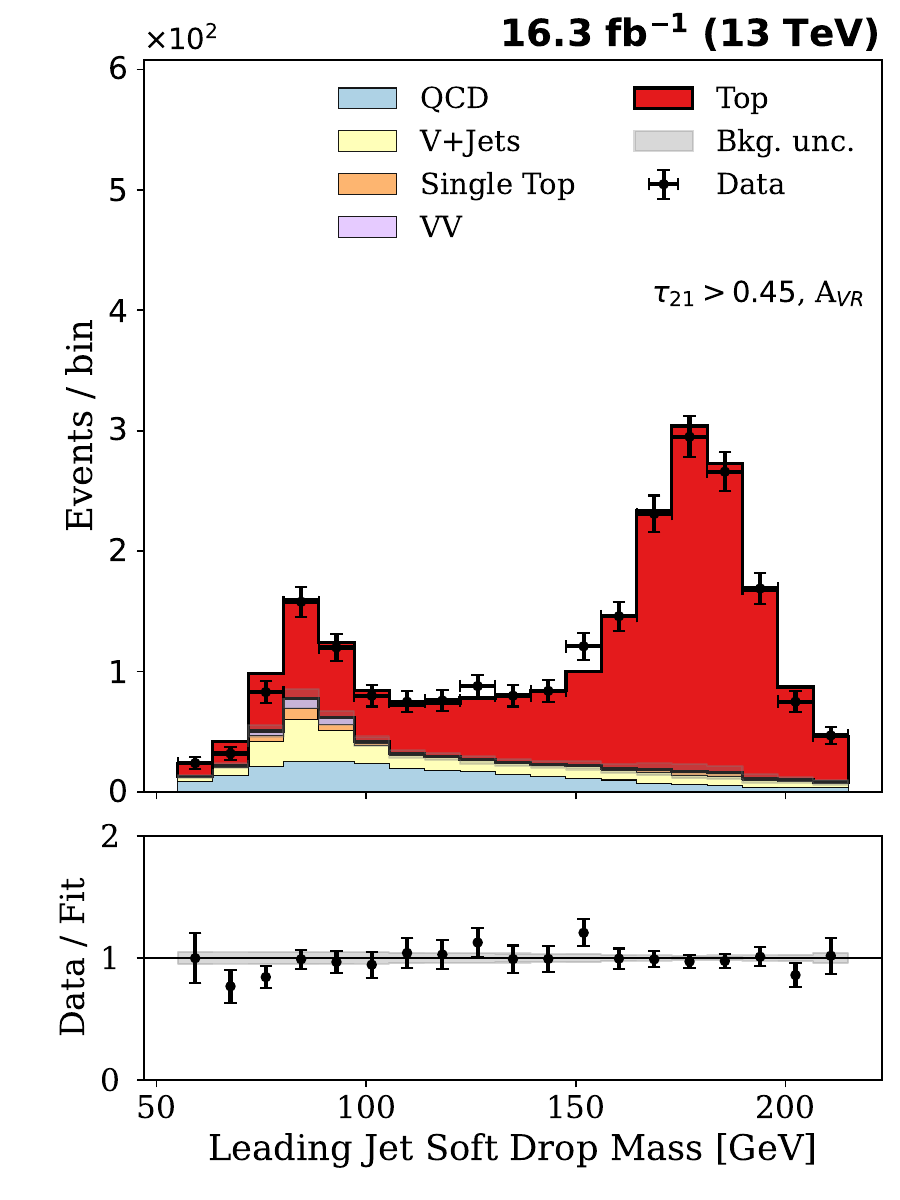}
        \includegraphics[width=.23\textwidth]{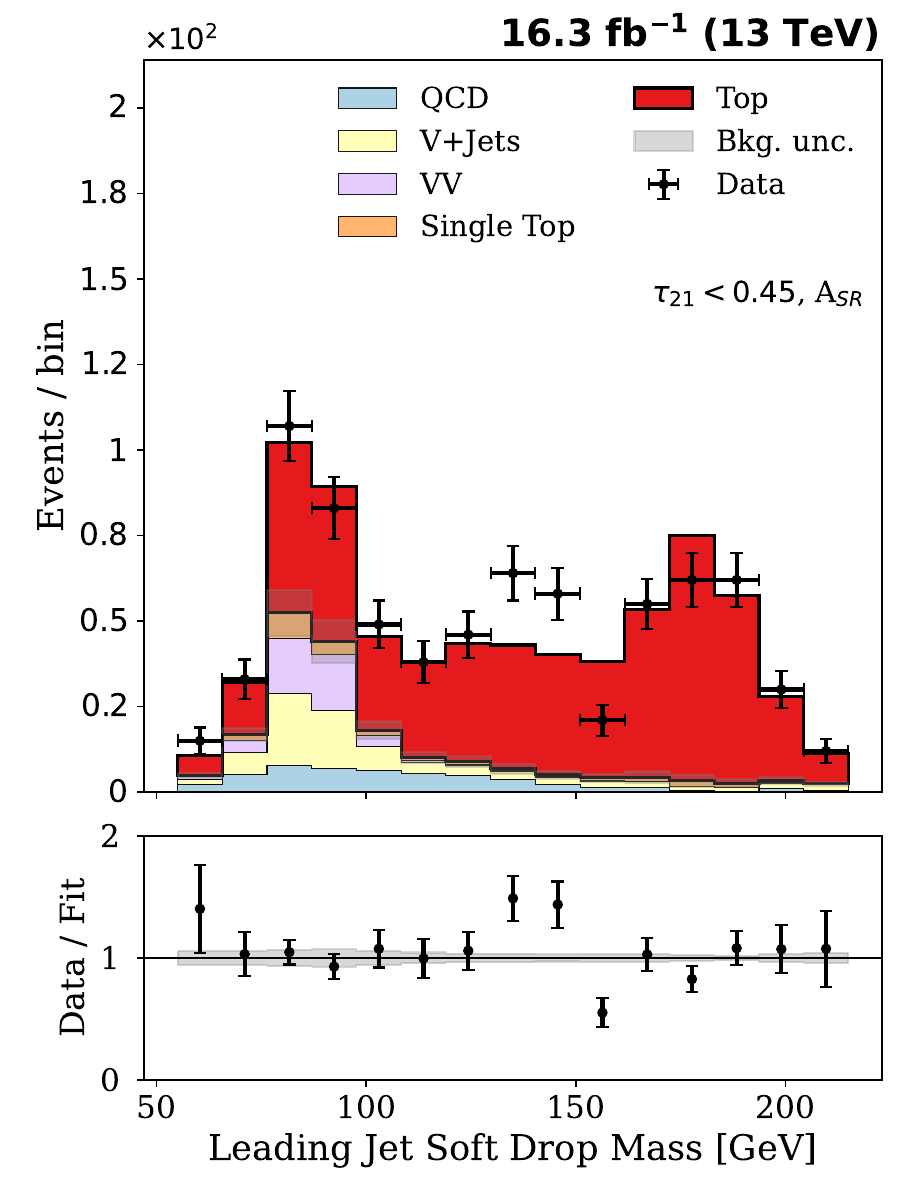}
    \caption{Leading jet soft drop mass where both jets are considered anomalous based on the \textsc{OmniLearned} large model score. The region where both jets have low $\tau_{21}$ values is shown on the right while the region where at least one jet fails the $\tau_{21}$ selection is shown on the left. Shaded regions represent the total background uncertainty.}
    \label{fig:large_results}
\end{figure}

In both subjettiness categories, the dominant process selected by the anomaly score is the pair production of top quarks, which is found to be well described by the simulations within uncertainties.  The significance obtained for t$\bar{\mathrm{t}}$ production using the Asymptotic approximation~\cite{Cowan:2010js} is well above 10, qualitatively consistent with our previous findings on the Aspen Open Dataset and other studies using CMS Open Data to rediscover t$\bar{\mathrm{t}}$ production in other phase space regions~\cite{Knapp:2020dde}. The QCD multijet background, which corresponds to 97.5\% of all measured events after basic event selection, contributes approximately 25\% and 30\% of the events in the main anomalous regions with low and high $\tau_{21}$ subjettiness, respectively. Here is important to reiterate that the anomaly score  does not use \textsc{OmniLearned} labels associated to known physics processes, but relies only on subjettiness information, making the selected events far from trivial.

\begin{figure}[ht]
    \centering
        \includegraphics[width=.23\textwidth]{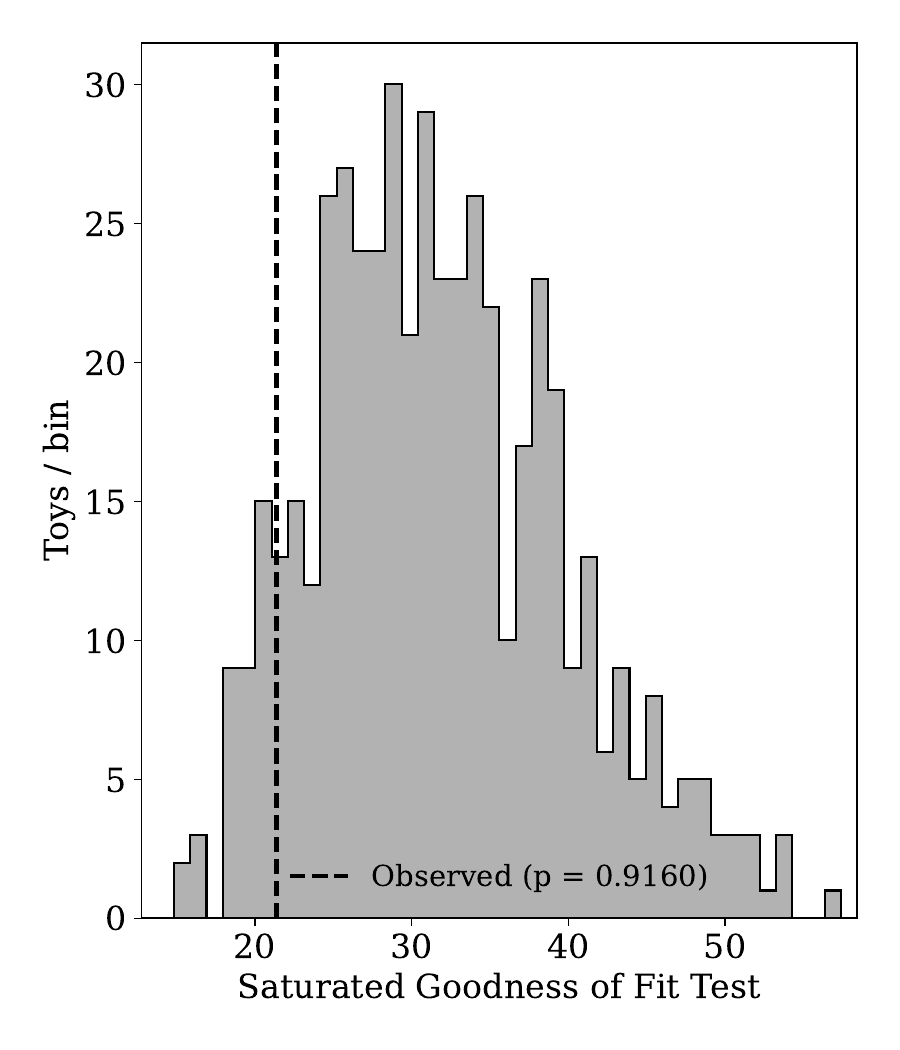}
        \includegraphics[width=.23\textwidth]{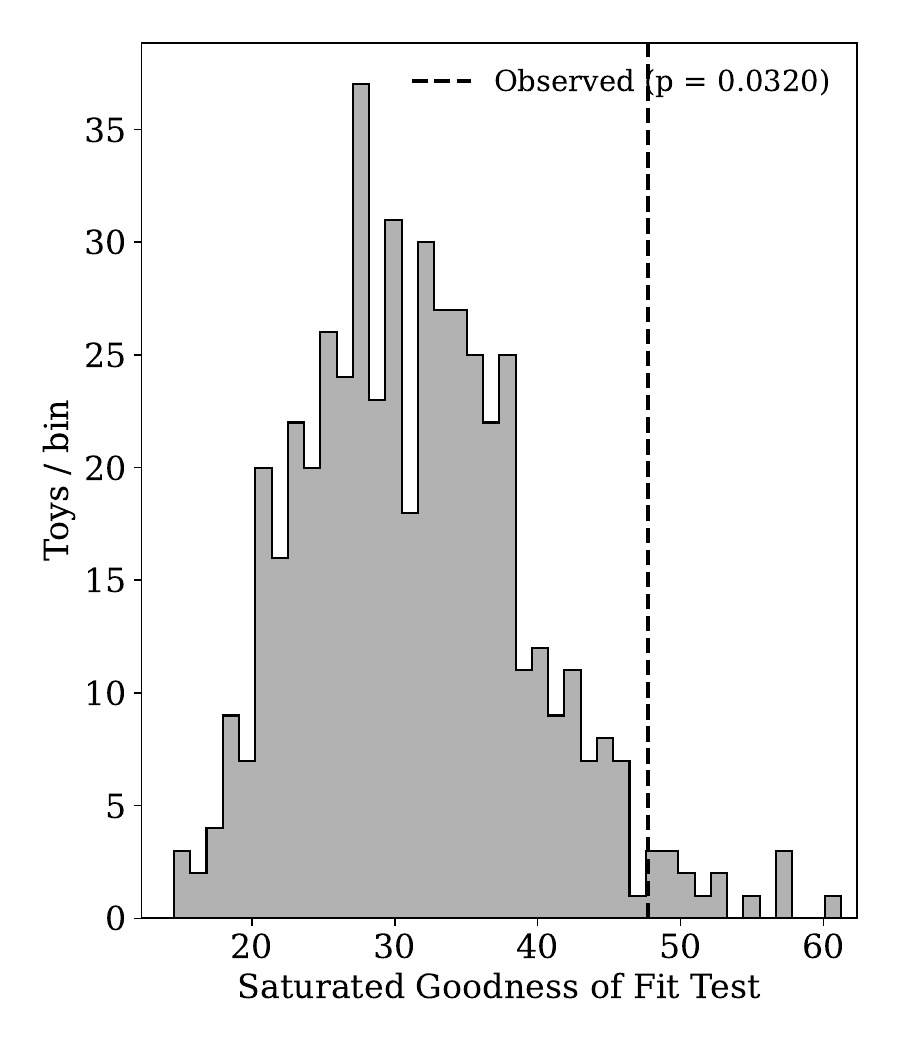}
    \caption{Goodness-of-fit test based on the saturated test statistic evaluated only in the anomalous regions of interest for the small (left) and large (right) \textsc{OmniLearned} model anomaly scores.}
    \label{fig:gof_bkg}
\end{figure}

Next, we repeat the same analysis strategy, but replace the anomaly detection score from the small \textsc{OmniLearned} model with that of the large model, with roughly 250 times more trainable parameters, but otherwise trained using the same data and loss functions as the small model. We first validate the ABCD assumption using simulated QCD events. The comparison is shown in Fig.~\ref{fig:abcd_l}. Again, we observe good agreement between the prediction and the distribution of QCD events within the statistical uncertainties of the available simulations. Next, we proceed using the same fit strategy as before, with fit results shown in Fig.~\ref{fig:large_results}.

\begin{figure}[ht]
    \centering
        \includegraphics[width=.23\textwidth]{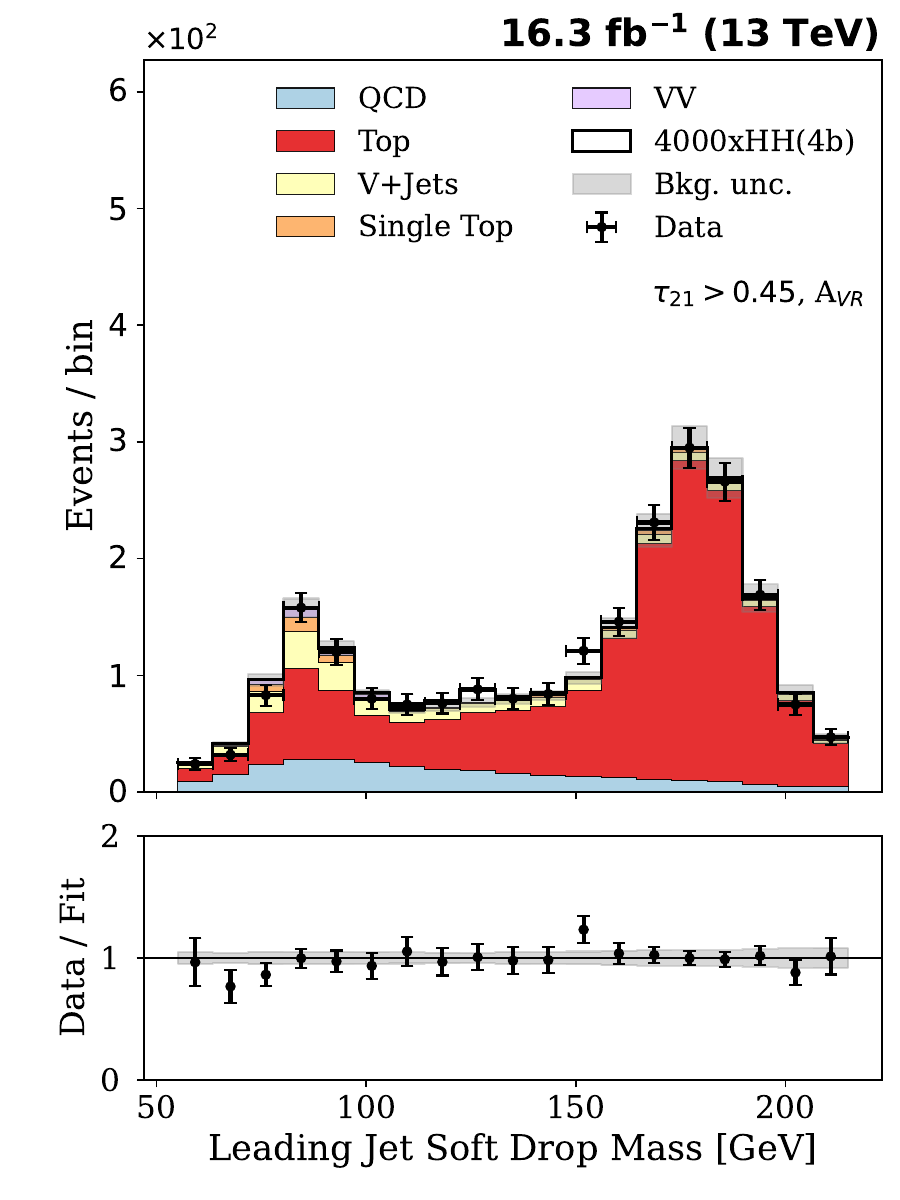}
        \includegraphics[width=.23\textwidth]{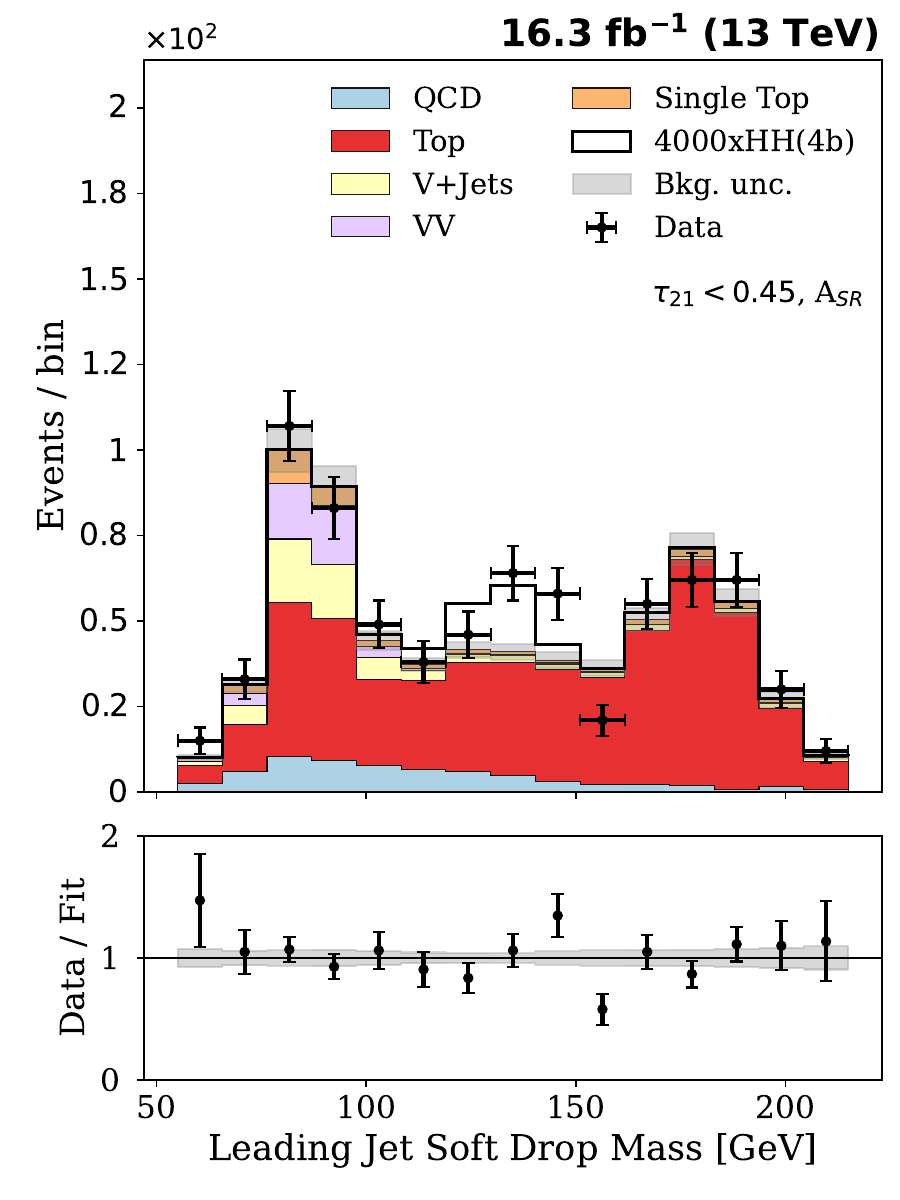}
    \caption{Leading jet soft drop mass where both jets are considered anomalous based on the \textsc{OmniLearned} large model score and the HH production is considered as the signal of interest. The region where both jets have low $\tau_{21}$ values is shown on the right while the region where at least one jet fails the $\tau_{21}$ selection is shown on the left. Shaded regions represent the total background uncertainty.}
    \label{fig:large_results_s}
\end{figure}
\begin{figure}[ht]
    \centering
        \includegraphics[width=.22\textwidth]{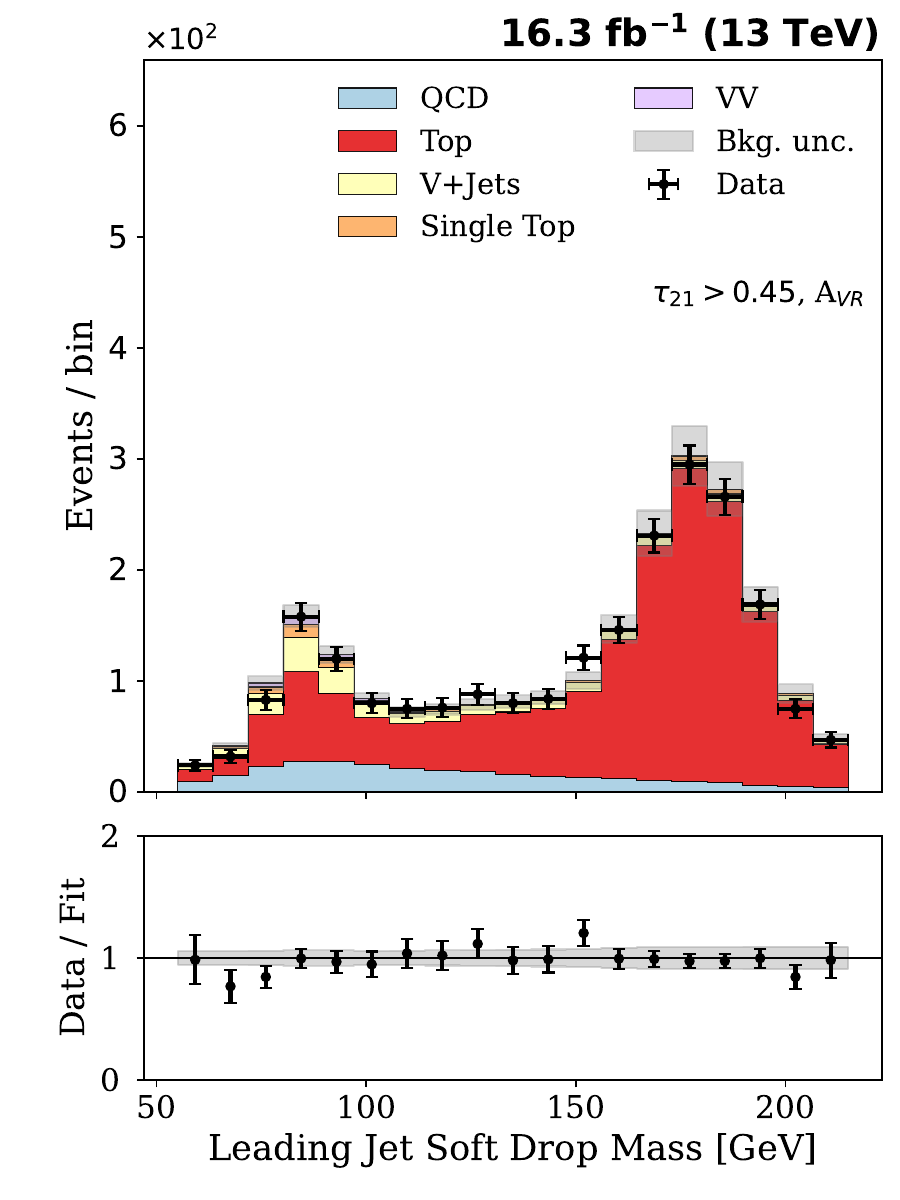}
        \includegraphics[width=.22\textwidth]{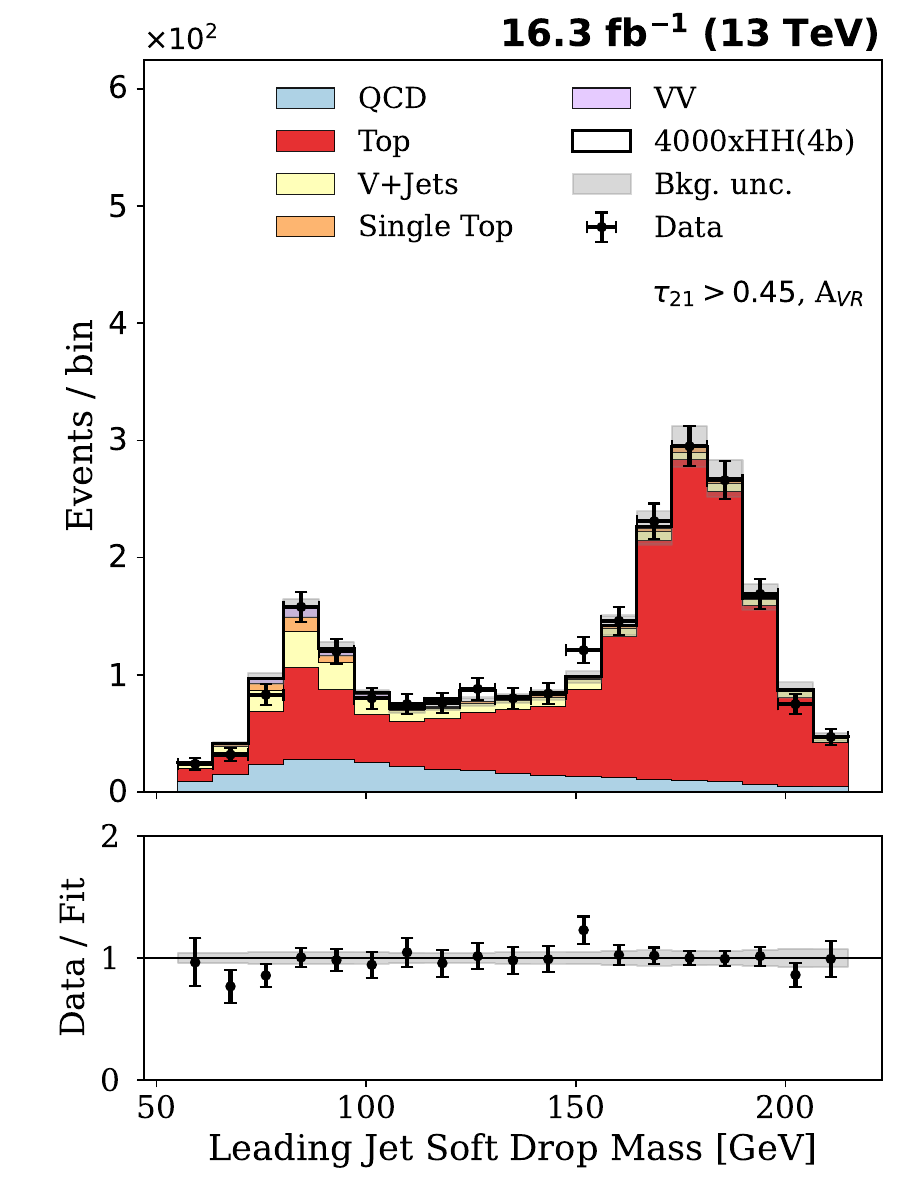}
        \includegraphics[width=.22\textwidth]{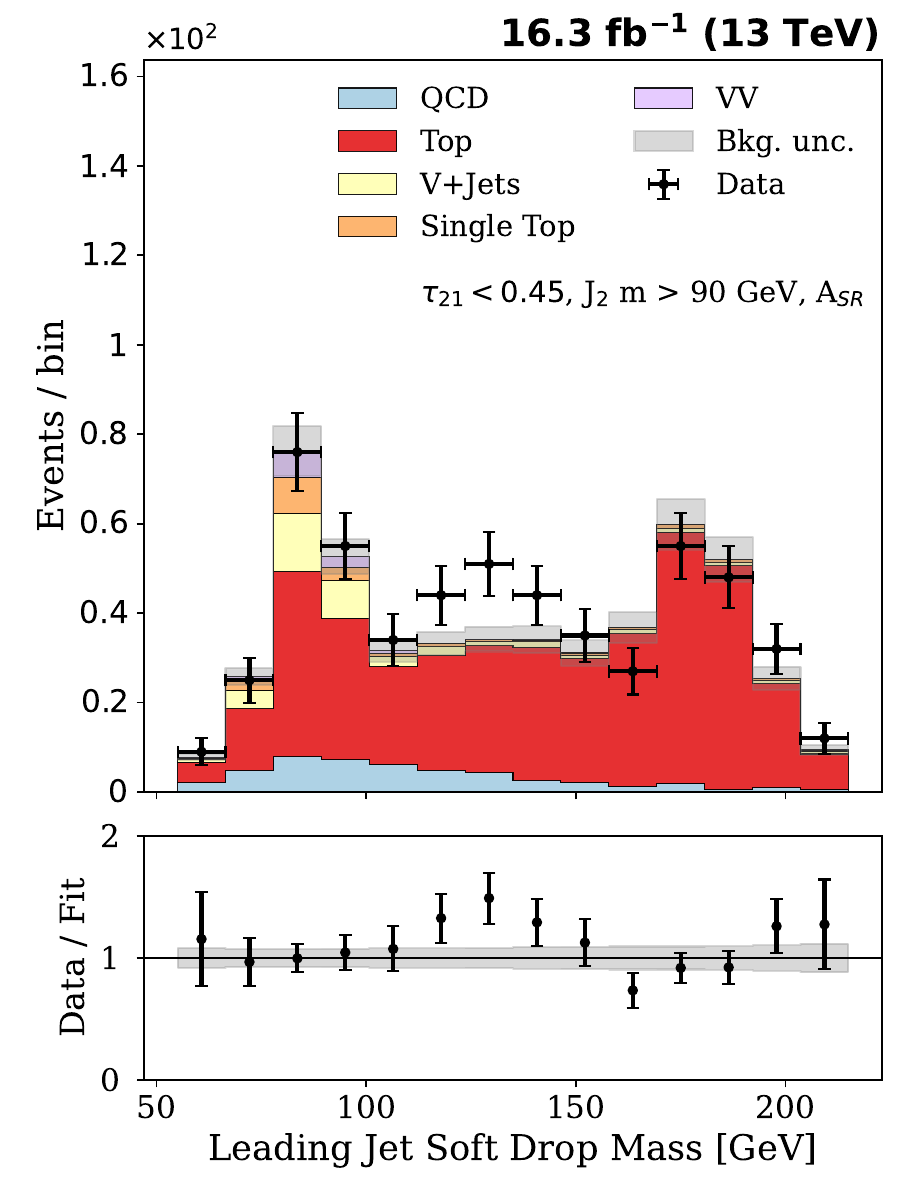}
        \includegraphics[width=.22\textwidth]{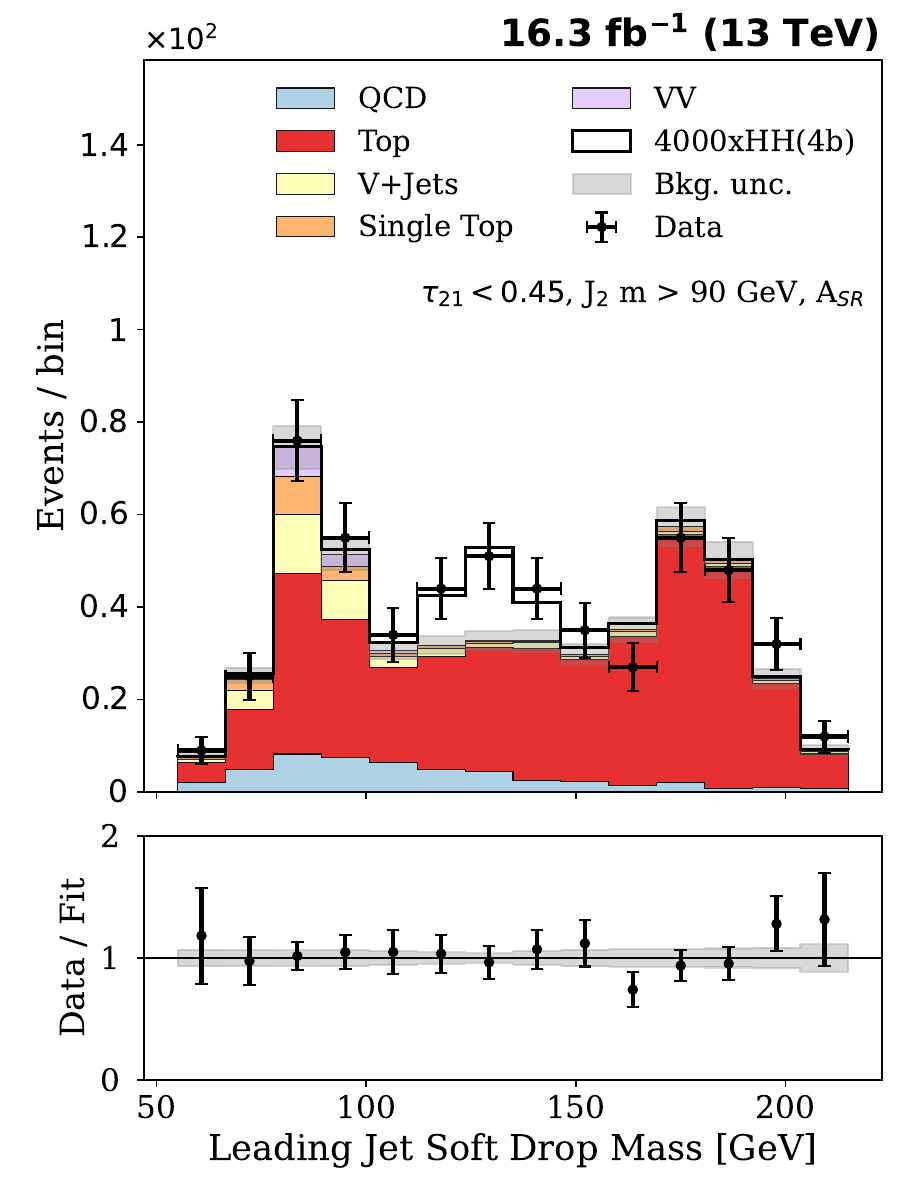}
        \includegraphics[width=.22\textwidth]{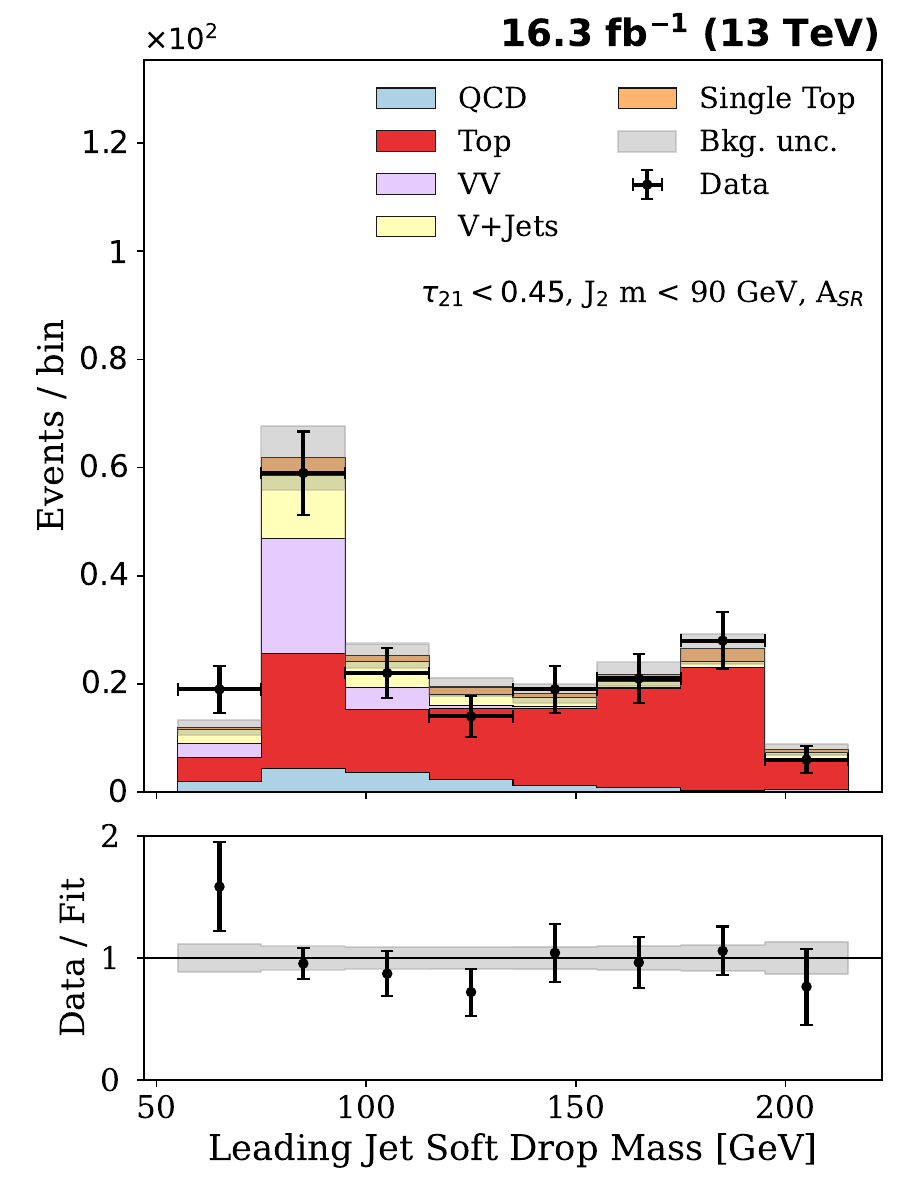}                
        \includegraphics[width=.22\textwidth]{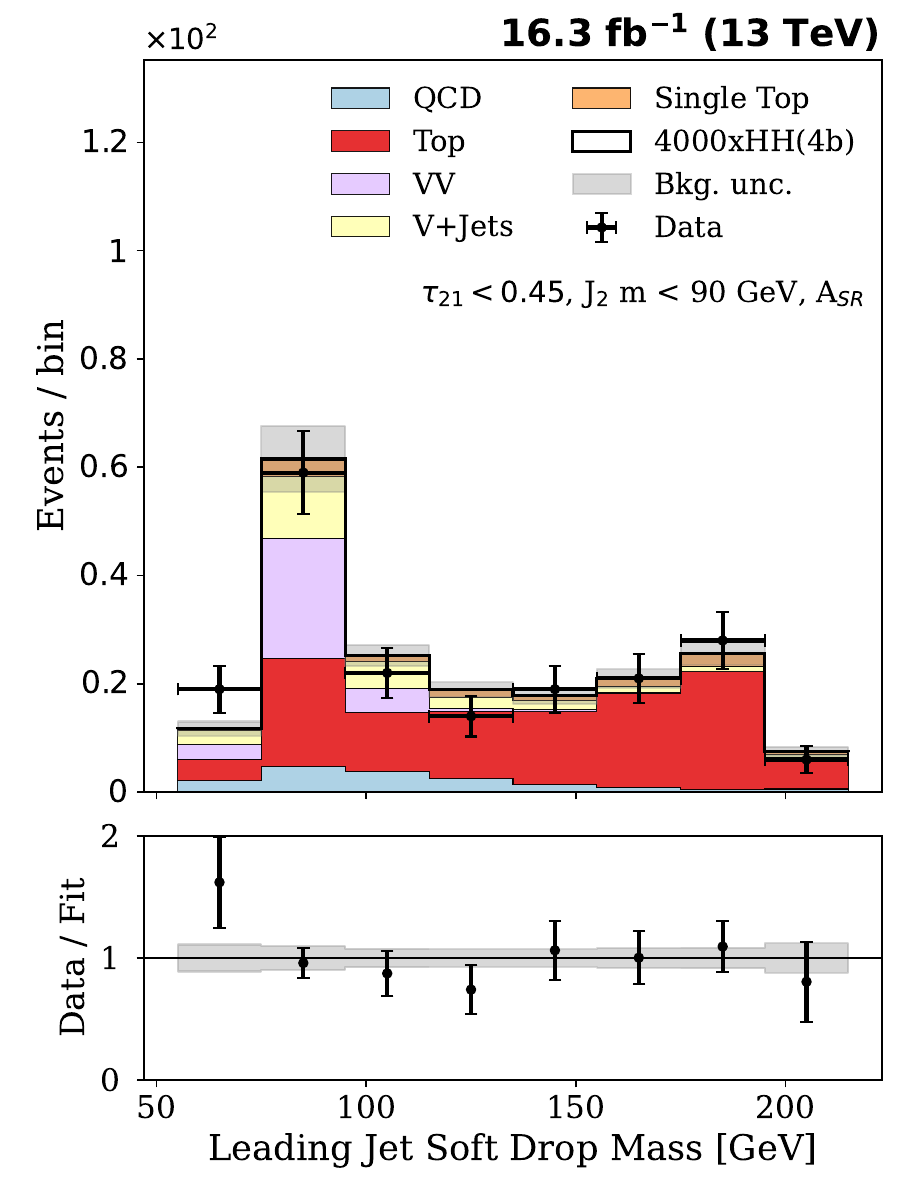}
       
    \caption{Leading jet soft drop mass where both jets are considered anomalous based on the \textsc{OmniLearned} large model score. The region where both jets have low $\tau_{21}$ values is split in the regions where the mass of the subleading jet is above (middle) and below (bottom) 90 GeV.  The region where at least one jet fails the $\tau_{21}$ selection is shown at the top. Results of the background-only fit are shown on the left while results considering a HH signal component are shown on the right. Shaded regions represent the total background uncertainty.}
    \label{fig:large_results_split_sr12}
\end{figure}

Notably, despite the only change being the use of a larger network model, the event composition selected by the anomaly detector changes considerably. In particular, while t$\bar{\mathrm{t}}$ remains the dominant process, we observe more events located at lower mass values. Additionally, the background description in the region where both jets have $\tau_{21} < 0.45$ shows a significant disagreement between the data and the fit results. We assess the compatibility between the fit model and the data in these anomalous regions by performing a goodness-of-fit (GOF) test based on the saturated test statistic~\cite{Cousins:2013saturated}, where the distribution of the test statistic is determined by sampling 500 toy datasets and the resulting distribution is compared to the observation to determine the p-value. Since by construction the ABCD control regions, apart from the main anomalous regions, are expected to agree with the data, we include them to constrain the normalization parameters during the fit, but calculate the GOF  only in the two main anomalous regions. The GOF test statistic distributions compared to the observed value are shown in Fig.~\ref{fig:gof_bkg} for both the small and large \textsc{OmniLearned} model scores. While the GOF test for the small model yield reasonable p-values, the large model shows an inconsistency with the data and expected physics processes as evidenced by the relatively small p-value.

\begin{figure}[ht]
    \centering
        \includegraphics[width=.23\textwidth]{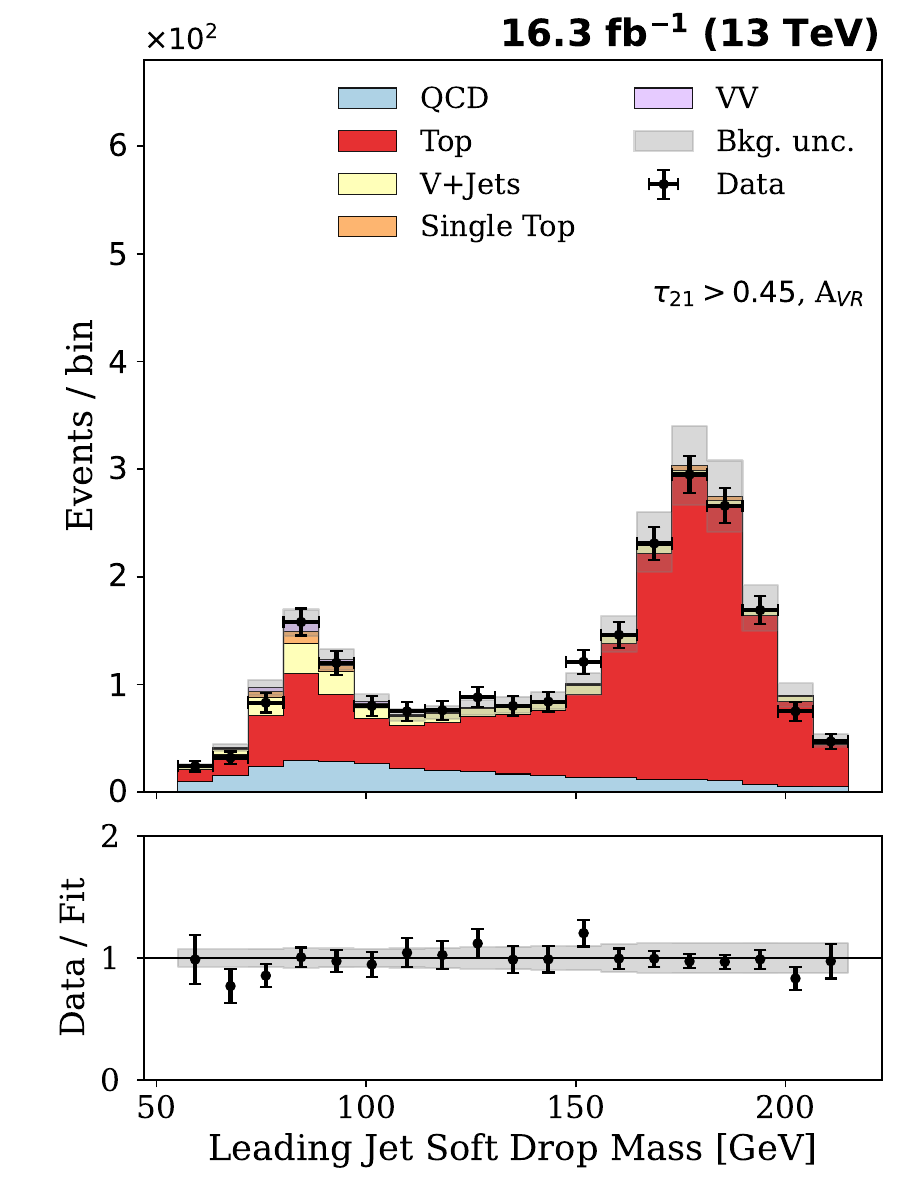}
        \includegraphics[width=.23\textwidth]{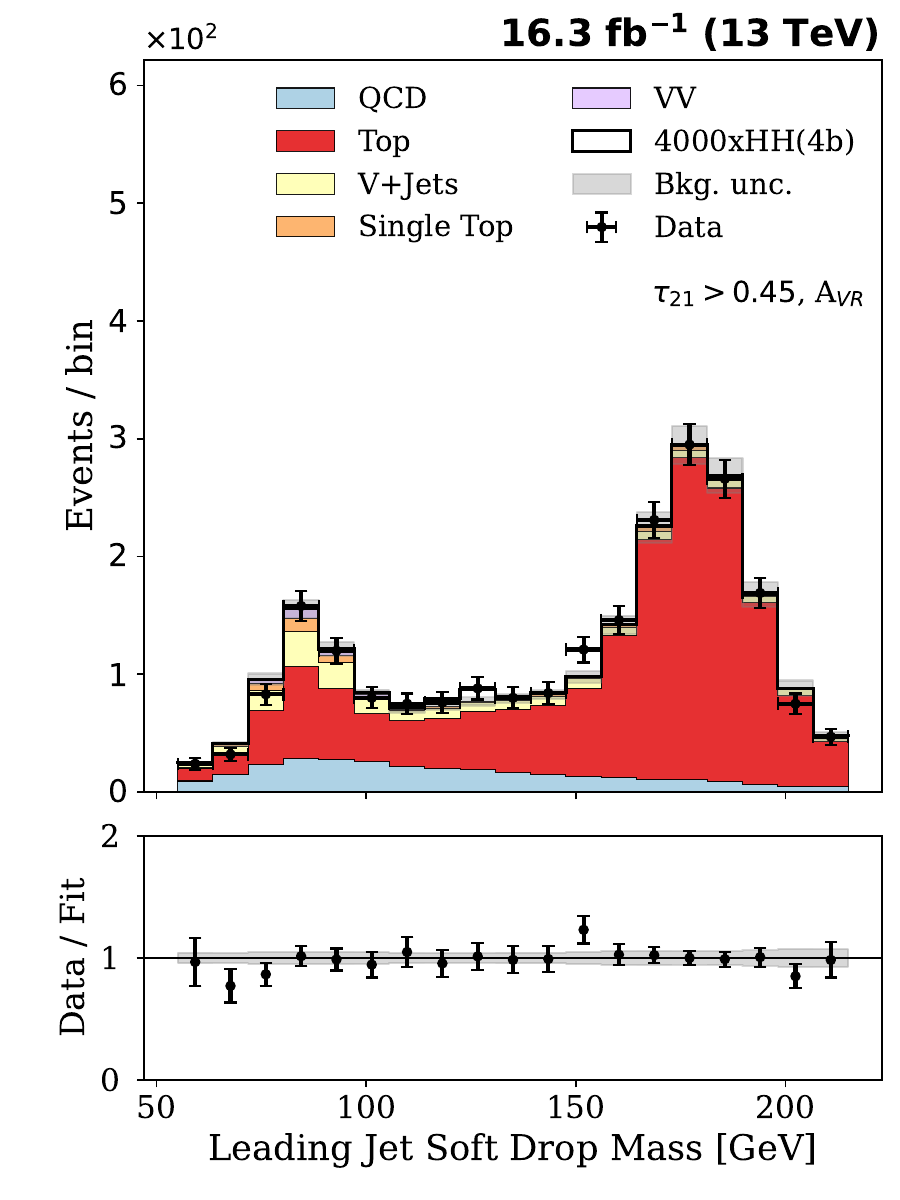}
        \includegraphics[width=.23\textwidth]{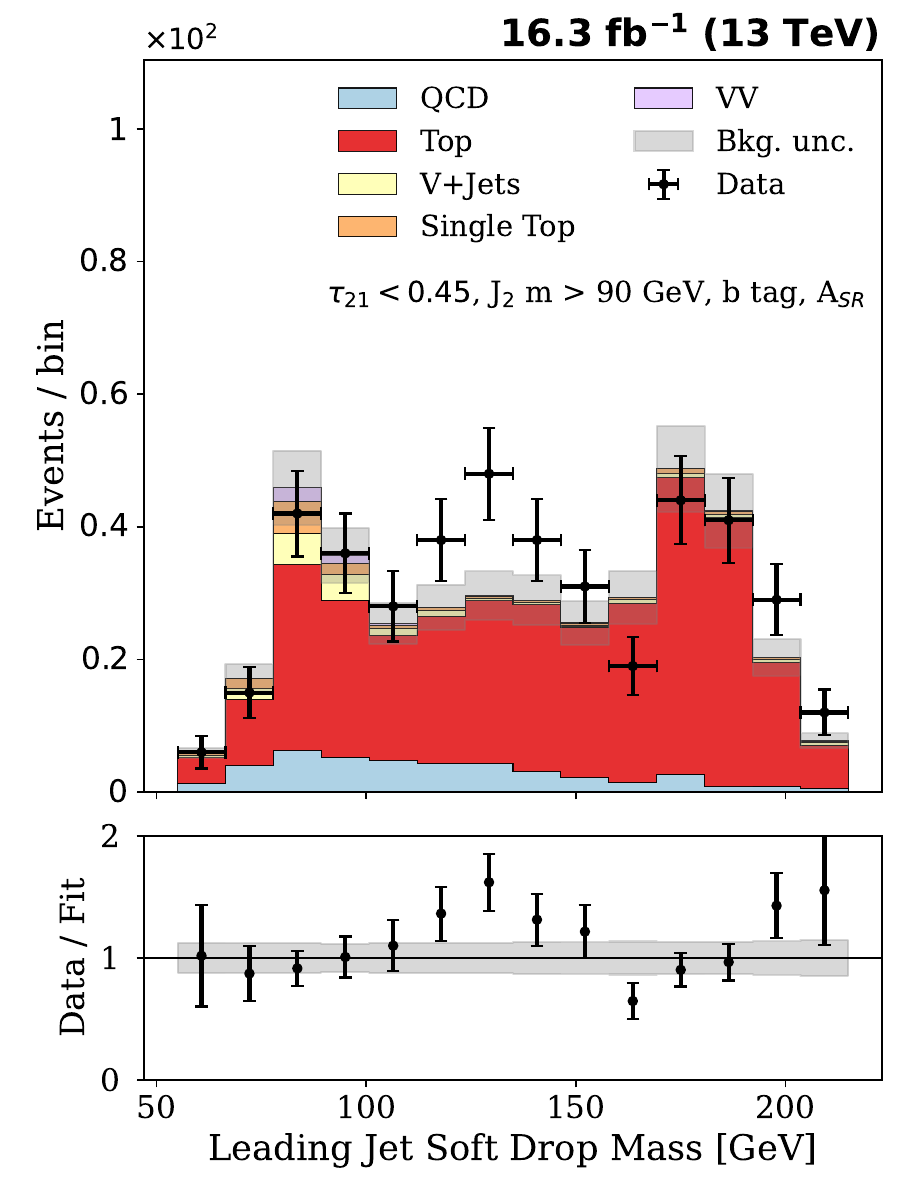}
        \includegraphics[width=.23\textwidth]{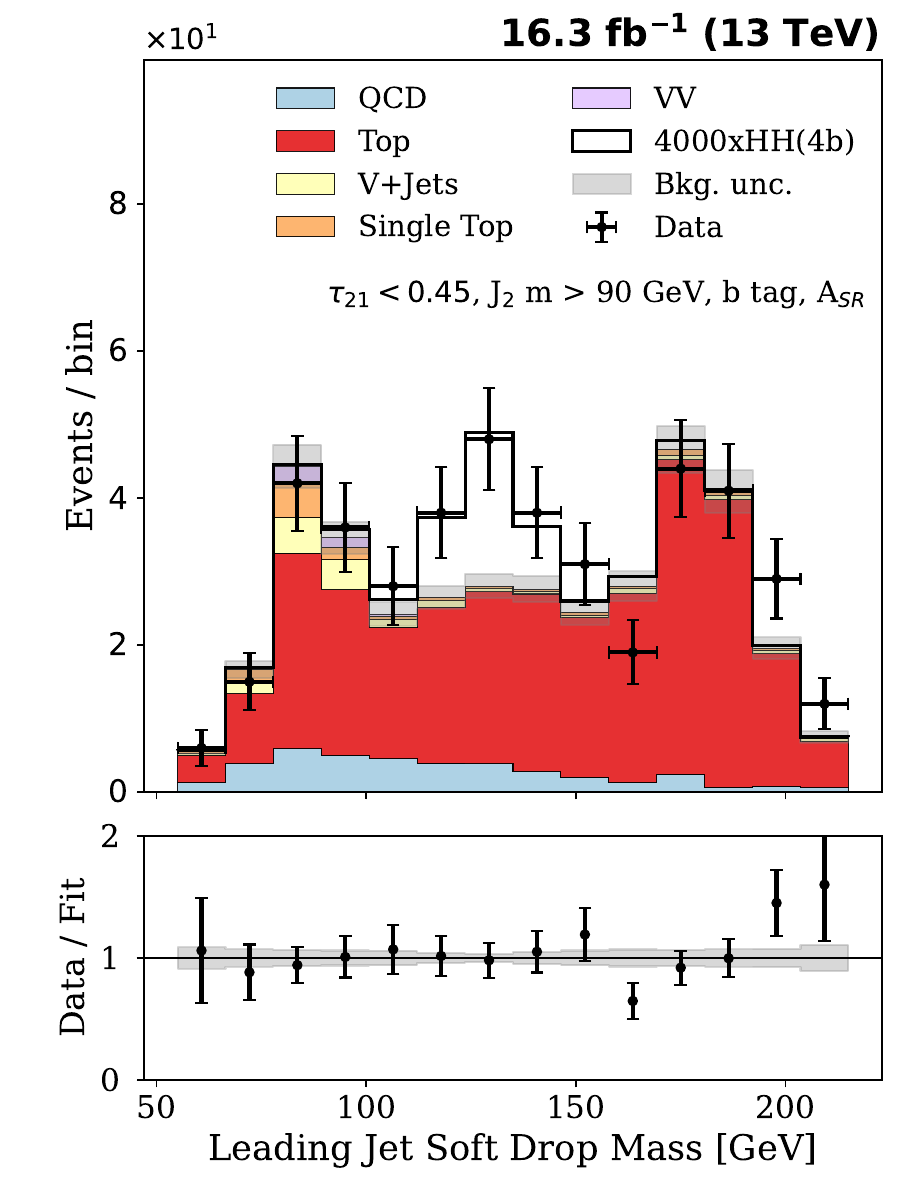}
        \includegraphics[width=.23\textwidth]{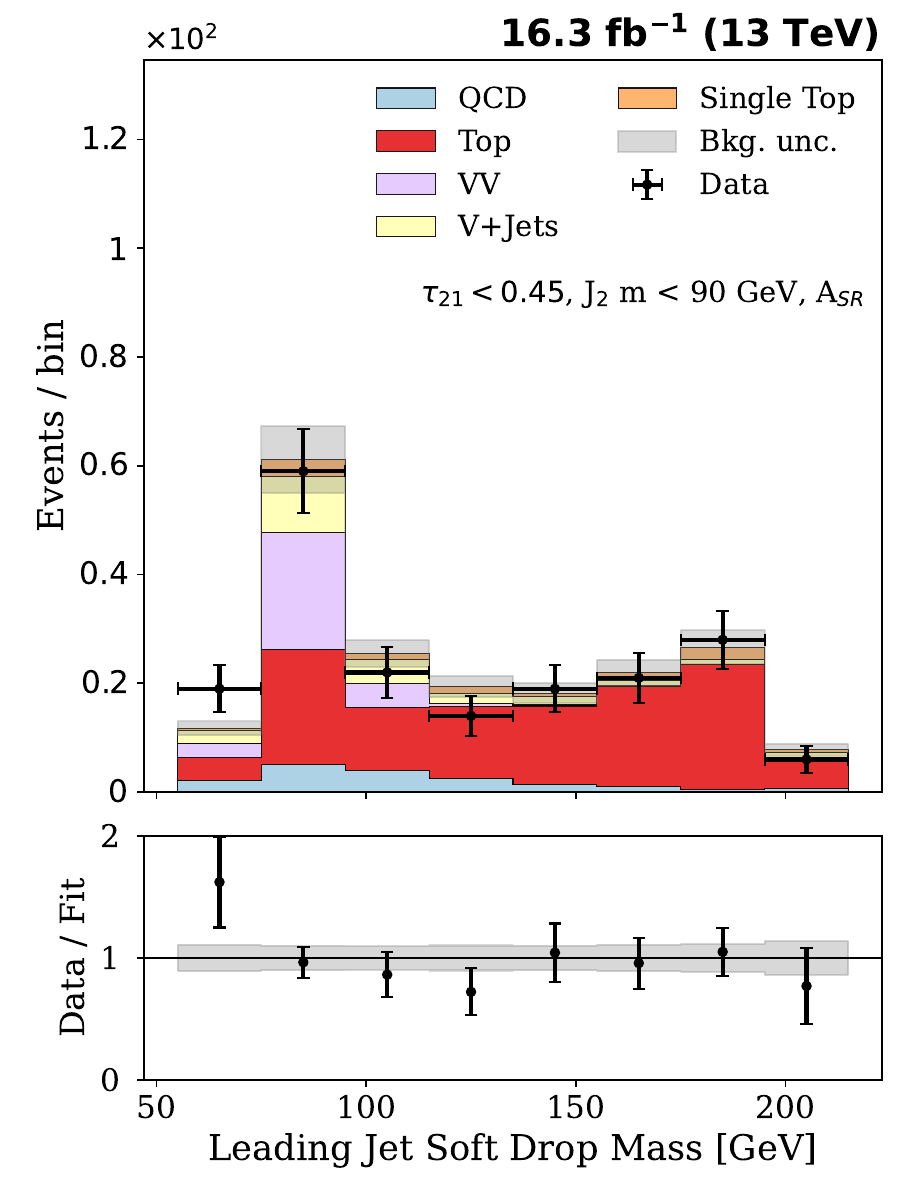}                
        \includegraphics[width=.23\textwidth]{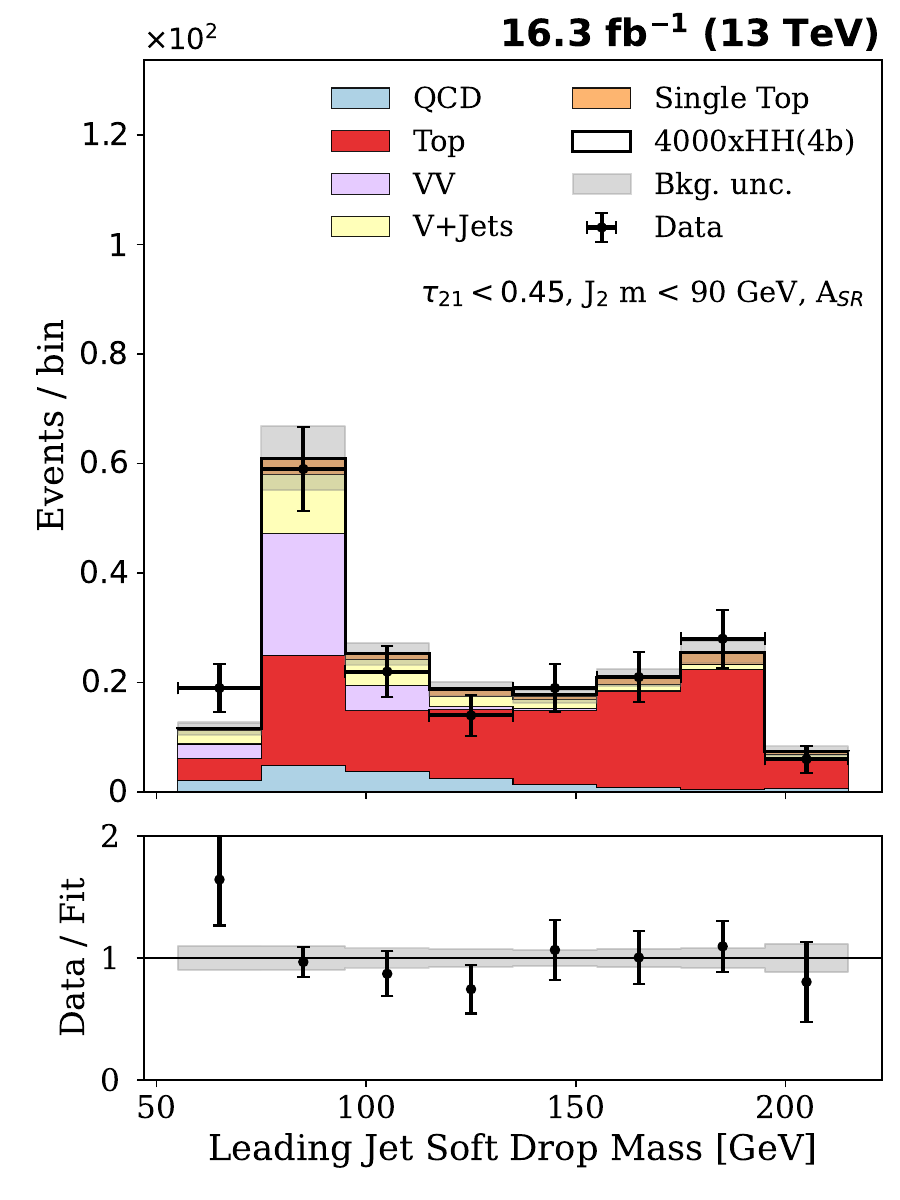}
       
    \caption{Leading jet soft drop mass where both jets are considered anomalous based on the \textsc{OmniLearned} large model score. The region where both jets have low $\tau_{21}$ values is split in the regions where the mass of the subleading jet is above (middle) and below (bottom) 90 GeV, with the region where the subleading jet mass is above 90 GeV to also be required to have at least one jet b-tagged.  The region where at least one jet fails the $\tau_{21}$ selection is shown on the top. Results of the background-only fit are shown on the left while results considering a HH signal component are shown on the right.  Shaded regions represent the total background uncertainty.}
    \label{fig:large_results_split_sr23}
\end{figure}

Additionally, in the region where both jets have $\tau_{21} < 0.45$, we observe an unexpected structure at soft drop mass values of around 130 GeV. To further investigate the properties of the selected events, we consider a different signal hypothesis, namely the pair production of Higgs bosons (HH). The HH simulations, also available from the CMS Open Data release, are performed at NLO with trilinear Higgs coupling set to 1.0 and using \PythiaEight for showering and hadronization. Even though the expected contribution from the Standard Model HH is expected to be less than a single event after selection, we investigate the compatibility of the data with the HH template. The results obtained using the HH sample as a signal, while t$\bar{\mathrm{t}}$ is now considered part of the background processes, are shown in Fig.~\ref{fig:large_results_s}. The fit requires the Standard Model HH decaying to pairs of b-quarks signal to be scaled by a factor of around 4000, far exceeding the current limits on HH production~\cite{CMS:2025ngq}\footnote{We stress that the HH simulation is used here only as a shape proxy for a signal with similar kinematic properties. We do not claim evidence for this process, which would be in direct contradiction with existing experimental constraints.}. Compared to the background-only hypothesis, the Asymptotic significance is 3.3, indicating a preference for the signal template over the background-only model. Since we only include the soft drop mass of the leading jet in the fit, we investigate whether the subleading jet mass has an impact on the observed distribution. We further split the region with jets satisfying $\tau_{21} < 0.45$ into two sub-regions: one where the subleading jet soft drop mass is below 90 GeV and one where it is above. We fit all regions simultaneously, performing the ABCD background estimation for QCD independently per region. The results we obtain are shown in Fig.~\ref{fig:large_results_split_sr12} for both the background-only fit and the HH signal plus background fit. While the fit results in the region with subleading jet soft drop mass below 90 GeV are more consistent with data, the region where the soft drop mass is above 90 GeV seems to highlight even further the disagreement between the data and expected physics processes, favoring the presence of the HH-like signal and resulting in an observed significance of 3.84 compared to the background-only hypothesis.

\begin{figure}[ht]
    \centering
        \includegraphics[width=.23\textwidth]{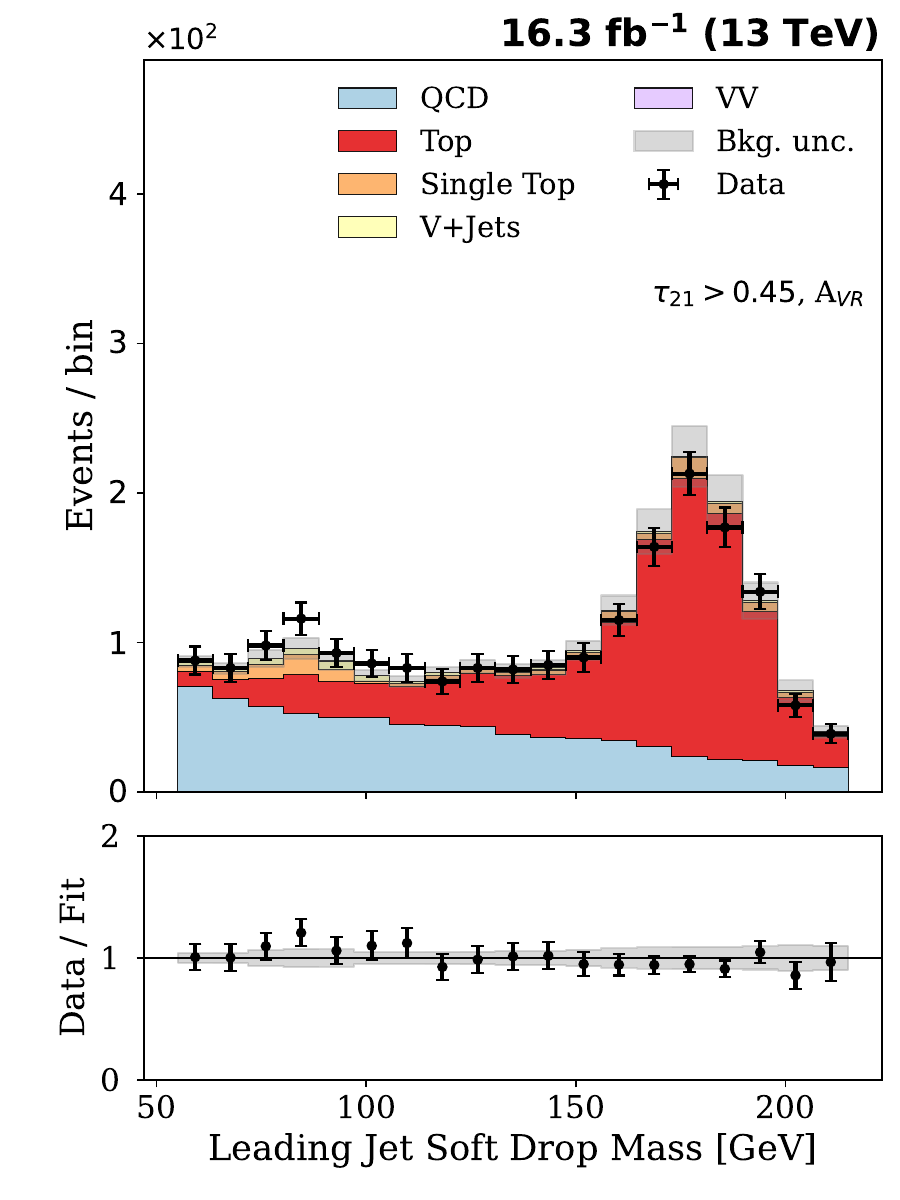}
        \includegraphics[width=.23\textwidth]{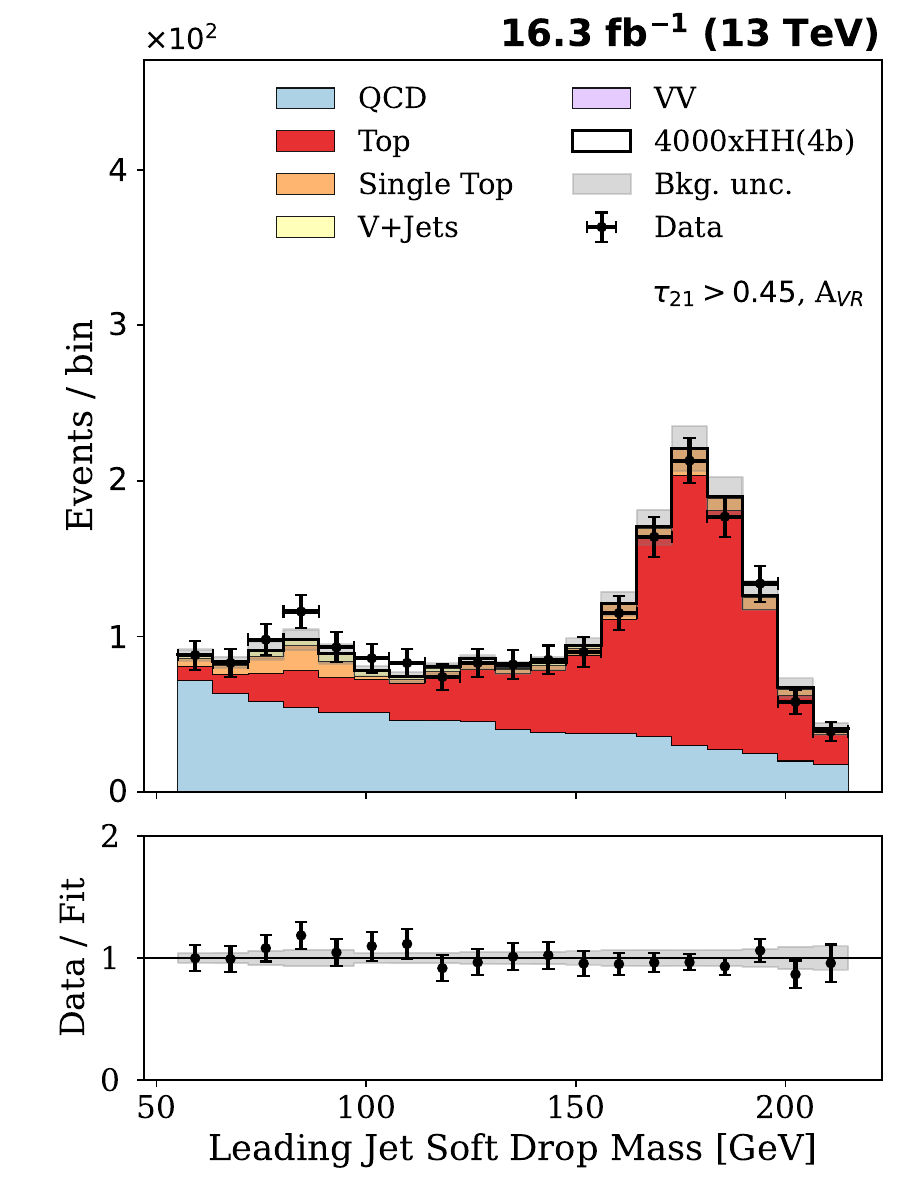}
        \includegraphics[width=.23\textwidth]{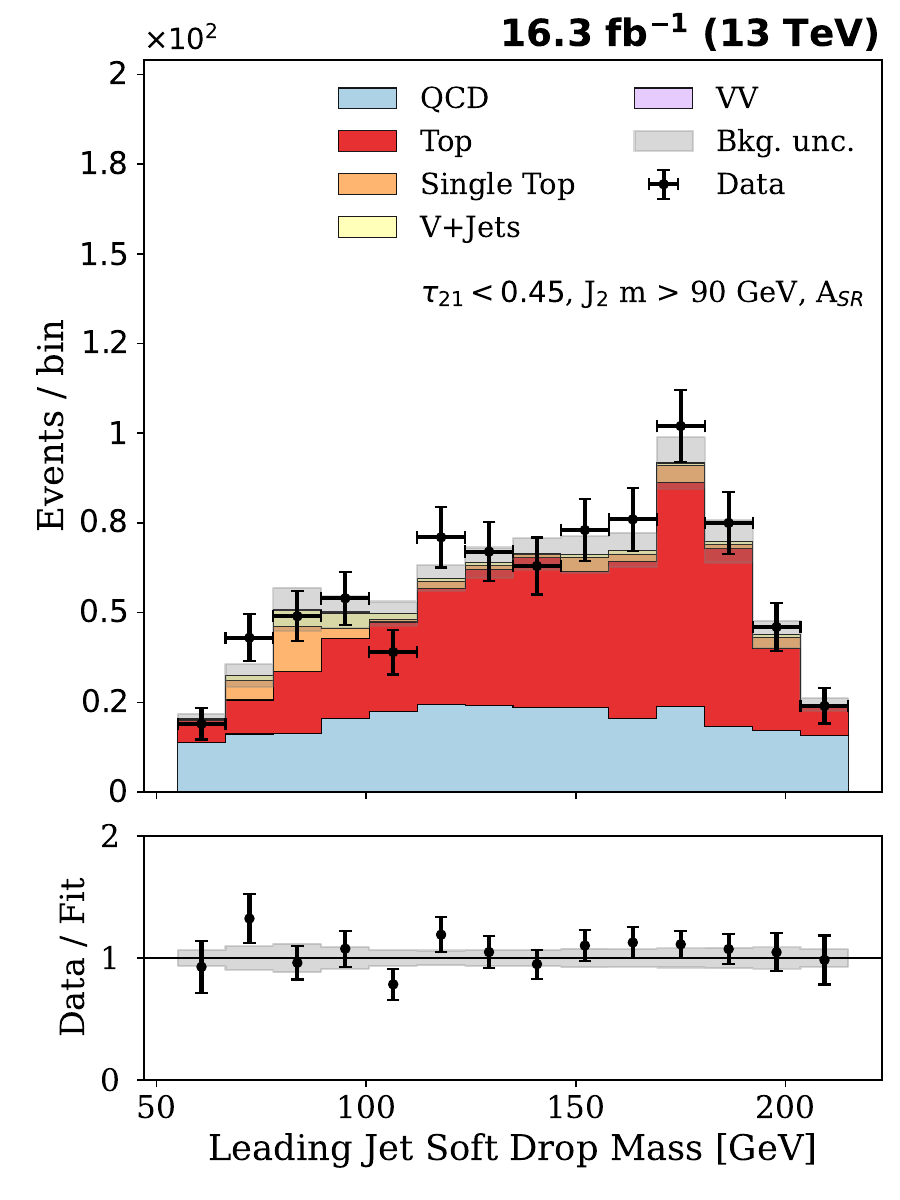}
        \includegraphics[width=.23\textwidth]{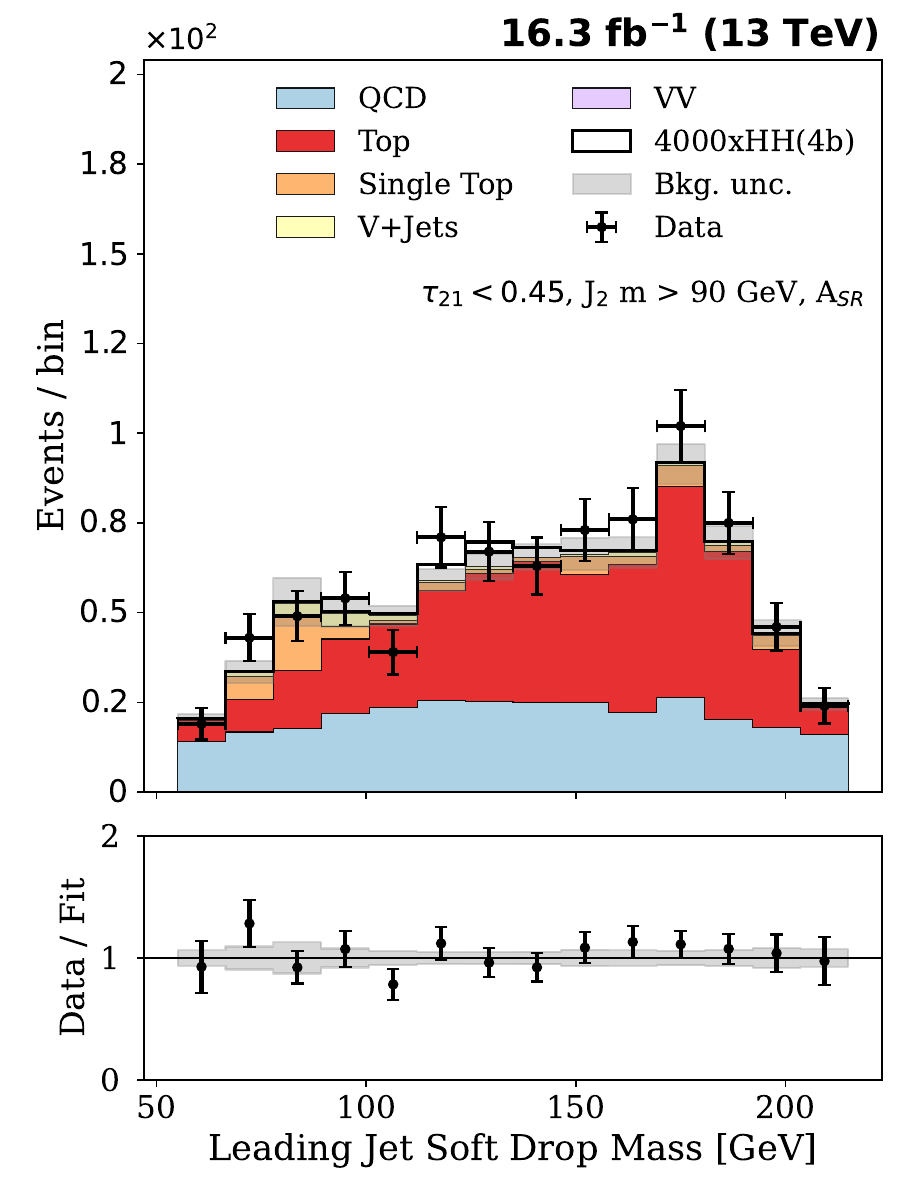}
        \includegraphics[width=.23\textwidth]{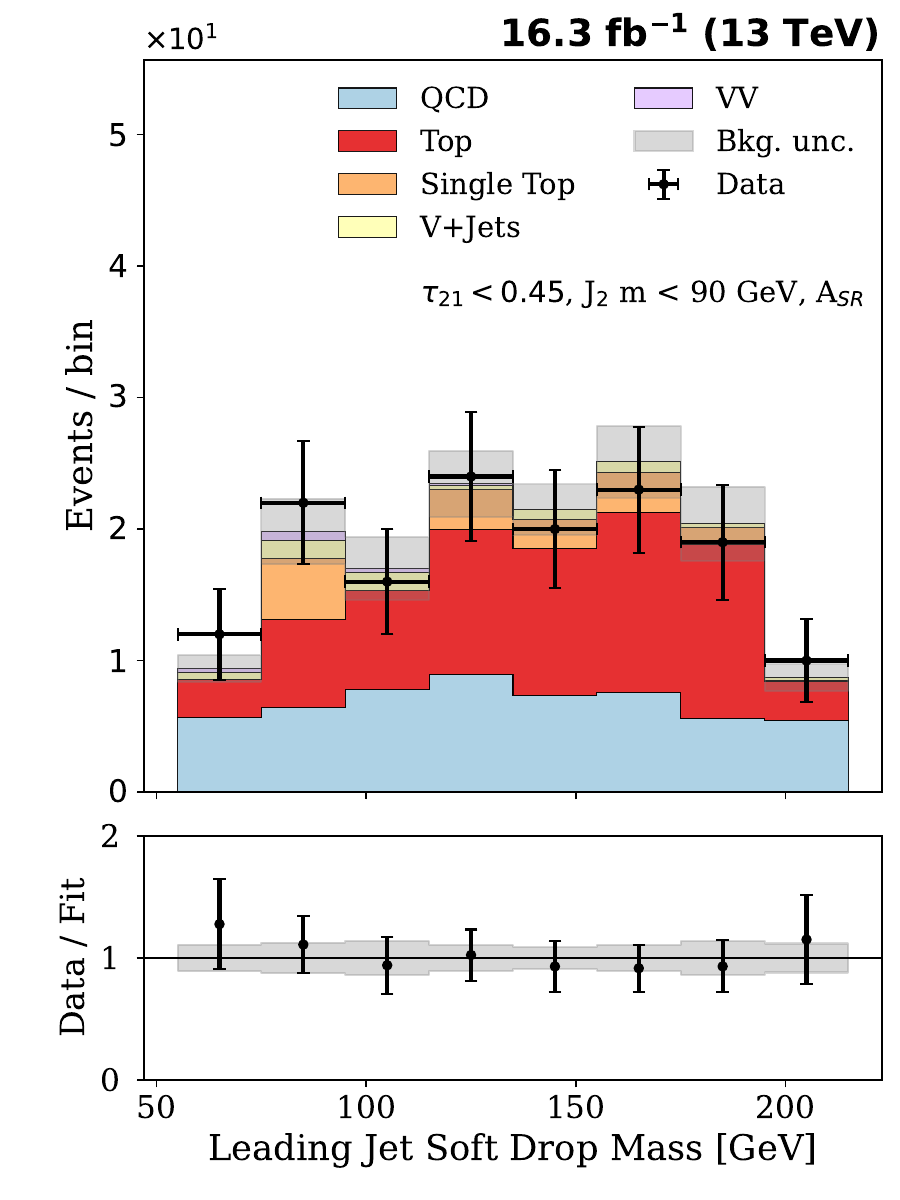}         
        \includegraphics[width=.23\textwidth]{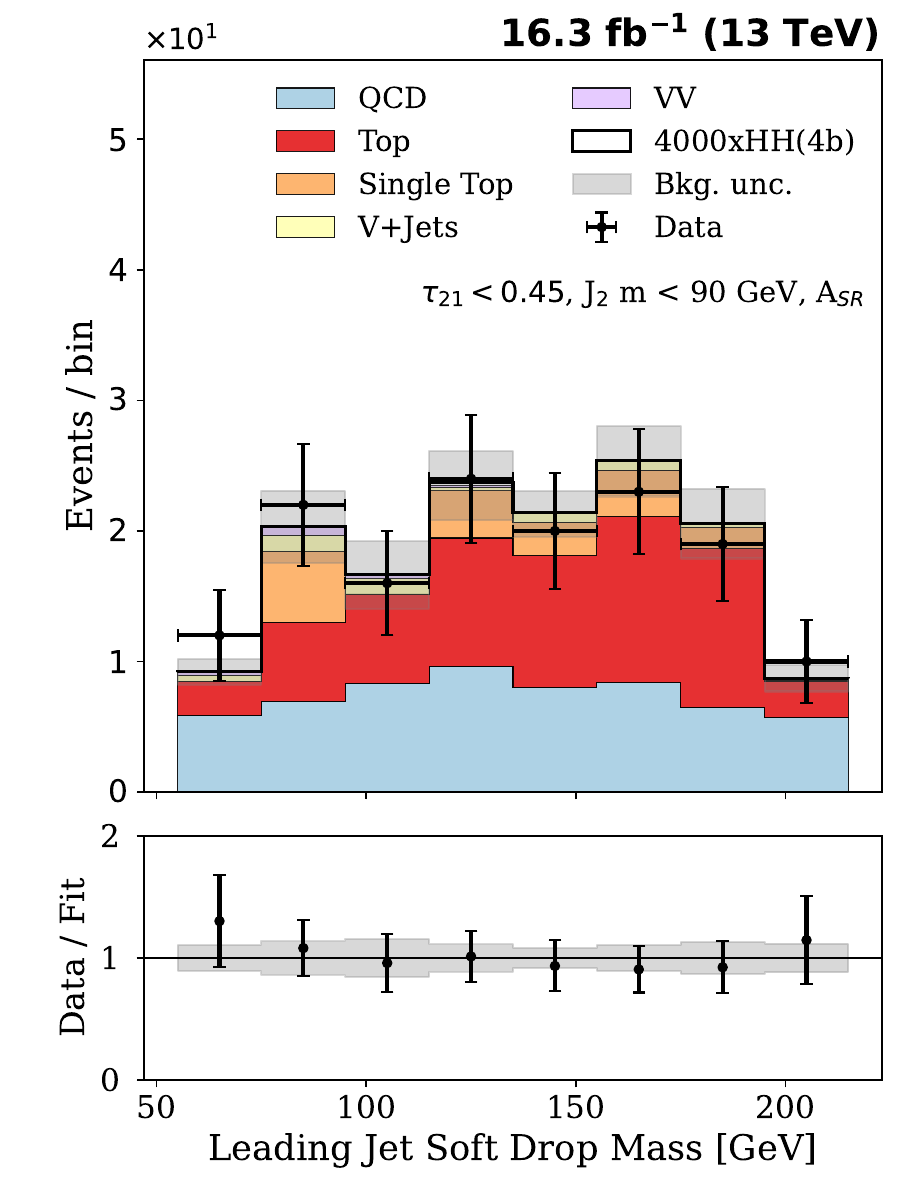}
       
    \caption{Leading jet soft drop mass where both jets are considered anomalous based on the CMS X(bb) tagger. The region where both jets have low $\tau_{21}$ values is split in the regions where the mass of the subleading jet is above (middle) and below (bottom) 90 GeV.  The region where at least one jet fails the $\tau_{21}$ selection is shown on the top. Results of the background-only fit are shown on the left while results considering a HH signal component are shown on the right.  Shaded regions represent the total background uncertainty.}
    \label{fig:htag_results}
\end{figure}

Next, we investigate whether the flavor composition of the selected jets has any distinguishing features. We extend the previous fit strategy by  requiring that, in the region where the subleading jet has mass above 90 GeV, at least one of the jets is b-tagged using the \textsc{DeepCSV}~\cite{CMS:2017wtu} algorithm and satisfying the medium working point, defined by an expected b-tag efficiency of 0.6 for true b-quark initiated jets and background rejection for light-quark initiated jet types of 0.01.  The results of the fit with this additional selection are displayed in Fig.~\ref{fig:large_results_split_sr23}. The additional b-tagging requirement further improves the compatibility of the data with the HH signal and changes the observed significance from 3.84 to 4.25 using the Asymptotic calculation\footnote{While we have not looked in many places, we do note that we have now made a series of selections that would affect the global p-value.}.

\begin{figure}[ht]
    \centering
        \includegraphics[width=.44\textwidth]{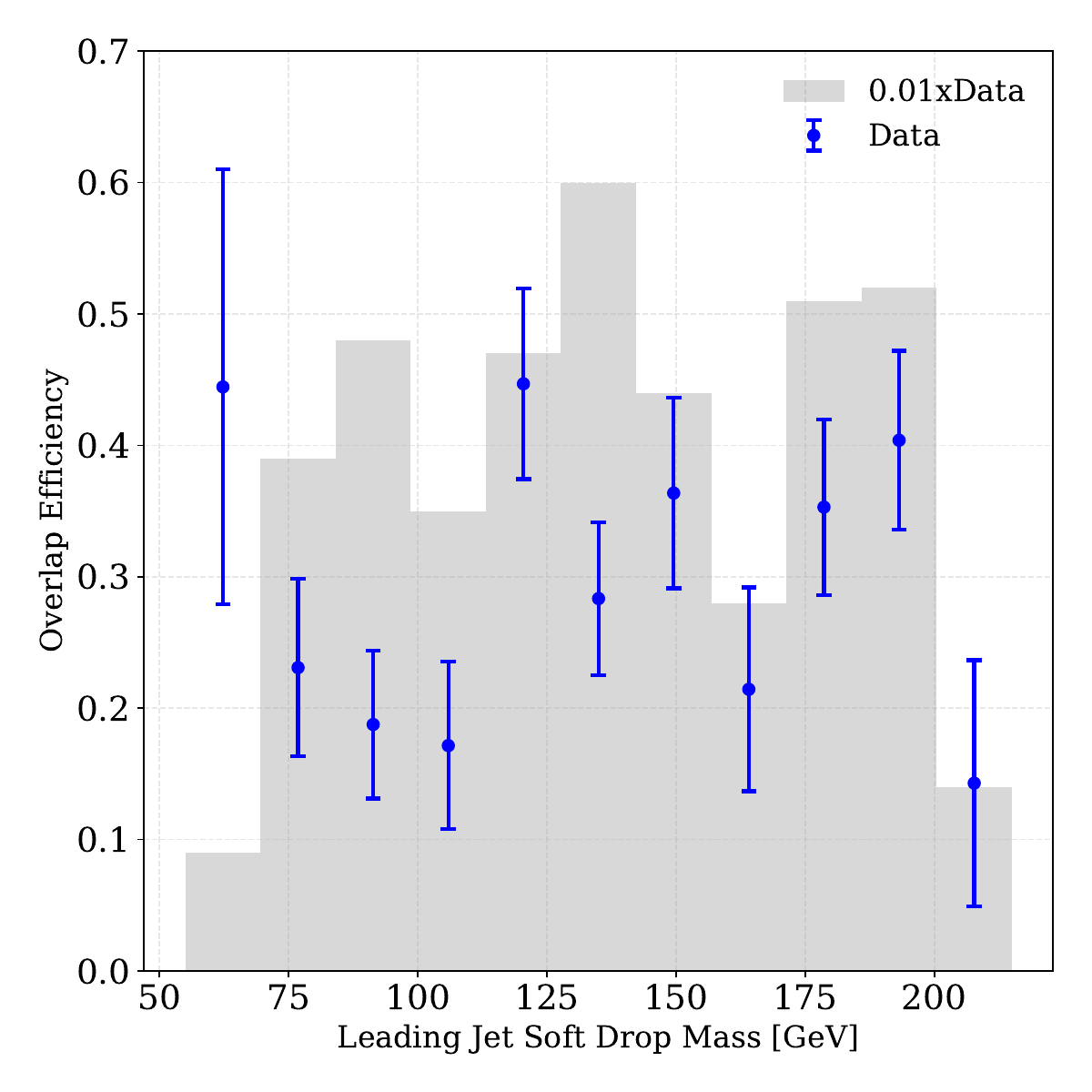}
               
    \caption{Fraction of events selected to be anomalous by both the \textsc{OmniLearned} and X(bb) from all events selected as anomalous by the \textsc{OmniLearned} score as a function of the leading jet soft drop mass. Overlaid is the histogram of the distribution of events selected as anomalous by \textsc{OmniLearned}.}
    \label{fig:overlap}
\end{figure}

To summarize, we find improved data compatibility when including an additional signal process whose leading jet soft drop mass shows a localized distribution around 130 GeV, subleading jet have soft drop mass above 90 GeV, and at least one of the boosted jets is b-tagged. However, given these properties, searches for boosted HH production~\cite{CMS:2022gjd} should also be sensitive to the discrepancy we observe. We investigate the sensitivity of our current analysis workflow when considering a different strategy, closer to searchers performed for HH production. We select anomalous events using a generic $X\rightarrow bb$ tagger, also available in the CMS Open Dataset samples~\cite{CMS:2025kje}, and trained on simulated jets distributed uniformly in mass and transverse momentum, thus minimizing the dependence of the jet selection based on the mass of the original particle. This score is then divided by the same tagger's QCD score to maximize the separation against QCD, consistent with our current strategy. Apart from this change, all other analysis choices are maintained, including the requirement that the subleading jet soft drop mass to be above 90 GeV, but removing the b-tagging requirements, since they are redundant with the new anomaly score. The results are shown in Fig~\ref{fig:htag_results}. In this case, the observed significance is 0.99, consistent with the background model. As a further cross-check, upper limits on the HH production cross section at 95\% confidence level based on the CLs criterion~\cite{Junk:1999kv,Read:2002hq} obtained using the asymptotic formulas are 296 relative to the SM prediction, qualitatively in agreement with limits for HH production given the significantly smaller dataset, phase space region, and absence of dedicated optimization for a HH search. We also quantify the fraction of events selected as anomalous by \textsc{OmniLearned} is also selected as anomalous by the X(bb) tagger. The results of the overlap fraction are shown in Fig.~\ref{fig:overlap}. Although the kinematic distribution of the anomalies selected by \textsc{OmniLearned} seems consistent with HH production, only 20-40\% of the events selected by \textsc{OmniLearned} are also selected by the X(bb) tagger, possibly indicating different jet substructure properties.

\begin{figure}[ht]
    \centering
        \includegraphics[width=.23\textwidth]{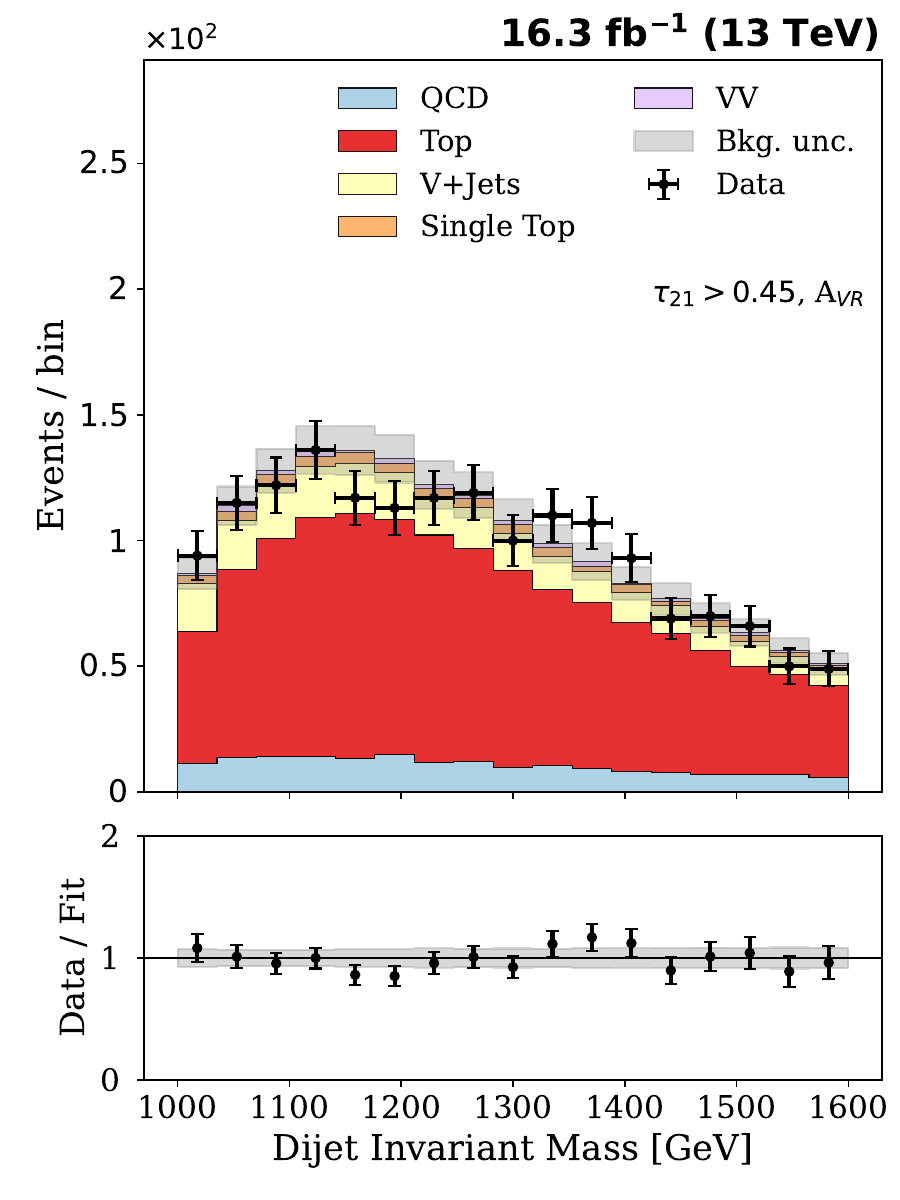}
        \includegraphics[width=.23\textwidth]{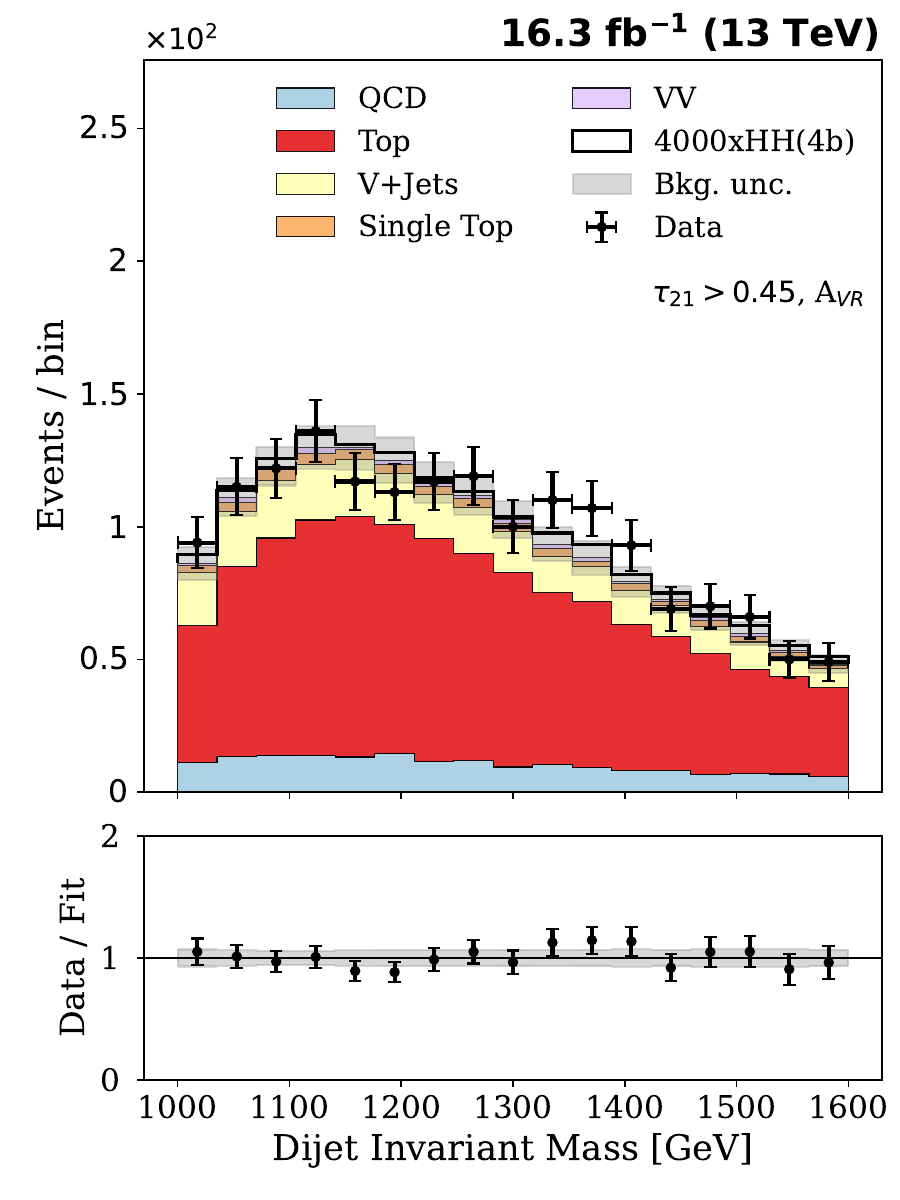}
        \includegraphics[width=.23\textwidth]{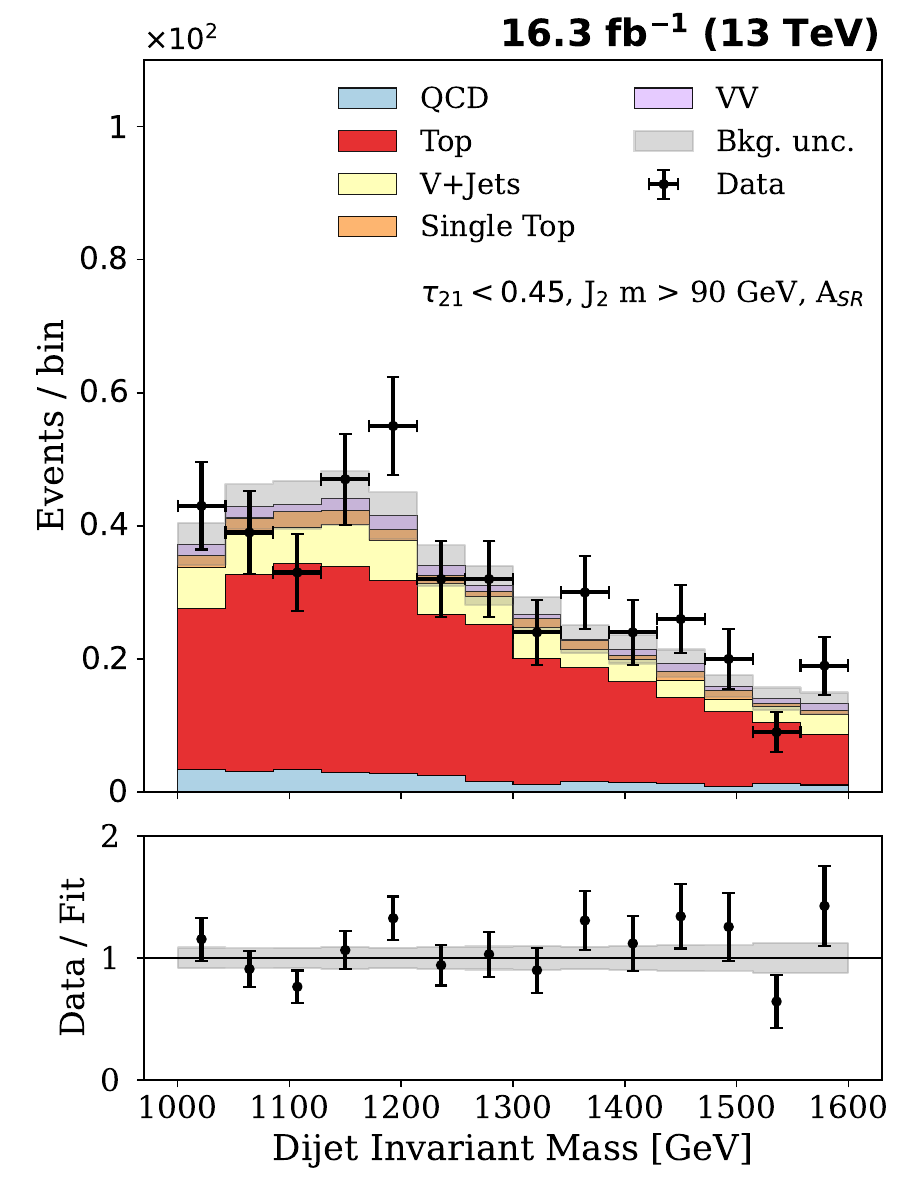}
        \includegraphics[width=.23\textwidth]{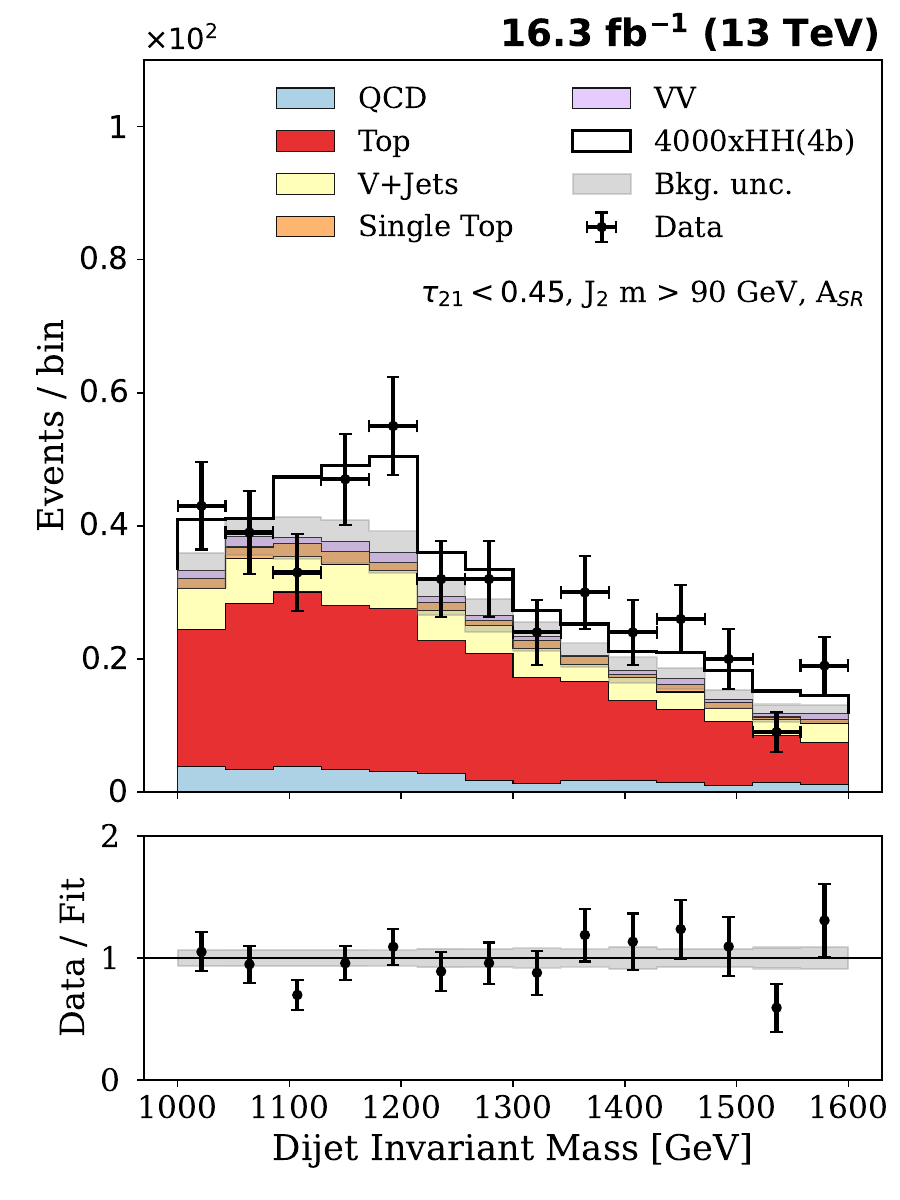}
        \includegraphics[width=.23\textwidth]{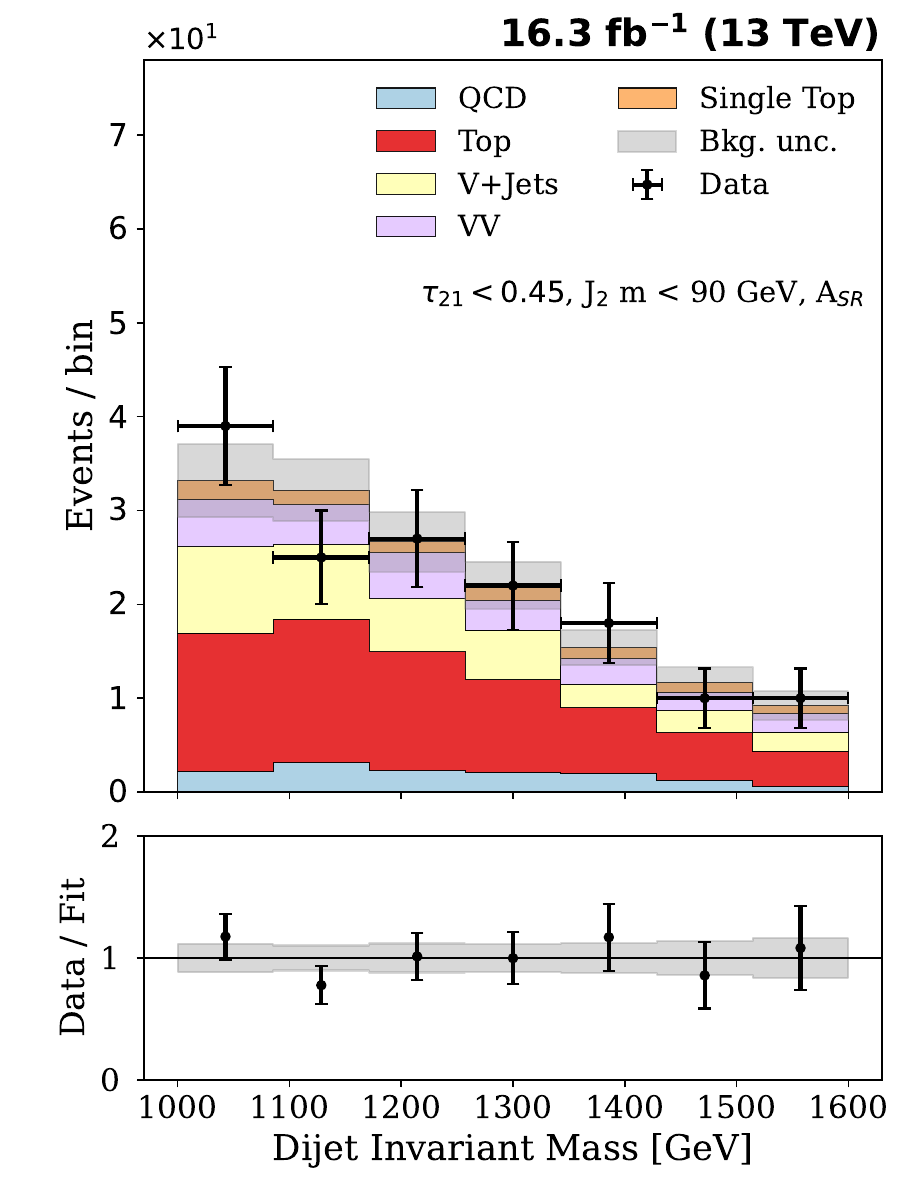}
         \includegraphics[width=.23\textwidth]{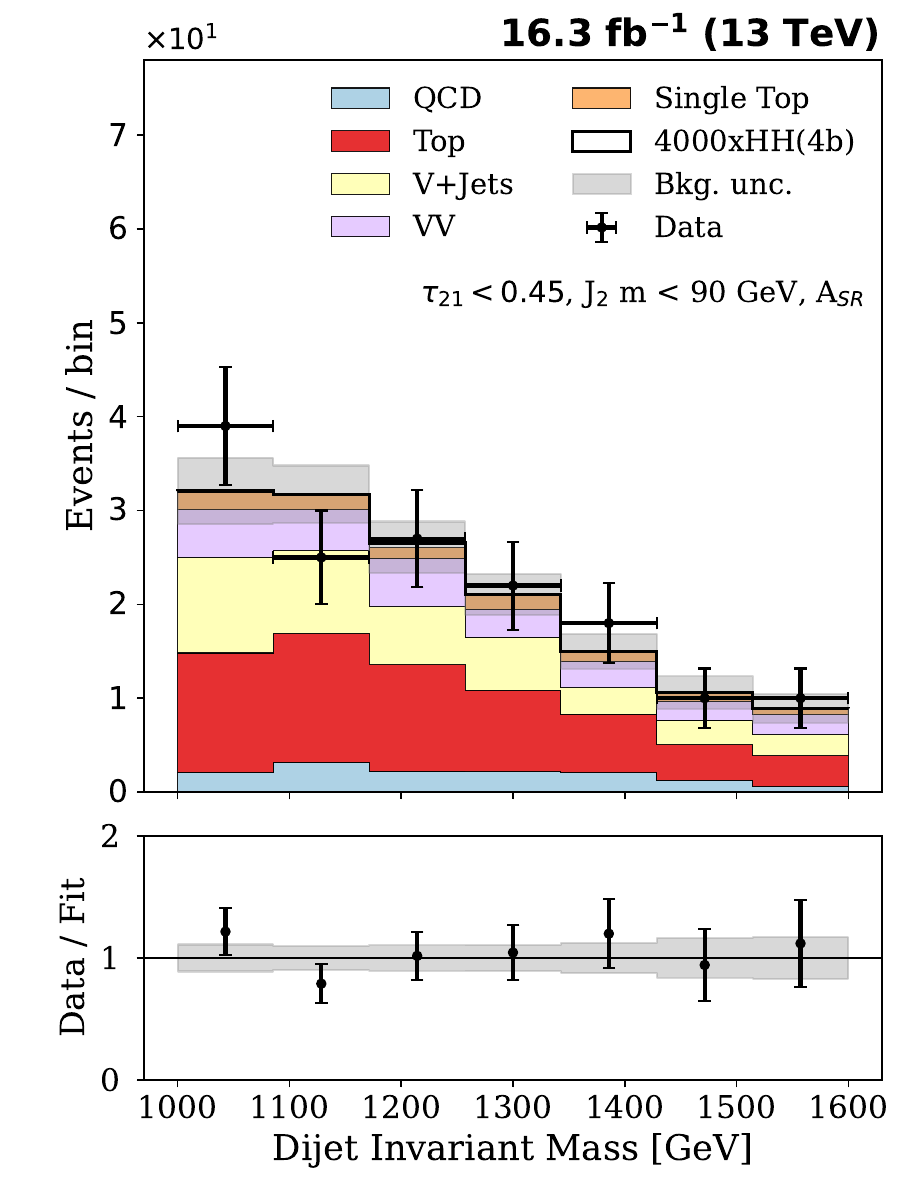}
       
    \caption{Distribution of the dijet invariant mass where both jets are considered anomalous based on the \textsc{OmniLearned} model score. The region where both jets have low $\tau_{21}$ values is split in the regions where the mass of the subleading jet is above (middle) and below (bottom) 90 GeV.  The region where at least one jet fails the $\tau_{21}$ selection is shown on the top. Results of the background-only fit are shown on the left while results considering a HH signal component are shown on the right.  Shaded regions represent the total background uncertainty.}
    \label{fig:large_results_dimass}
\end{figure}

Next, we investigate whether an additional observable beyond the leading jet soft drop mass also shows an unexpected distribution. In particular, we look at the distribution of the invariant mass of the two leading jets. We apply the same fit strategy as before,  using the large \textsc{OmniLearned} model to define the anomaly score. The results of the fit are shown in Fig.~\ref{fig:large_results_dimass}. The dijet mass distribution in the region with high $\tau_{21}$ values shows a smooth distribution while the region where both jets have low  $\tau_{21}$ values shows higher fluctuations. The HH signal hypothesis improves the description of the data. However, we caution that this improvement may be artificial, as the limited HH simulation sample size introduces statistical fluctuations in the signal template which happen to coincide with the data fluctuations. Furthermore, in the absence of a genuine resonant process, the dijet invariant mass from HH production is not expected to exhibit localized structures. Here is also important to compare with similar methods. Particularly, in~\cite{CMS:2024nsz,CMS:2025sch}, several anomaly detection methods are used to search for new physics based on the dijet mass and jet soft drop mass distributions. However, none of the anomaly detection methods employed define the anomaly score based on the same categories we use based on different jet substructure signatures. While different anomaly score can be sensitive to the same new physics process, we demonstrate that even the same anomaly score, trained on a significantly bigger model, can lead to different anomalous events to be selected. Additionally, in~\cite{CMS:2024nsz,CMS:2025sch}, anomalous events are required to have dijet mass above 1455 GeV, which as seem from Fig.~\ref{fig:large_results_dimass}, would reject the majority of the events selected by this work.

\begin{figure}[ht]
    \centering
        \includegraphics[width=.43\textwidth]{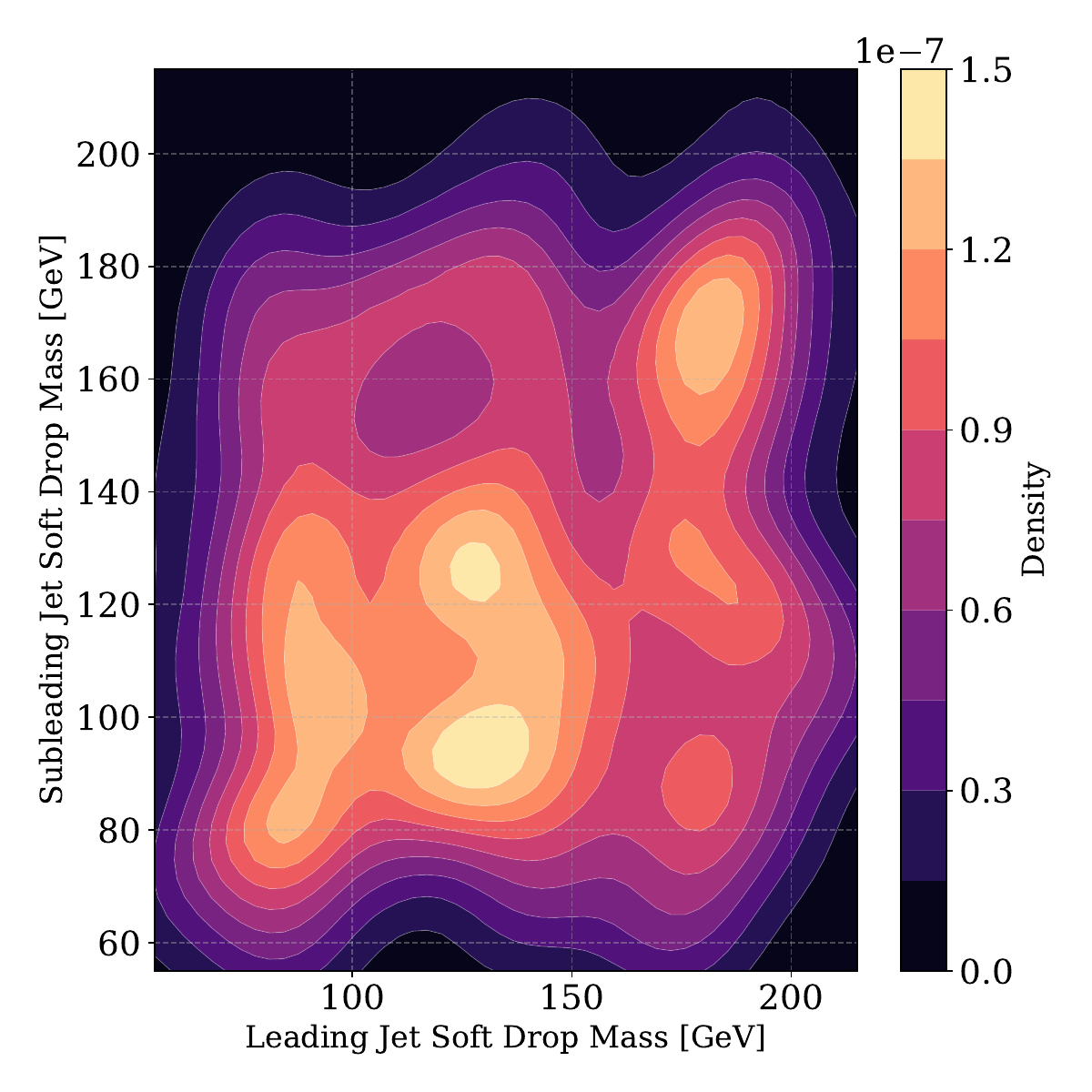}
        \includegraphics[width=.43\textwidth]{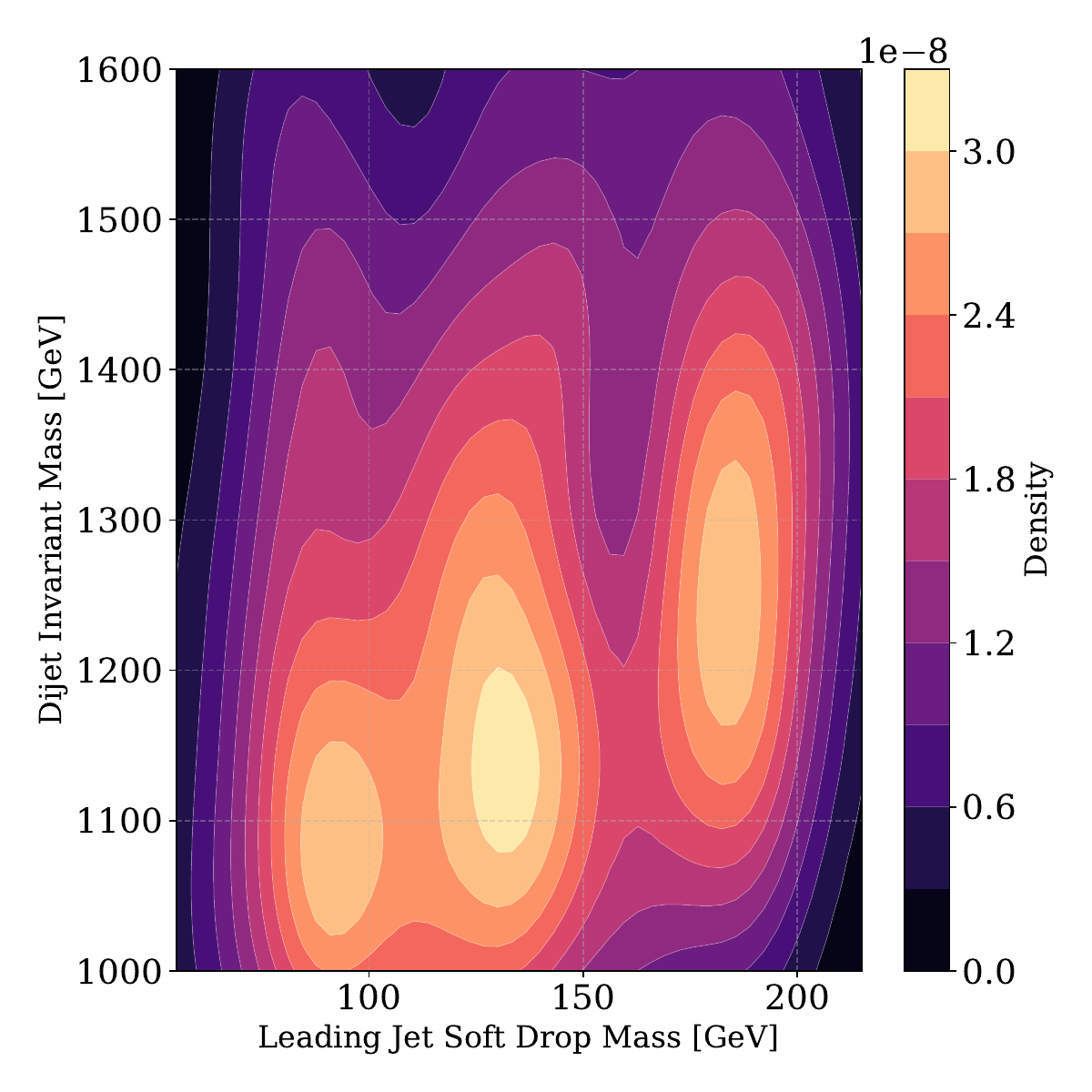}
       
    \caption{Two-dimensional pdf estimated using a KDE with a Gaussian kernel for the soft drop masses of both leading and subleading jet (left) and between the leading jet soft drop mass and the invariant mass of the dijet system.}
    \label{fig:kde}
\end{figure}

Finally, we carry out a qualitative study of the correlations between the soft drop masses of the two jets, as well as between the leading jet soft drop mass and the dijet invariant mass. We use a kernel density estimator (KDE) with a Gaussian kernel to estimate the probability density function (PDF) of the data in the high-anomaly, low $\tau_{21}$, b-tagged region. When estimating the joint PDF of the leading jet soft drop mass and the dijet invariant mass, we additionally require the subleading jet soft drop mass to be above 90 GeV. Results are shown in  in Fig.~\ref{fig:kde}. The estimated joint PDF of the soft drop masses of the two jets shows several localized excesses. Some are compatible with known Standard Model processes, such as both jets having soft drop mass around 90 GeV, consistent with the presence of two W- or Z-bosons, or both jets having soft drop mass around 170 GeV, compatible with top quark pair production.  The region where the leading jet has soft drop mass around 130 GeV shows a localized excess where the subleading jet mass ranges from 90 to 135 GeV, consistent with our previous observation that requiring the subleading jet mass above 90 GeV enhances the significance of the excess in the leading jet soft drop mass. Regarding the dijet invariant mass, events with a leading jet soft drop mass around 130 GeV are distributed broadly around 1150 GeV, while those with a leading jet soft drop mass around 90 (175) GeV are concentrated near a dijet mass of 1050 (1250) GeV, suggesting that the discrepancy observed in the dijet invariant mass distribution may not be directly correlated with the possible excess seen in the leading jet soft drop mass. We also perform additional validation of the fit results and investigate different observables in App.~\ref{app:val}.

\section{Conclusion and Outlook}
\label{sec:conclusions}

In this paper, we have performed an analysis of regions of phase space identified by the different anomaly scores defined by the OmniLearned foundation model.  Using data published by the CMS experiment, we show that the large OmniLearned model is able to rediscover the top quark as an anomaly, just like the medium and small models of Ref.~\cite{Bhimji:2025isp}. We validate data-driven background estimation procedures that model the individual background components relevant for observables described by high hadronic activity regions.  The initial motivation for this work was the observation of a non-smoothly falling mass spectrum in the lower mass sideband from the large OmniLearned model. After performing an in-depth analysis, we find that Standard Model-only processes are not able to describe this feature of the data.  When including HH-like signal events, we find a better fit to the data than with no additional signal component.  While we do not believe the excess, if it is real, is Standard Model HH, this observation could provide a hint about the types of events selected by the anomaly score.  We further validate our findings by comparing our anomaly detection strategy with one where the anomaly detection score is replaced by a tagger trained on generic $X\rightarrow bb$ decays. The results in this obtained are compatible with the background-only distributions. Curiously, the overlap between selected events by both \textsc{OmniLearned} and the X(bb) tagger anomaly scores is small, possibly hinting to a different jet substructure signature.

We hope that this study provides a useful benchmark for anomaly detection followup.  All of the code and data used to prepare the results in this paper are public and we invite further scrutiny of the events and of our methods. Ultimately, our results are limited by the size of the available data and we would be exciting to see how this region of phase space looks like with (much) more data. 

\section*{Code Availability}

The code for \textsc{OmniLearned} can be found at~\cite{vinicius_mikuni_2026_18489564} while all results of this paper can be reproduced using the software described in \url{https://github.com/ViniciusMikuni/OmniLearnedAD}.

\section*{Acknowledgments}
We thank Oz Amram, Caroline Collard, Nathaniel Craig, Mia Liu,  Benedikt Maier, Maximilian Swiatlowski, Jesse Thaler, and Pietro Vischia for helpful discussions and feedback on the results of this work. VM is supported by JST EXPERT-J, Japan Grant Number JPMJEX2509.
BN is supported by the U.S. Department of Energy (DOE), Office of Science under contract DE-AC02-76SF00515.  This research used resources of the National Energy Research Scientific Computing Center, a DOE Office of Science User Facility supported by the Office of Science of the U.S. Department of Energy under Contract No. DE-AC02-05CH11231 using NERSC awards HEP-ERCAP0021099 and HEP-ERCAP0028249.

\appendix

\section{QCD Background Control Regions}
\label{app:abcd}

To extract the signal component we fit simultaneously multiple regions to constrain the background normalization in the regions with high anomaly score. In this appendix we show the results obtained by the fits in the other regions, primarily used to estimate the QCD contribution for the ABCD estimation. In these regions, the almost perfect agreement between data and background model is by construction, since the initial guess of the QCD yield in each bin is defined by the difference between the observed data in the region and the contribution from all other simulated processes. This initial normalization is then corrected during the fit simultaneously with the ABCD calculation. In the subsections below we show the results obtained in each of these regions. Since the results are very similar between background only fit and signal plus background, we show only the signal plus background results.

\subsection{Top Quark as a Signal}

\begin{figure}[ht]
    \centering
        \includegraphics[width=.23\textwidth]{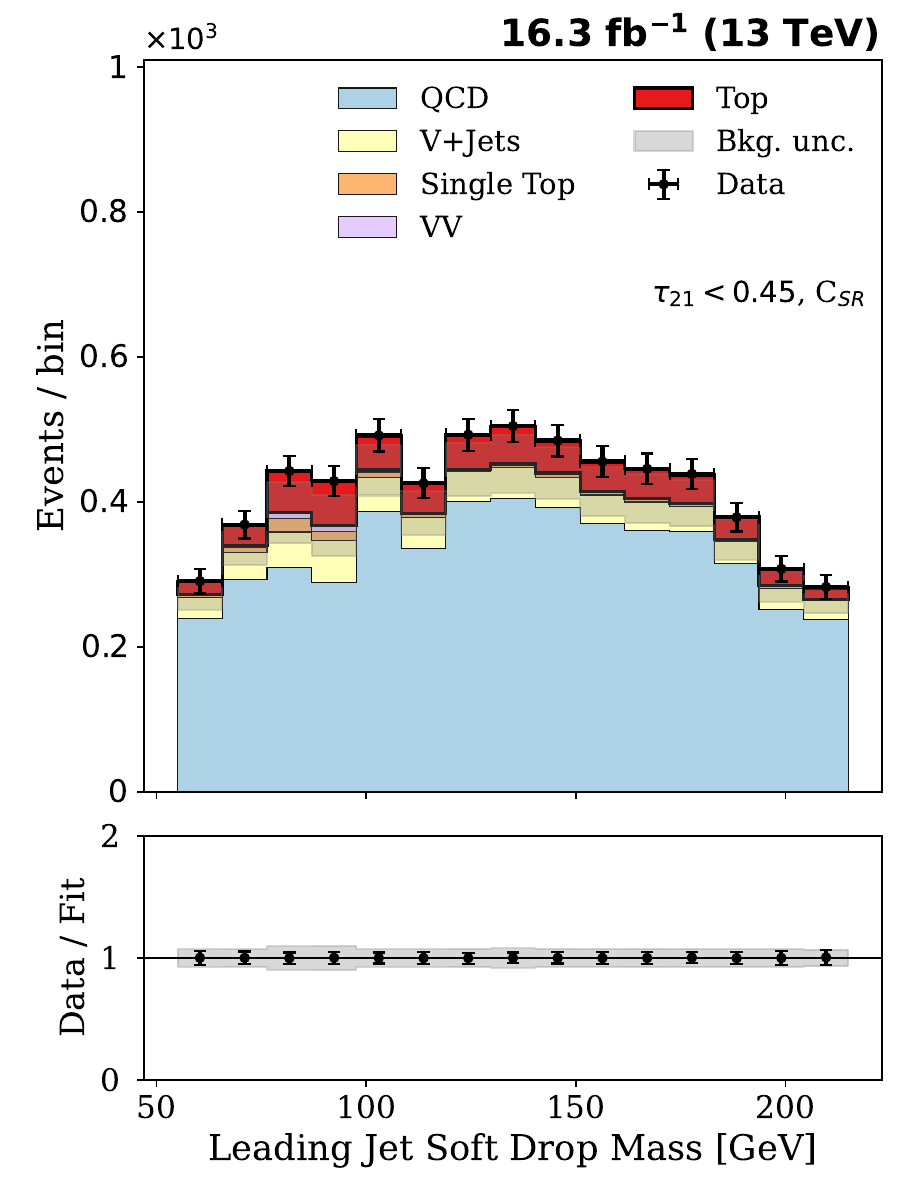}
        \includegraphics[width=.23\textwidth]{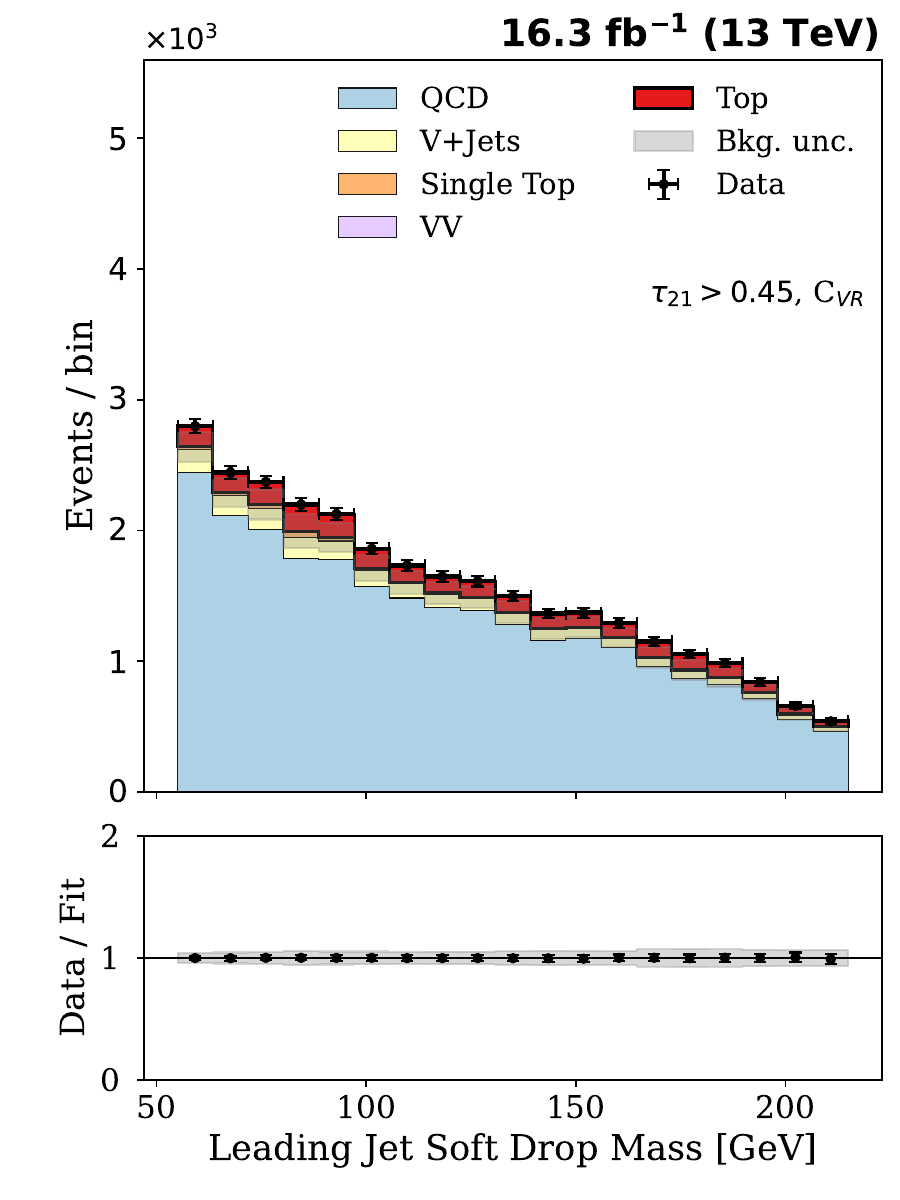}
        \includegraphics[width=.23\textwidth]{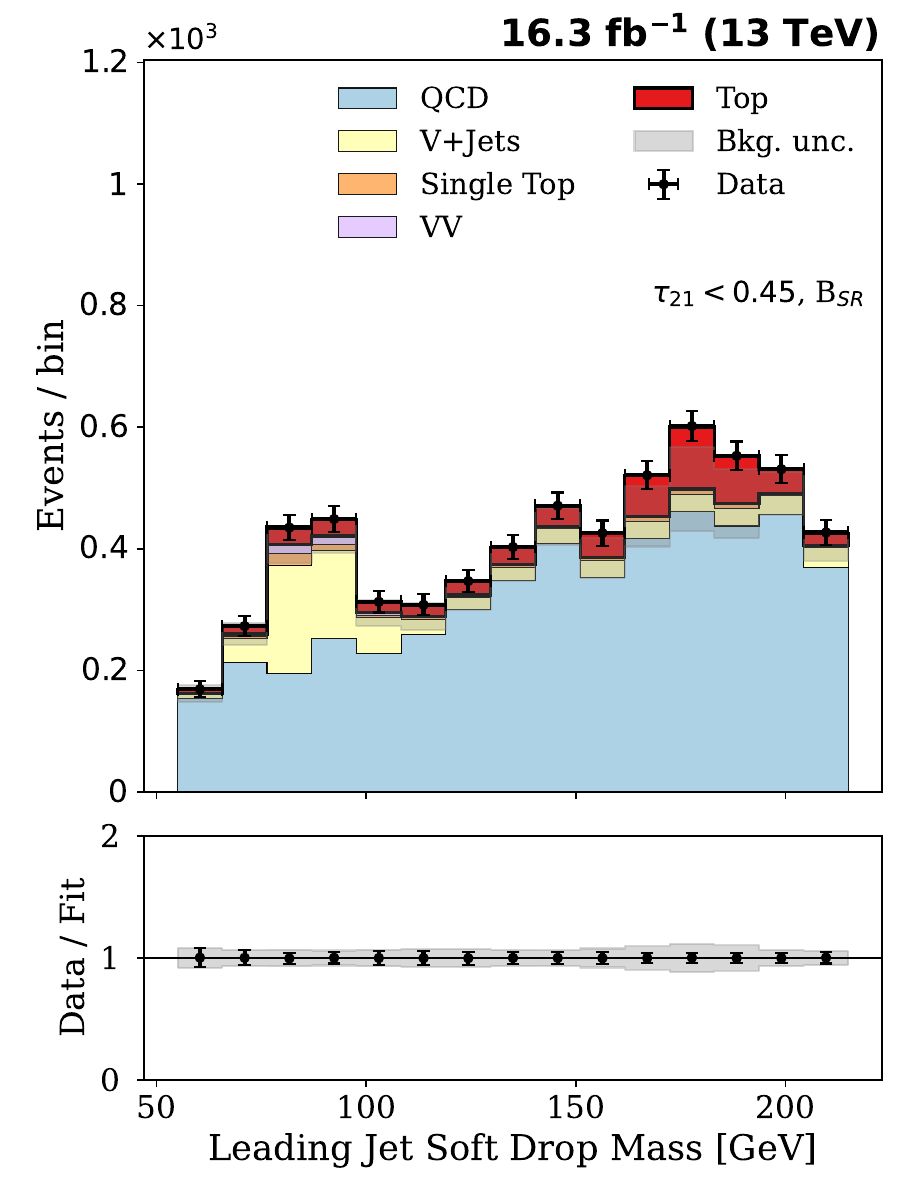}
        \includegraphics[width=.23\textwidth]{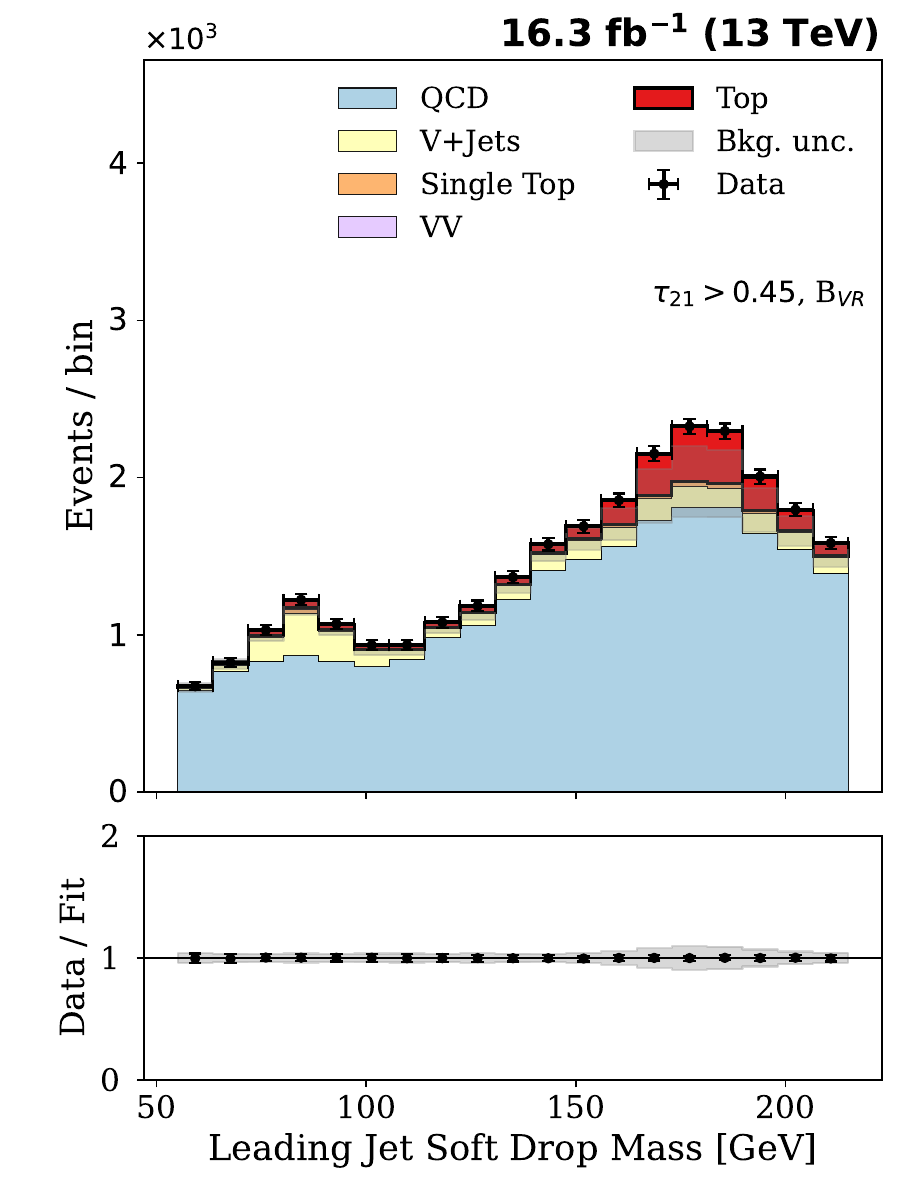}
        \includegraphics[width=.23\textwidth]{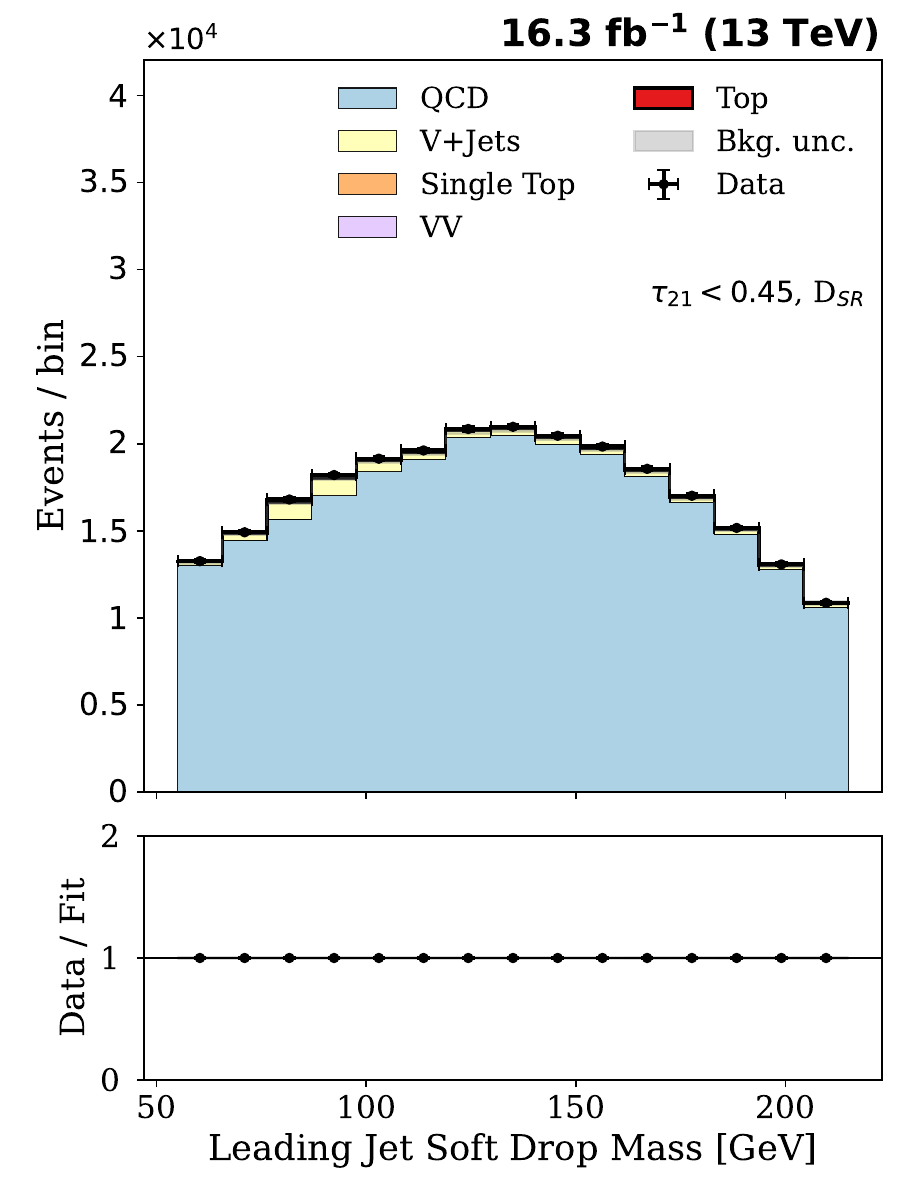}
        \includegraphics[width=.23\textwidth]{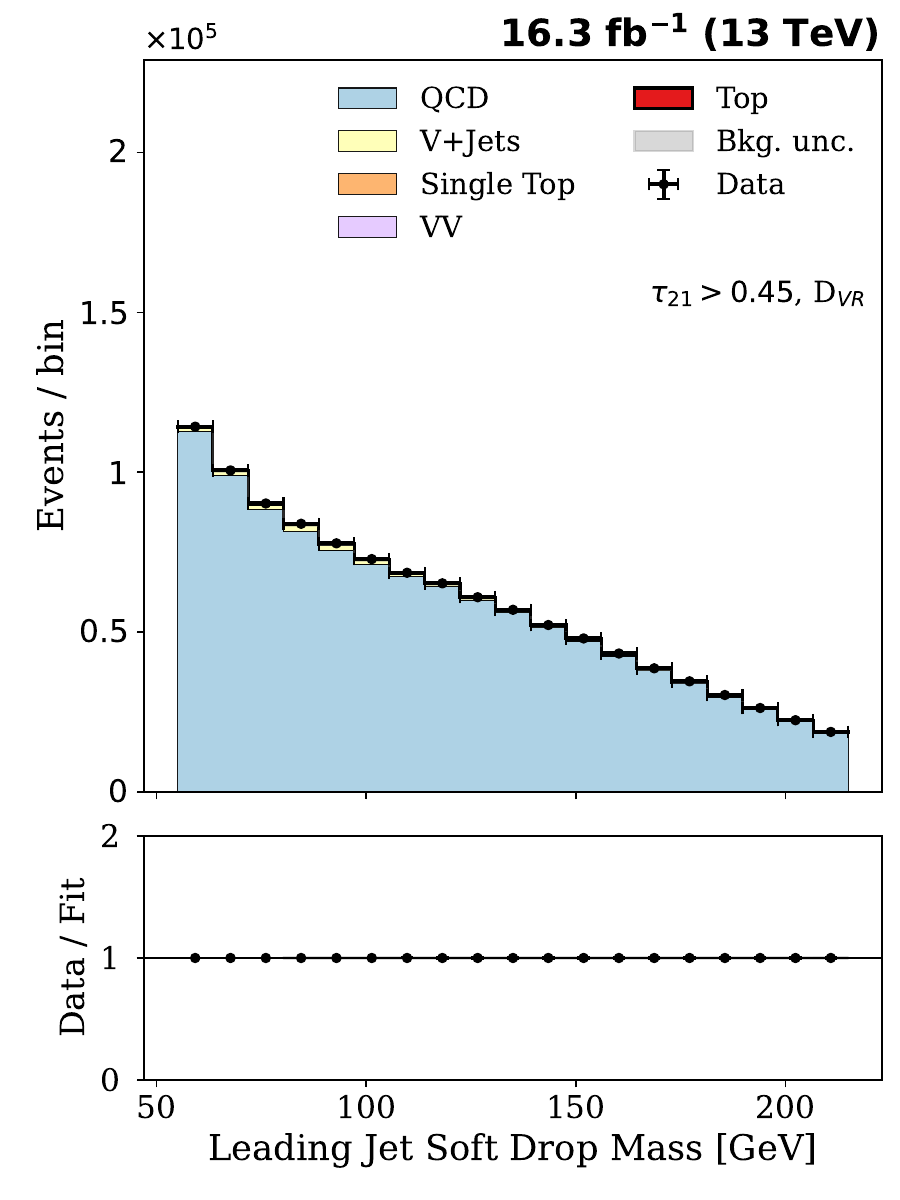}
        
    \caption{Leading jet soft drop mass where both jets are considered anomalous based on the \textsc{OmniLearned} small model score. The region where both jets have low $\tau_{21}$ values is shown at the left while the region where at least one jet fails the $\tau_{21}$ selection is shown at the right. The different regions used for the ABCD calculation are shown as rows. Shaded regions represent the total background uncertainty.}
    \label{fig:top_small_abcd}
\end{figure}

\begin{figure}[ht]
    \centering
        \includegraphics[width=.23\textwidth]{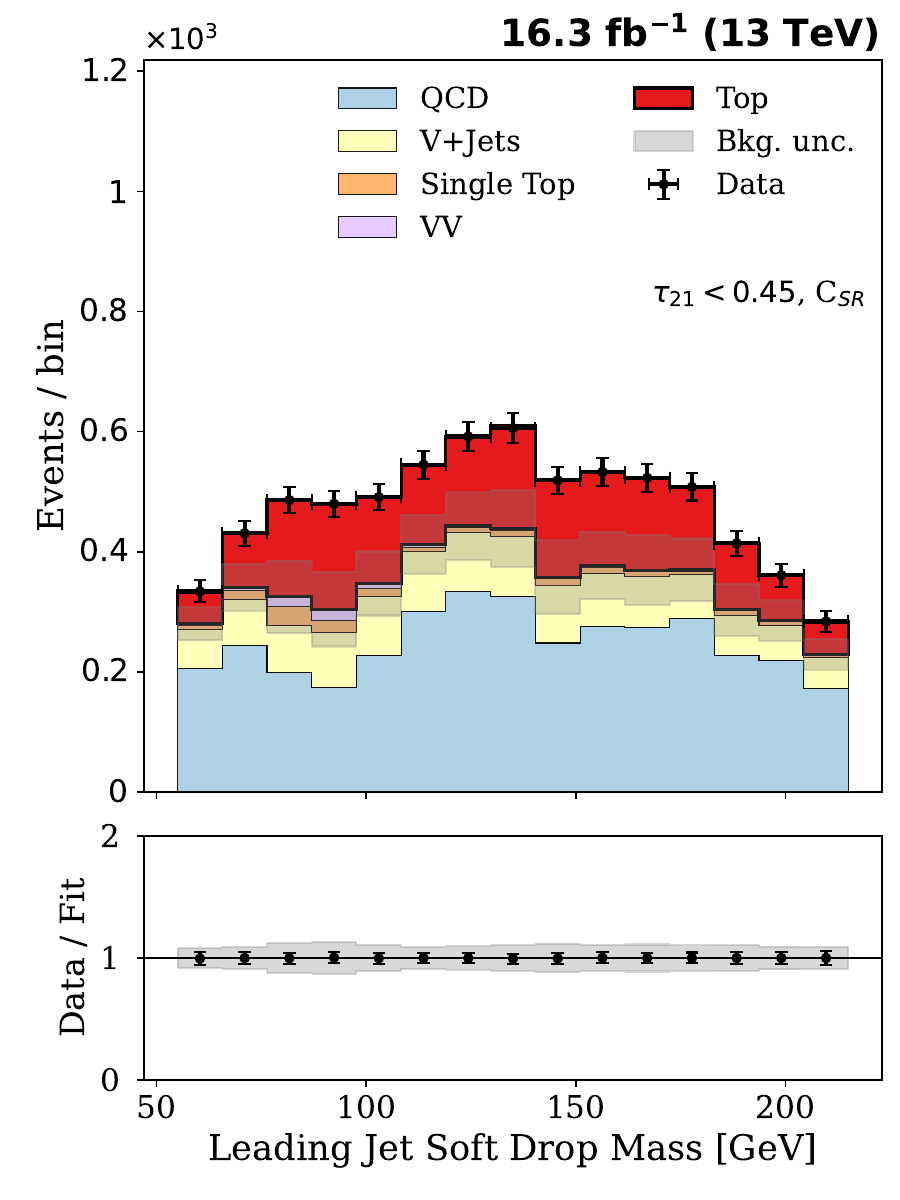}
        \includegraphics[width=.23\textwidth]{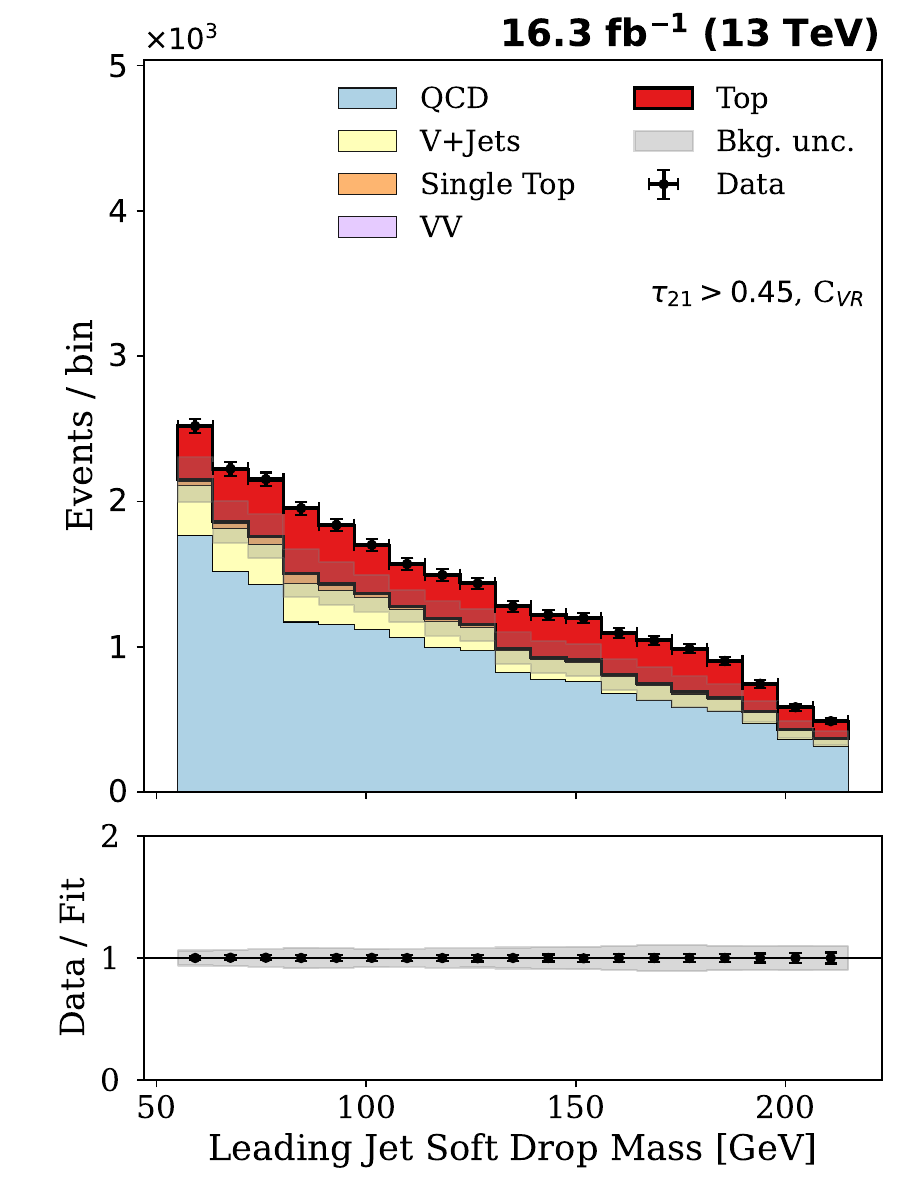}
        \includegraphics[width=.23\textwidth]{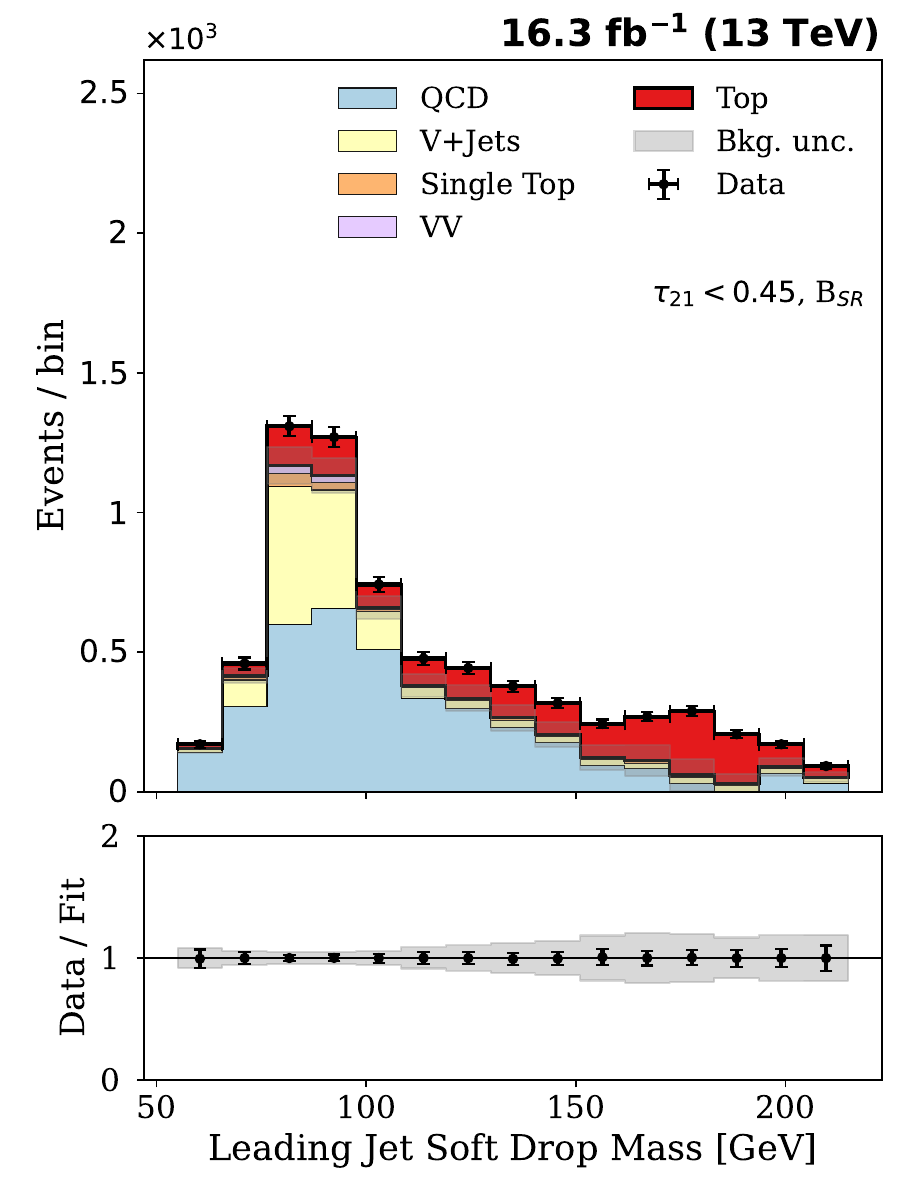}
        \includegraphics[width=.23\textwidth]{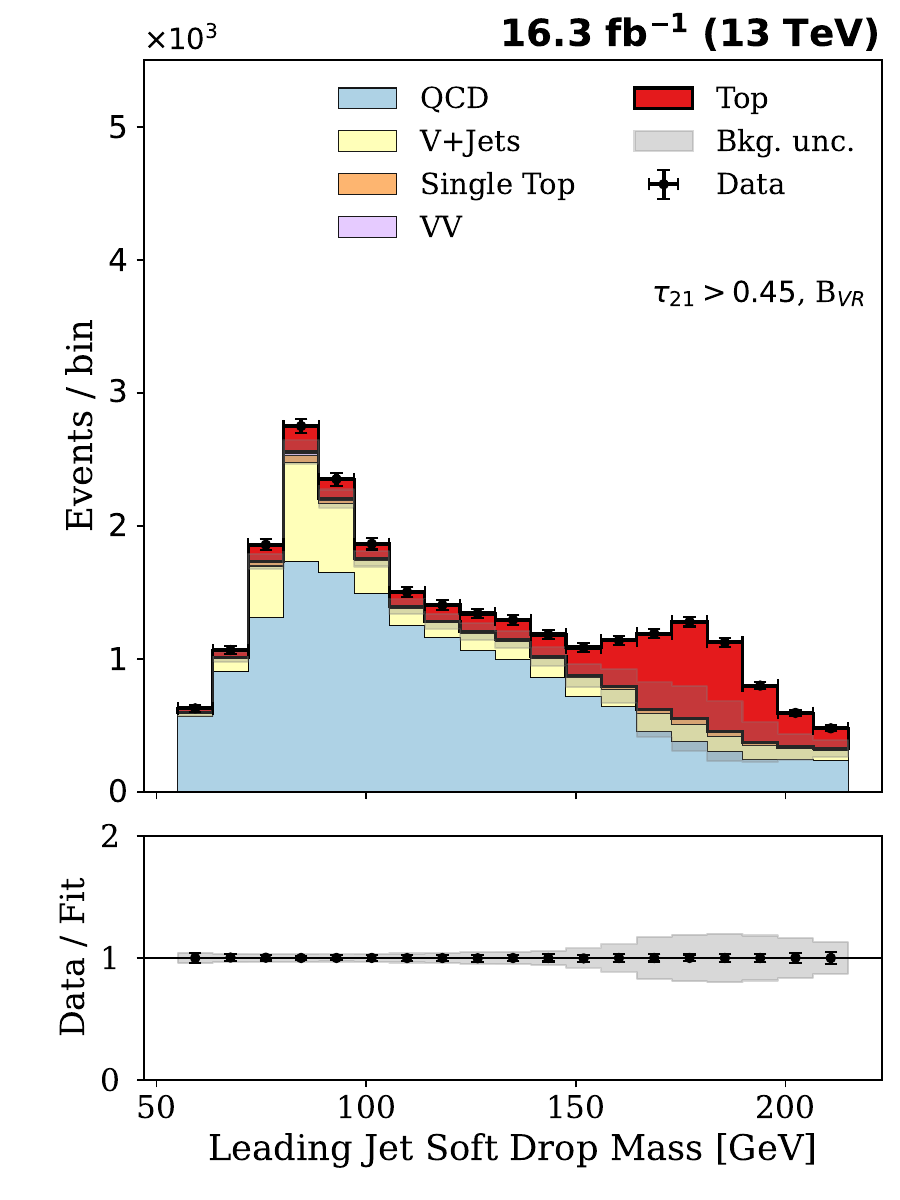}
        \includegraphics[width=.23\textwidth]{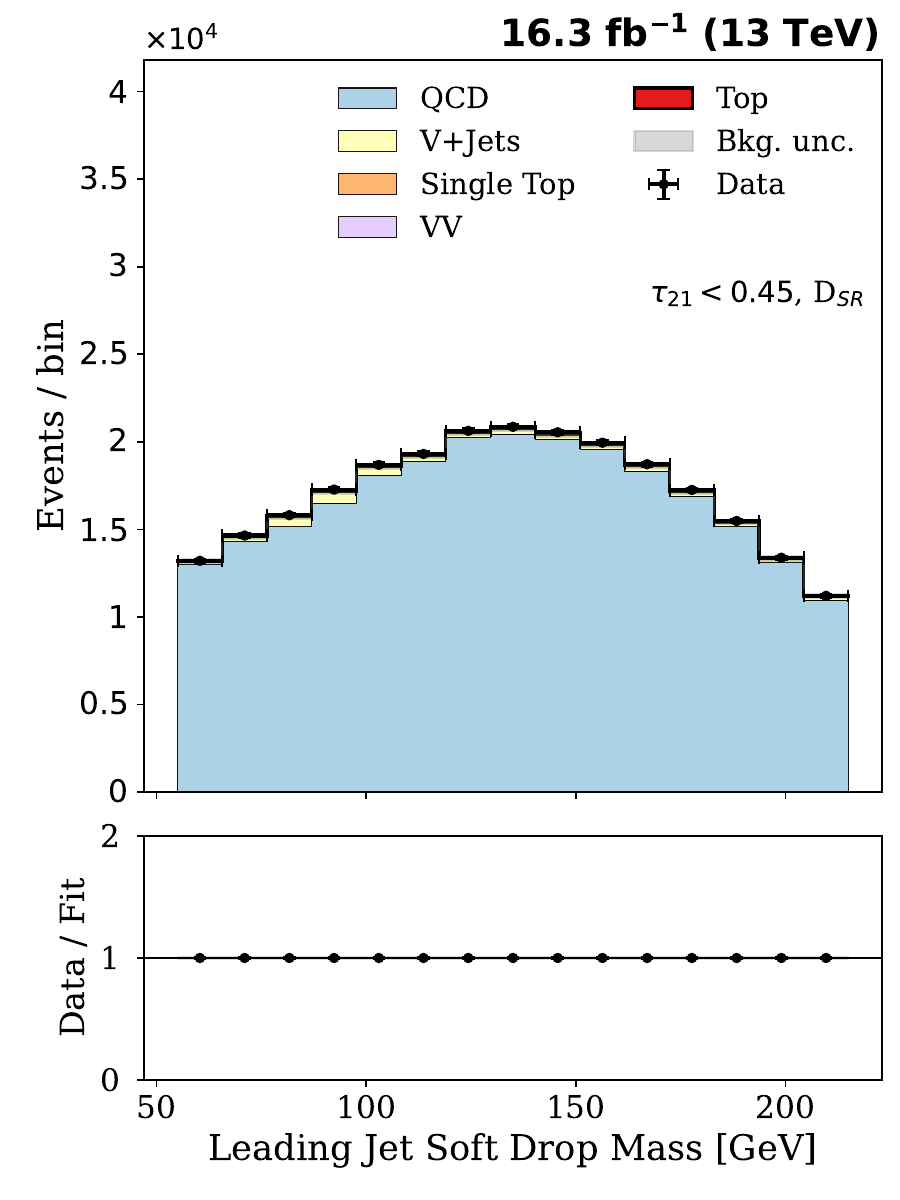}                
        \includegraphics[width=.23\textwidth]{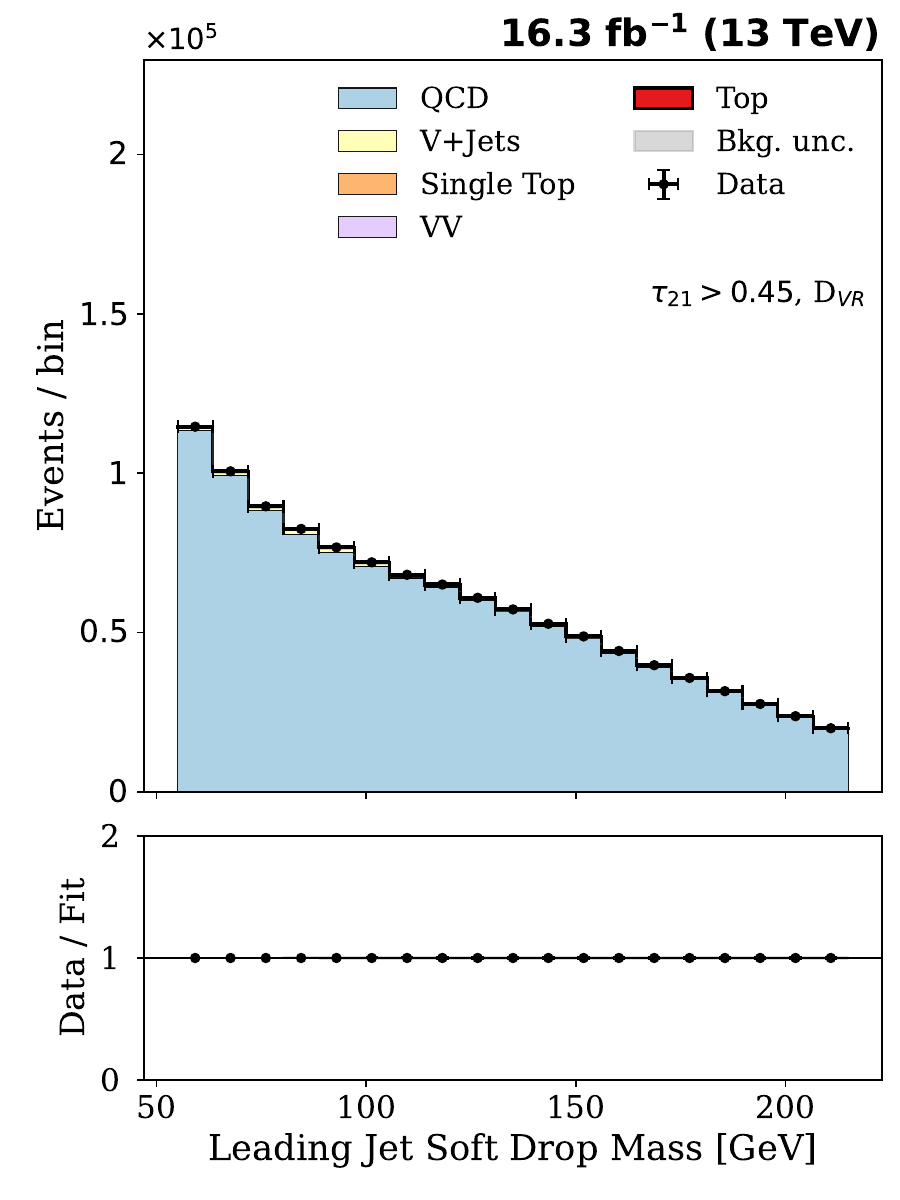}
        
    \caption{Leading jet soft drop mass where both jets are considered anomalous based on the \textsc{OmniLearned} small model score. The region where both jets have low $\tau_{21}$ values is shown at the left while the region where at least one jet fails the $\tau_{21}$ selection is shown at the right. The different regions used for the ABCD calculation are shown as rows. Shaded regions represent the total background uncertainty.}
    \label{fig:top_large_abcd}
\end{figure}

Regions where top quark pair production are considered the main signal are shown in Figs.~\ref{fig:top_small_abcd} and~\ref{fig:top_large_abcd} for the small and large \textsc{OmniLearned} scores, respectively. 

\subsection{Dihiggs as a Signal}

\begin{figure}[ht]
    \centering
        \includegraphics[width=.23\textwidth]{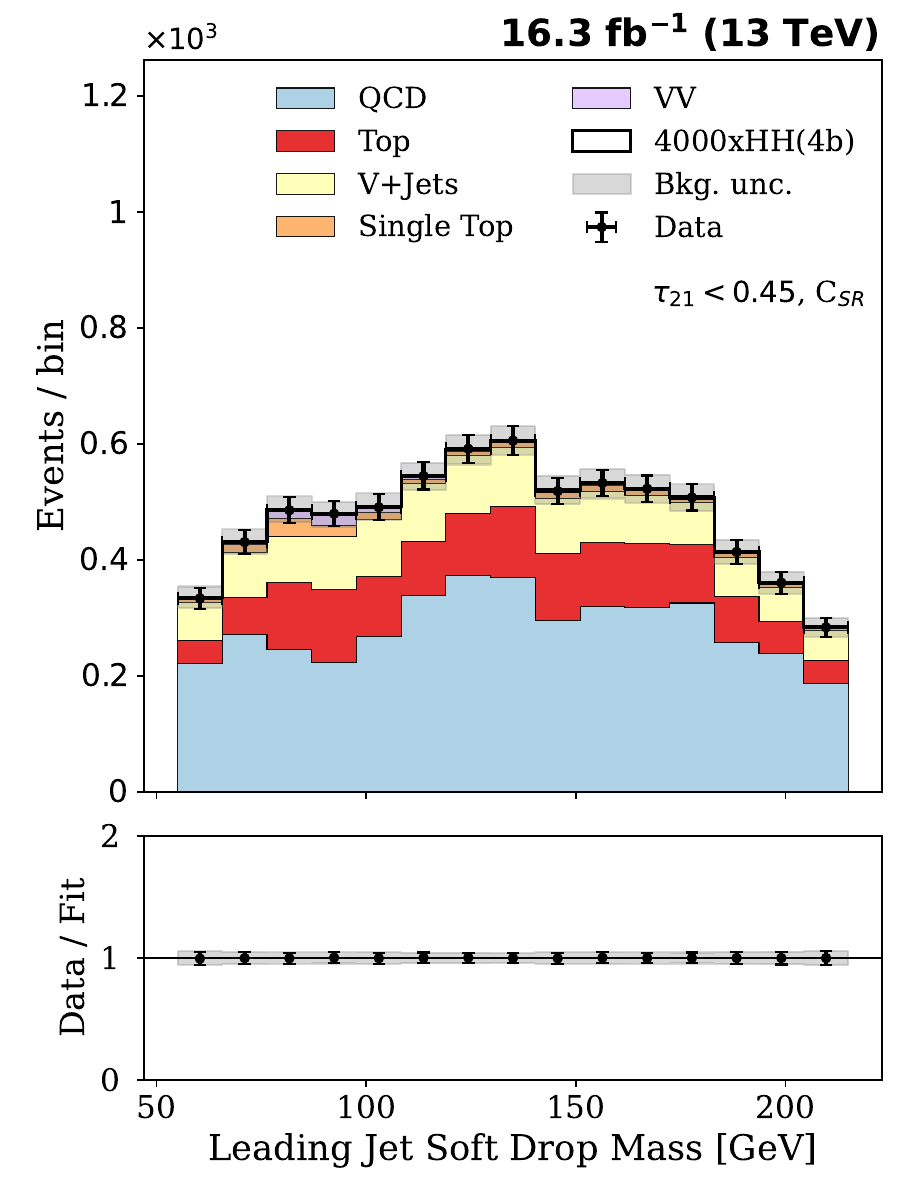}
        \includegraphics[width=.23\textwidth]{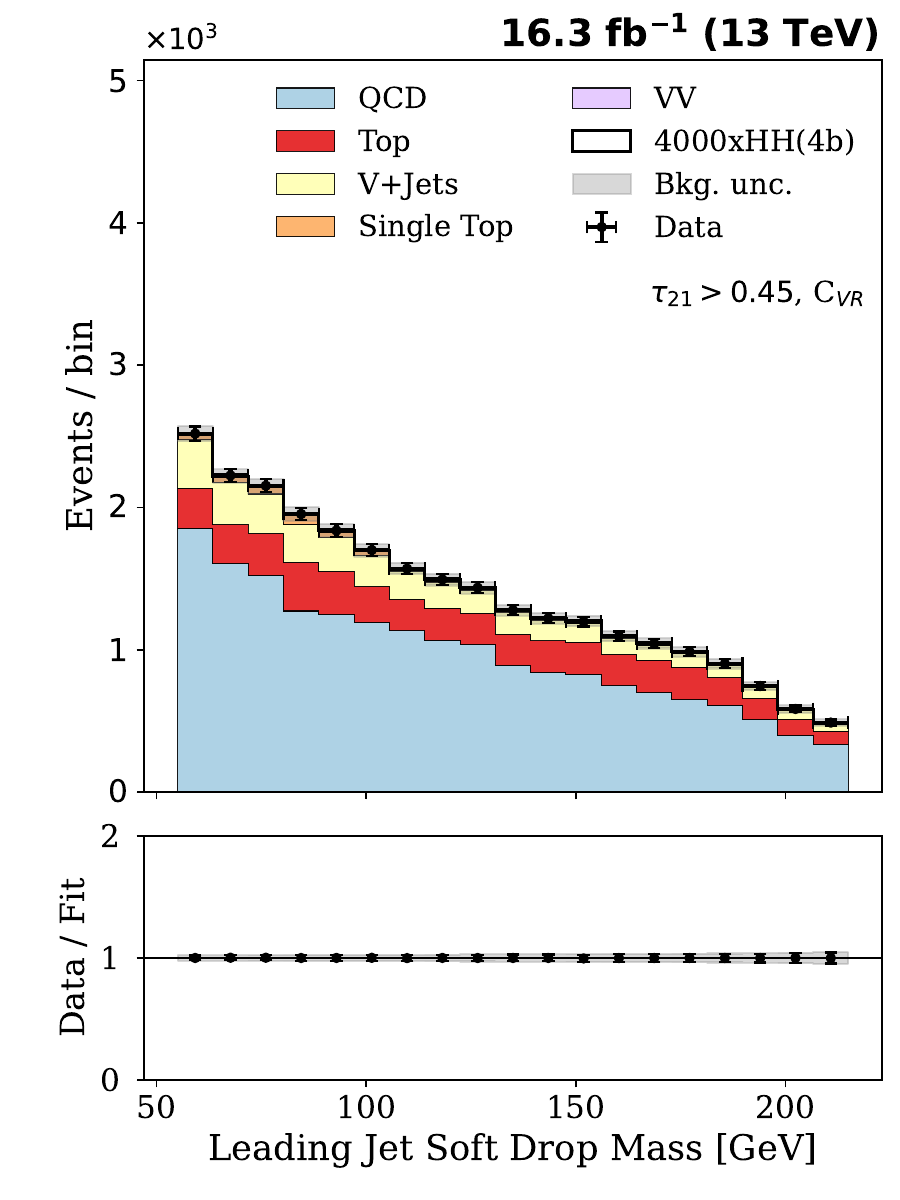}
        \includegraphics[width=.23\textwidth]{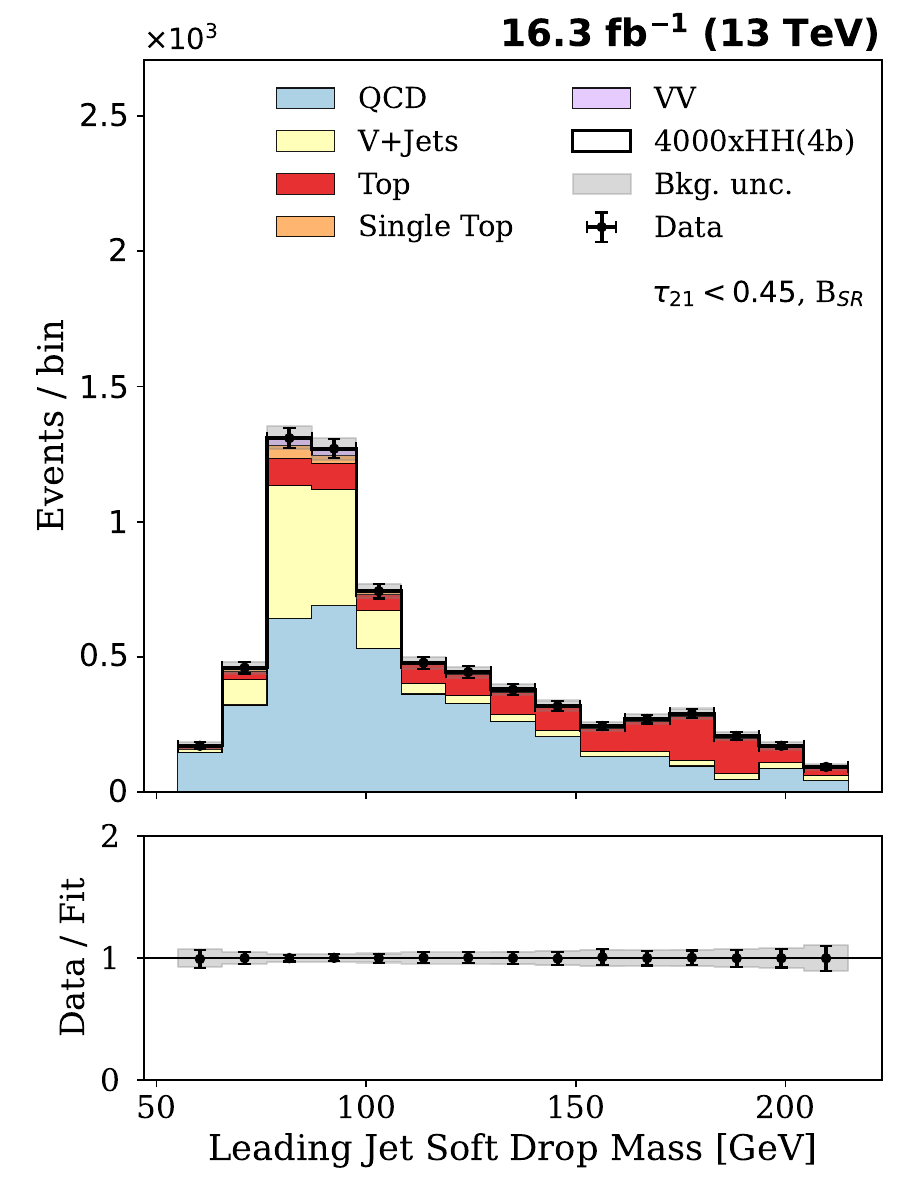}
        \includegraphics[width=.23\textwidth]{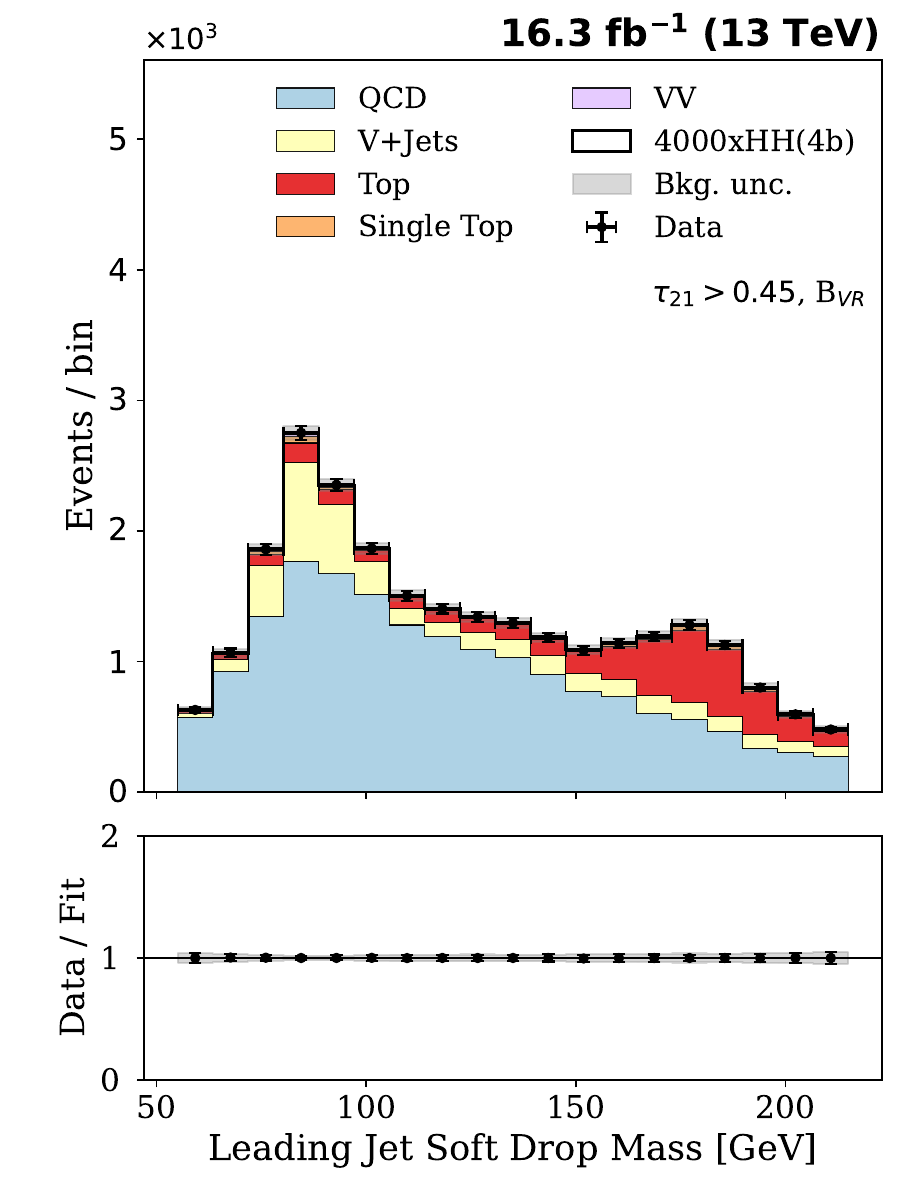}
        \includegraphics[width=.23\textwidth]{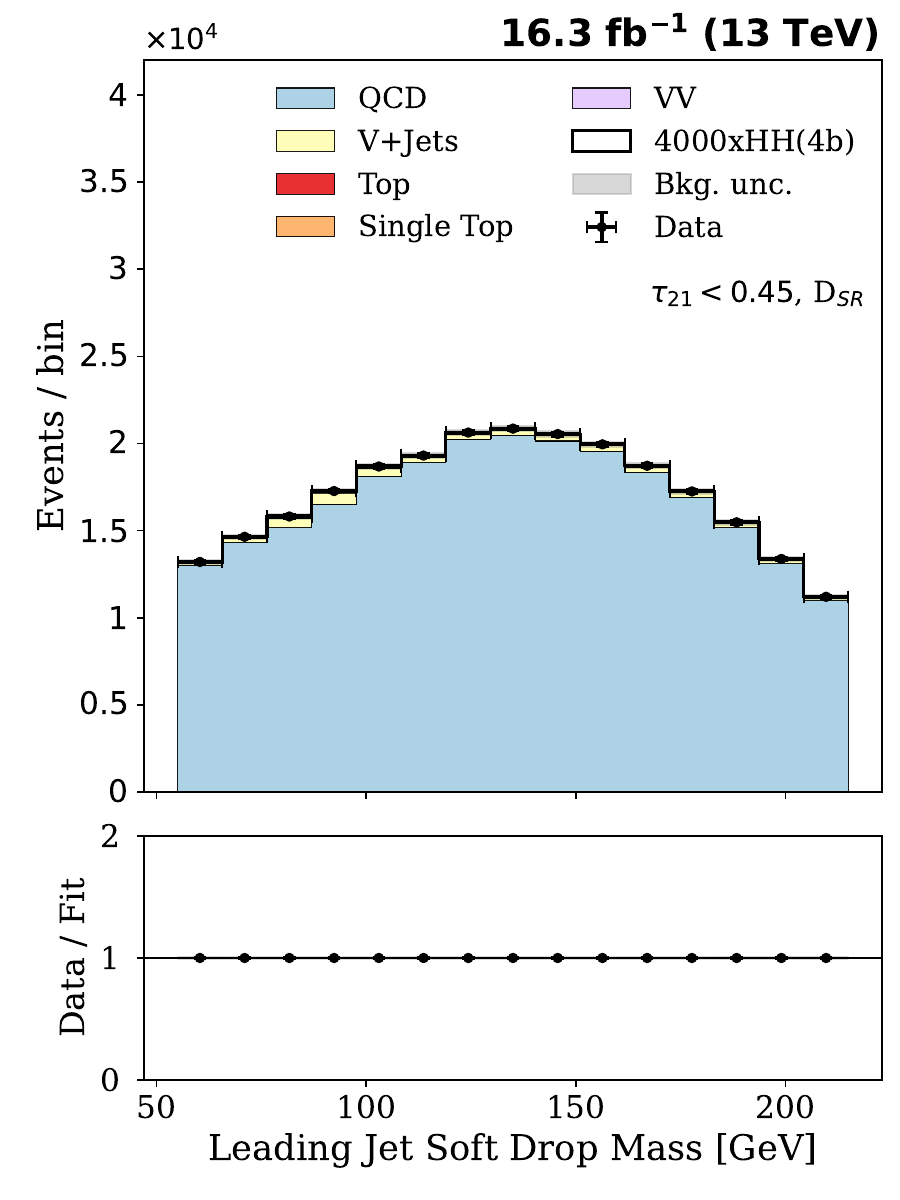}                
        \includegraphics[width=.23\textwidth]{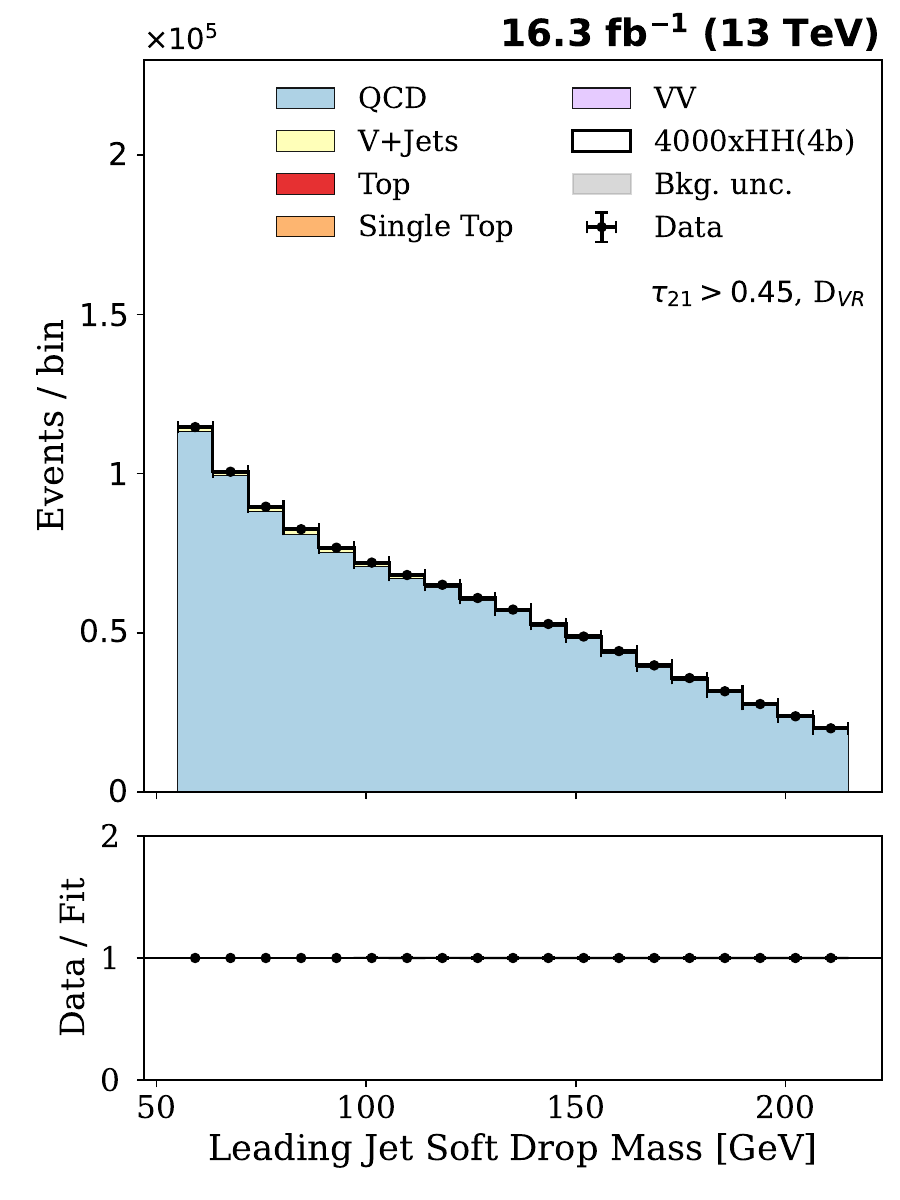}
        
    \caption{Leading jet soft drop mass where both jets are considered anomalous based on the \textsc{OmniLearned} large model score. The region where both jets have low $\tau_{21}$ values is shown at the left while the region where at least one jet fails the $\tau_{21}$ selection is shown at the right. The different regions used for the ABCD calculation are shown as rows. Shaded regions represent the total background uncertainty.}
    \label{fig:hh_baseline}
\end{figure}

\begin{figure*}[ht]
    \centering
        \includegraphics[width=.31\textwidth]{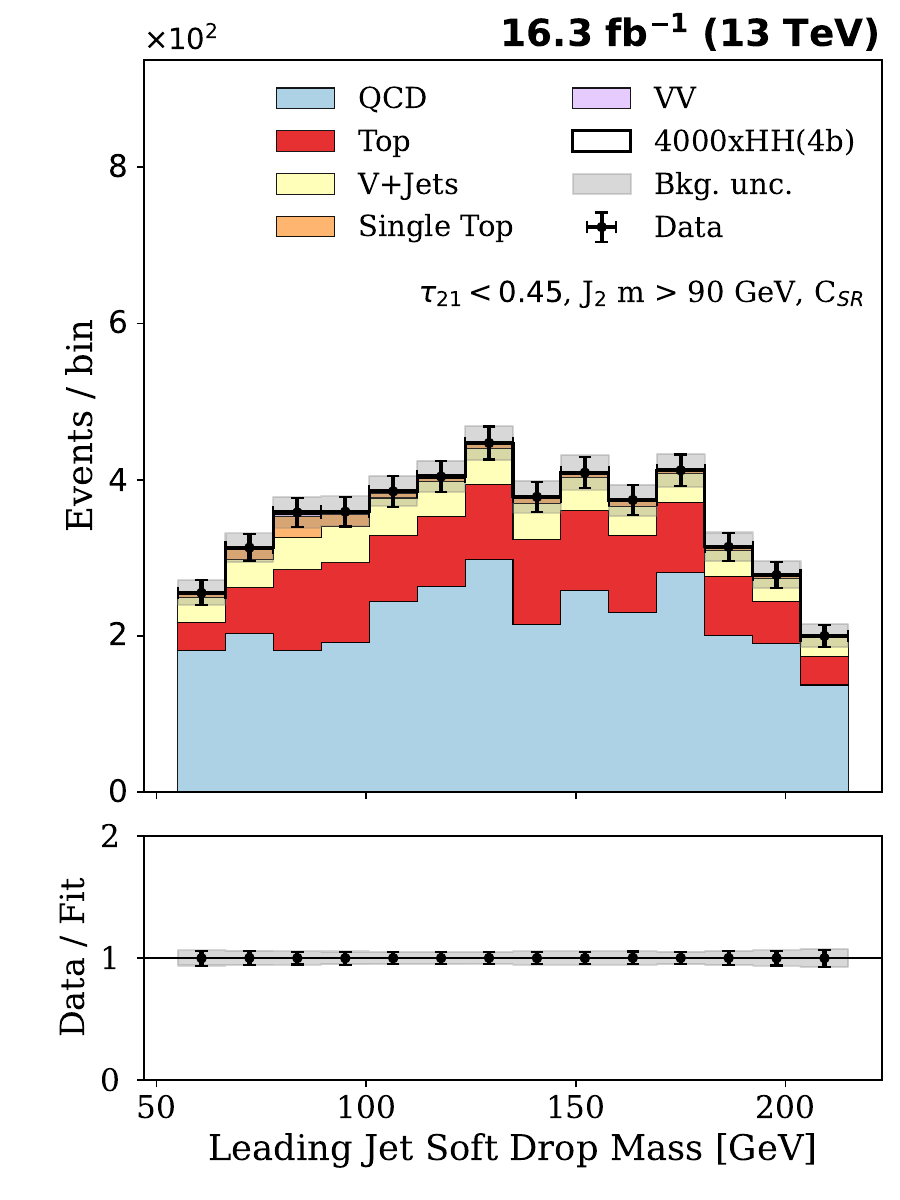}
        \includegraphics[width=.31\textwidth]{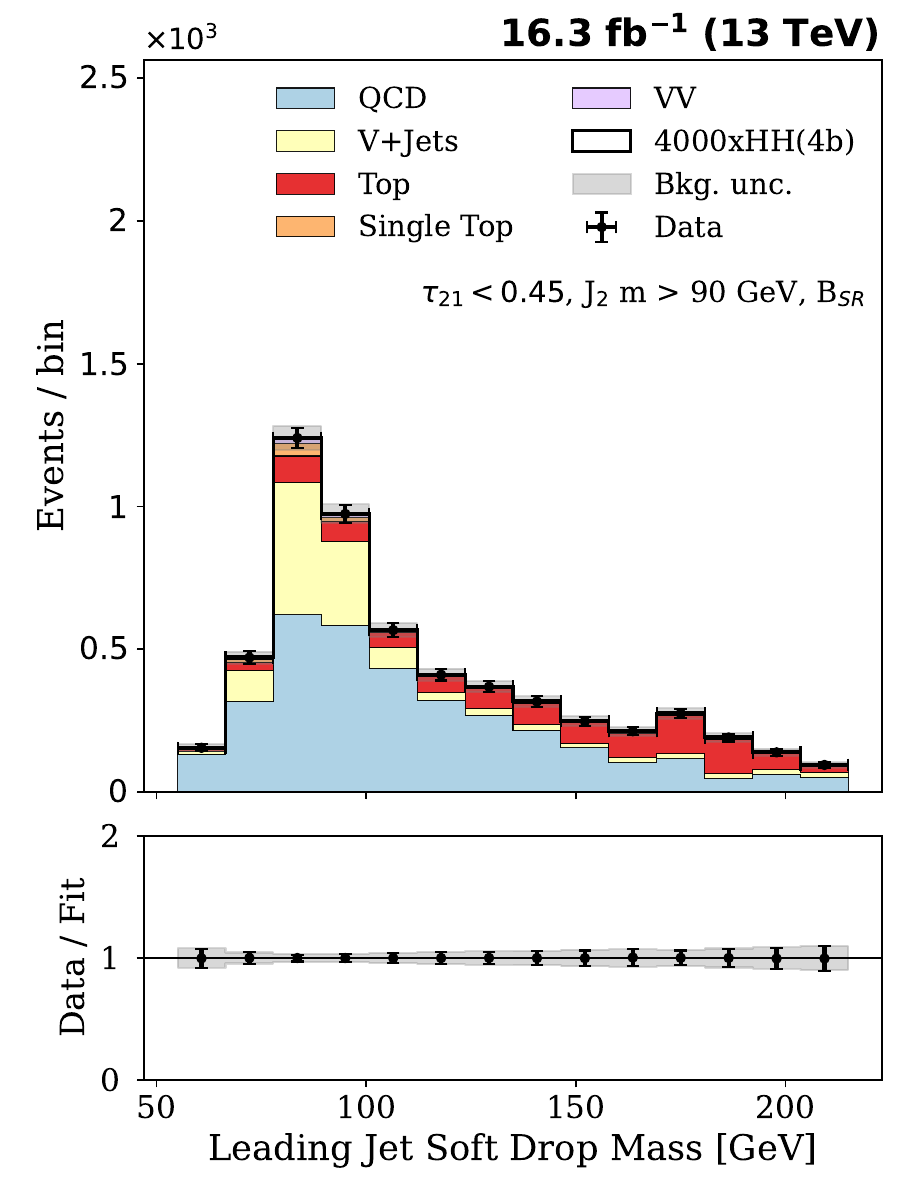}
        \includegraphics[width=.31\textwidth]{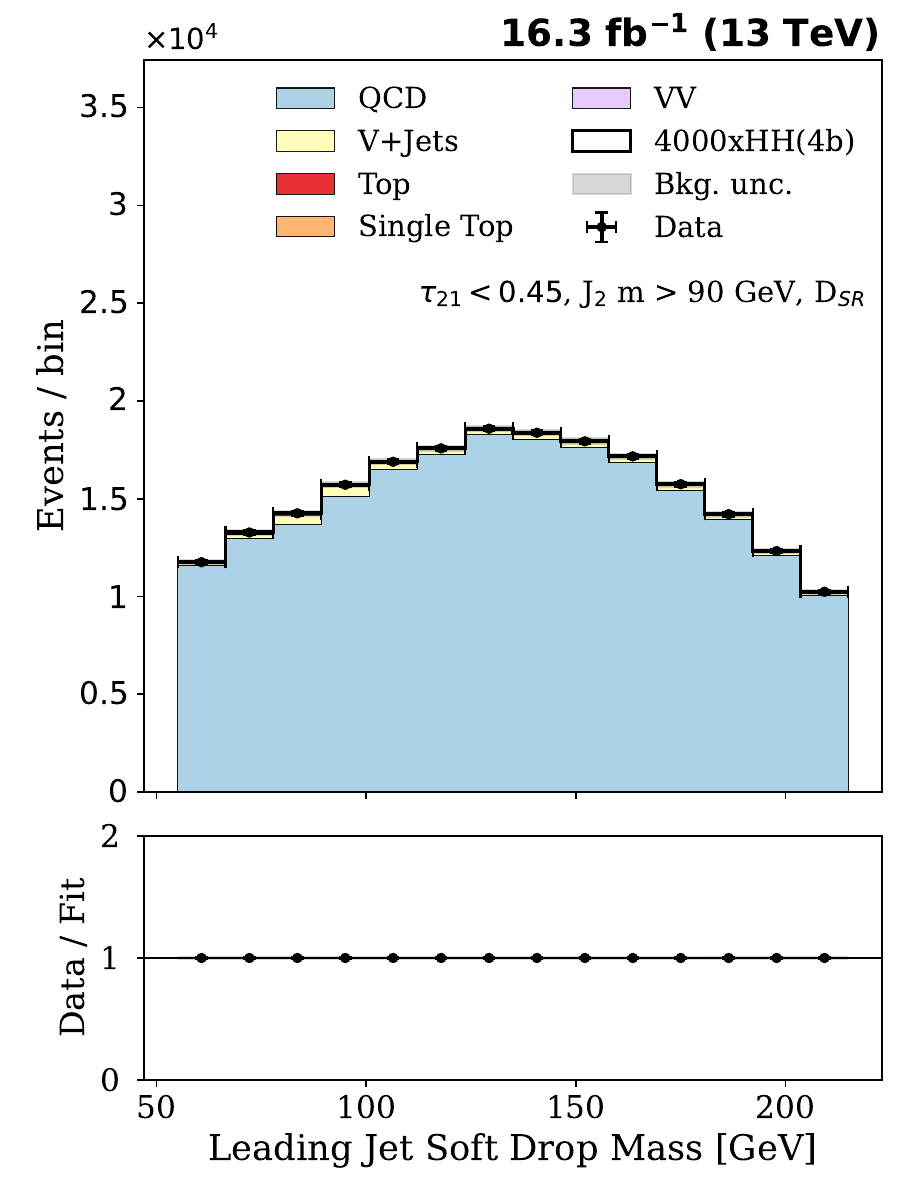}
        \includegraphics[width=.31\textwidth]{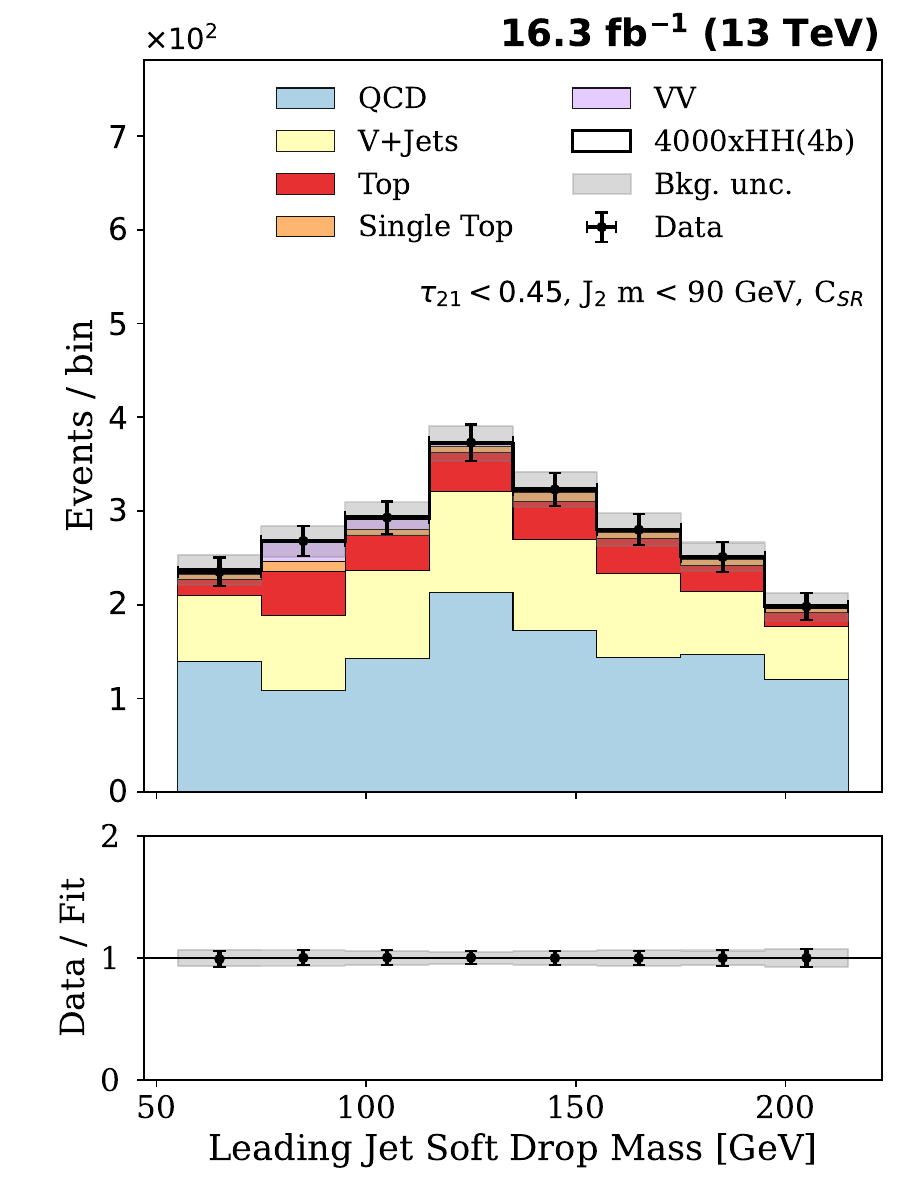}
        \includegraphics[width=.31\textwidth]{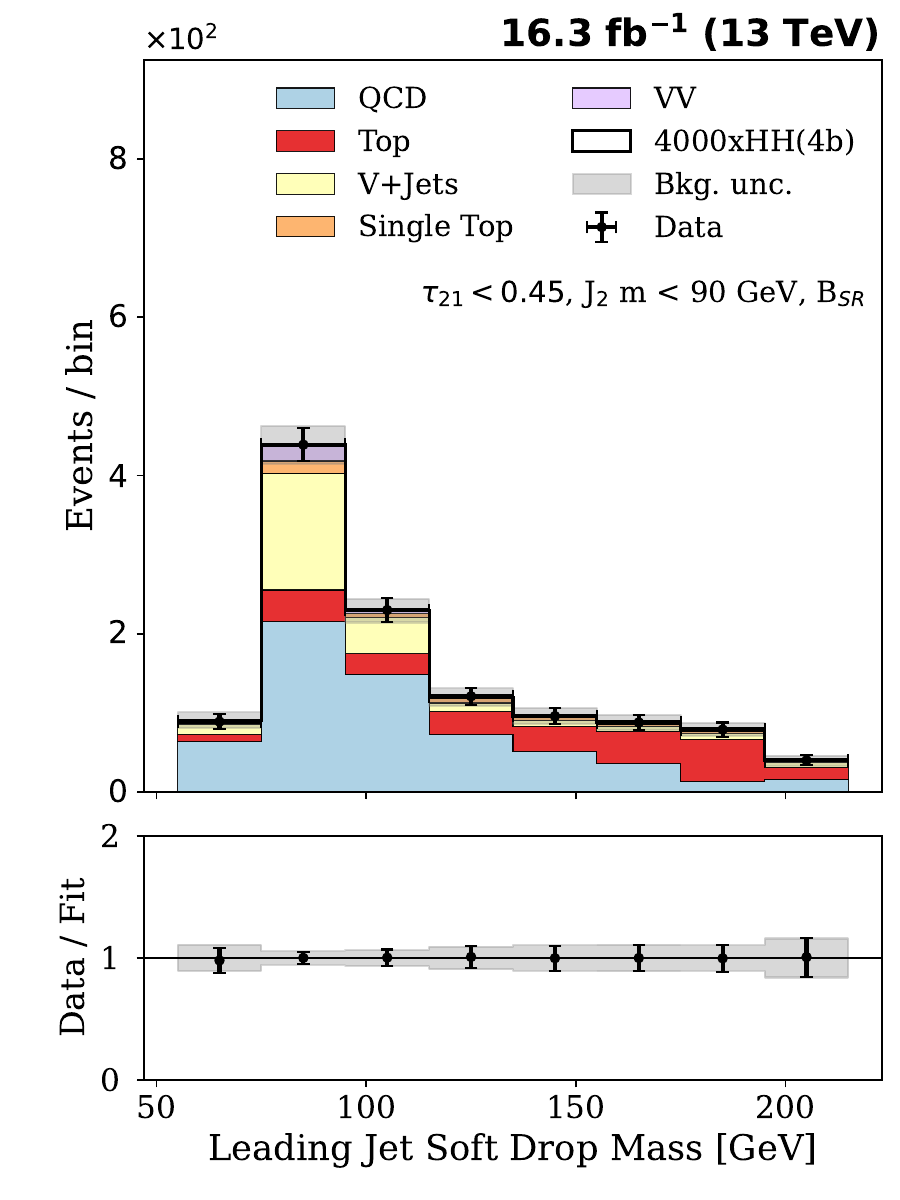}
        \includegraphics[width=.31\textwidth]{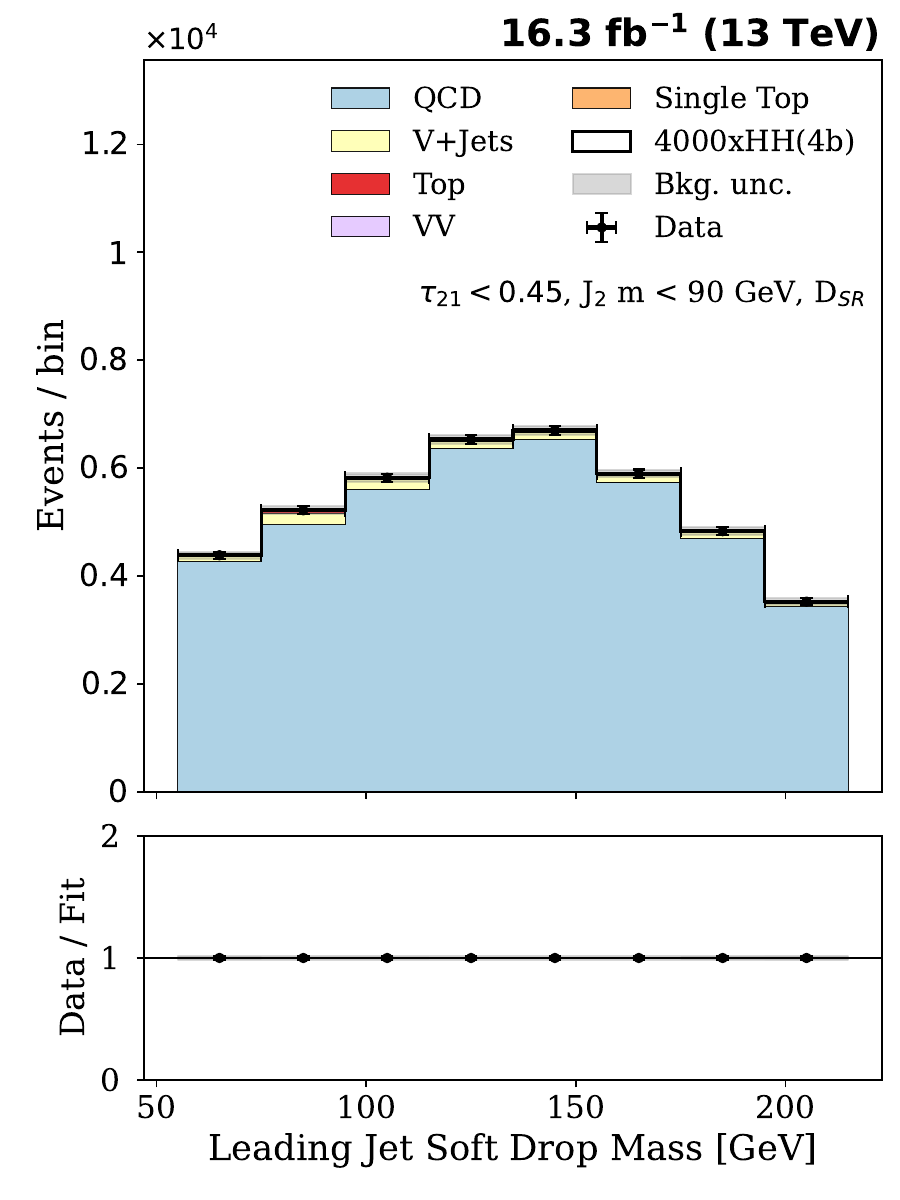}
        \includegraphics[width=.31\textwidth]{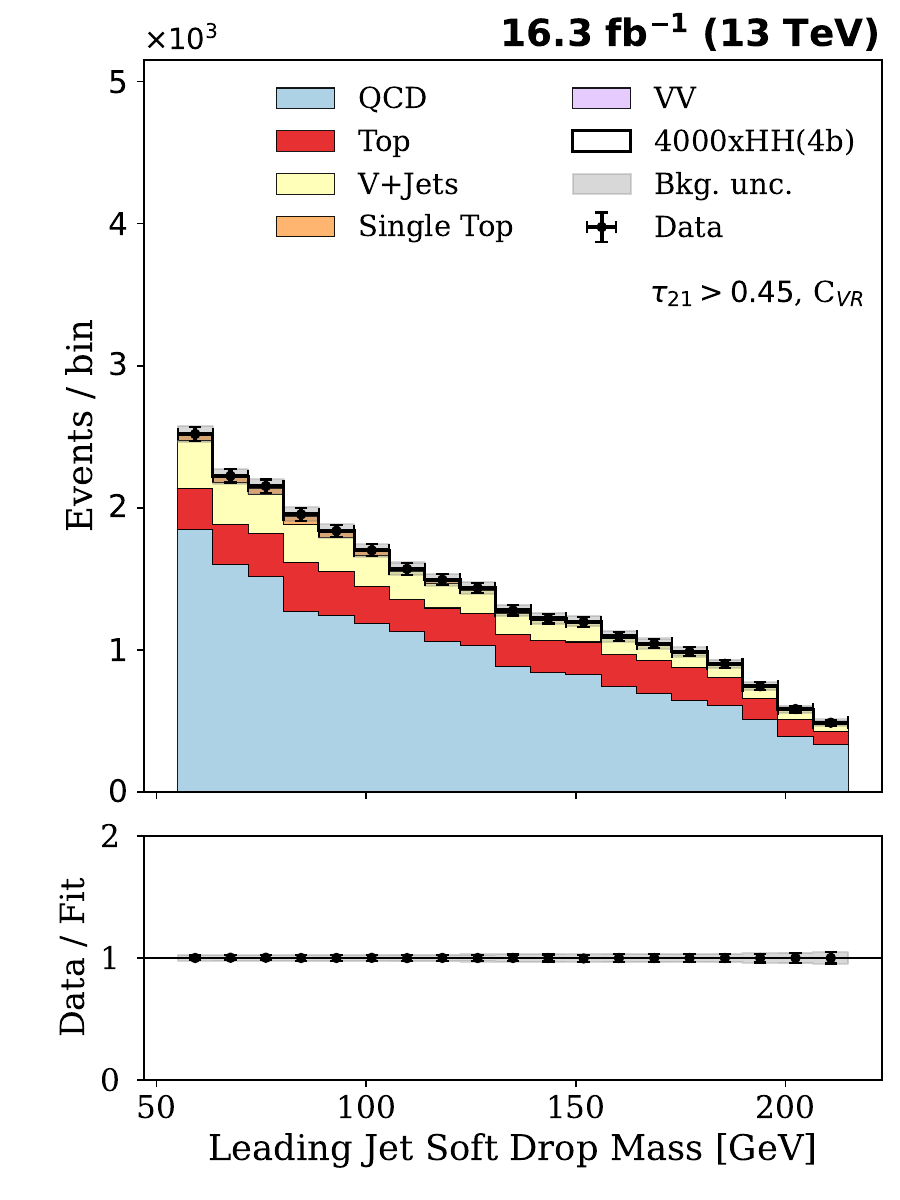}
        \includegraphics[width=.31\textwidth]{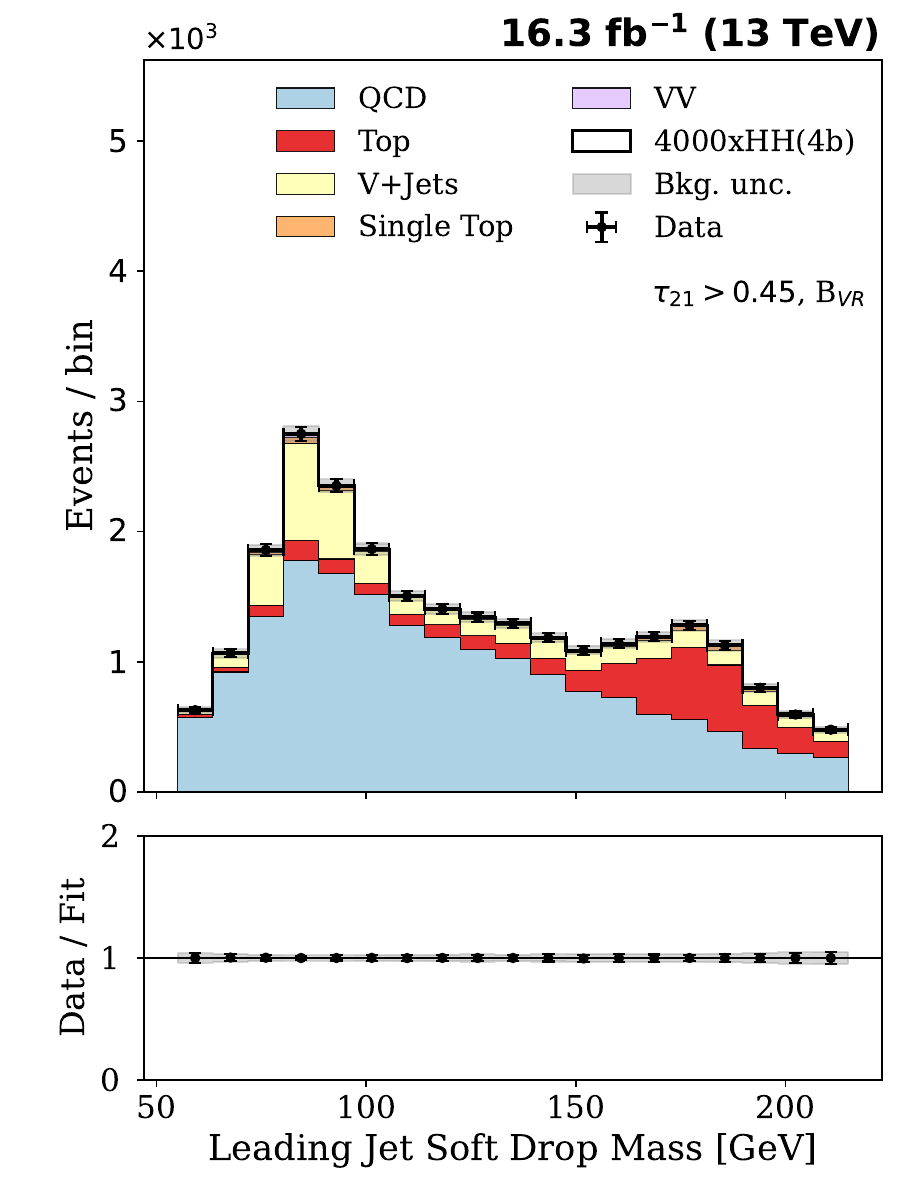}
        \includegraphics[width=.31\textwidth]{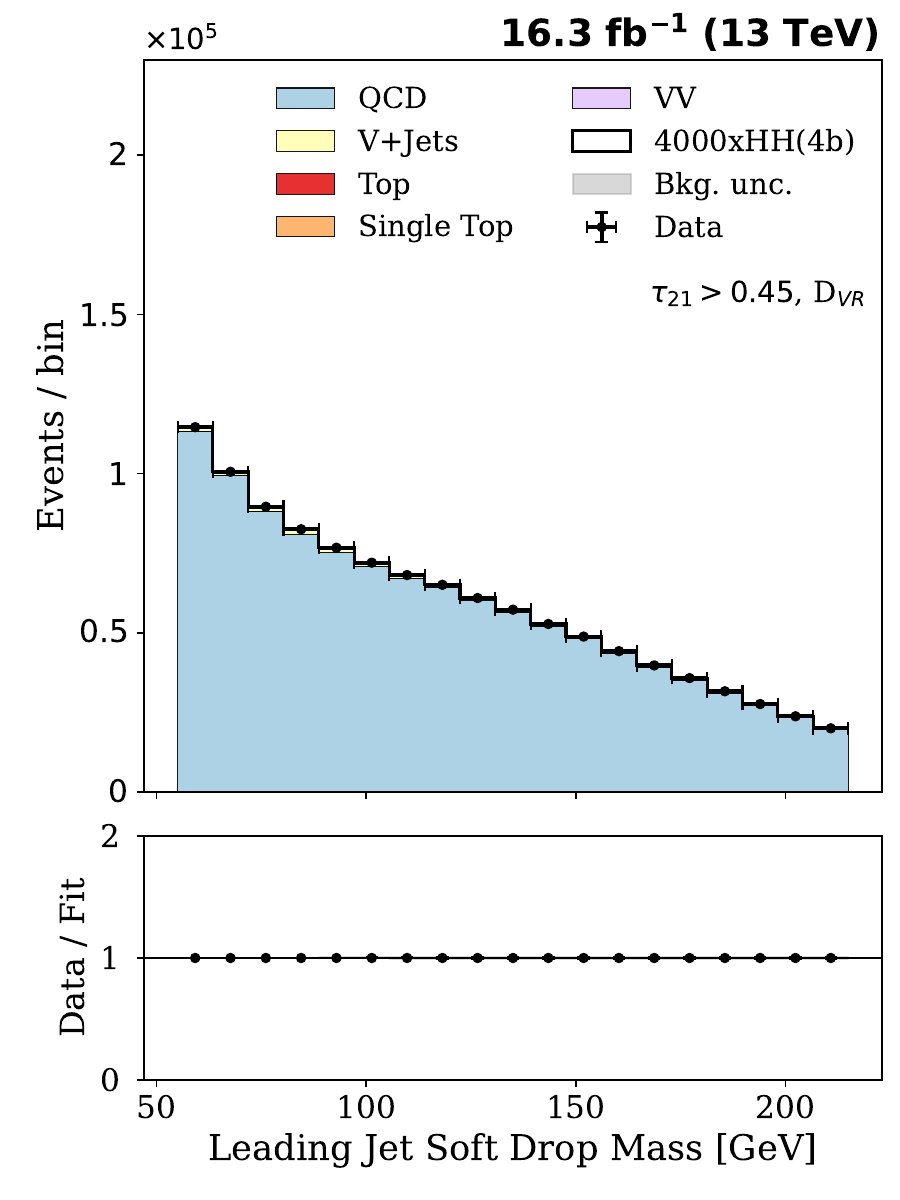}
        
    \caption{Leading jet soft drop mass where both jets are considered anomalous based on the \textsc{OmniLearned} large model score. The region where both jets have low $\tau_{21}$ values and subleading jet mass is above 90 GeV  is shown at the top while the region where the subleading jet mass is below 90 GeV is shown at the middle. The region where at least one jet fails the $\tau_{21}$ selection is shown at the bottom. The different regions used for the ABCD calculation are shown as columns. Shaded regions represent the total background uncertainty.}
    \label{fig:hh_sr12}
\end{figure*}

\begin{figure*}[ht]
    \centering
        \includegraphics[width=.31\textwidth]{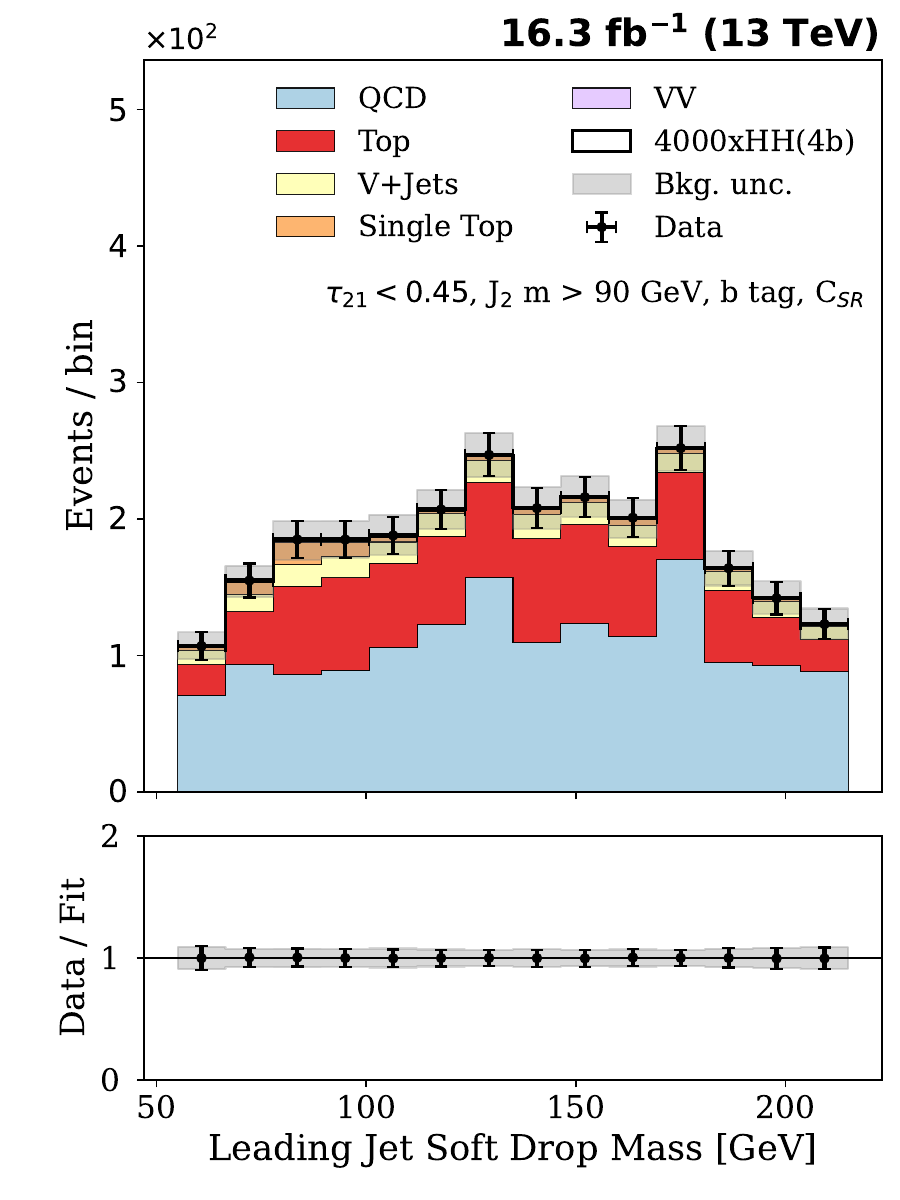}
        \includegraphics[width=.31\textwidth]{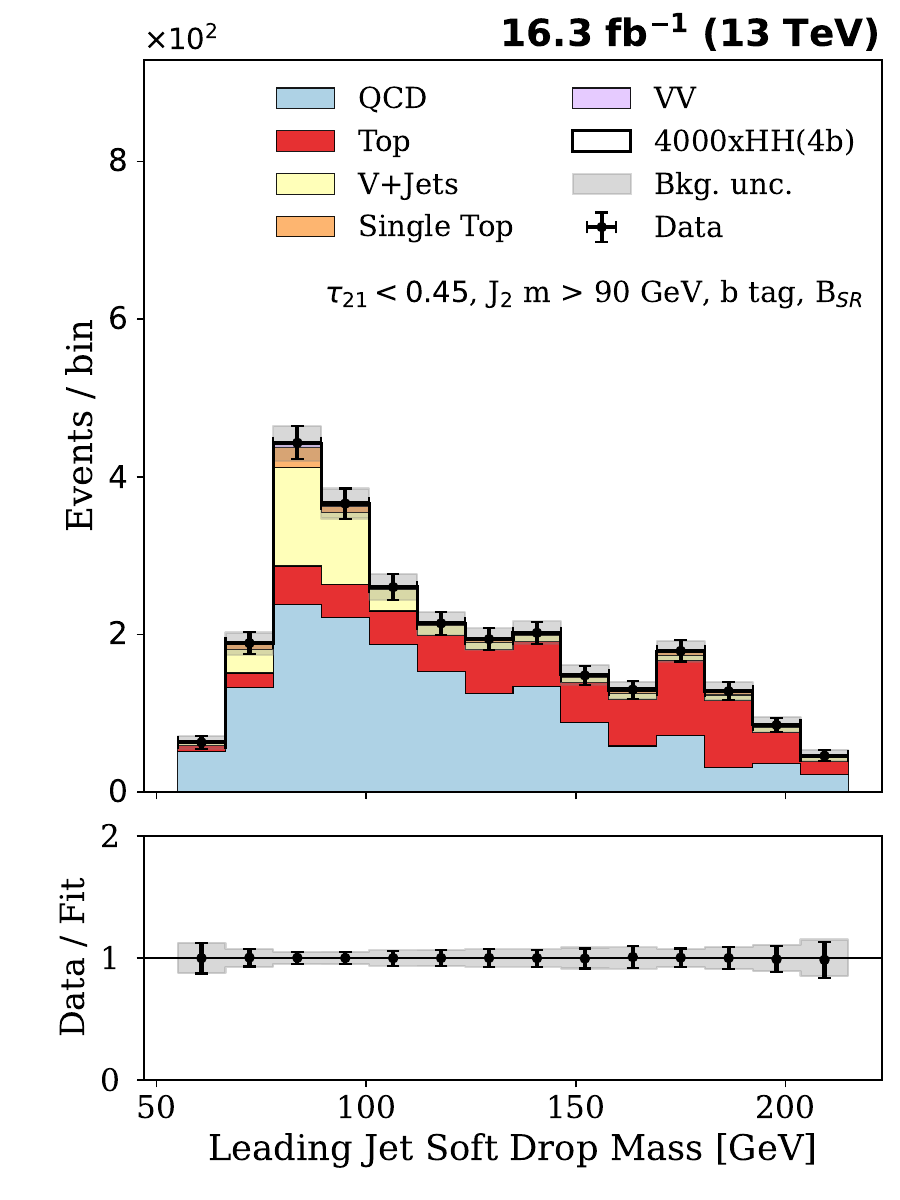}
        \includegraphics[width=.31\textwidth]{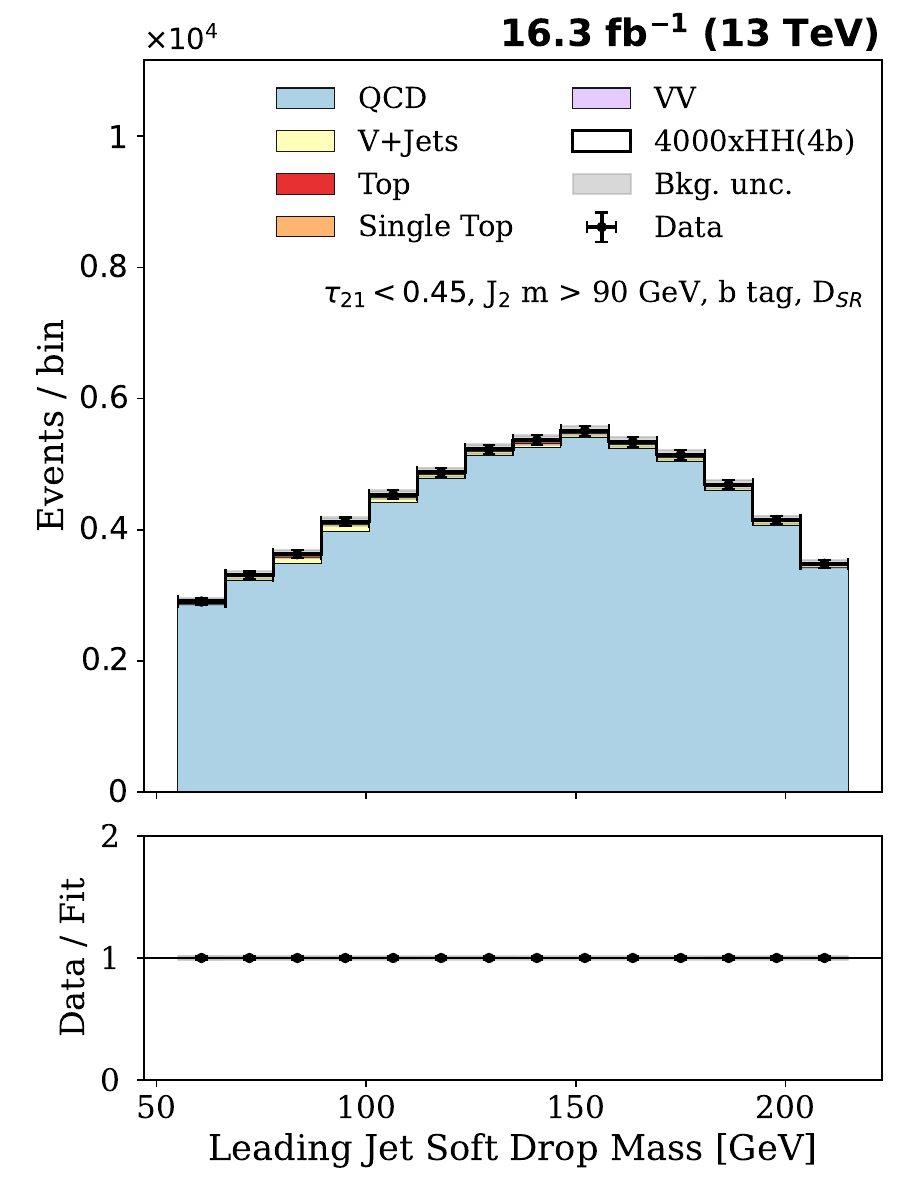}
        \includegraphics[width=.31\textwidth]{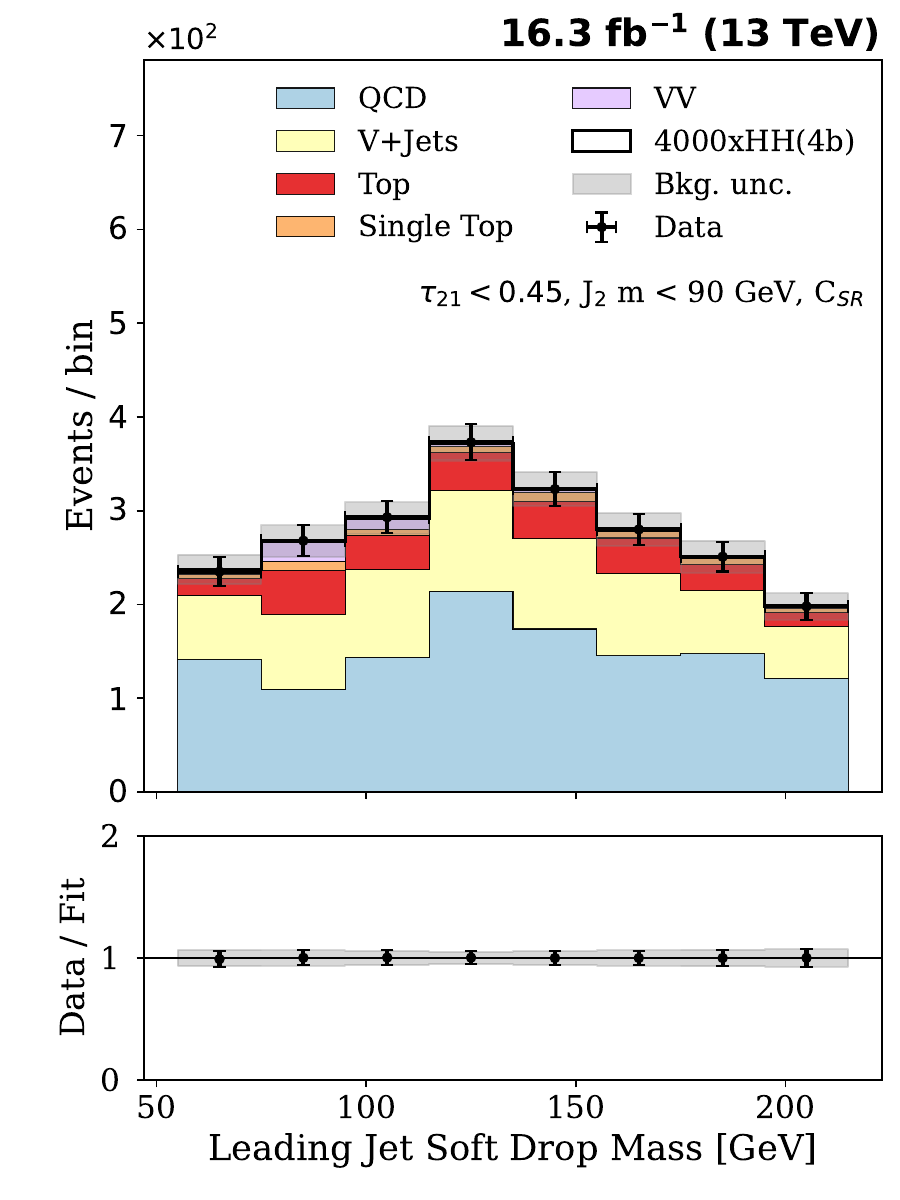}
        \includegraphics[width=.31\textwidth]{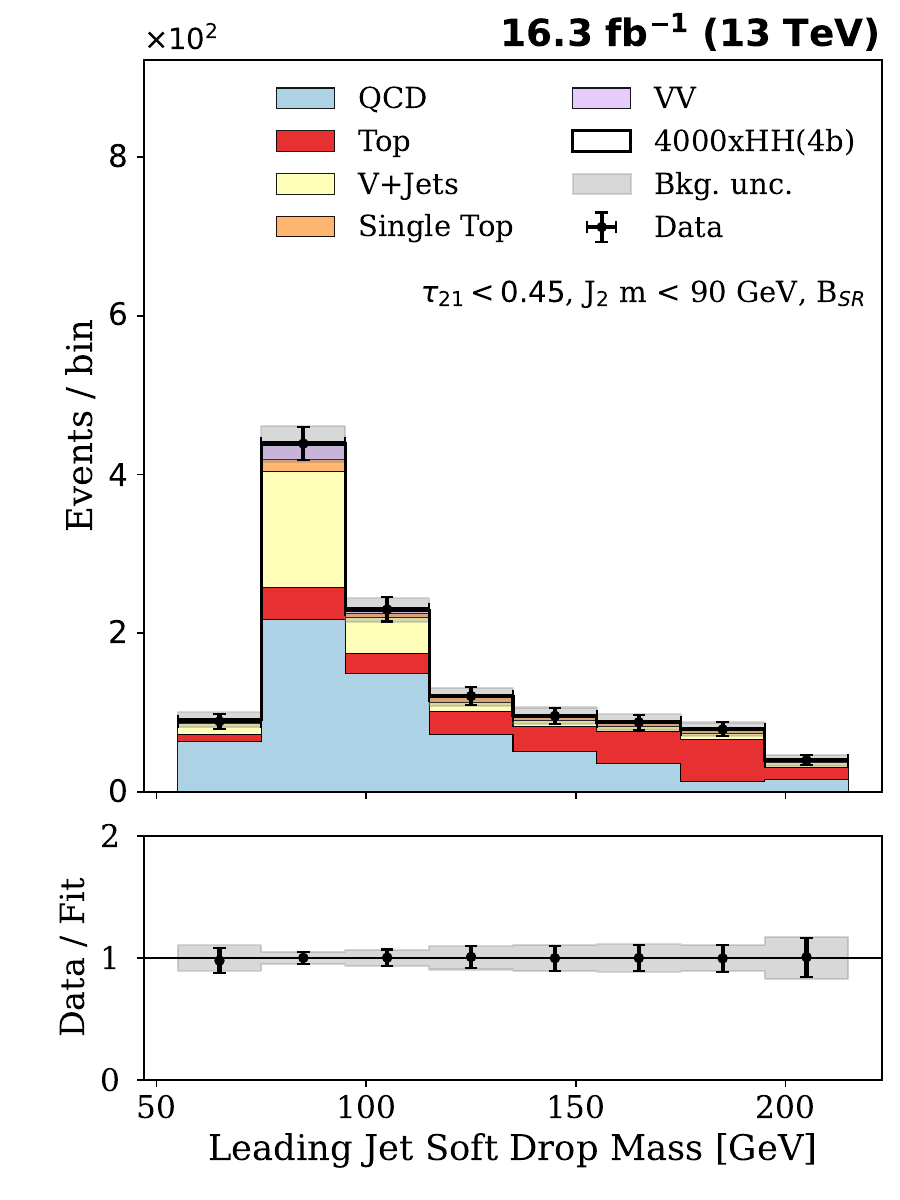}
        \includegraphics[width=.31\textwidth]{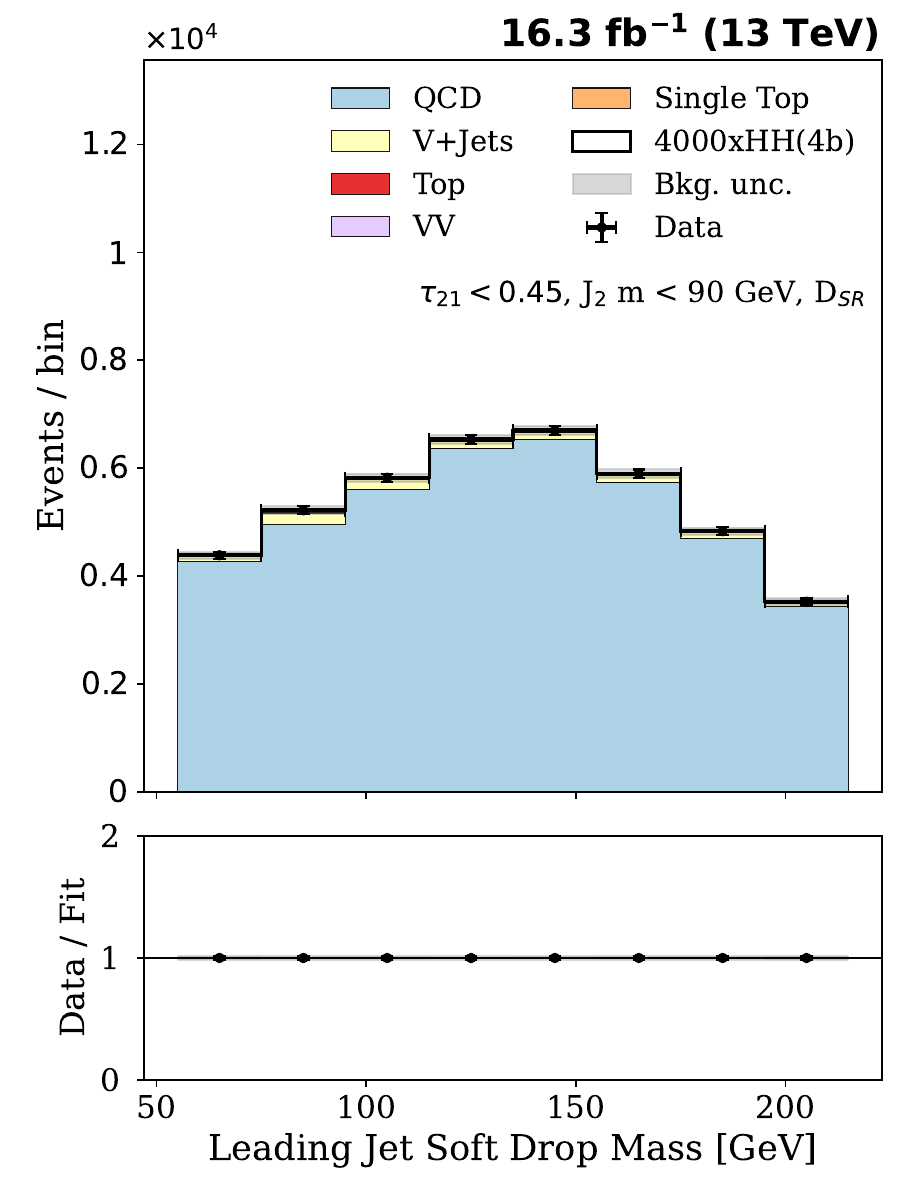}
        \includegraphics[width=.31\textwidth]{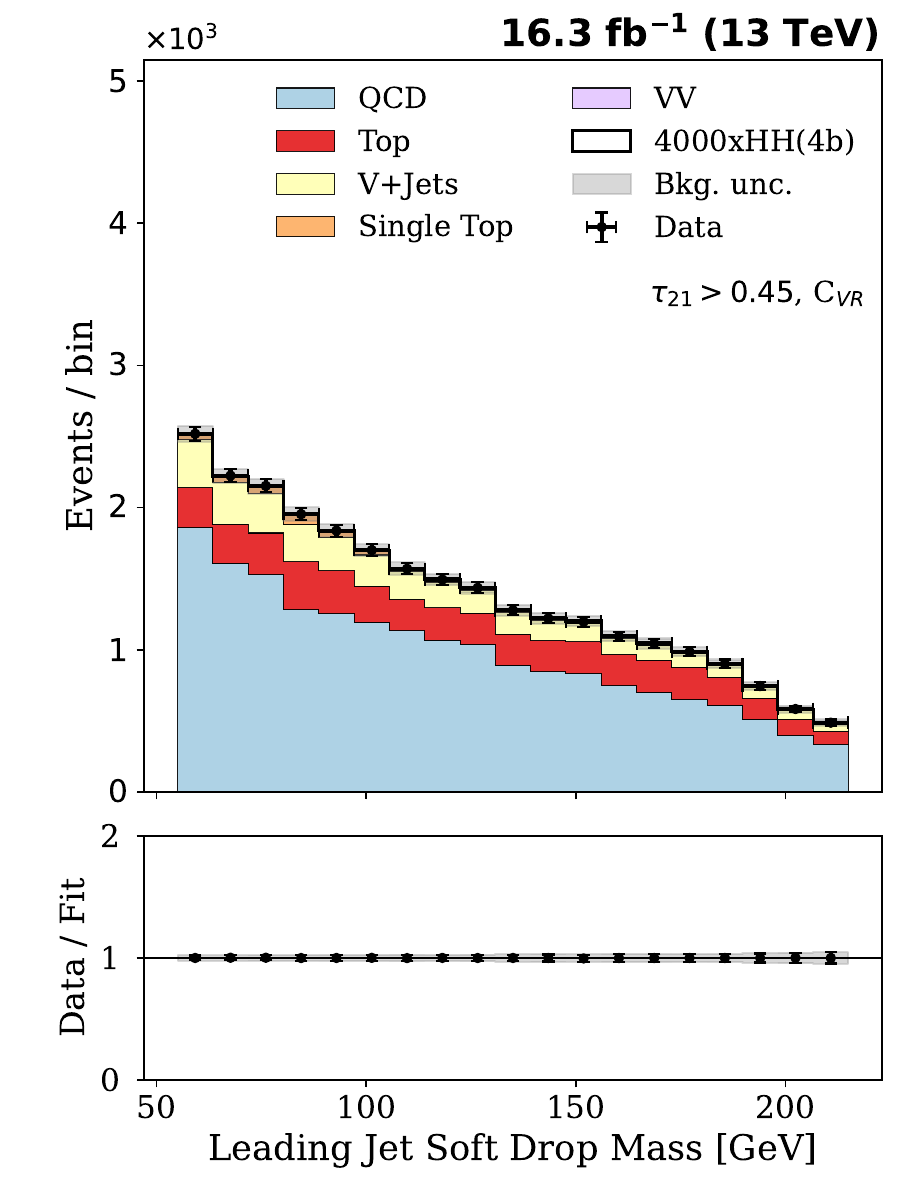}
        \includegraphics[width=.31\textwidth]{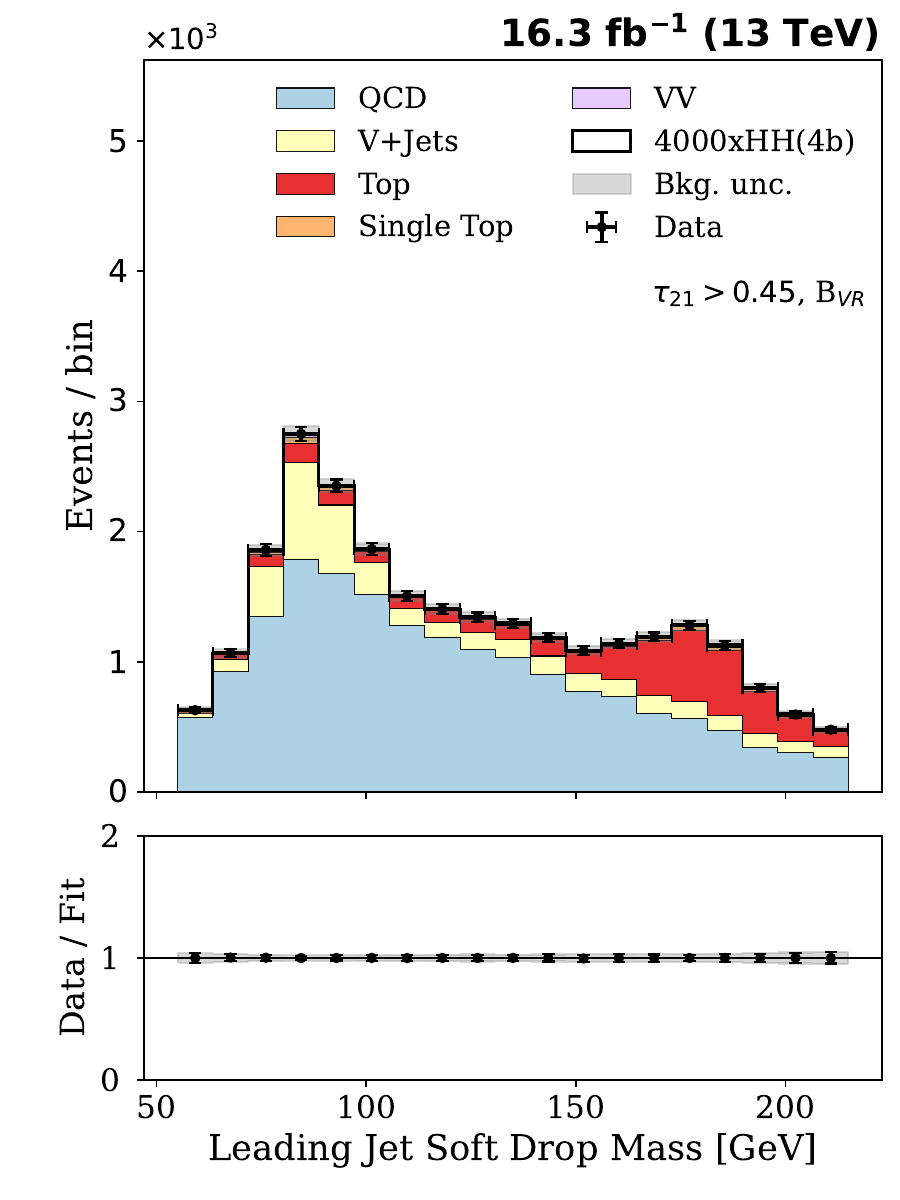}
        \includegraphics[width=.31\textwidth]{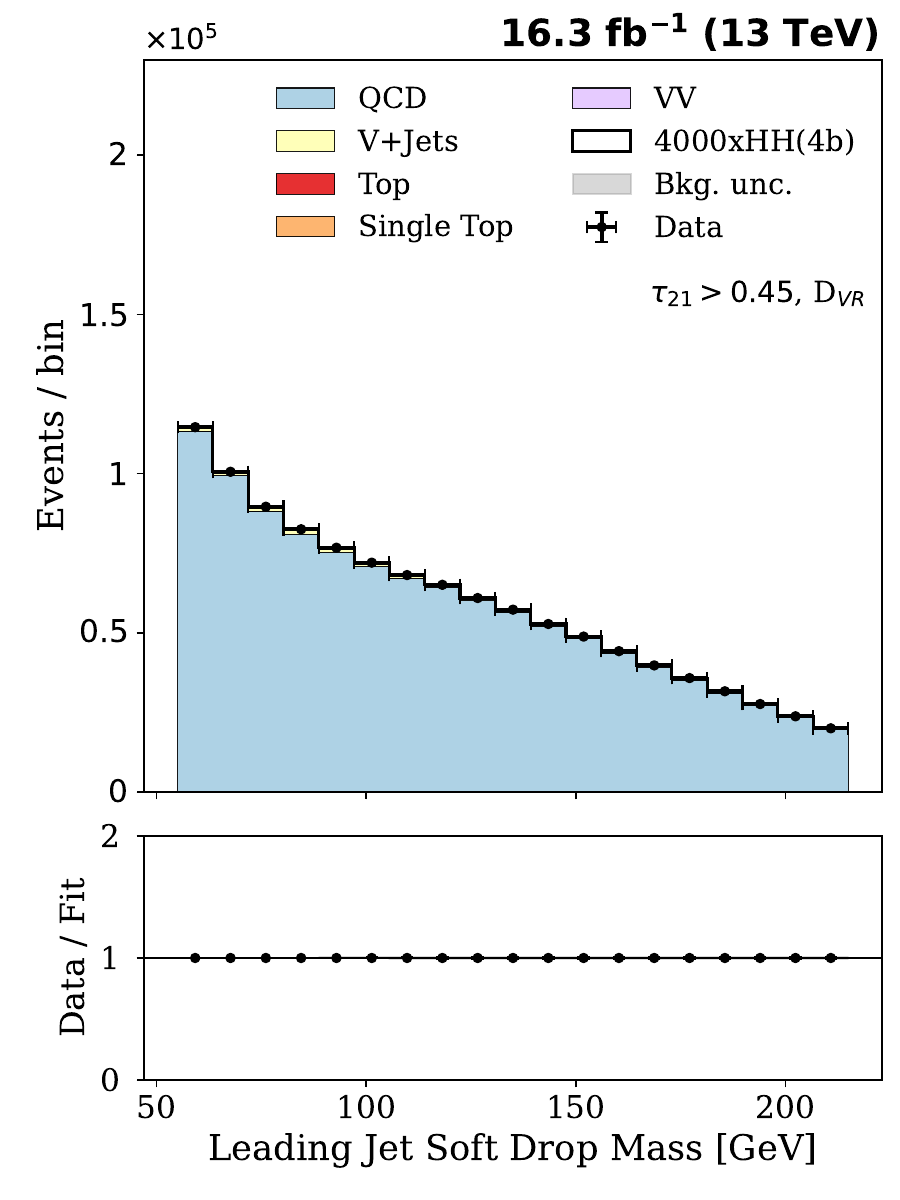}
        
    \caption{Leading jet soft drop mass where both jets are considered anomalous based on the \textsc{OmniLearned} large model score. The region where both jets have low $\tau_{21}$ values at least one jet is b-tagged, and subleading jet mass is above 90 GeV  is shown at the top while the region where the subleading jet mass is below 90 GeV is shown at the middle. The region where at least one jet fails the $\tau_{21}$ selection is shown at the bottom. The different regions used for the ABCD calculation are shown as columns. Shaded regions represent the total background uncertainty.}
    \label{fig:hh_sr32}
\end{figure*}

Results for different selections of the most anomalous regions using the HH sample as the signal are shown in Figs~\ref{fig:hh_baseline},~\ref{fig:hh_sr12}, and~\ref{fig:hh_sr32} for the baseline selection, selection requiring the subleading jet with mass above 90 GeV, and selection requiring at least one b-tagged jet on top of the previous selection, respectively.

\subsection{X(bb) Anomaly Score}
\begin{figure*}[ht]
    \centering
        \includegraphics[width=.31\textwidth]{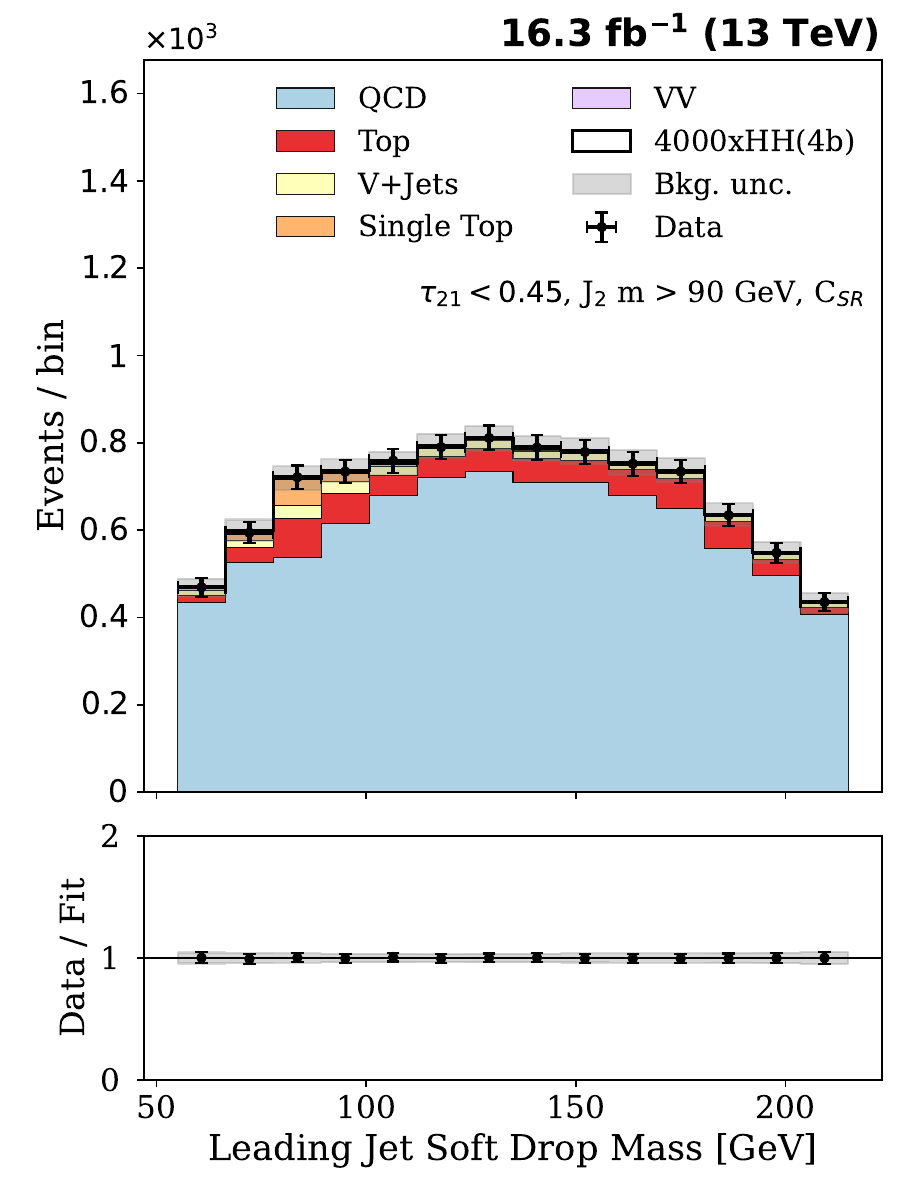}
        \includegraphics[width=.31\textwidth]{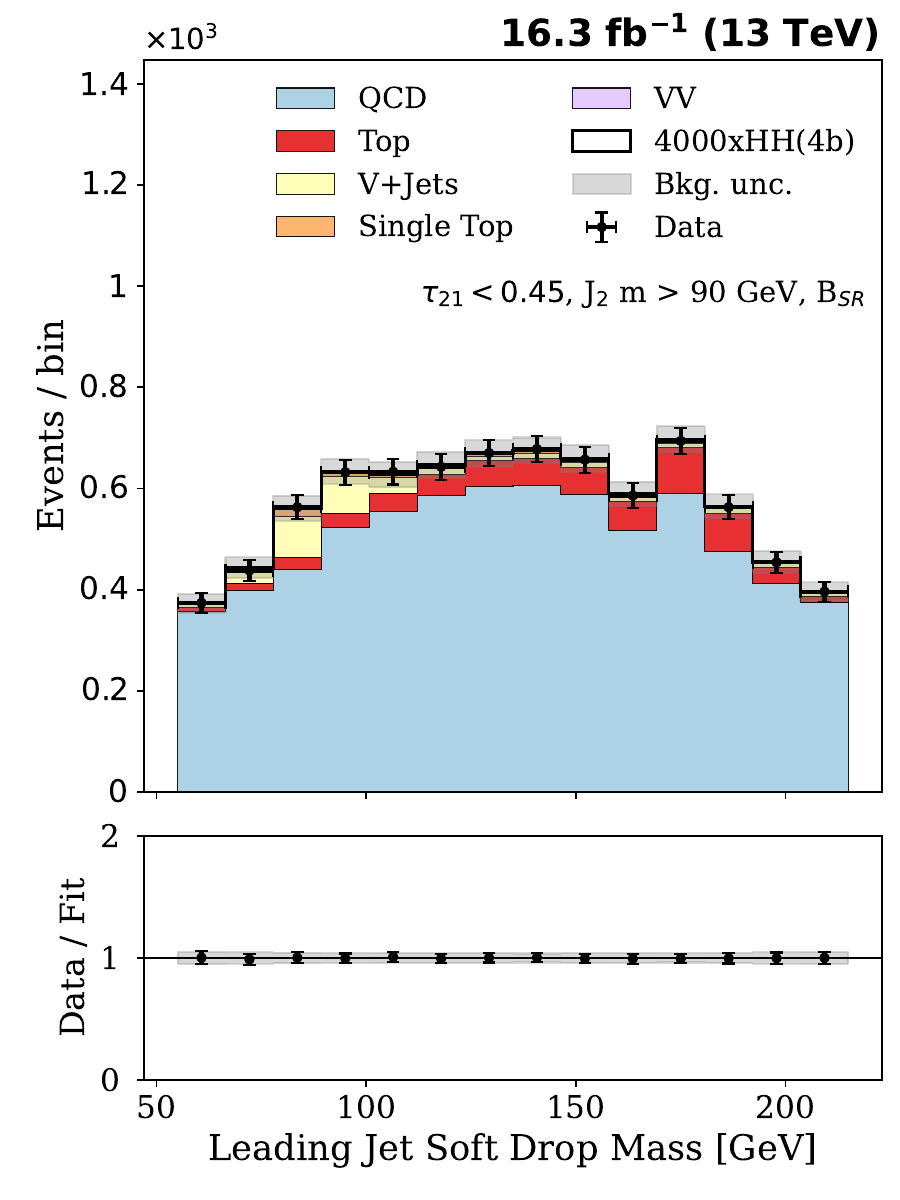}
        \includegraphics[width=.31\textwidth]{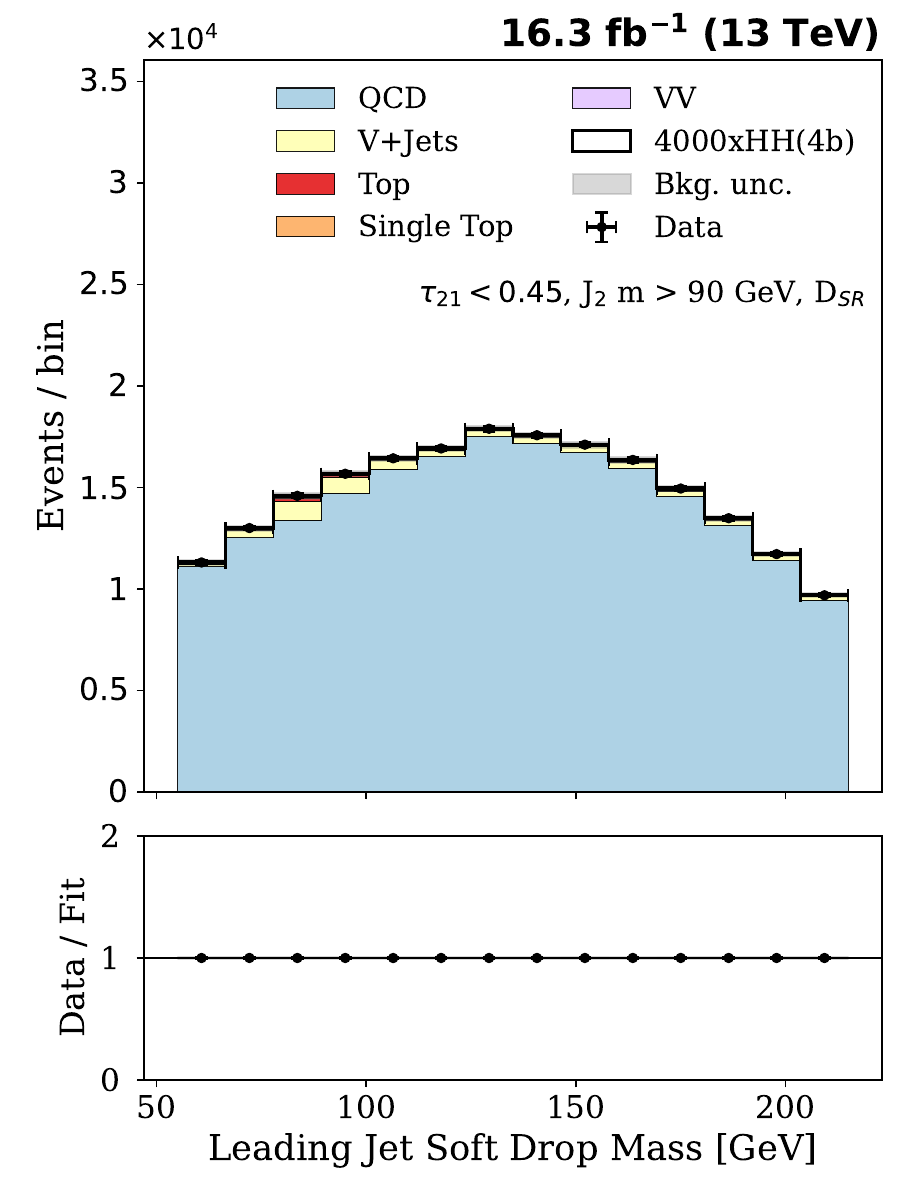}
        \includegraphics[width=.31\textwidth]{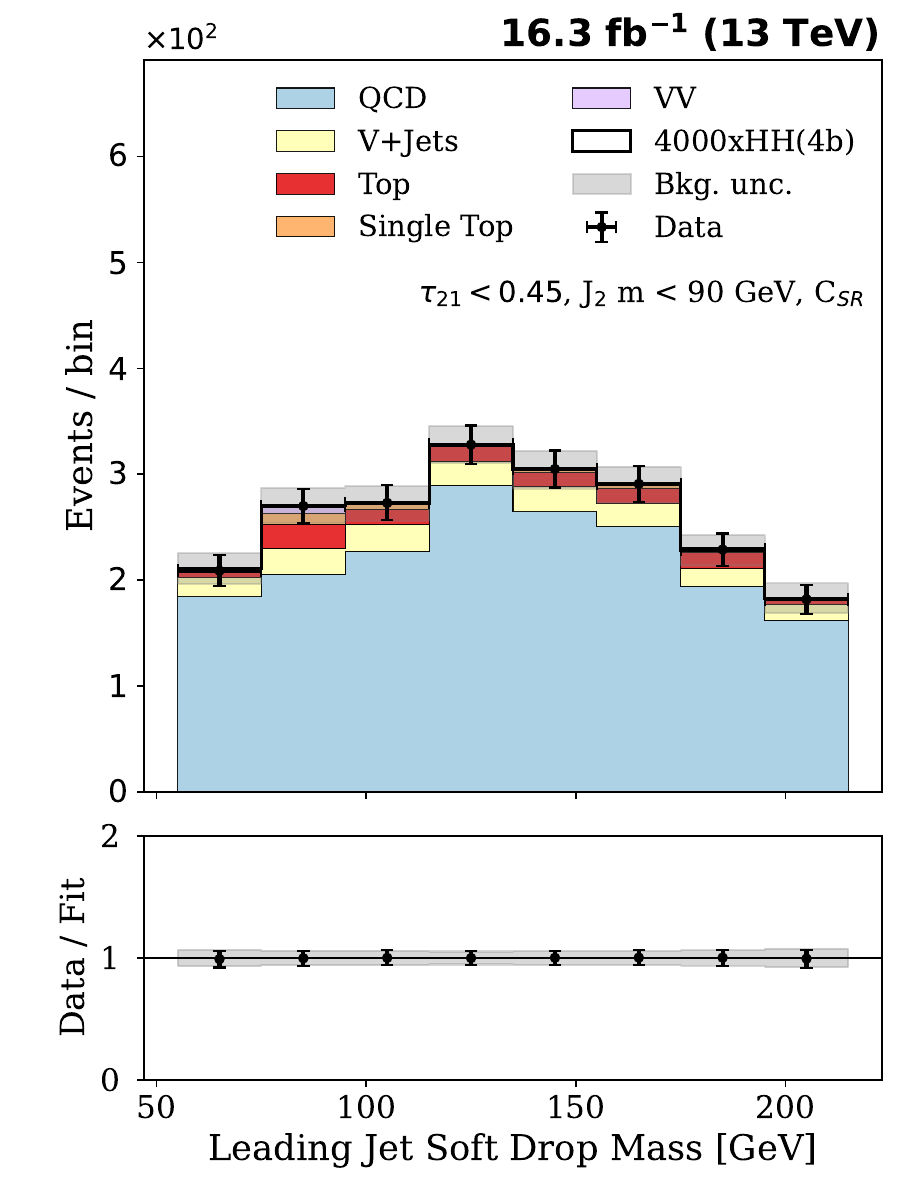}
        \includegraphics[width=.31\textwidth]{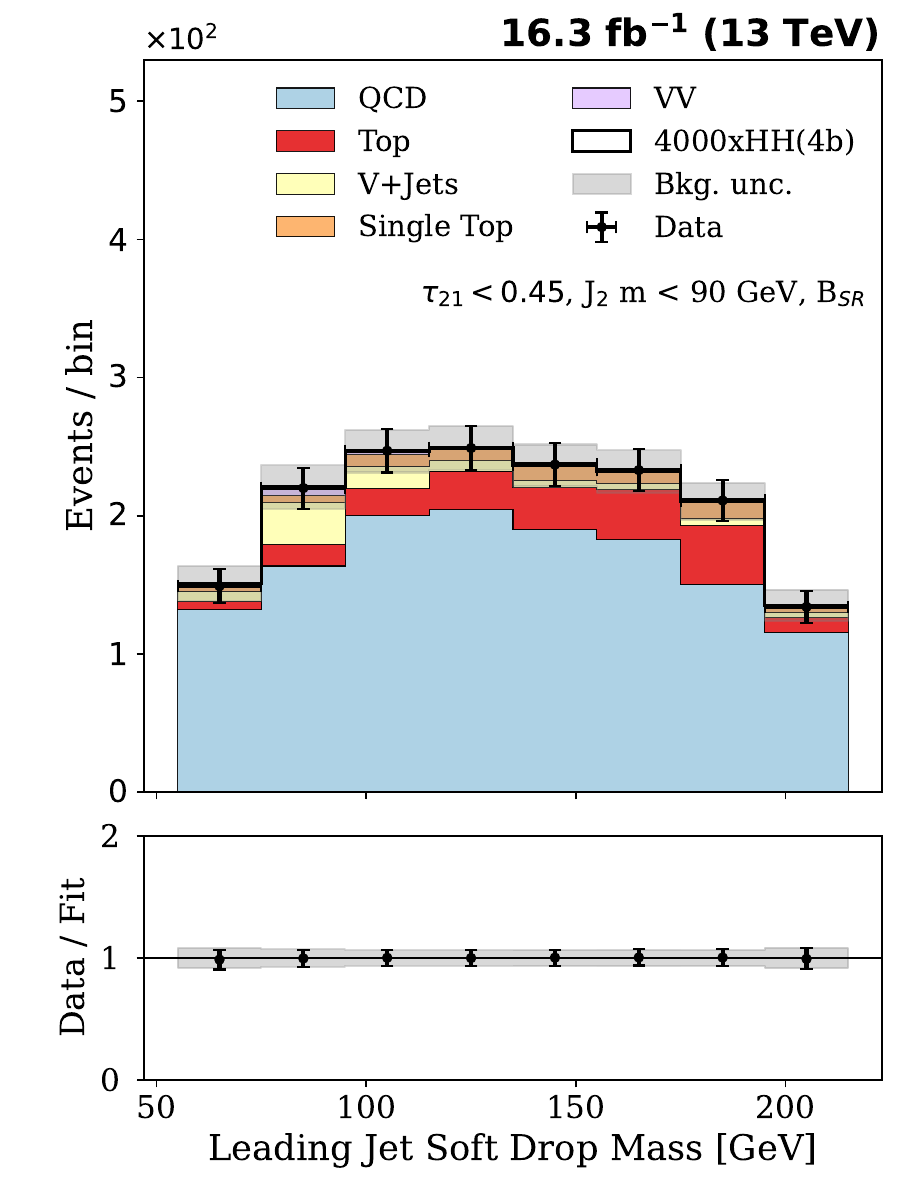}
        \includegraphics[width=.31\textwidth]{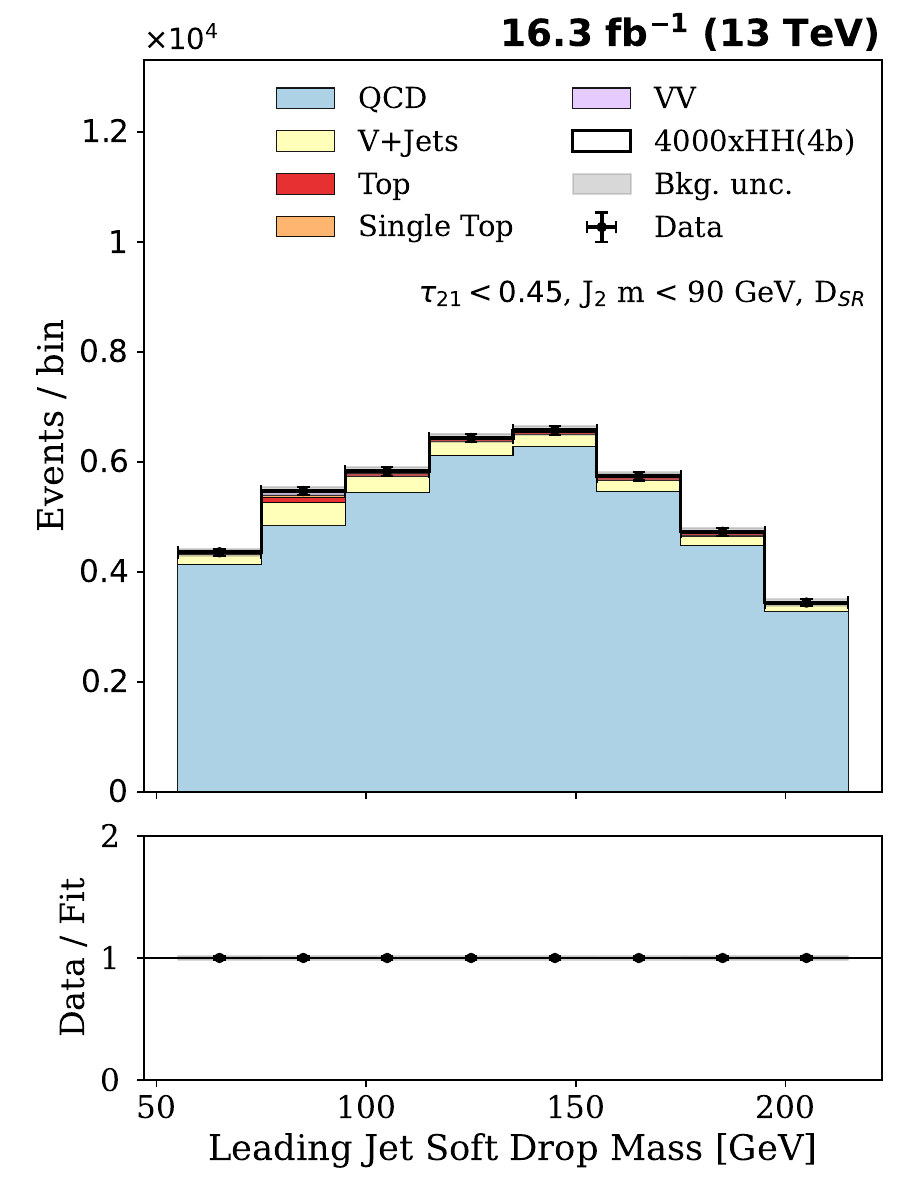}
        \includegraphics[width=.31\textwidth]{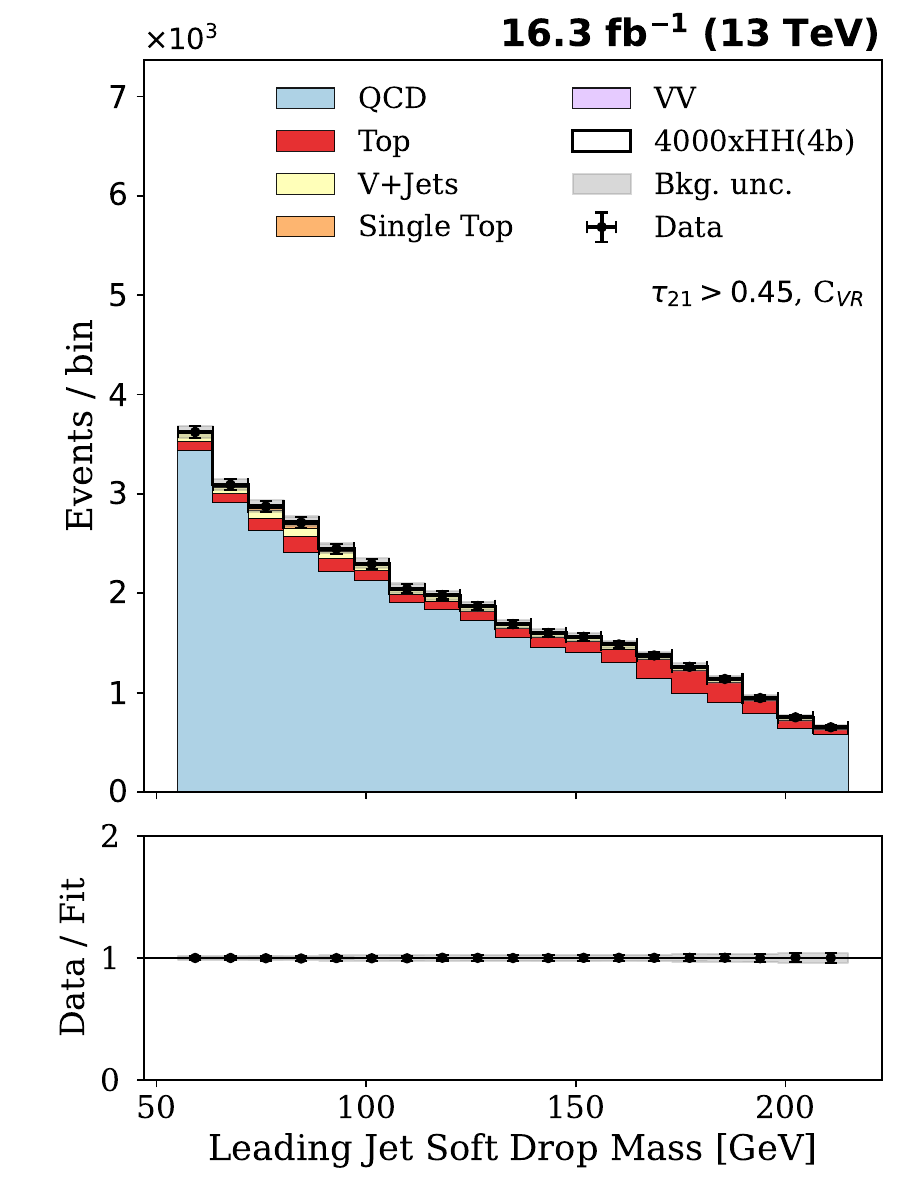}
        \includegraphics[width=.31\textwidth]{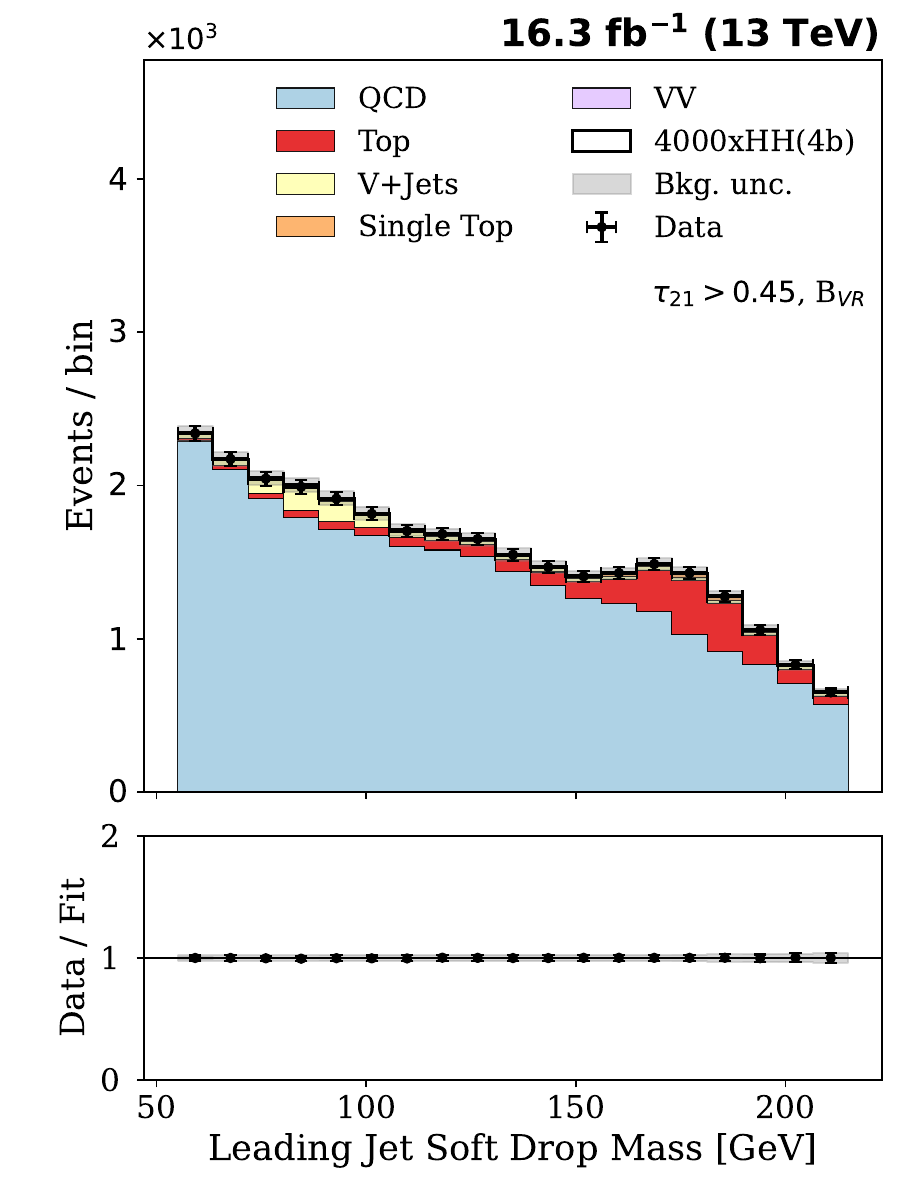}
        \includegraphics[width=.31\textwidth]{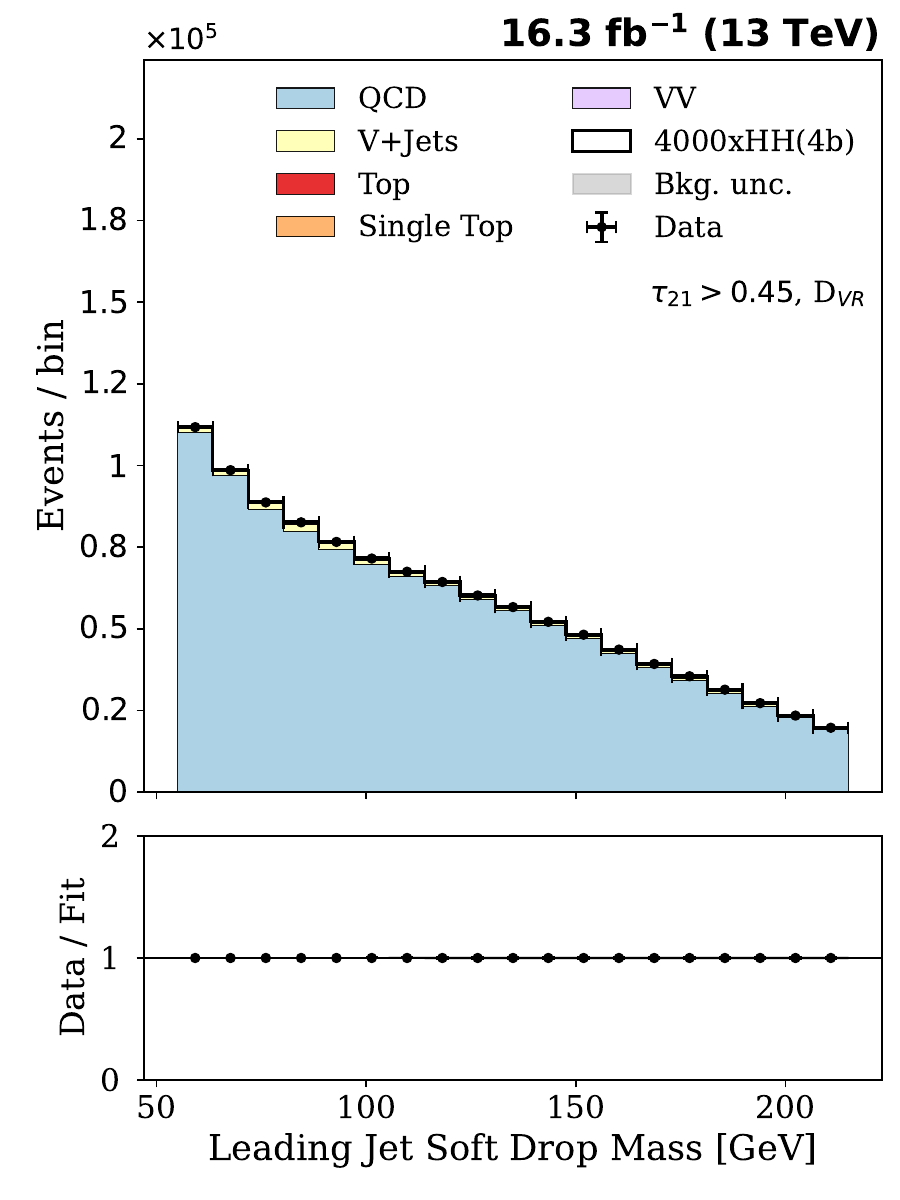}
        
    \caption{Leading jet soft drop mass where both jets are considered anomalous based on the X(bb) tagger score. The region where both jets have low $\tau_{21}$ values and subleading jet mass is above 90 GeV  is shown at the top while the region where the subleading jet mass is below 90 GeV is shown at the middle. The region where at least one jet fails the $\tau_{21}$ selection is shown at the bottom. The different regions used for the ABCD calculation are shown as columns. Shaded regions represent the total background uncertainty.}
    \label{fig:hh_htag}
\end{figure*}

Results replacing the \textsc{OmniLearned} anomaly score by the X(bb) tagger are shown in Fig.~\ref{fig:hh_htag}.

\subsection{Dijet Invariant Mass}
\begin{figure*}[ht]
    \centering
        \includegraphics[width=.31\textwidth]{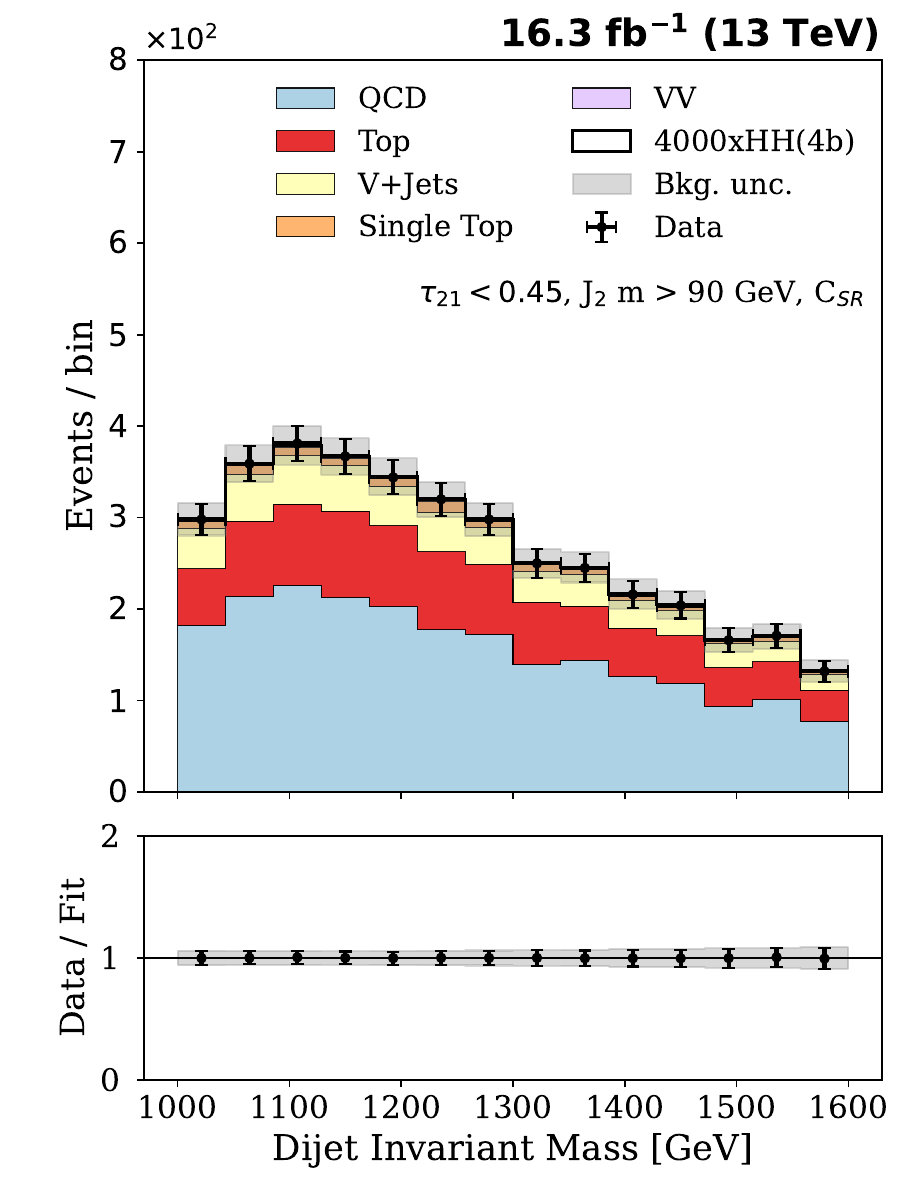}
        \includegraphics[width=.31\textwidth]{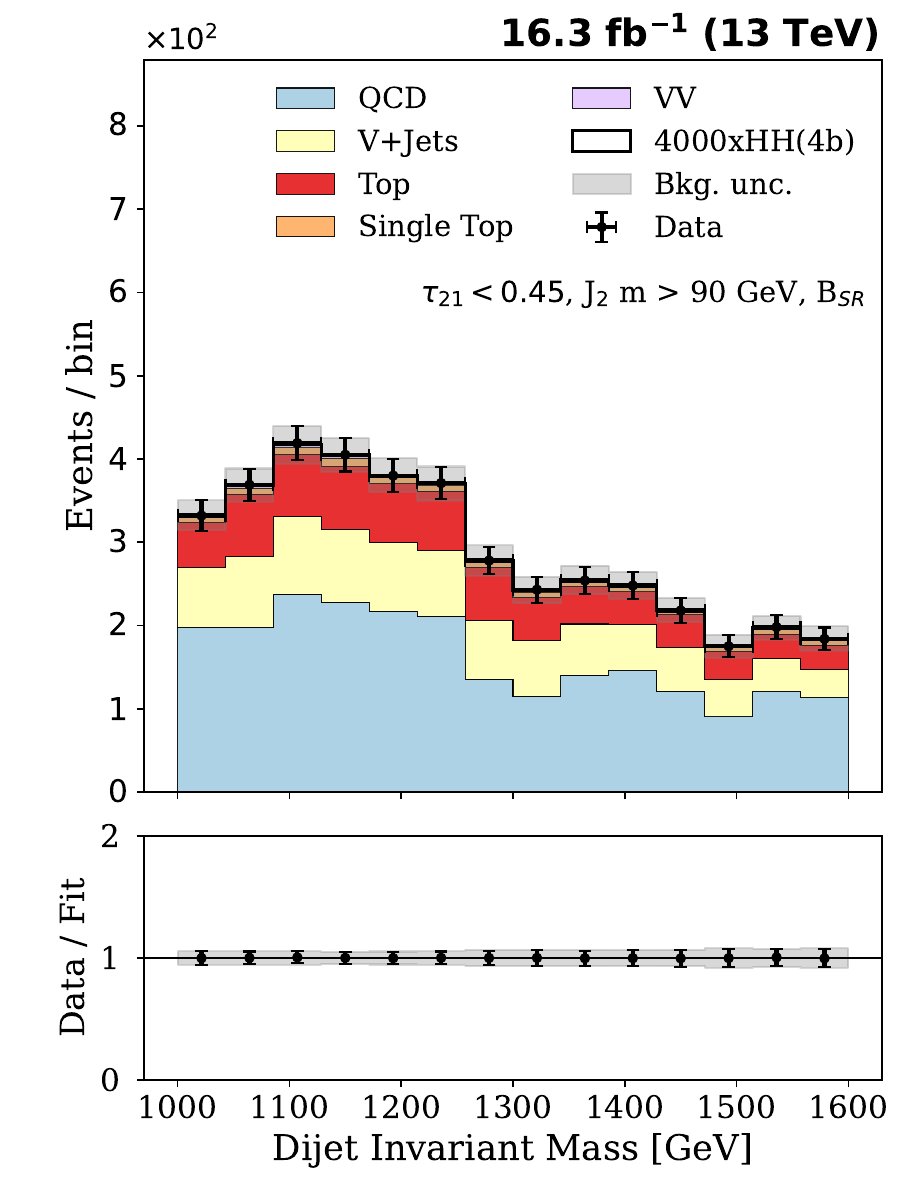}
        \includegraphics[width=.31\textwidth]{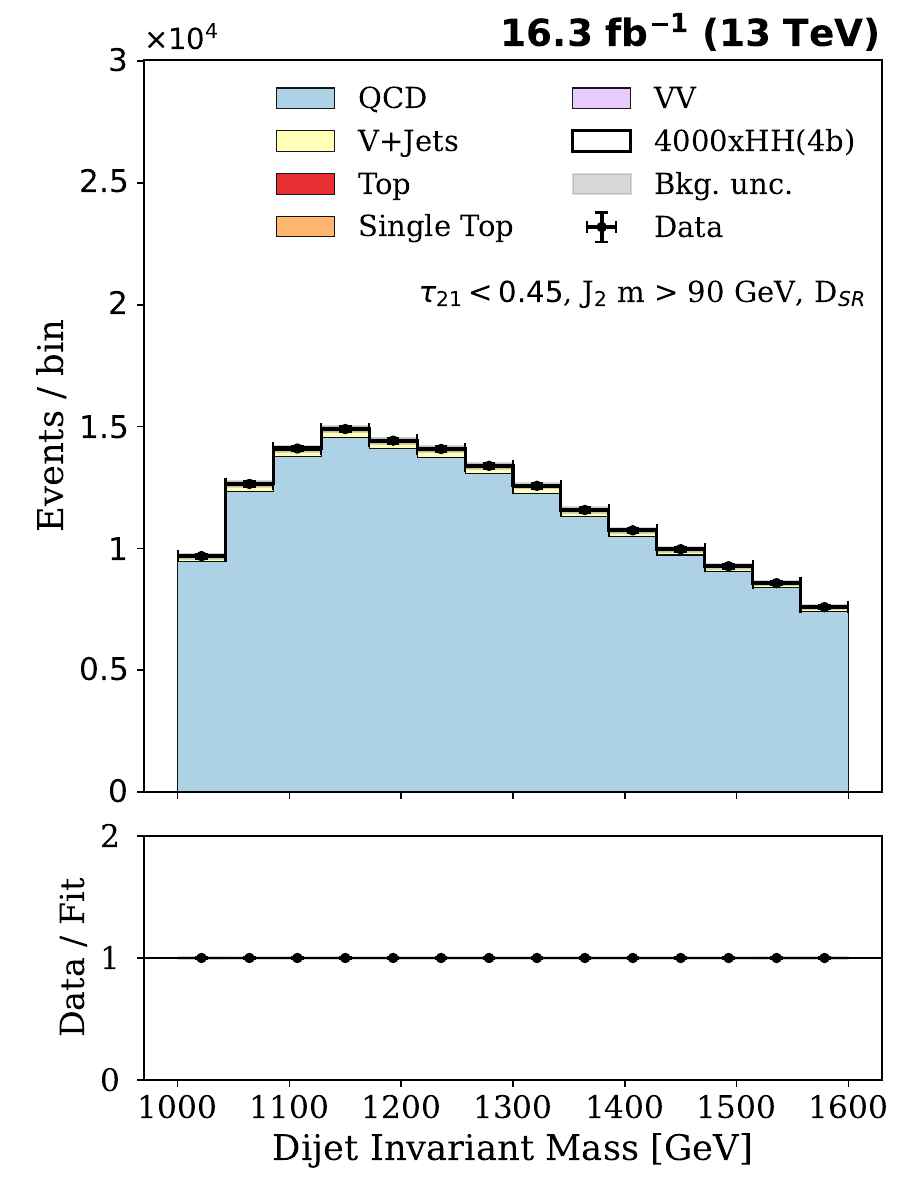}
        \includegraphics[width=.31\textwidth]{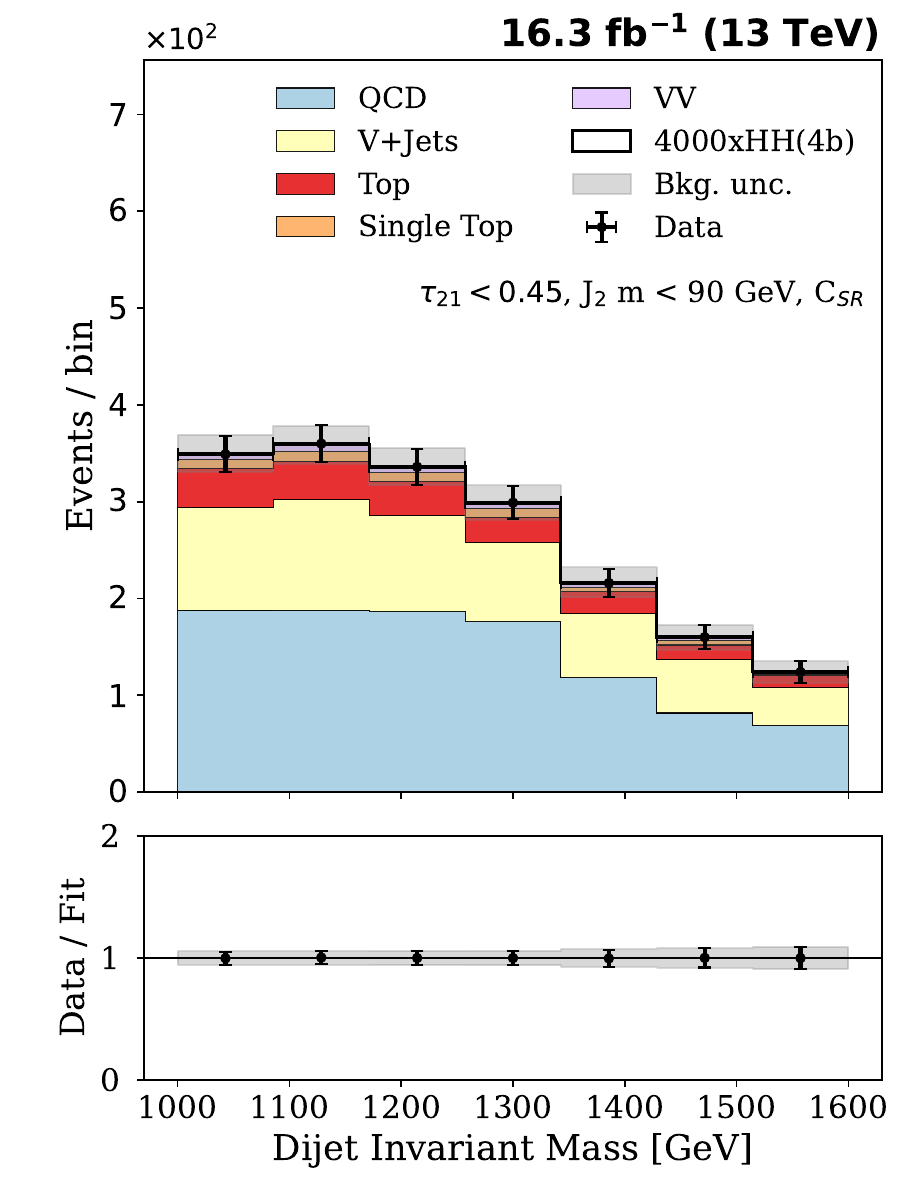}
        \includegraphics[width=.31\textwidth]{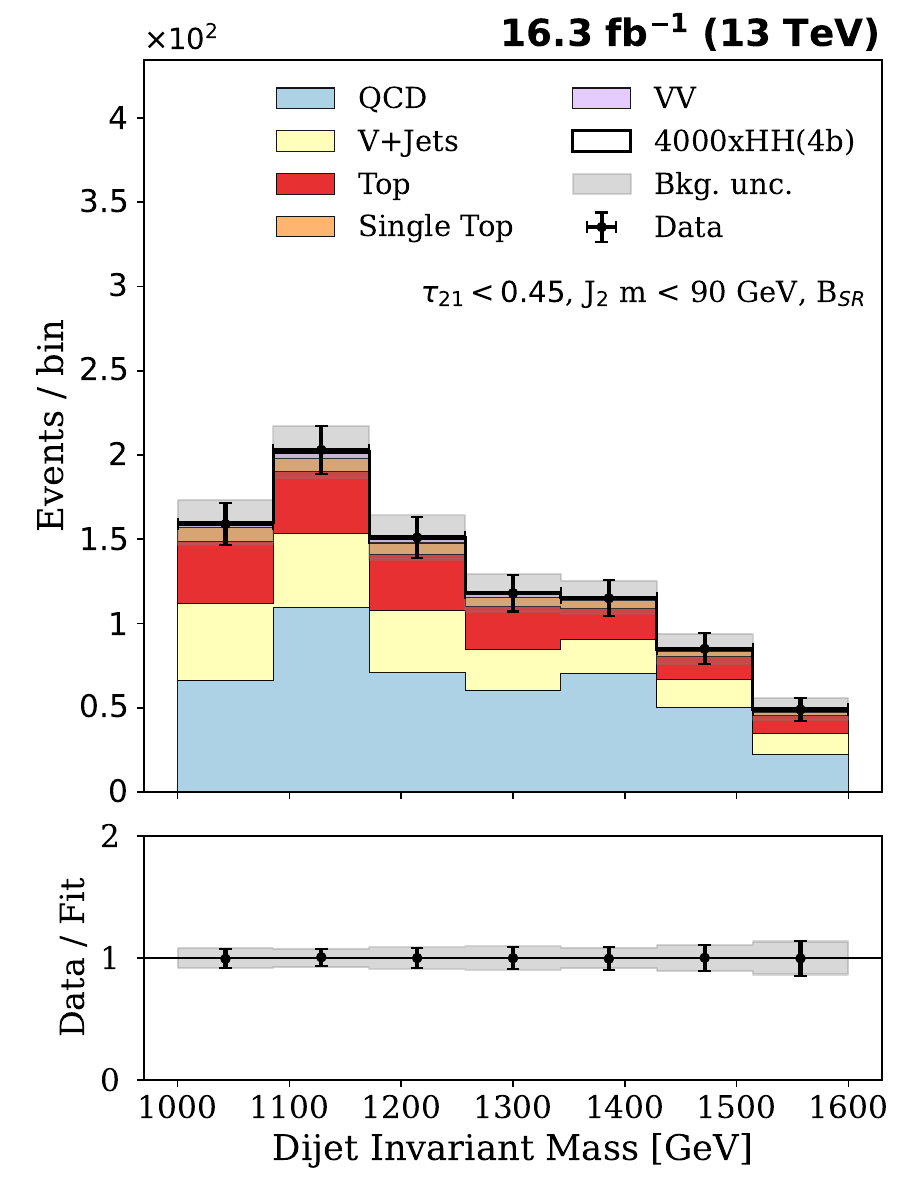}
        \includegraphics[width=.31\textwidth]{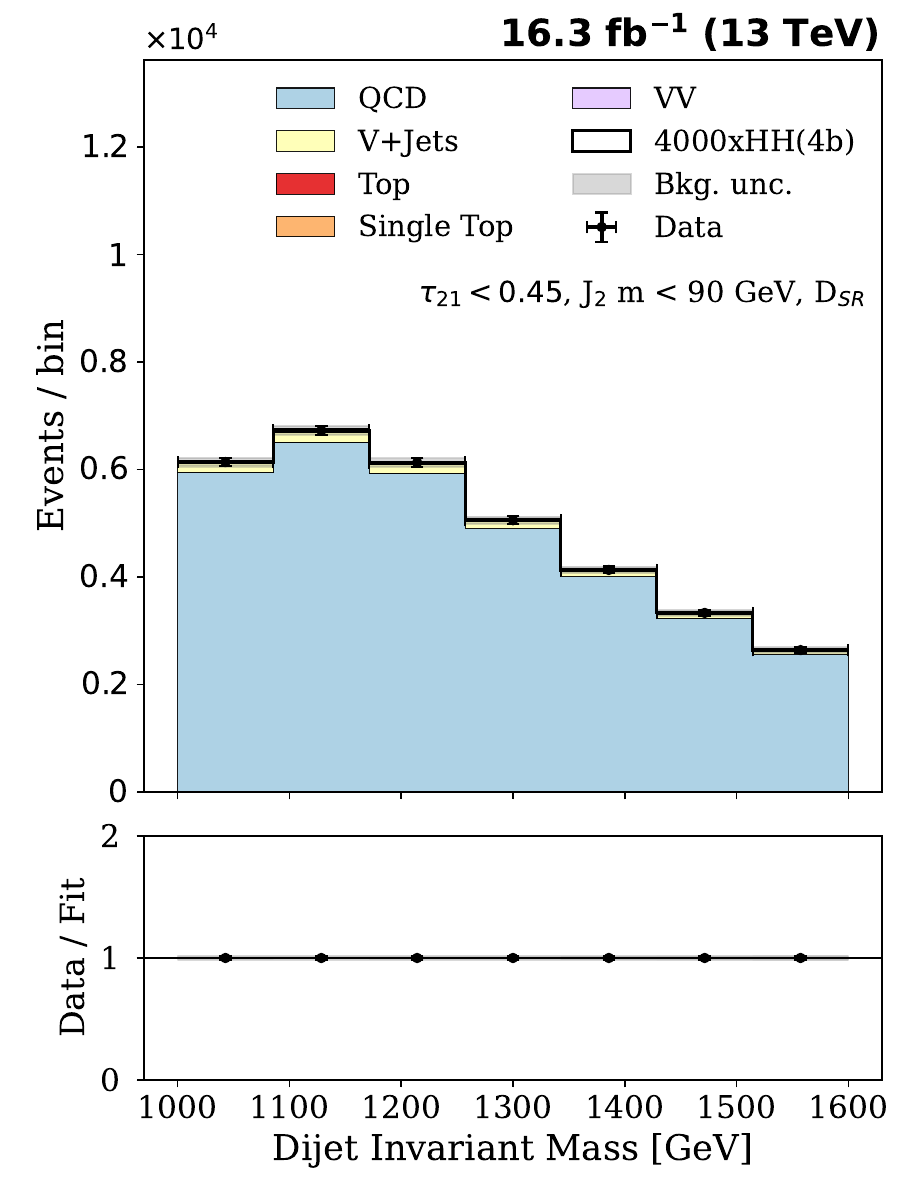}
        \includegraphics[width=.31\textwidth]{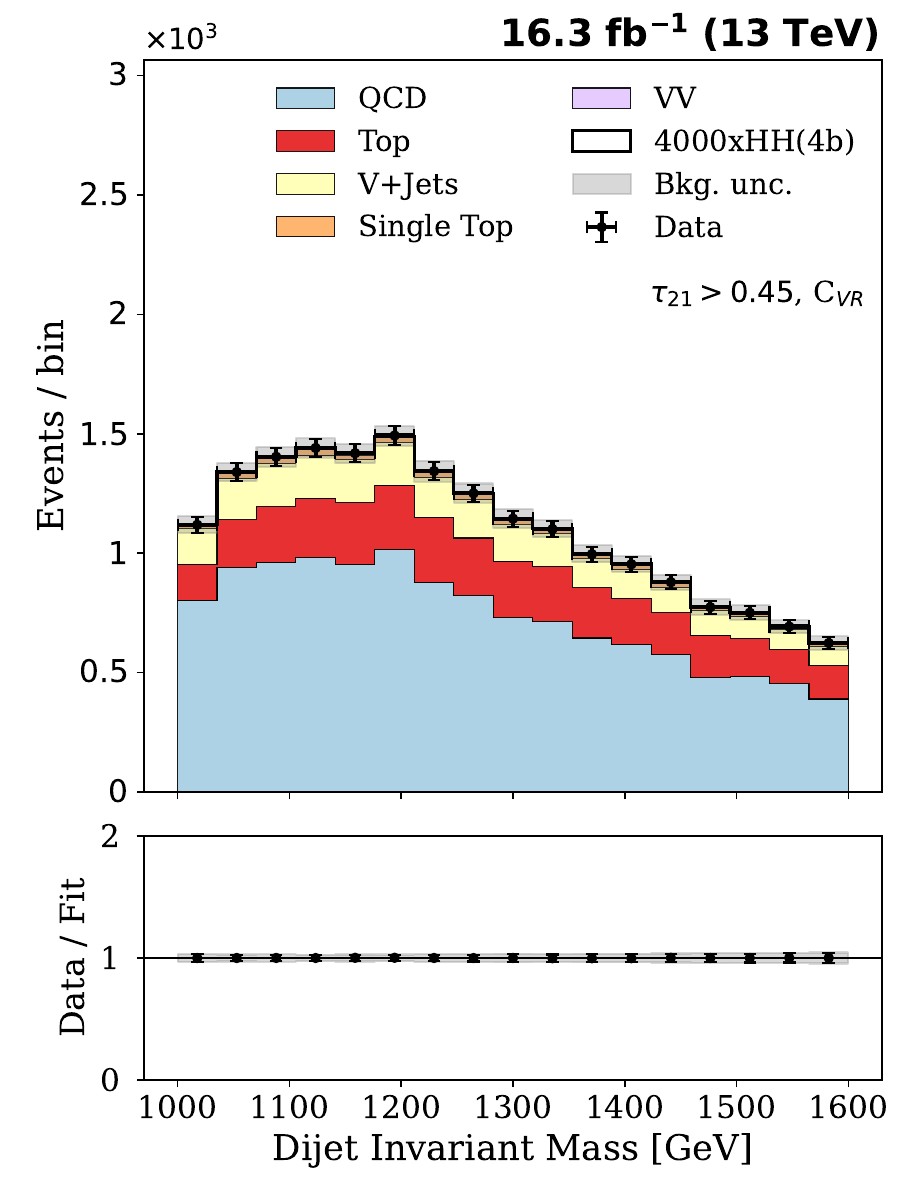}
        \includegraphics[width=.31\textwidth]{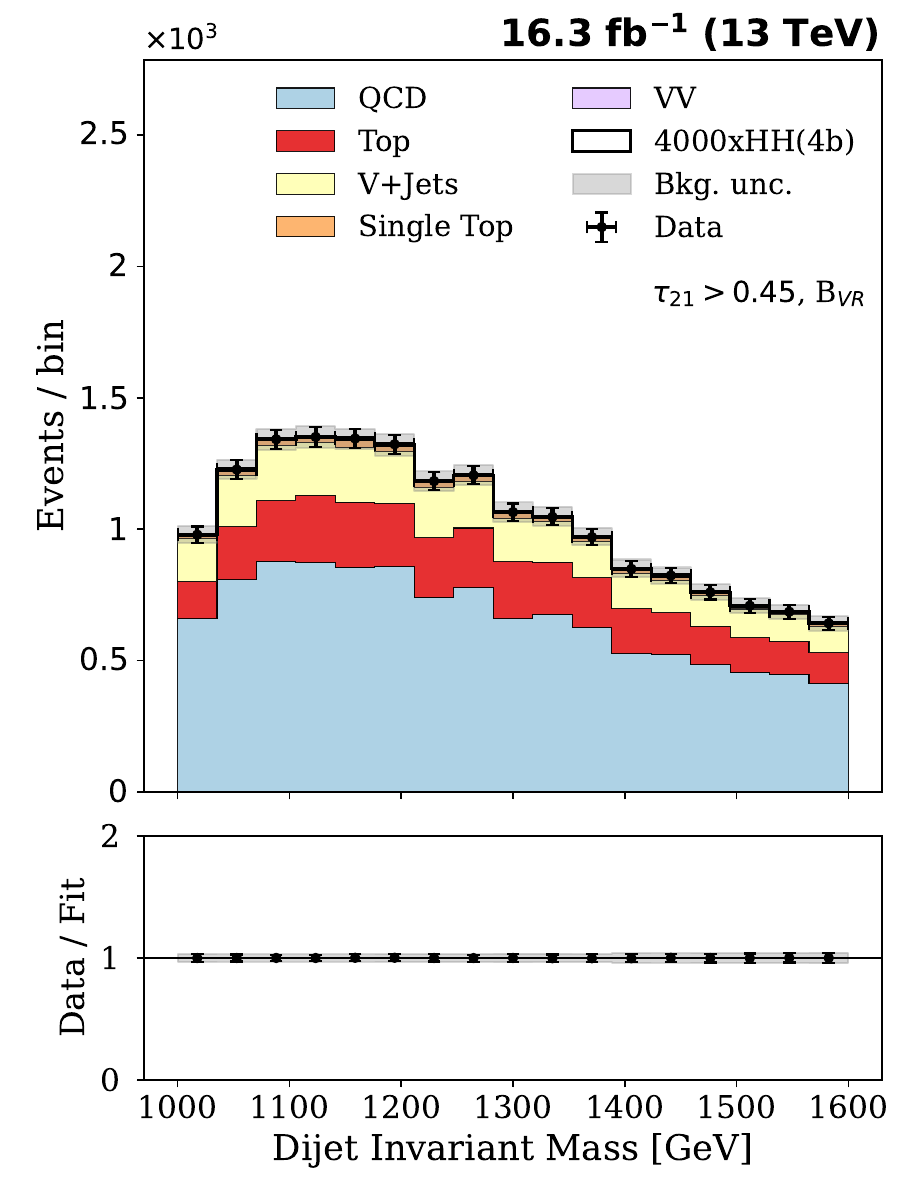}
        \includegraphics[width=.31\textwidth]{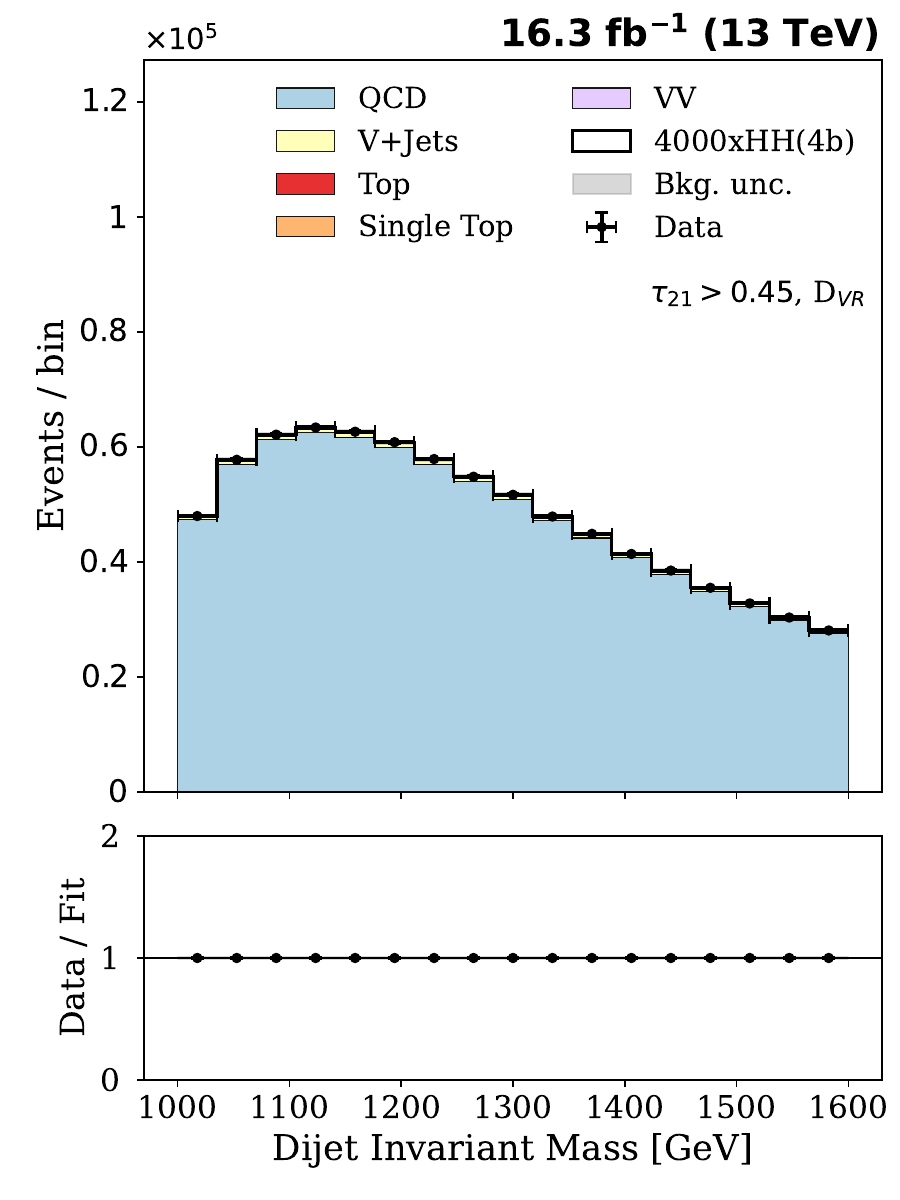}
        
    \caption{Distribution of the dijet invariant mass where both jets are considered anomalous based on the \textsc{OmniLearned} model score. The region where both jets have low $\tau_{21}$ values and subleading jet mass is above 90 GeV  is shown at the top while the region where the subleading jet mass is below 90 GeV is shown at the middle. The region where at least one jet fails the $\tau_{21}$ selection is shown at the bottom. The different regions used for the ABCD calculation are shown as columns. Shaded regions represent the total background uncertainty.}
    \label{fig:hh_dimass}
\end{figure*}

Results using the \textsc{OmniLearned} anomaly to fit the dijet invariant mass are shown in Fig.~\ref{fig:hh_dimass}.

\section{Additional Validation Results}
\label{app:val}
In this section we provide additional studies carried out to validate the analysis procedure described in this document and to investigate the results obtained.

\subsection{Jet Substructure Study}
\begin{figure}[ht]
    \centering
        \includegraphics[width=.23\textwidth]{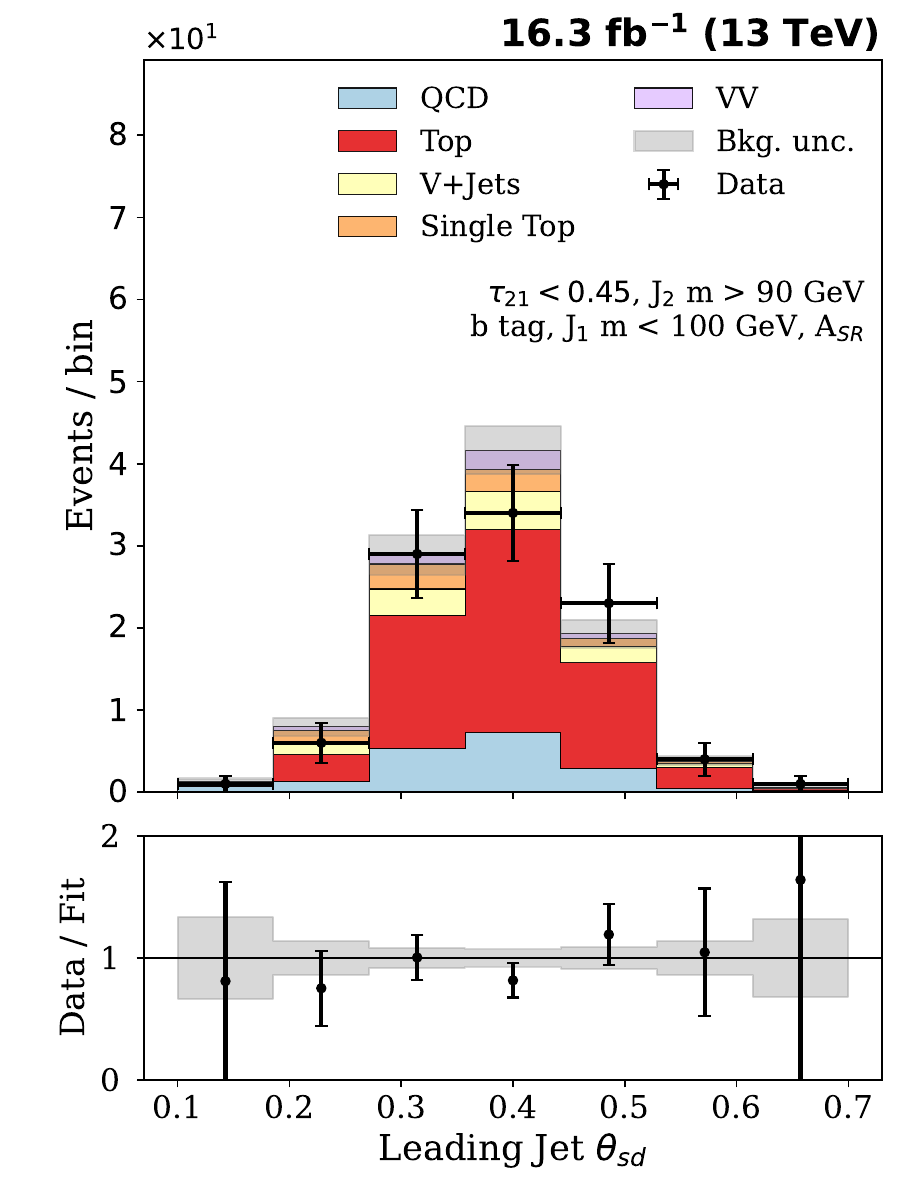}
        \includegraphics[width=.23\textwidth]{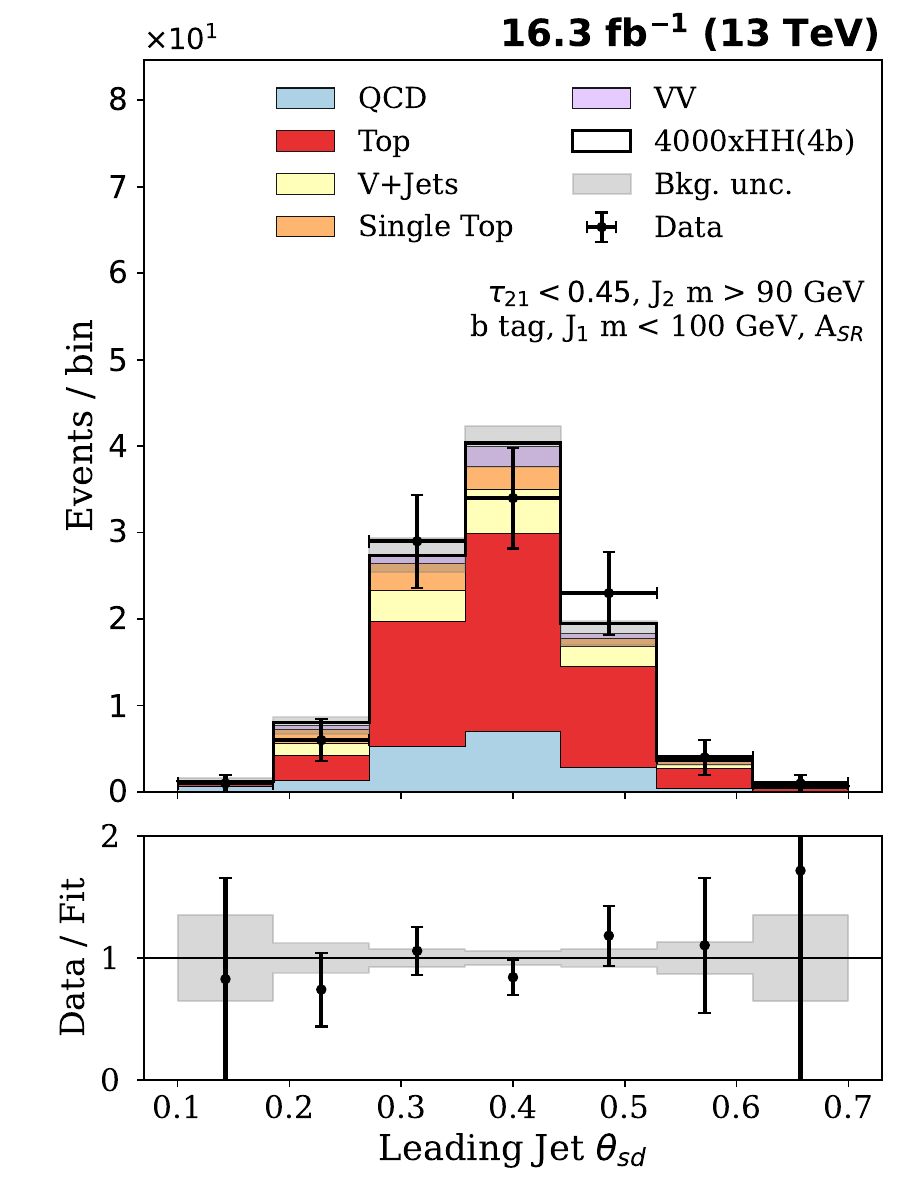}
        \includegraphics[width=.23\textwidth]{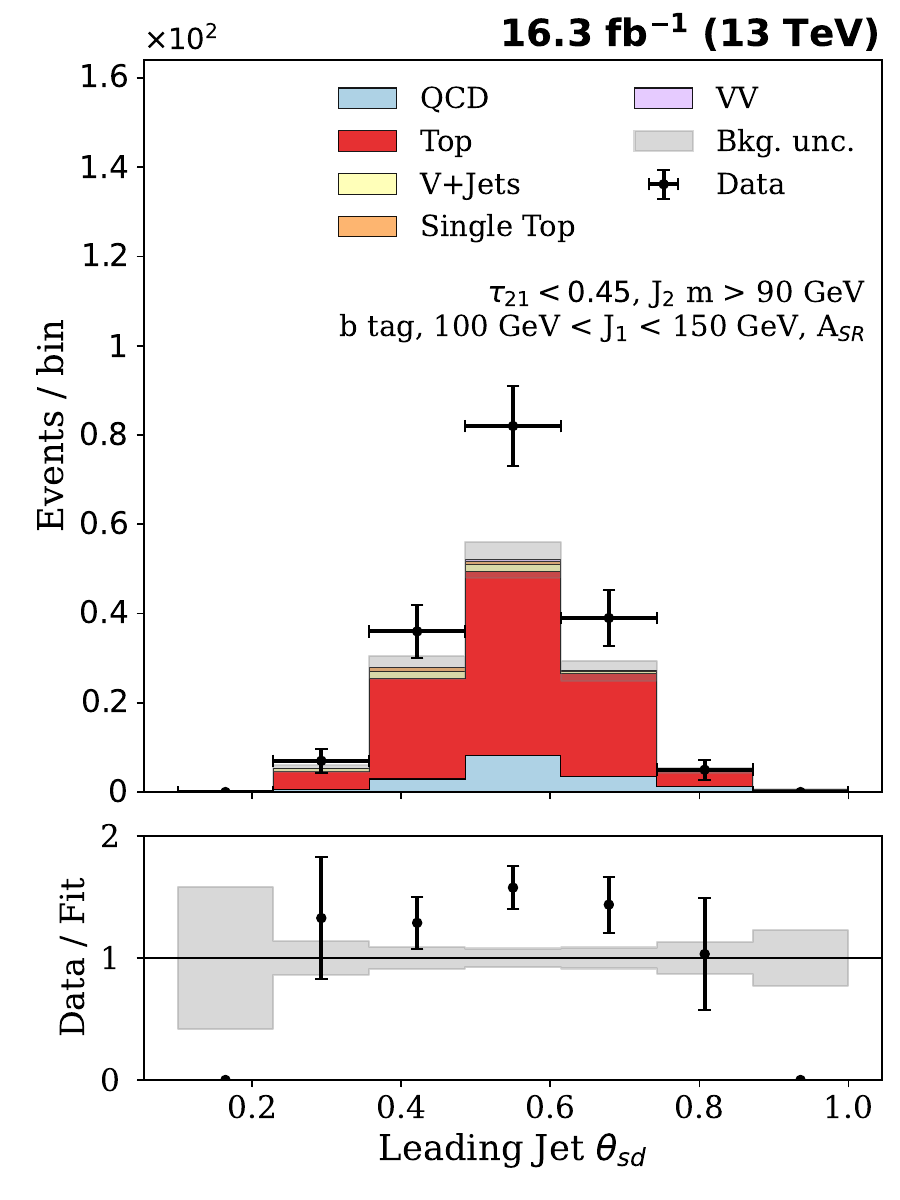}
        \includegraphics[width=.23\textwidth]{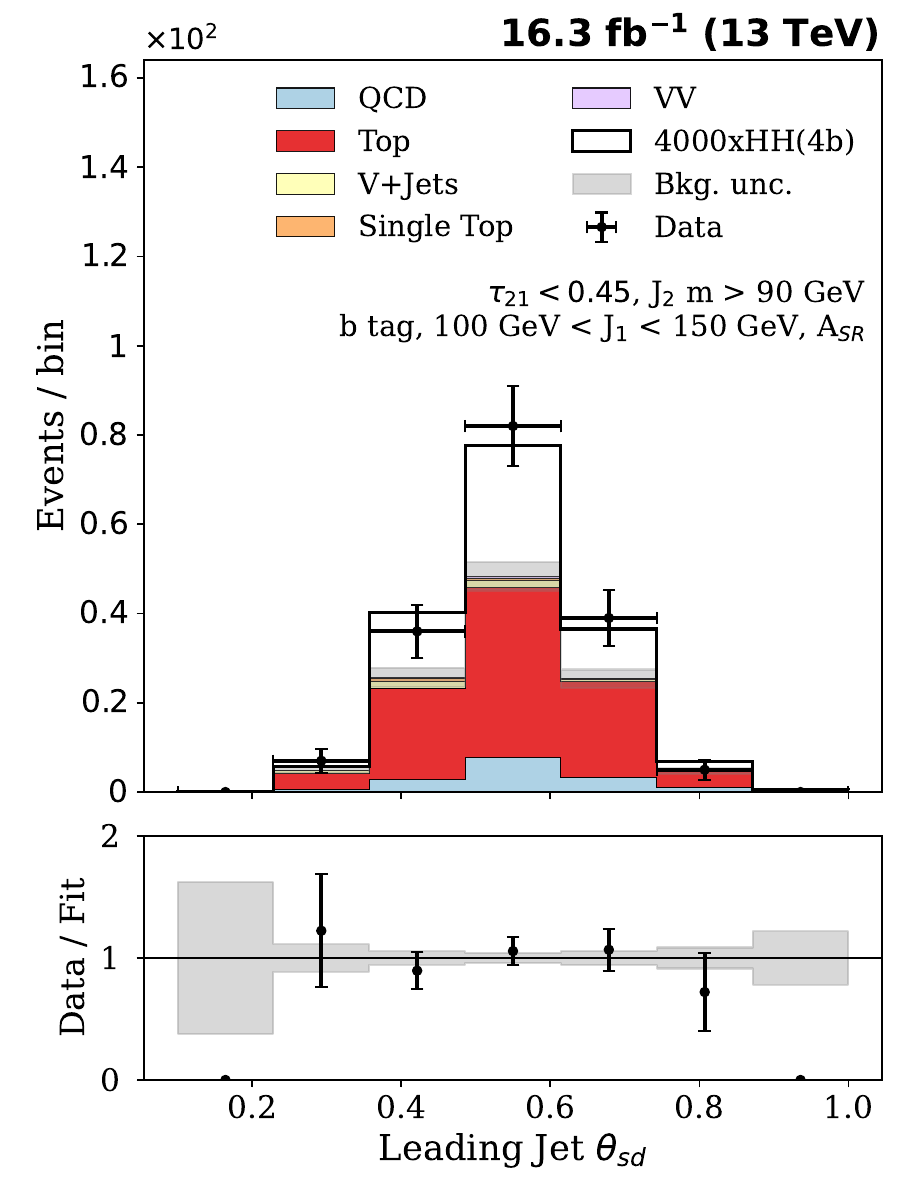}
        \includegraphics[width=.23\textwidth]{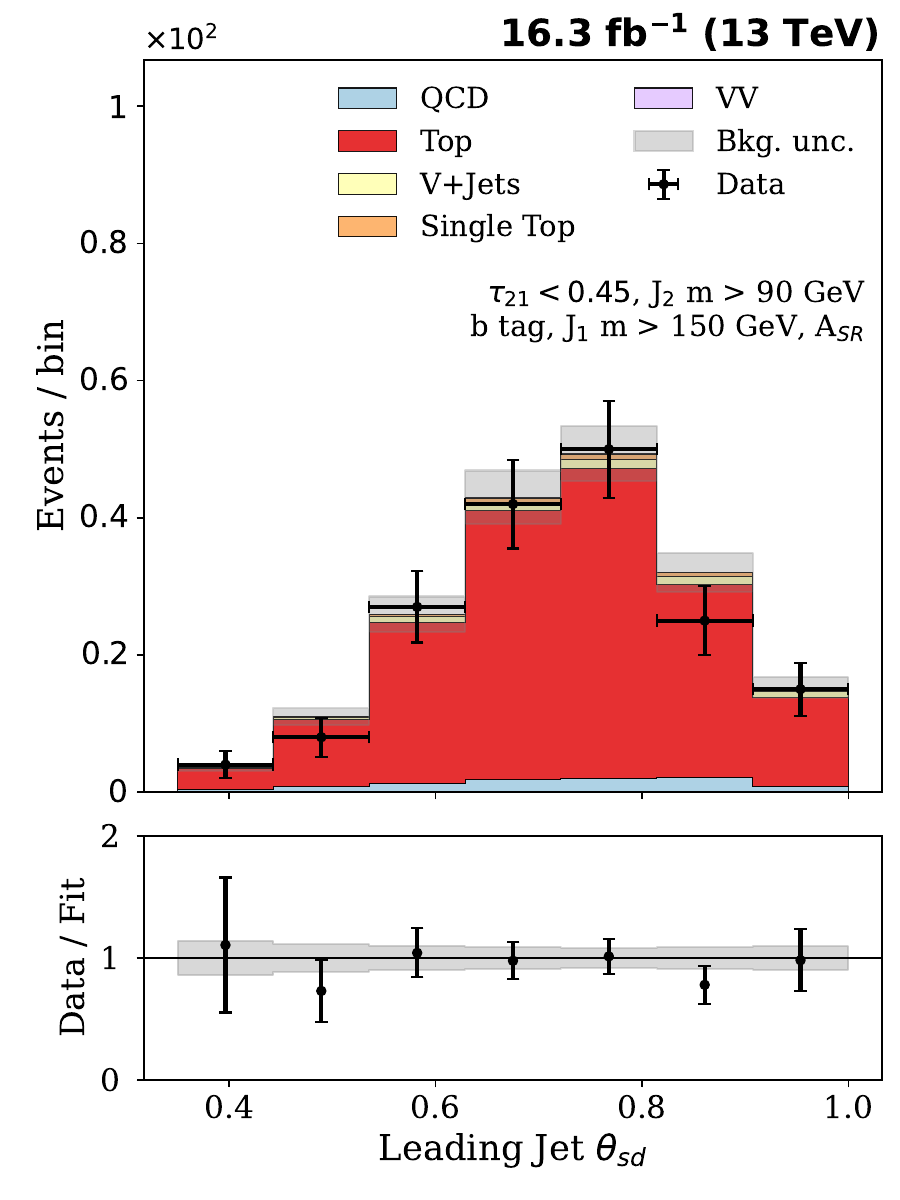}                
        \includegraphics[width=.23\textwidth]{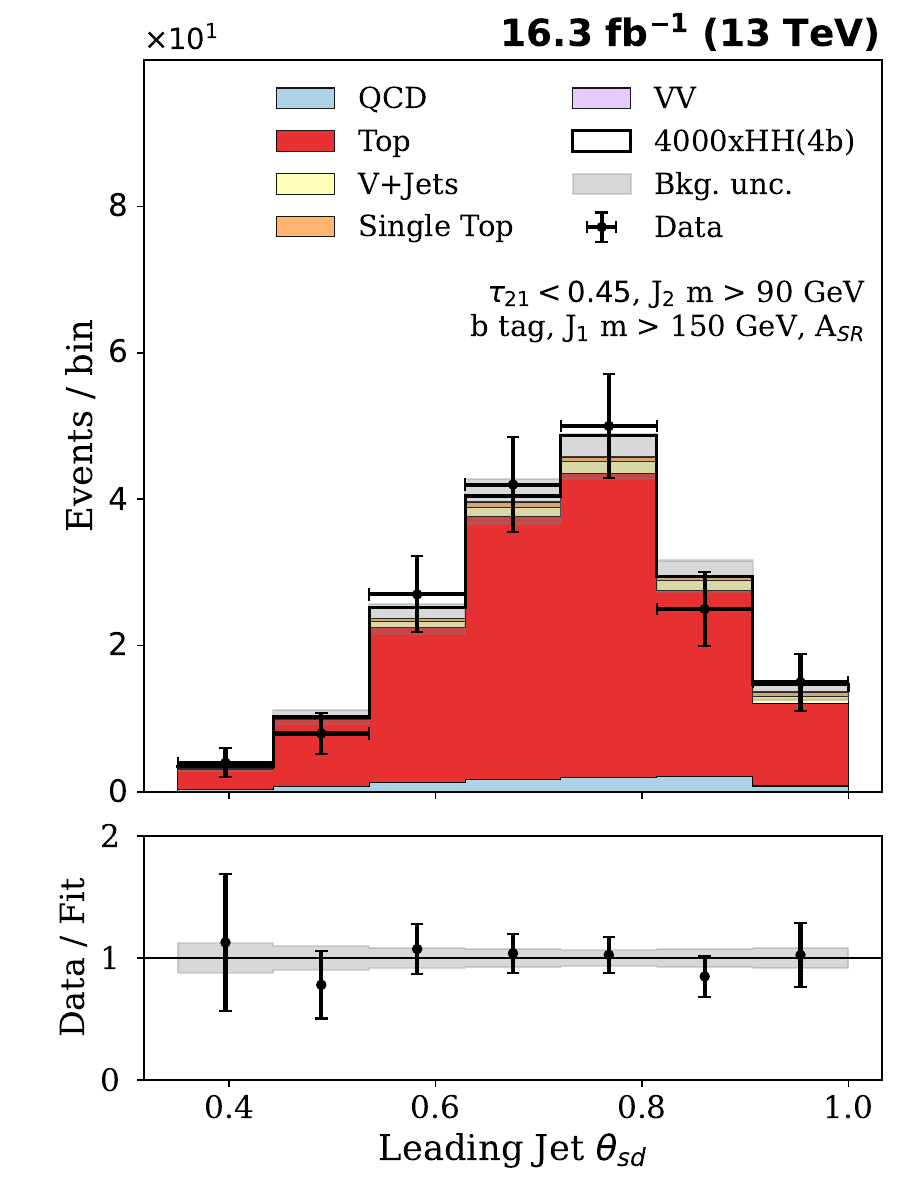}
        
    \caption{Leading jet $\theta$ for events where both jets are considered anomalous based on the \textsc{OmniLearned} large model score. Different mass intervals determined simultaneously in the fit are shown in the different rows while the background only results are on the left and the signal plus background results on the right. Shaded regions represent the total background uncertainty.}
    \label{fig:theta}
\end{figure}

\begin{figure}[ht]
    \centering
        \includegraphics[width=.23\textwidth]{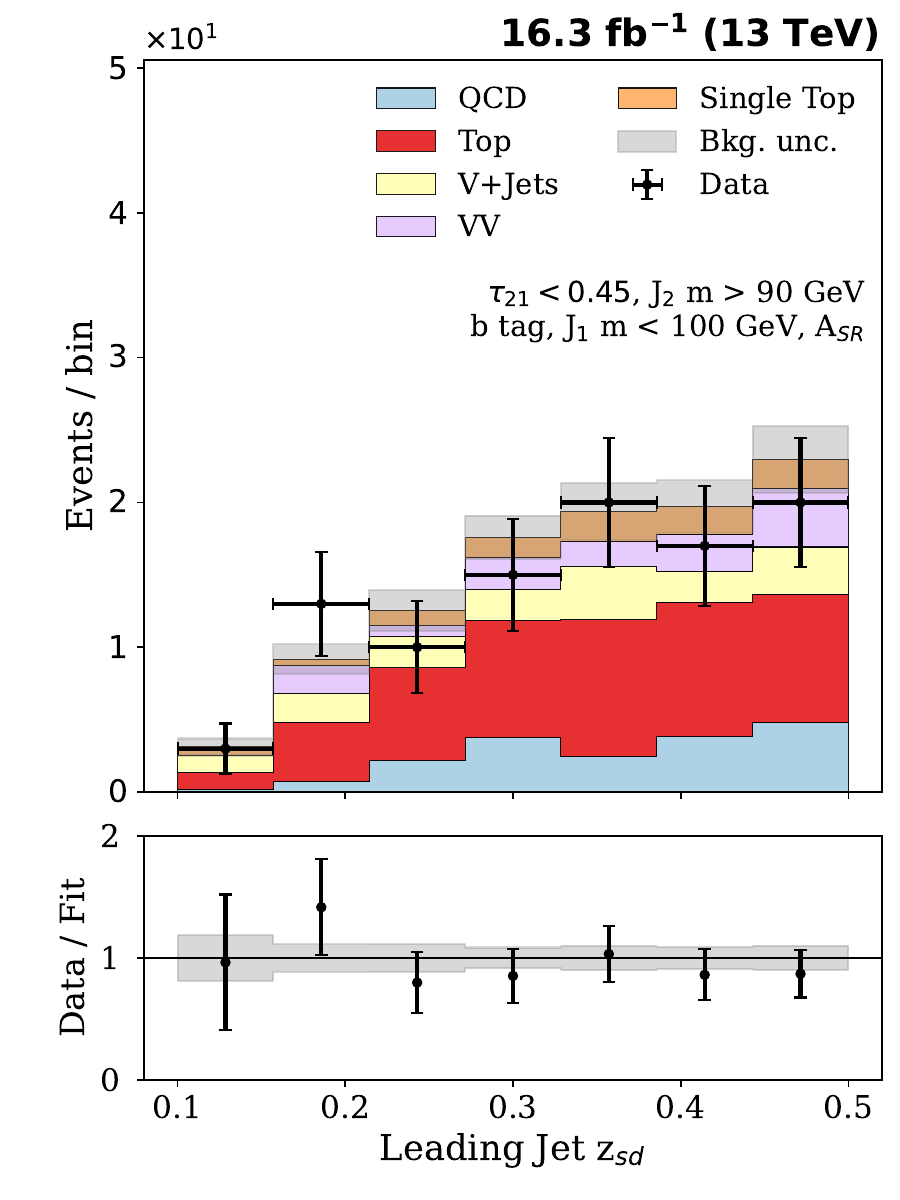}
        \includegraphics[width=.23\textwidth]{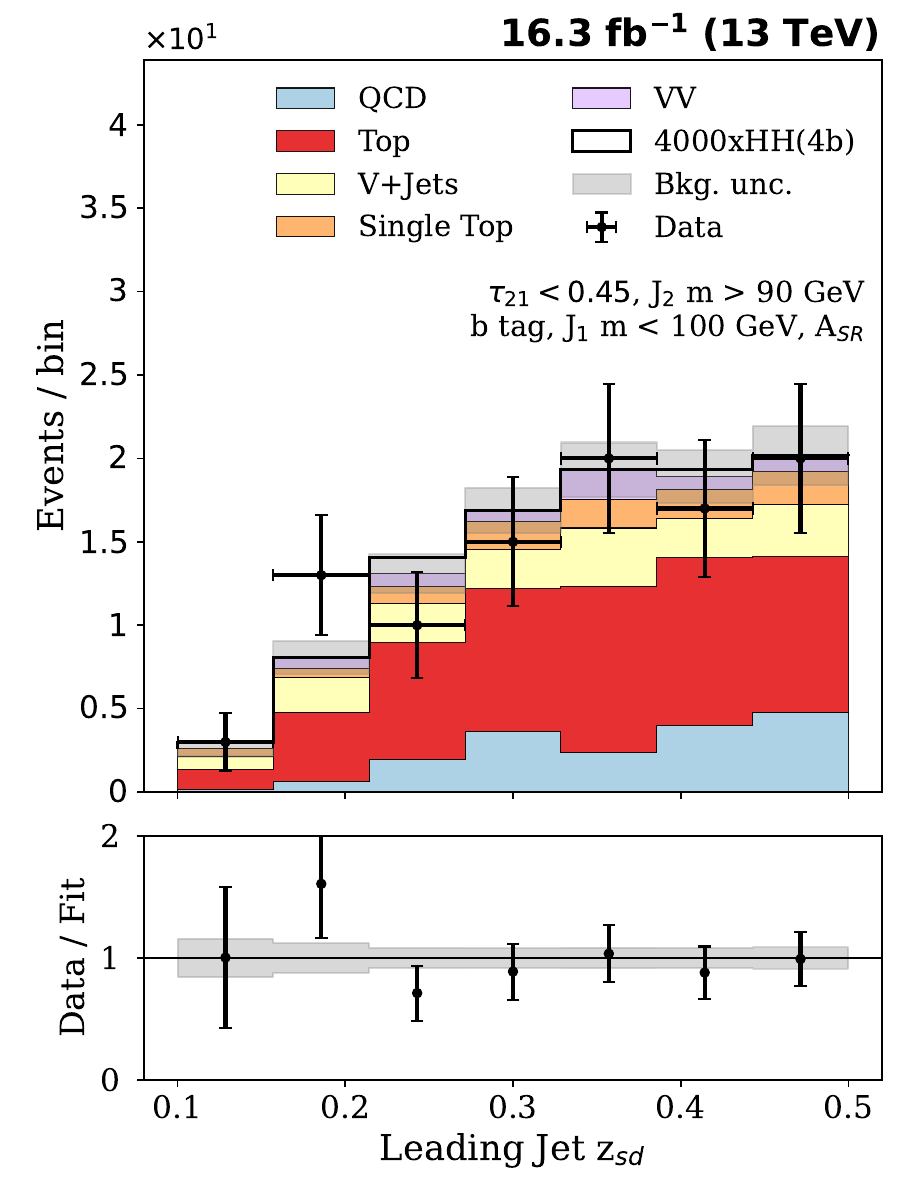}
        \includegraphics[width=.23\textwidth]{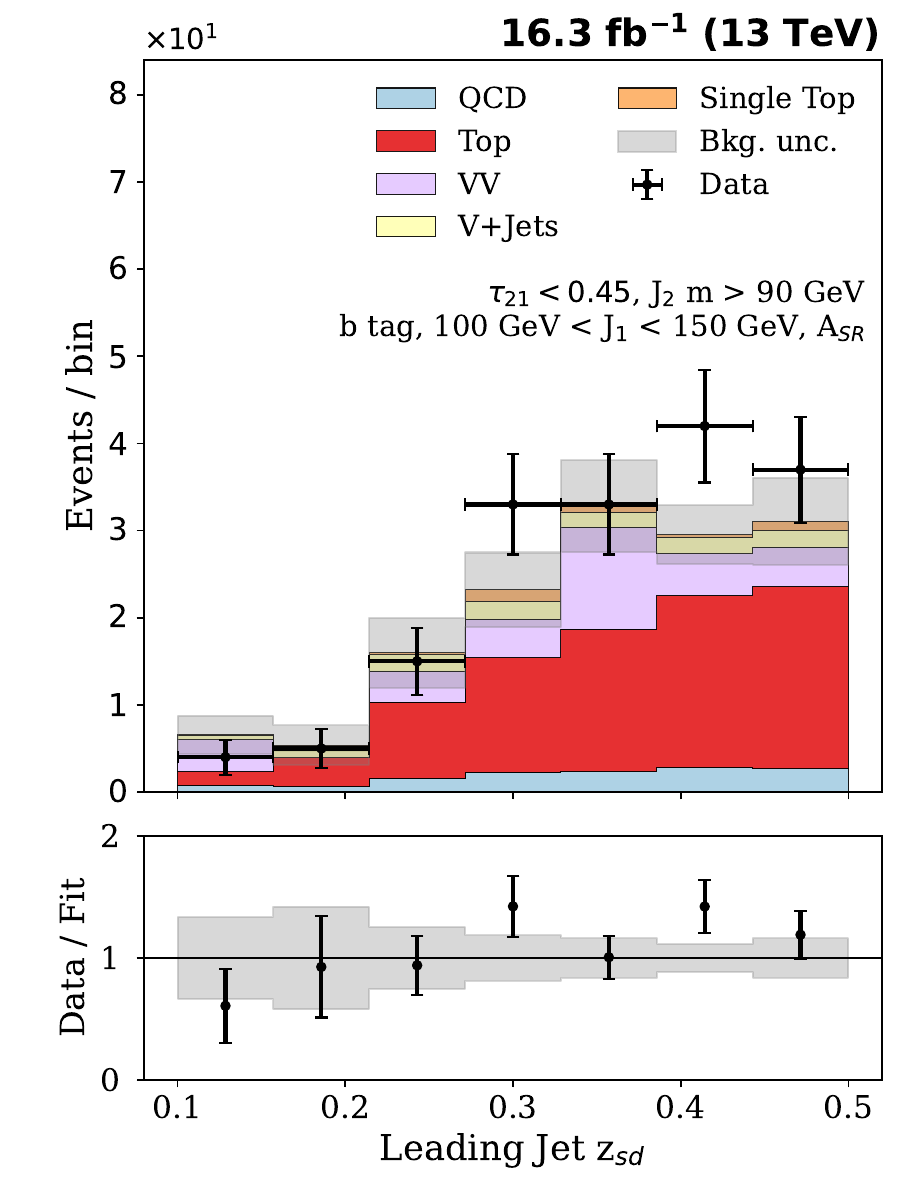}
        \includegraphics[width=.23\textwidth]{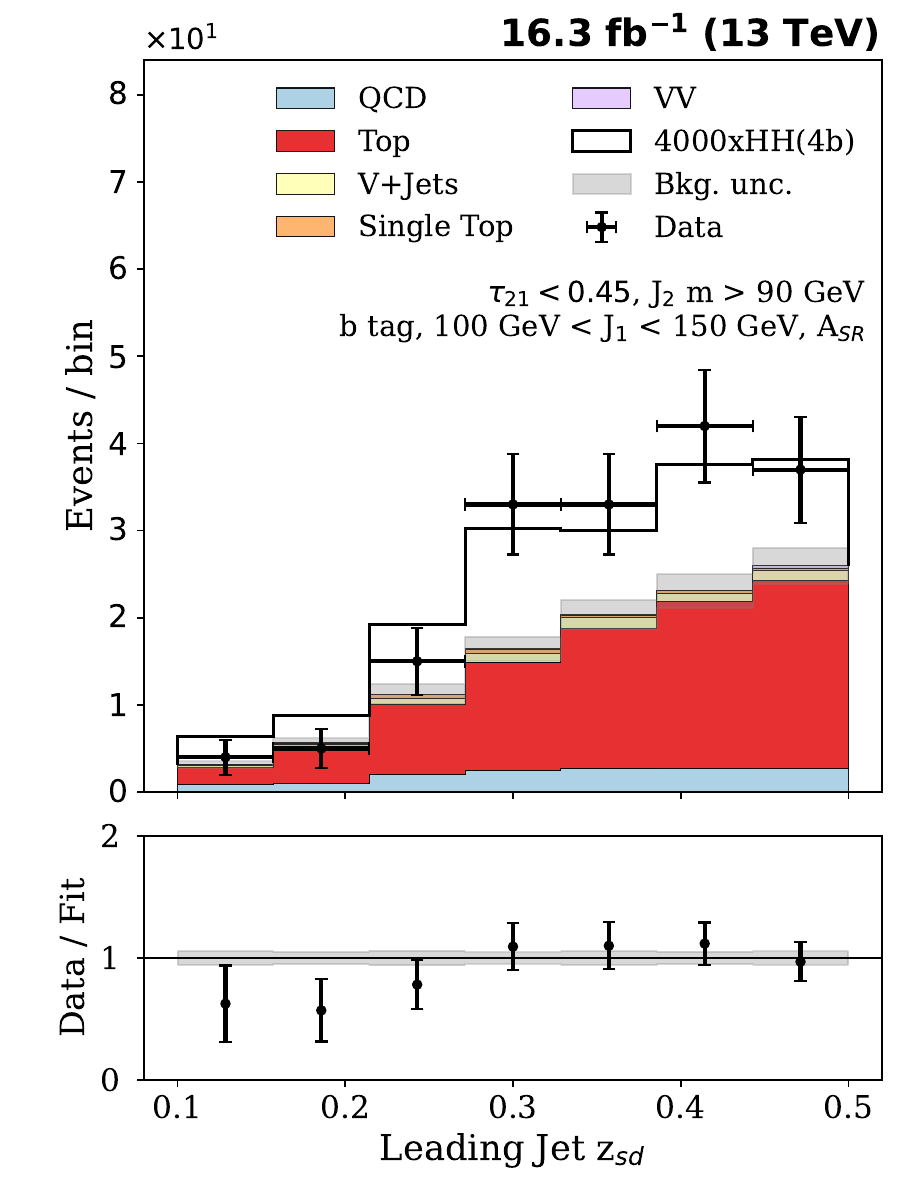}
        \includegraphics[width=.23\textwidth]{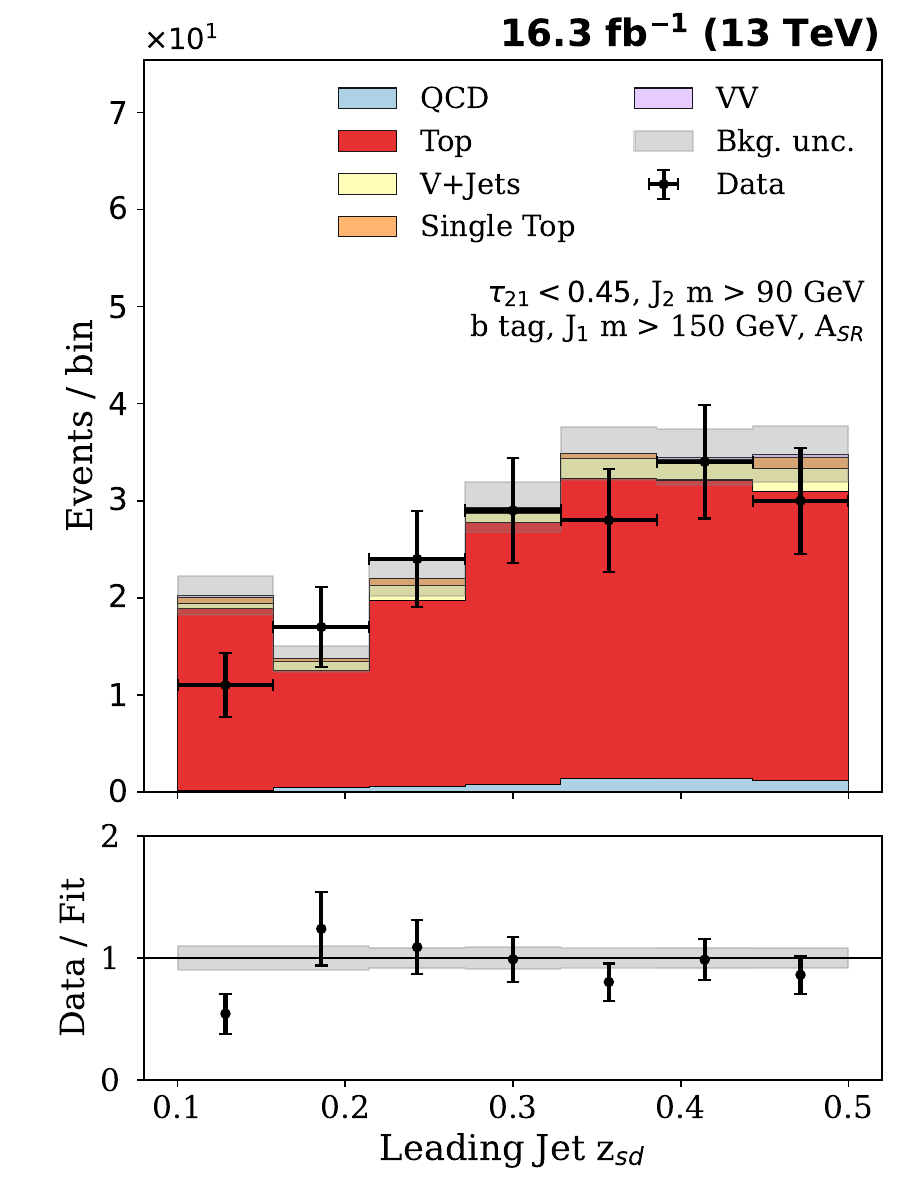}                
        \includegraphics[width=.23\textwidth]{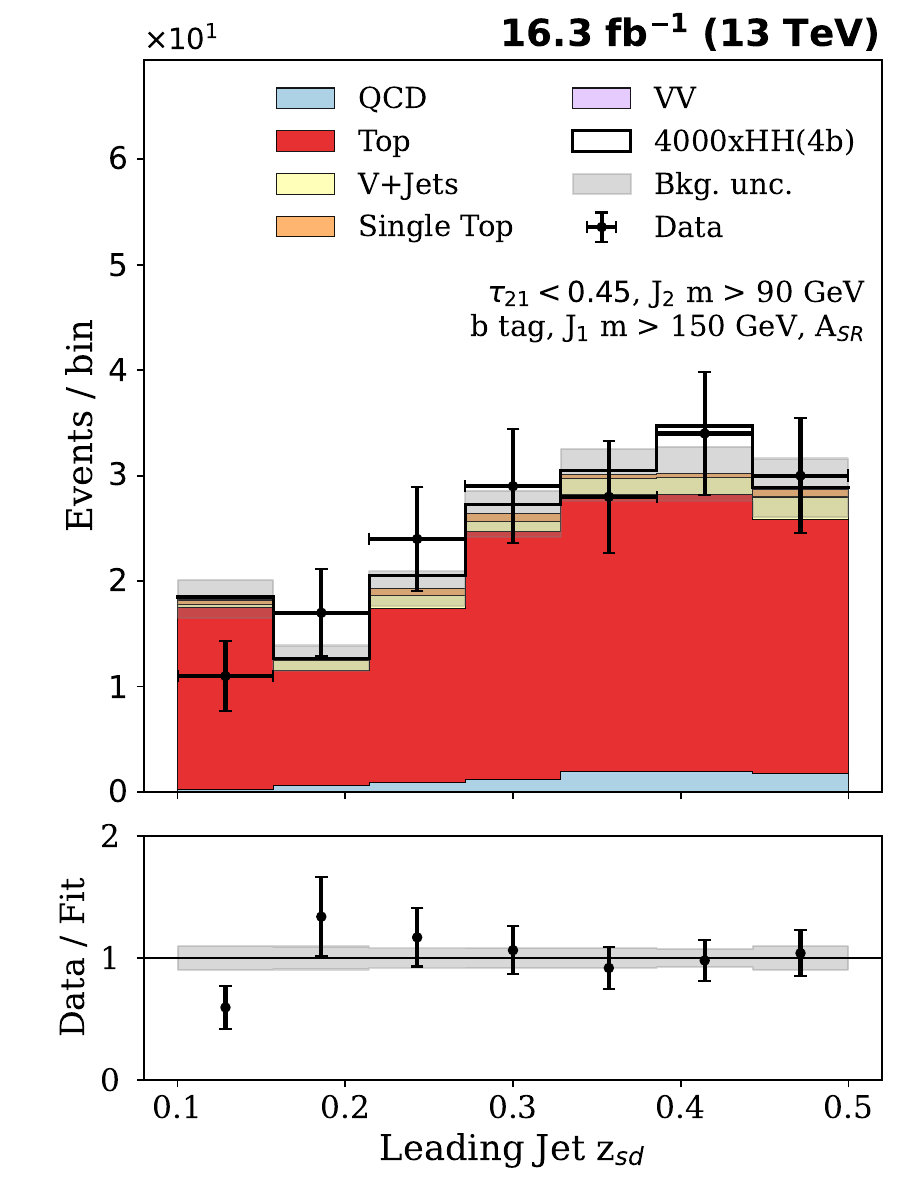}
        
    \caption{Leading jet $z$ for events where both jets are considered anomalous based on the \textsc{OmniLearned} large model score. Different mass intervals determined simultaneously in the fit are shown in the different rows while the background only results are on the left and the signal plus background results on the right. Shaded regions represent the total background uncertainty.}
    \label{fig:z}
\end{figure}

The soft drop mass is calculated based on the properties of the last two subjets after jet reclustering. In particular, the distribution of the momentum fraction $z$ and the angular observable $\theta$ are important quantities used during the calculation, where for a jet with radius $R$:
\begin{equation}
    z = \frac{\min(p_{T1},\, p_{T2})}{p_{T1} + p_{T2}},
    \qquad
    \theta = \frac{\Delta R_{12}}{R}
\end{equation}
for subjets with transverse momenta $p_{T1}$ and $p_{T2}$ separated by a distance $\Delta R_{12}$. In particular, the distributions of both $z$ and $\theta$ are useful probes to investigate if the anomalous events are compatible with 2-prong events, as expected from a HH sample. Conversely, simulation mis-modeling of these distributions could lead to unwanted artifacts to the soft drop mass distribution. We investigate the description of these quantities by repeating the full fit procedure carried out previously but replacing the observable of interest with $z$ or $\theta$. We use the most sensitive region for this study, namely the one that requires the subleading jet soft drop mass to be above 90 GeV and at least one of the jets need to be b tagged. The additional control regions with subleading jet mass below 90 GeV and high $\tau_{21}$ values is included in the fit to constrain the different background processes.  Additionally, we split the data into three non-overlapping regions based on the leading jet soft drop mass. The first category included jets with masses between 55 GeV and 100 GeV and is dominated by jets consisting of the decay products of W and Z bosons. The second category has leading jet masses between 100 GeV and 150 GeV and covers the region where the excess is observed. The last category selects jets with masses above 150 GeV and is dominated by jets consisting of the full decay products of top quarks. The results of the fit are shown in Figs.~\ref{fig:theta} and~\ref{fig:z} for $\theta$ and $z$ respectively. 

We observe an Asymptotic Significance of 4.25 for $z$ and of 4.05 for $\theta$, consistent with the results obtained using the soft drop mass as the main observable. Moreover, we observe the shape of the signal component to be compatible with the ones predicted by the HH sample.

\subsection{Split Top Quark Contribution}

\begin{figure}[ht]
    \centering
        \includegraphics[width=.23\textwidth]{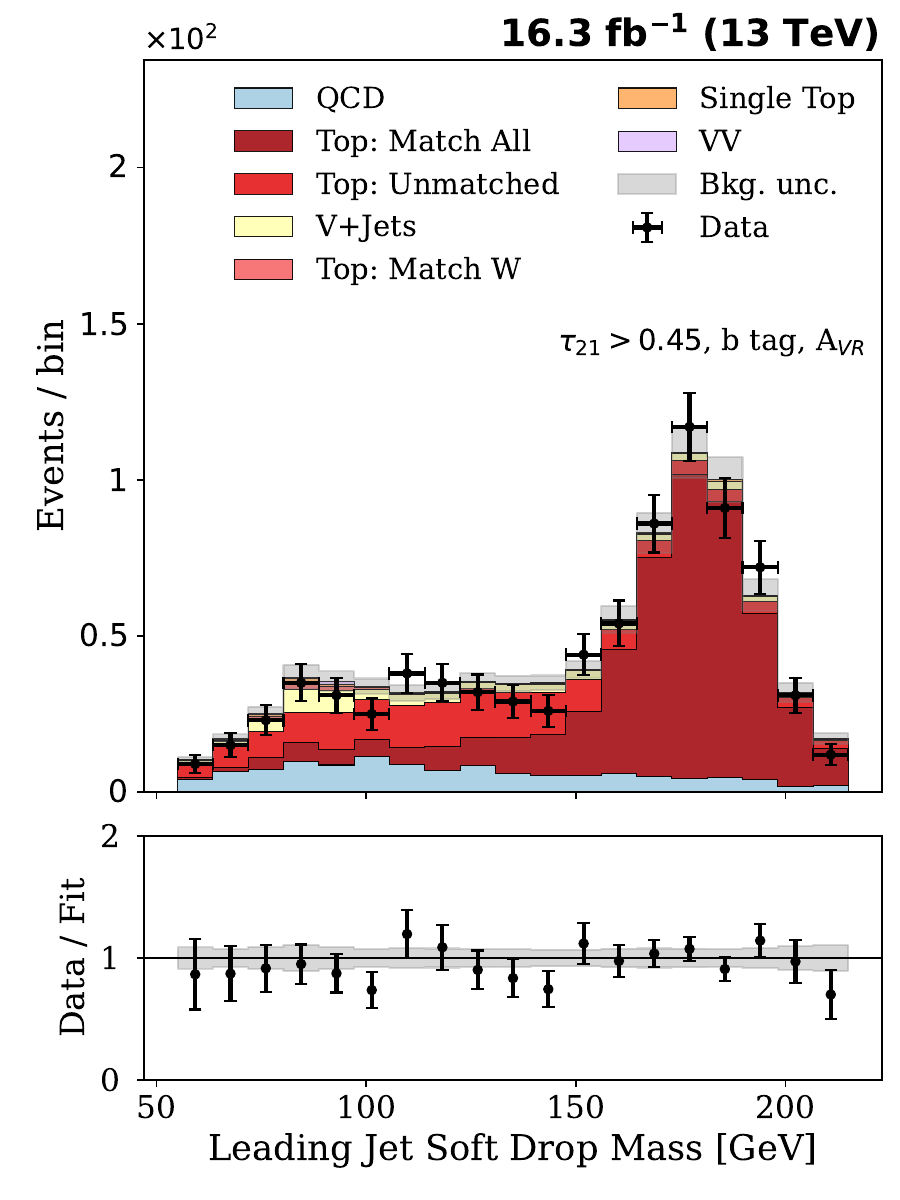}
        \includegraphics[width=.23\textwidth]{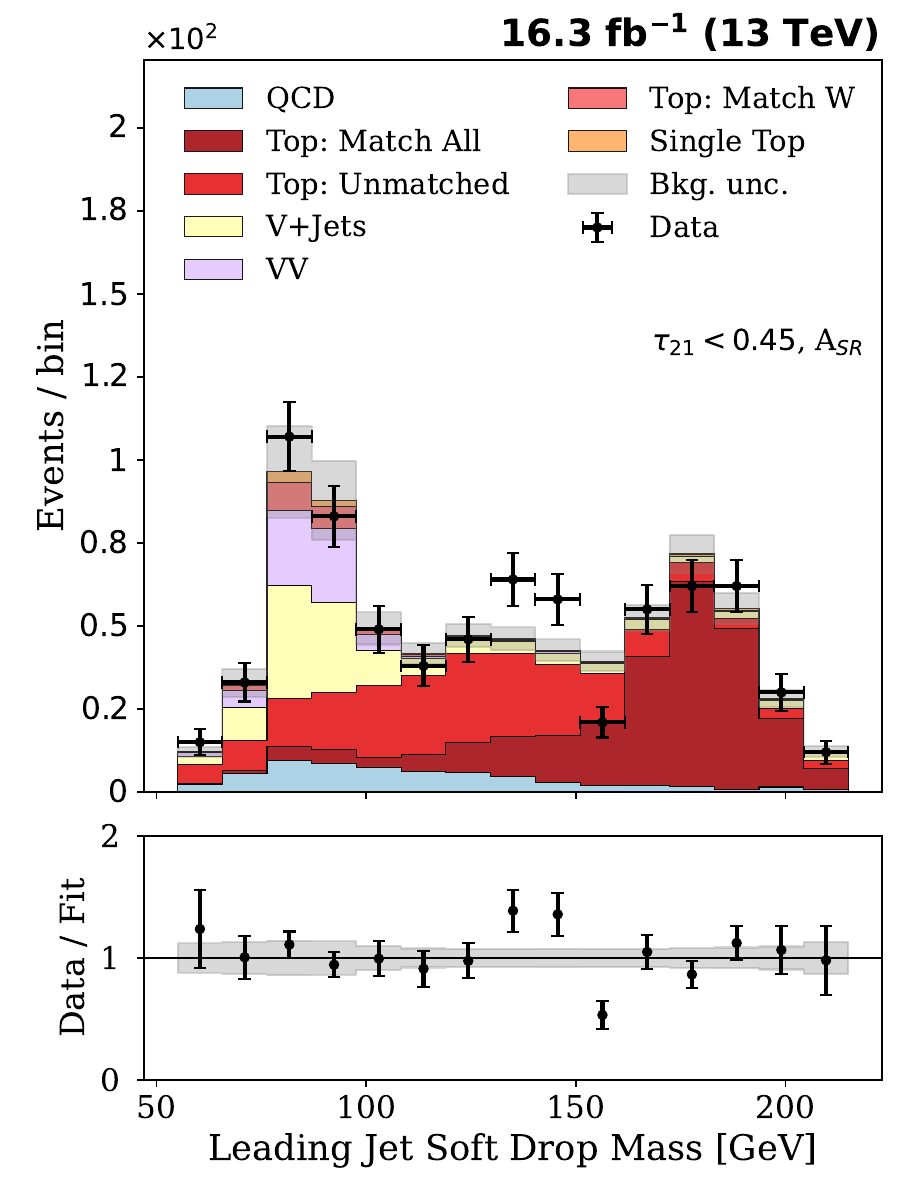}
        \includegraphics[width=.23\textwidth]{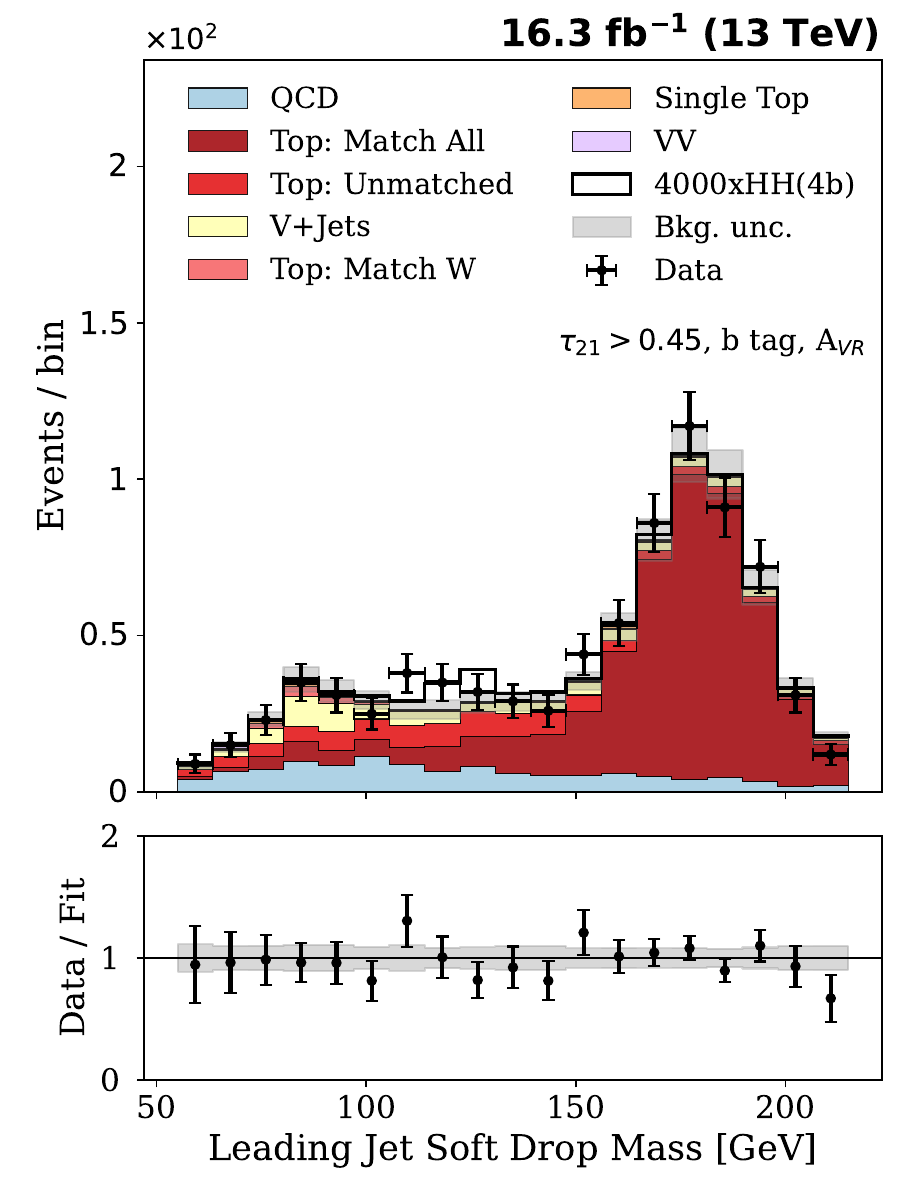}
        \includegraphics[width=.23\textwidth]{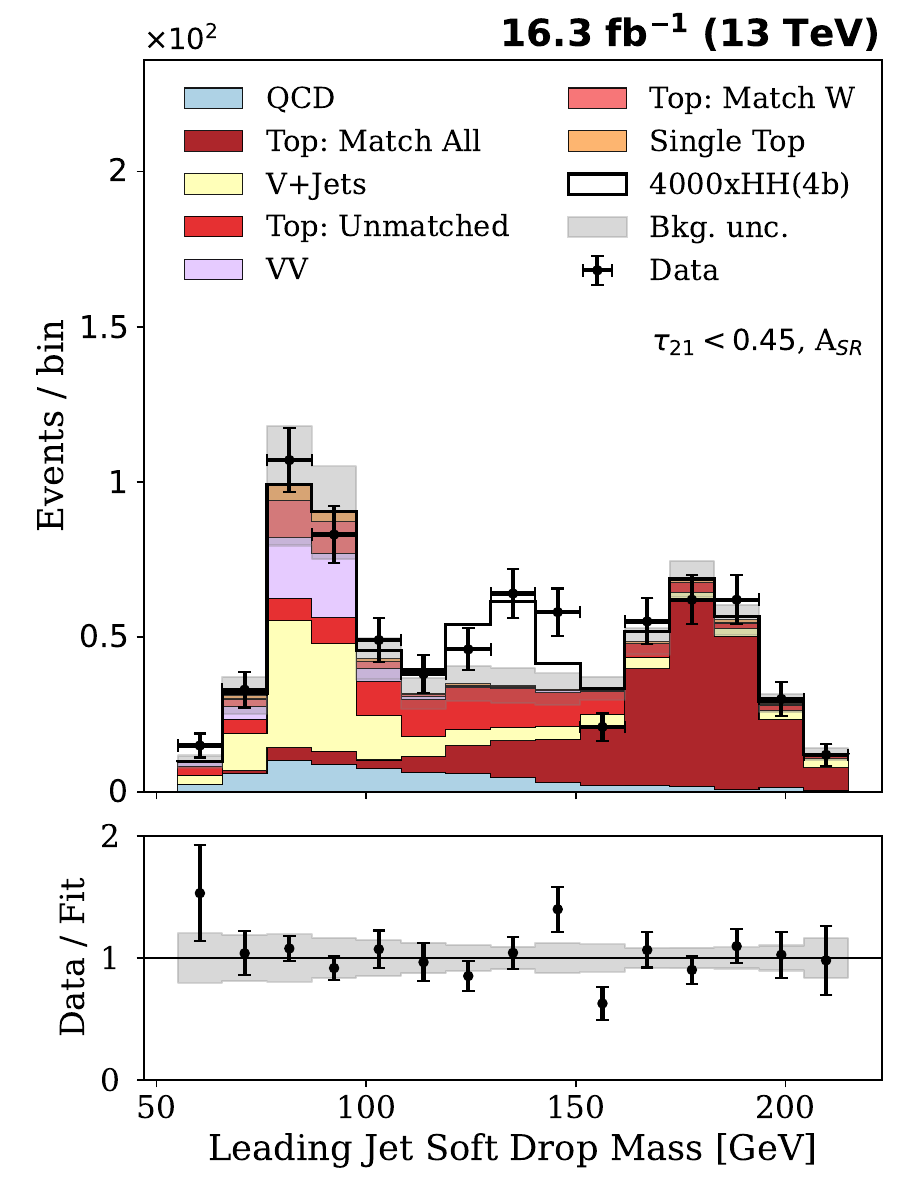}
    \caption{Leading jet soft drop mass where both jets are considered anomalous based on the \textsc{OmniLearned} large model score and the HH production is considered as the signal of interest. The region where both jets have low $\tau_{21}$ values is shown on the right while the region where at least one jet fails the $\tau_{21}$ selection is shown on the left. Shaded regions represent the total background uncertainty.}
    \label{fig:large_results_top_split}
\end{figure}

\begin{figure}[ht]
    \centering
        \includegraphics[width=.23\textwidth]{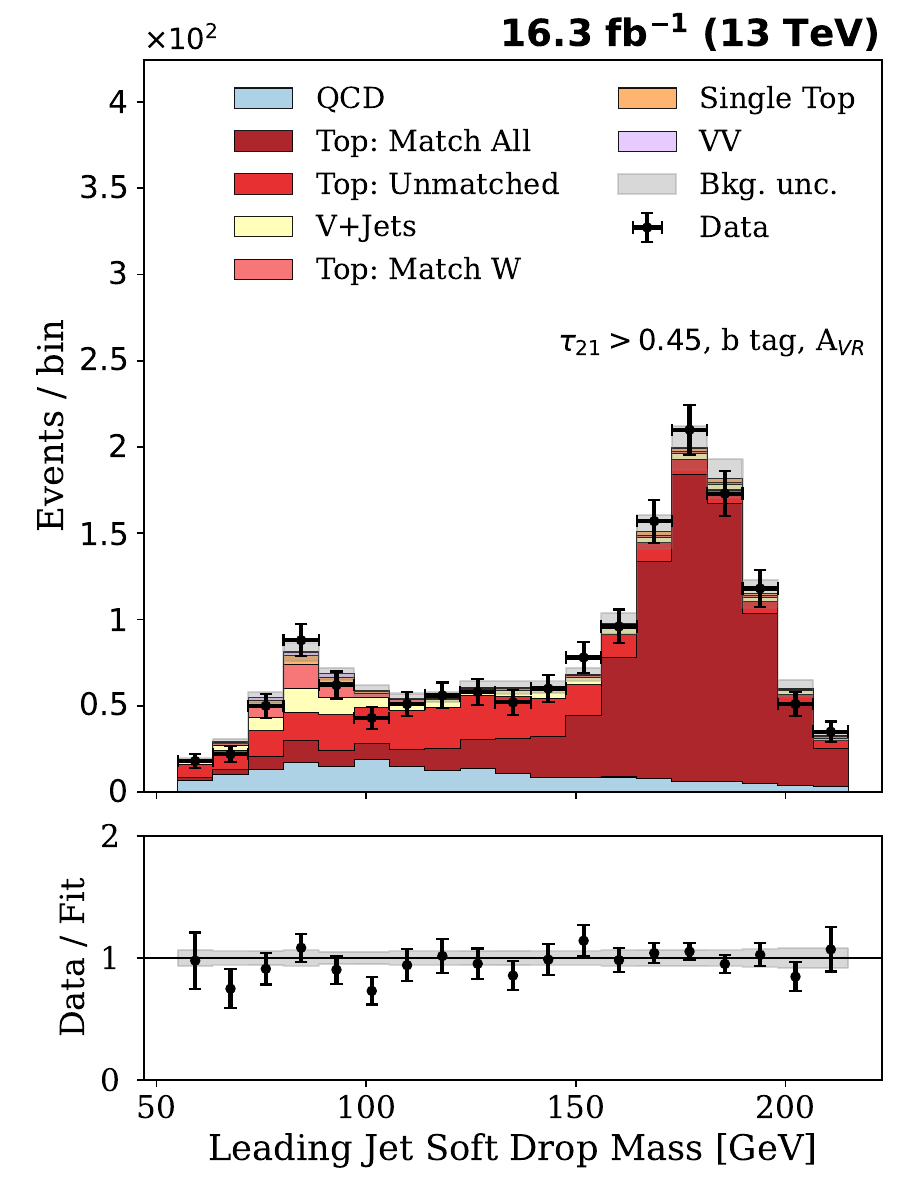}
        \includegraphics[width=.23\textwidth]{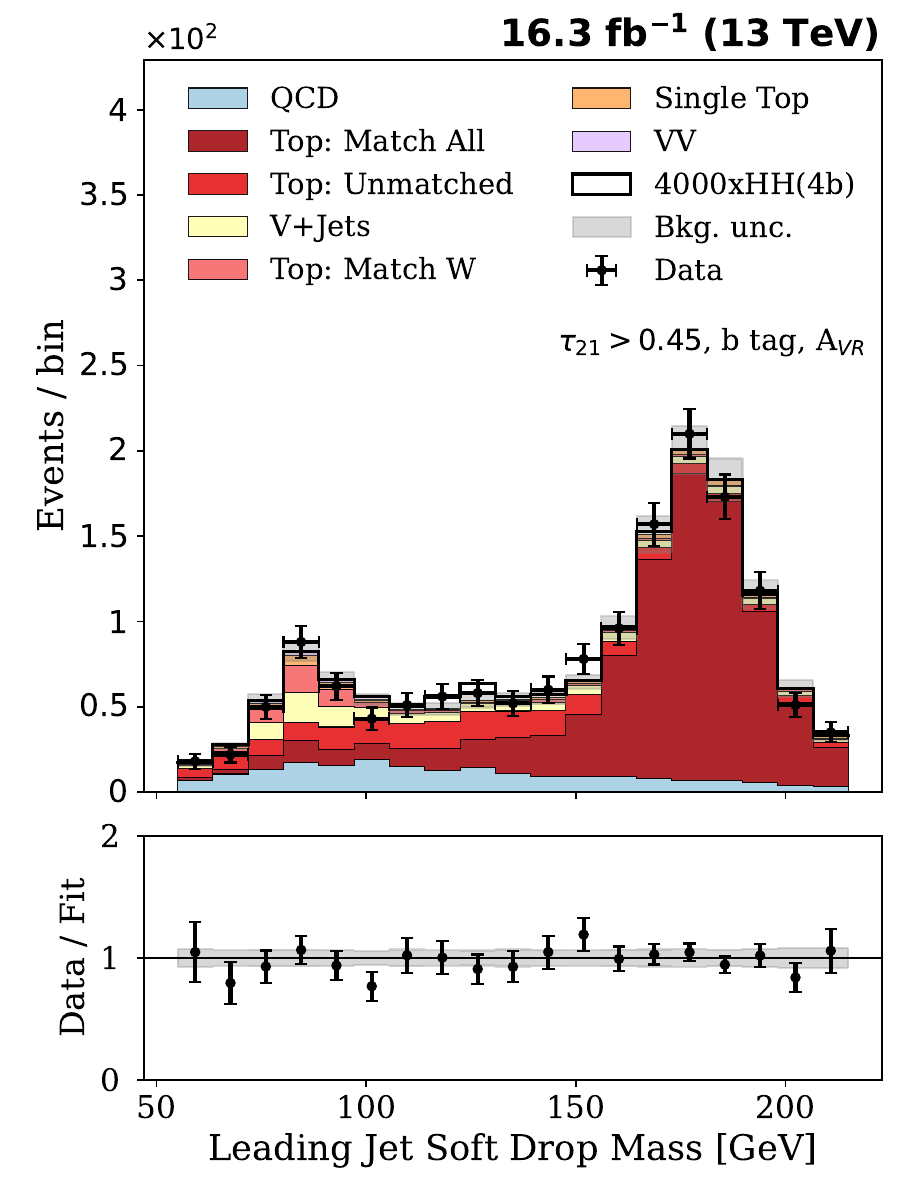}
        \includegraphics[width=.23\textwidth]{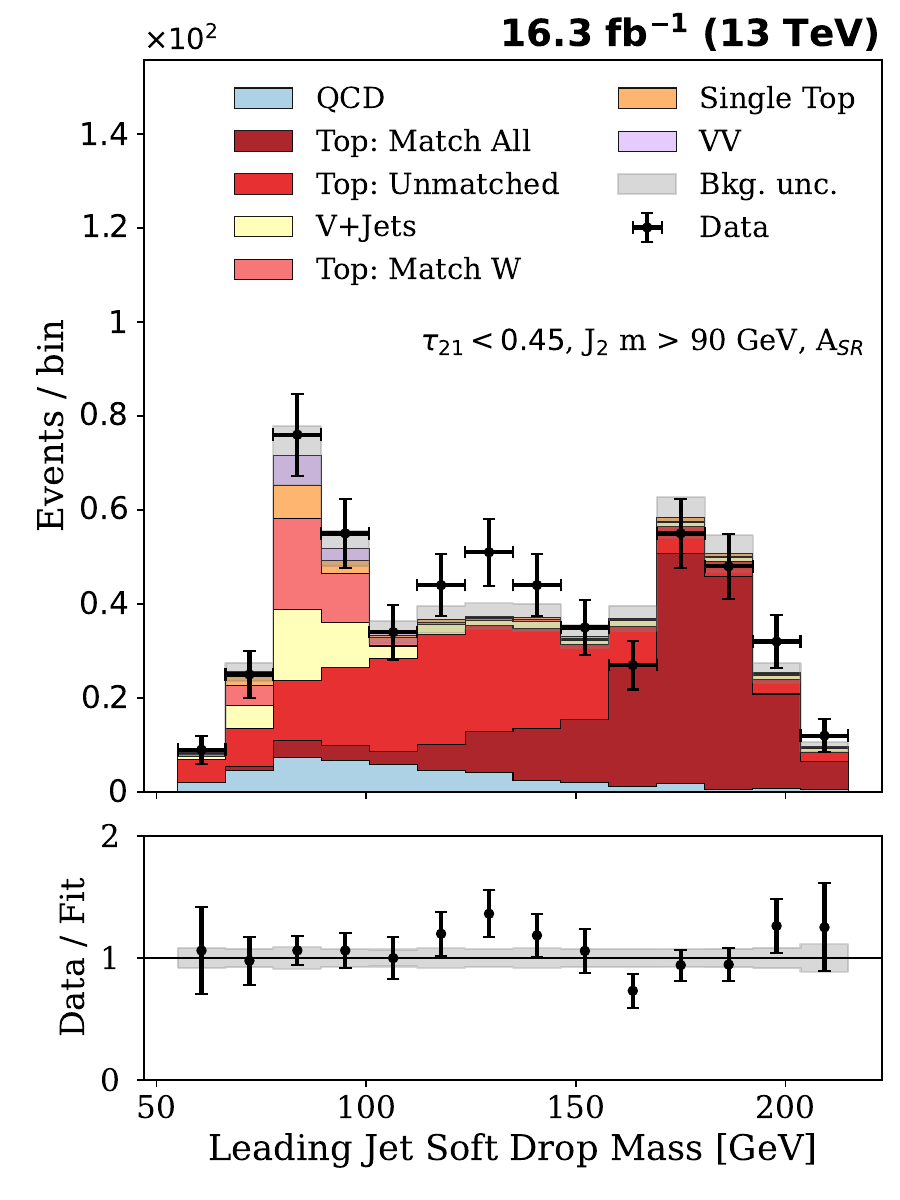}
        \includegraphics[width=.23\textwidth]{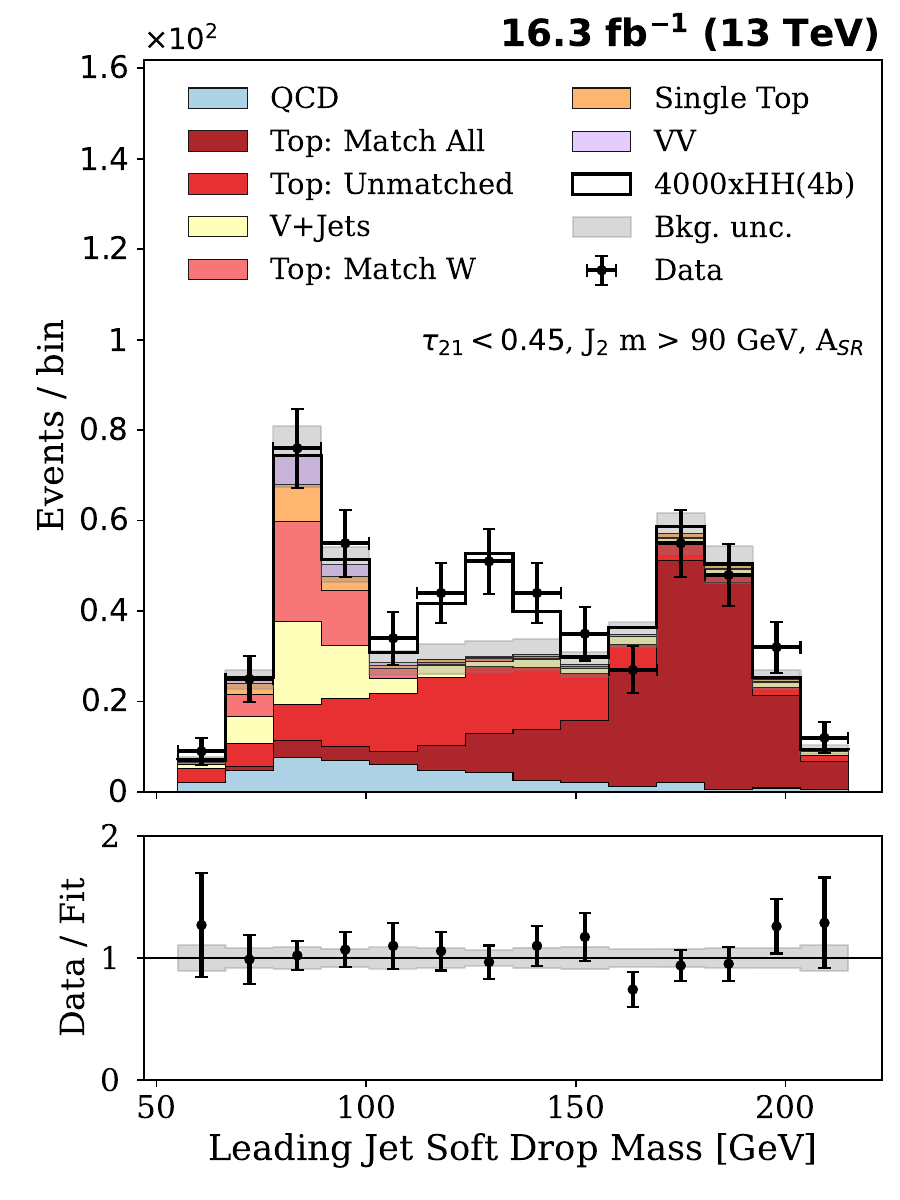}
        \includegraphics[width=.23\textwidth]{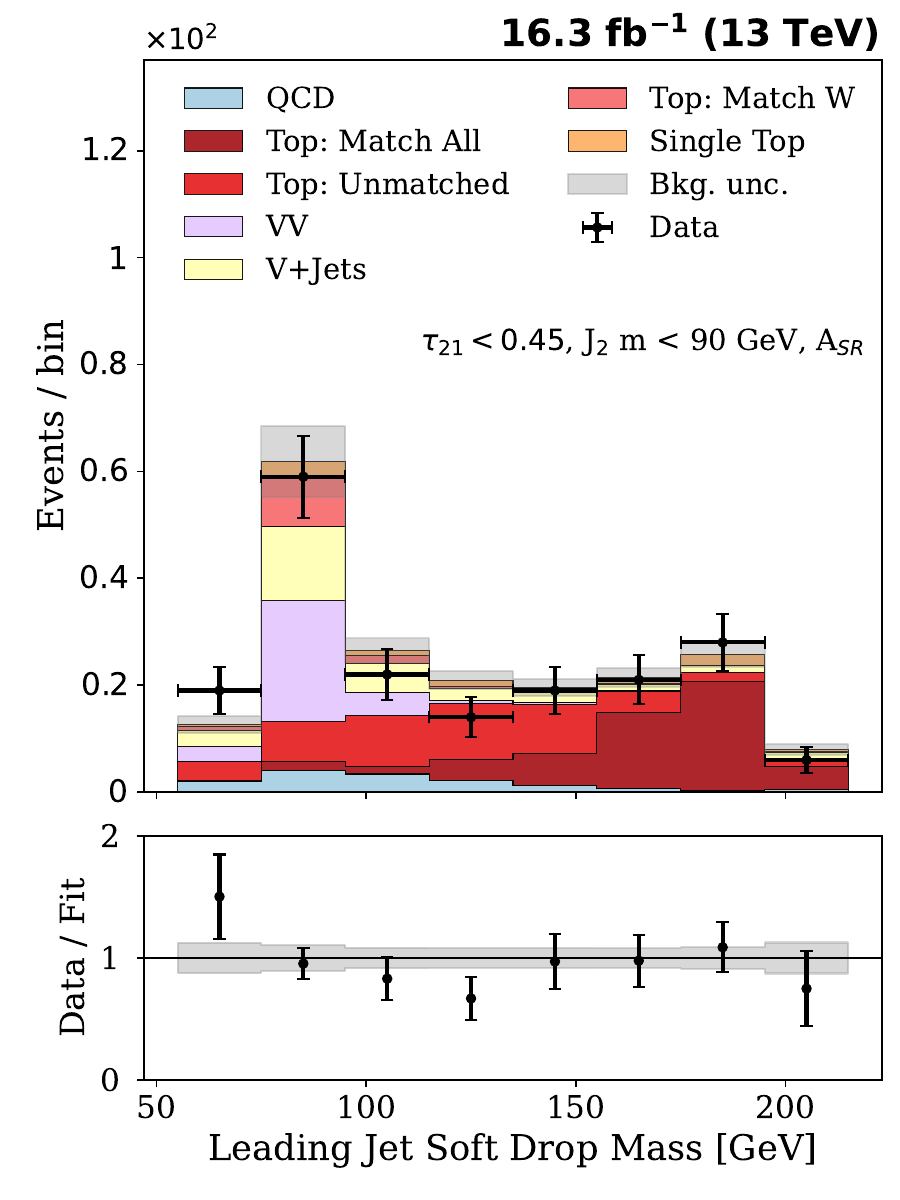}                
        \includegraphics[width=.23\textwidth]{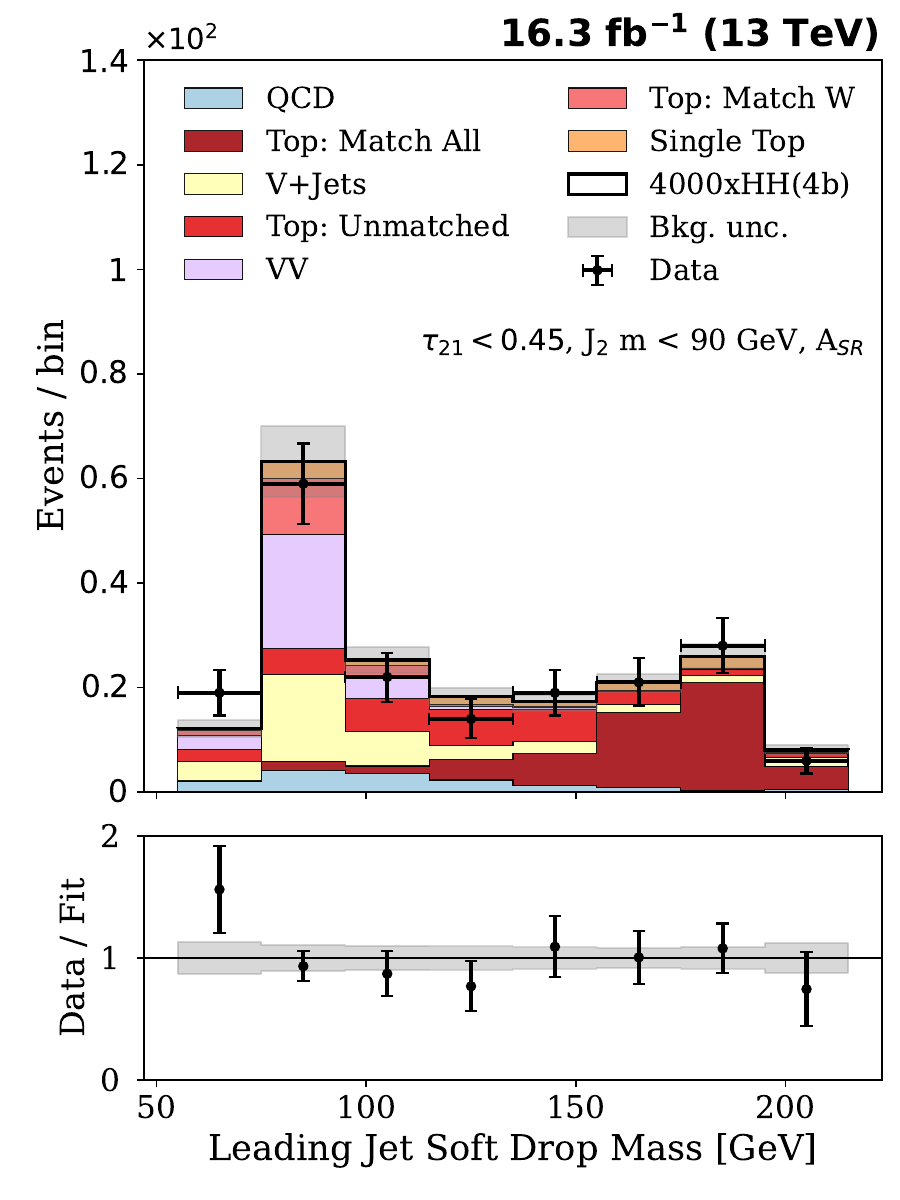}
       
    \caption{Leading jet soft drop mass where both jets are considered anomalous based on the \textsc{OmniLearned} large model score. The region where both jets have low $\tau_{21}$ values is split in the regions where the mass of the subleading jet is above (middle) and below (bottom) 90 GeV.  The region where at least one jet fails the $\tau_{21}$ selection and has at least one b-tagged jet is shown on the top. Results of the background-only fit are shown at the left while results considering a HH signal component are shown at the right. Shaded regions represent the total background uncertainty.}
    \label{fig:large_results_top_split_sr12}
\end{figure}

\begin{figure}[ht]
    \centering
        \includegraphics[width=.23\textwidth]{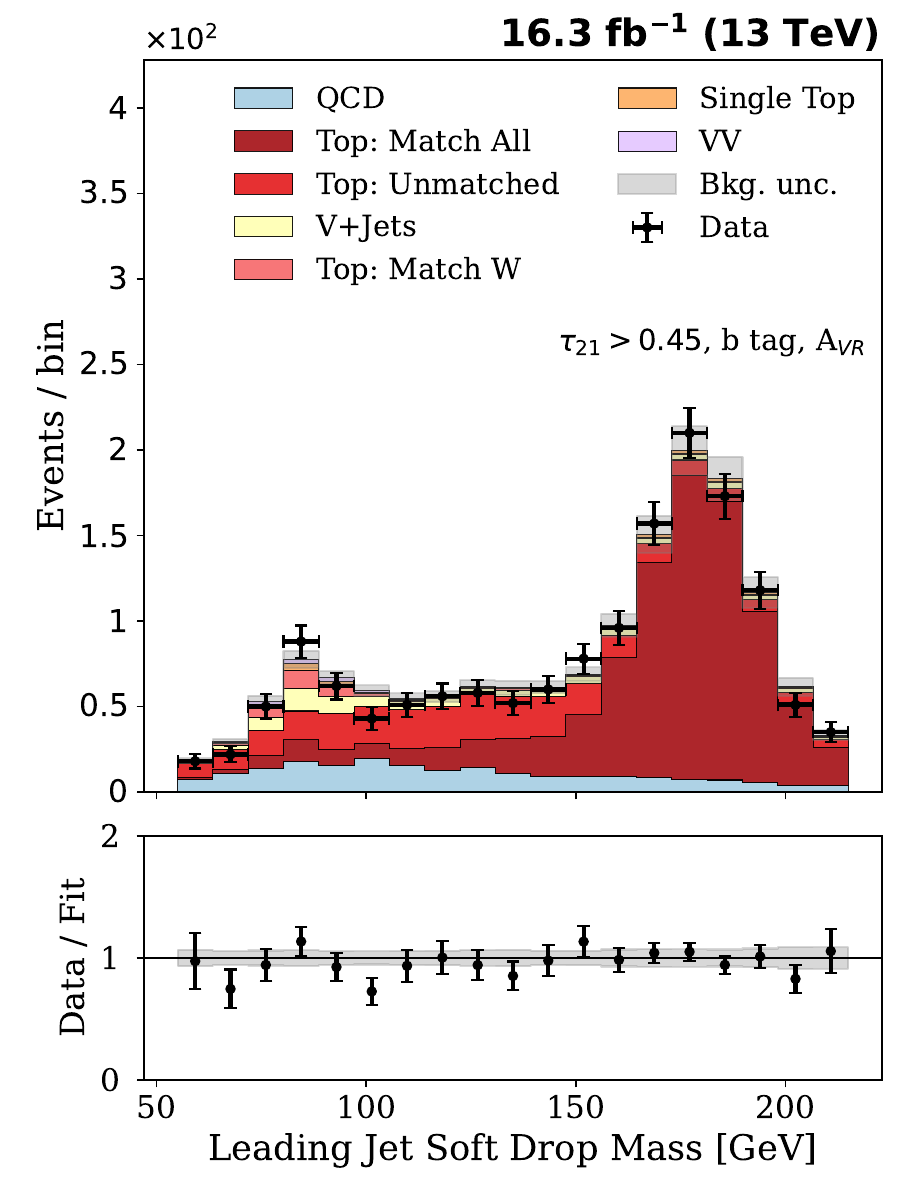}
        \includegraphics[width=.23\textwidth]{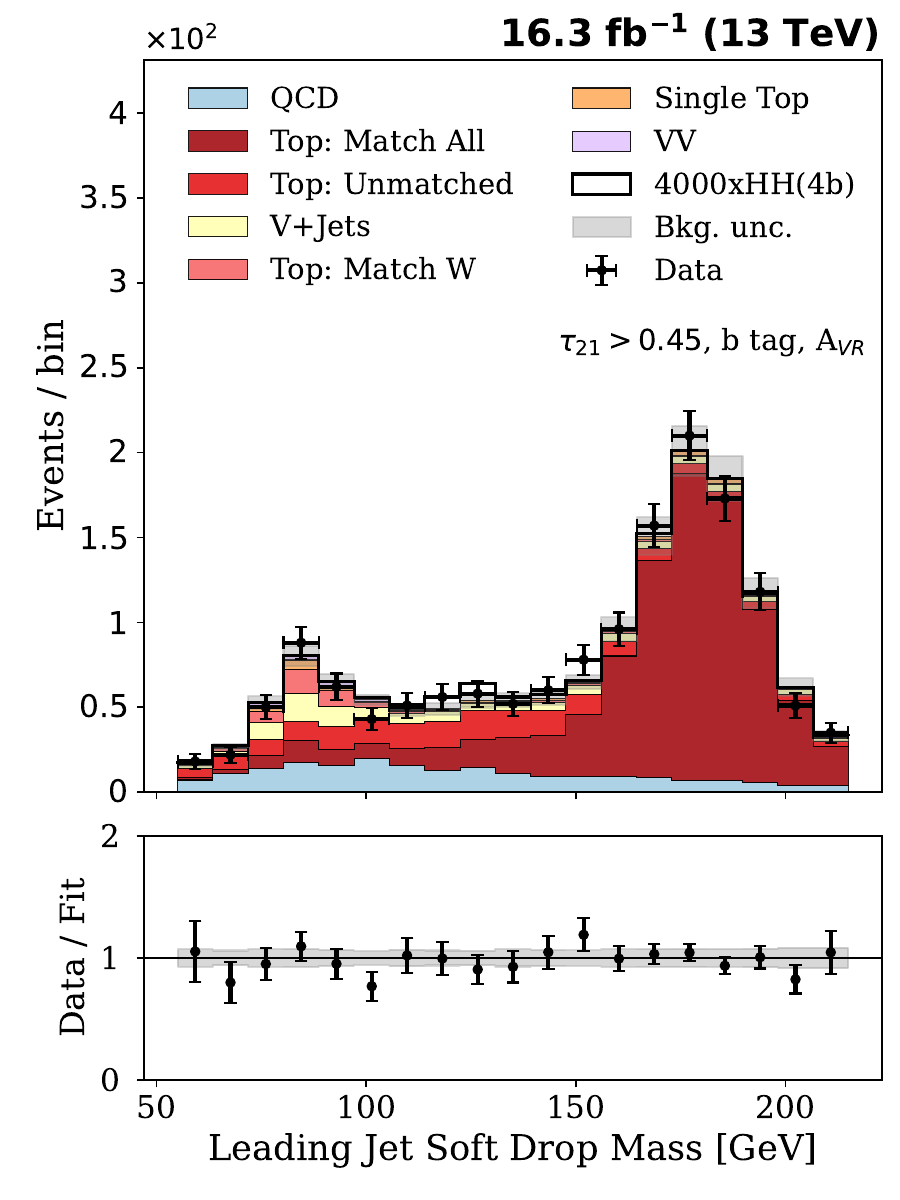}
        \includegraphics[width=.23\textwidth]{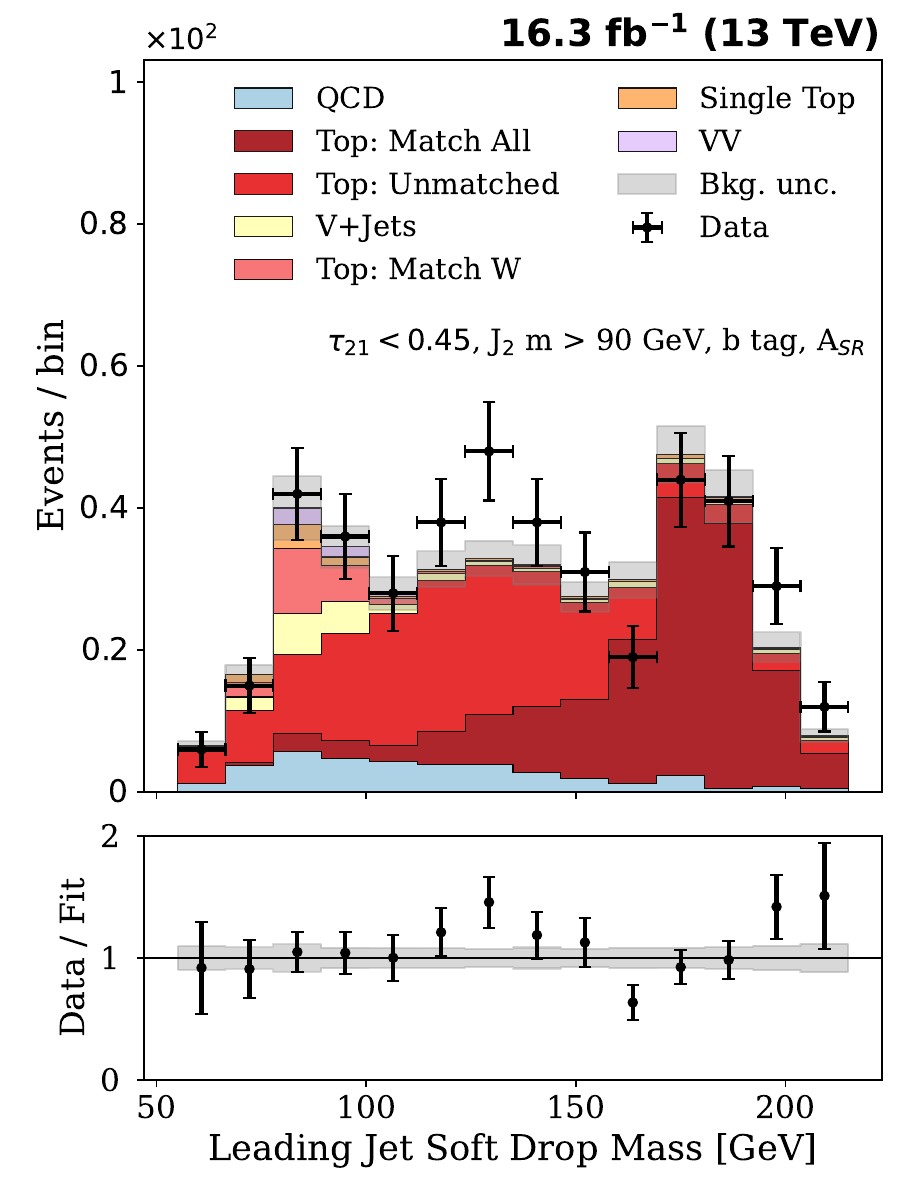}
        \includegraphics[width=.23\textwidth]{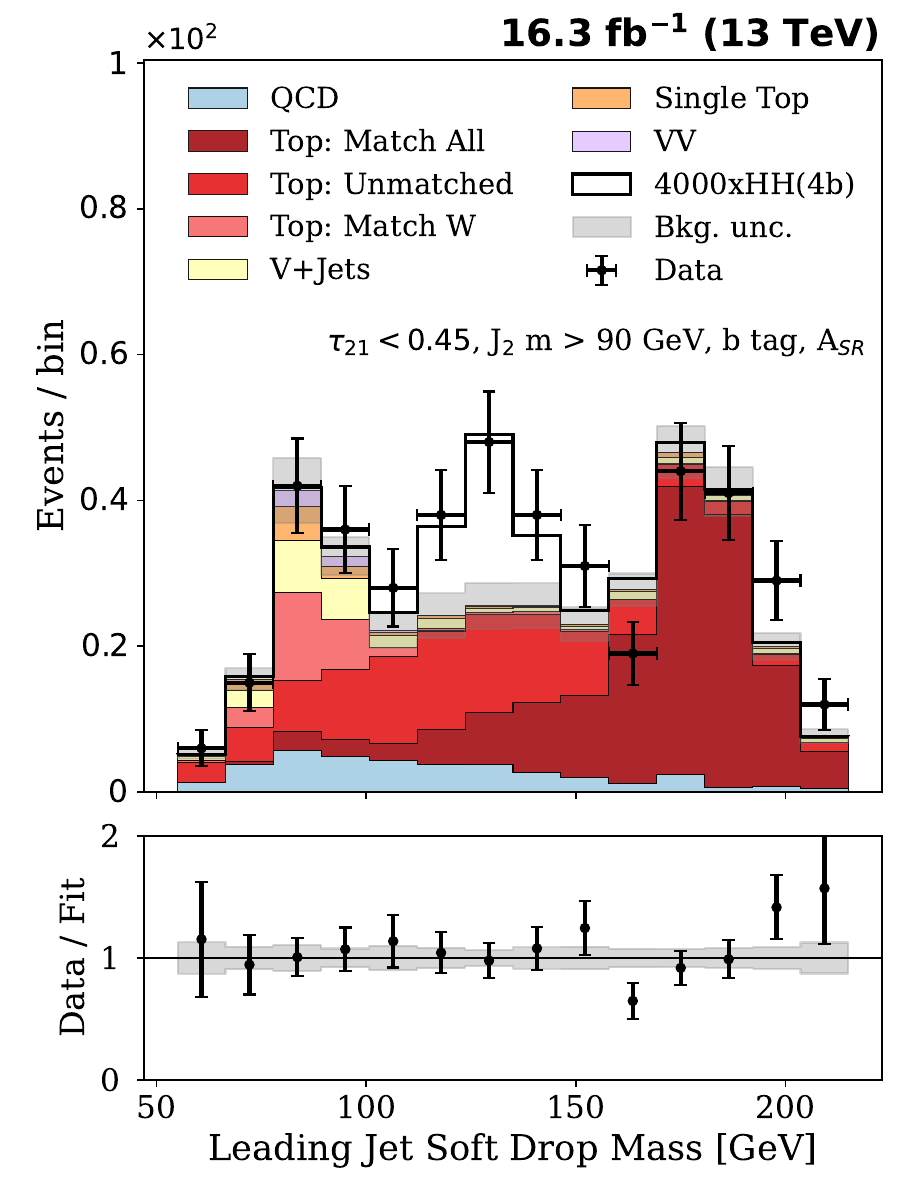}
        \includegraphics[width=.23\textwidth]{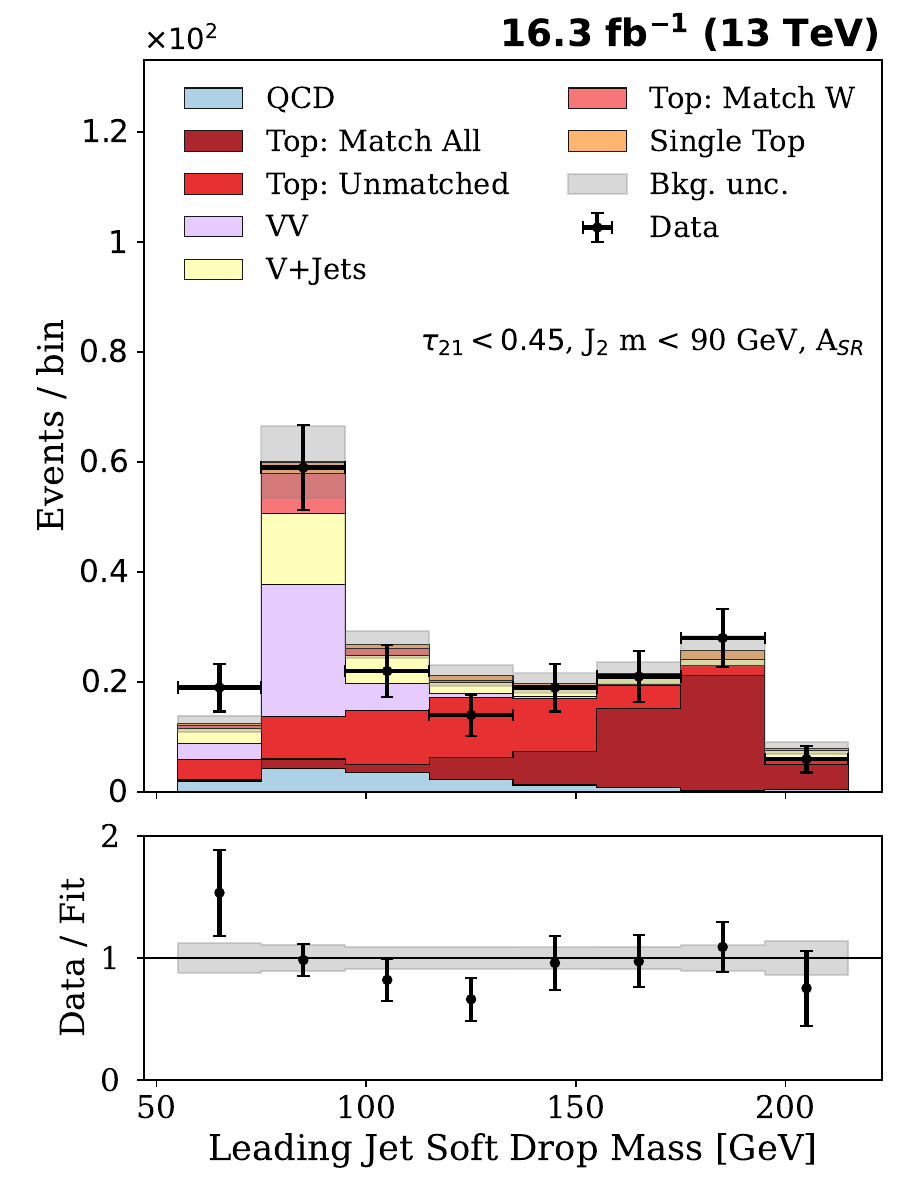}                
        \includegraphics[width=.23\textwidth]{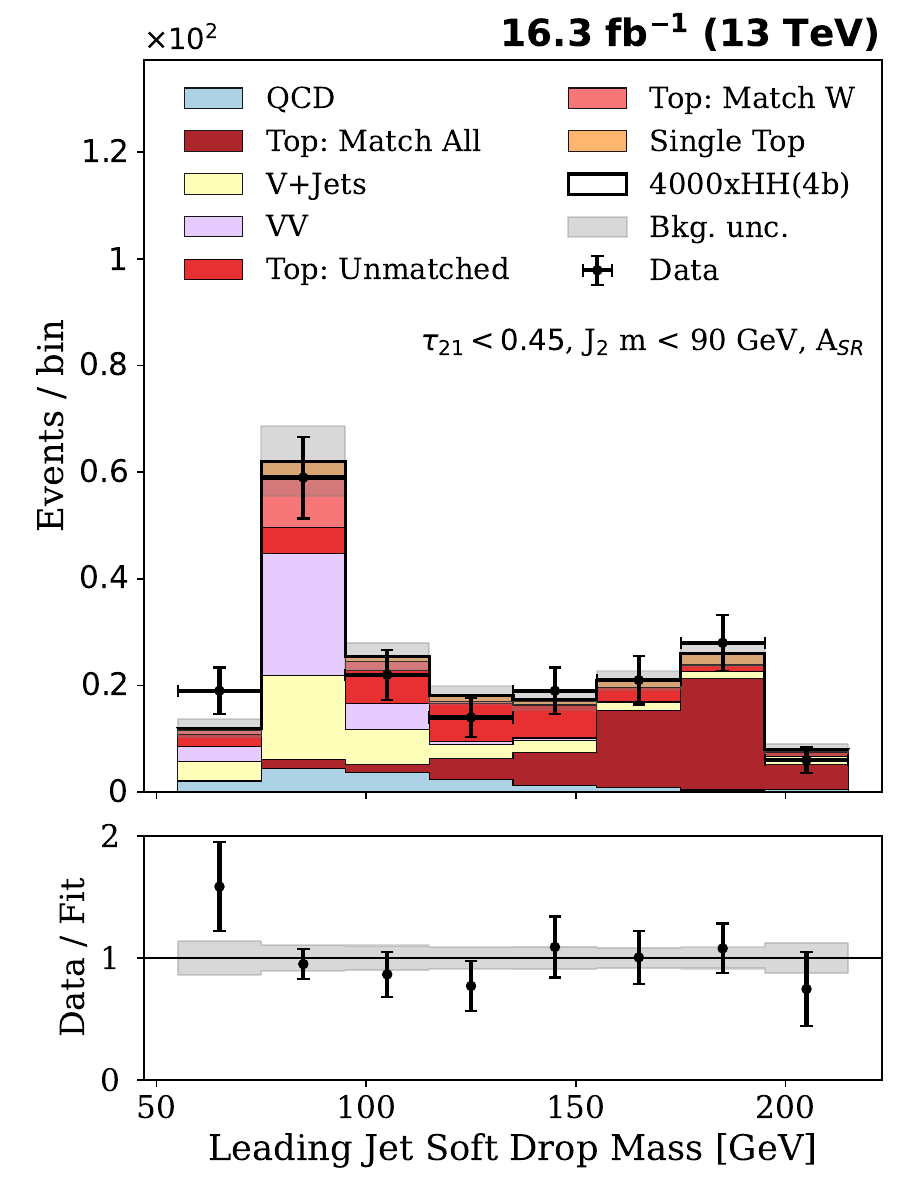}
       
    \caption{Leading jet soft drop mass where both jets are considered anomalous based on the \textsc{OmniLearned} large model score. The region where both jets have low $\tau_{21}$ values is split in the regions where the mass of the subleading jet is above (middle) and below (bottom) 90 GeV, with the region where the subleading jet mass is above 90 GeV to also be required to have at least one jet b-tagged.  The region where at least one jet fails the $\tau_{21}$ selection is shown on the top. Results of the background-only fit are shown at the left while results considering a HH signal component are shown at the right.  Shaded regions represent the total background uncertainty.}
    \label{fig:large_results_top_split_sr23}
\end{figure}

In this work we consider the top quark pair production as a single process. However, since the jet clustering and detector acceptance are not fully efficient, jets resulting from top quark decays can lead to different signatures and flavor composition. We study the impact of considering these effects by splitting the contributions of top quarks into three independent categories. The first category requires the reconstructed jet to be matched within a distance of 0.8 with all the generated decay products of the hadronic top quark, which includes the b hadron and the quarks from the W boson decay. The second category considers jets where only the W boson decay products are matched to the reconstructed jet. The last category considers the remaining jets from the top quark sample that were not selected by the previous categories. Similarly to the previous analysis strategy, we include a free normalization parameter, now independent for each category, that is determined during the fit. All the remaining systematic uncertainties are still considered correlated among the different top quark categories. Additionally, to enhance the constraining power of the validation region in the fit, we require at least one jet to be b-tagged in events that fail the $\tau_{21}$ selection. This additional requirement does not alter the composition of the main anomalous regions and is only used to better constrain the normalization factors associated with each top-quark category. The results obtained are shown in Figs.~\ref{fig:large_results_top_split},~\ref{fig:large_results_top_split_sr12}, and~\ref{fig:large_results_top_split_sr23}. The Asymptotic significances compared to the ones obtained by considering a single top quark category are listed in Tab.~\ref{tab:significance}.

\begin{table}[th]
    \centering
    \caption{Asymptotic significance obtained using the HH signal sample for different event selections and considering the top quark contribution either as a single component or split into multiple categories.}
    \label{tab:significance}
	\begin{tabular}{lccccc}
    \hline
          Selection &  Single & Split\\
            \hline
            Baseline & 3.30  & 2.73\\ 
            Baseline + J$_2(m) > 90$ GeV & 3.84 & 3.15 \\ 
            Baseline + J$_2(m) > 90$ GeV + b-tagged & 4.25 & 3.47 \\             
	\end{tabular}
\end{table}

While the observed significance decreases compared to the single category due to the additional normalization terms the fit results still favor the alternative hypothesis.

\begin{figure}[ht]
    \centering
        \includegraphics[width=.23\textwidth]{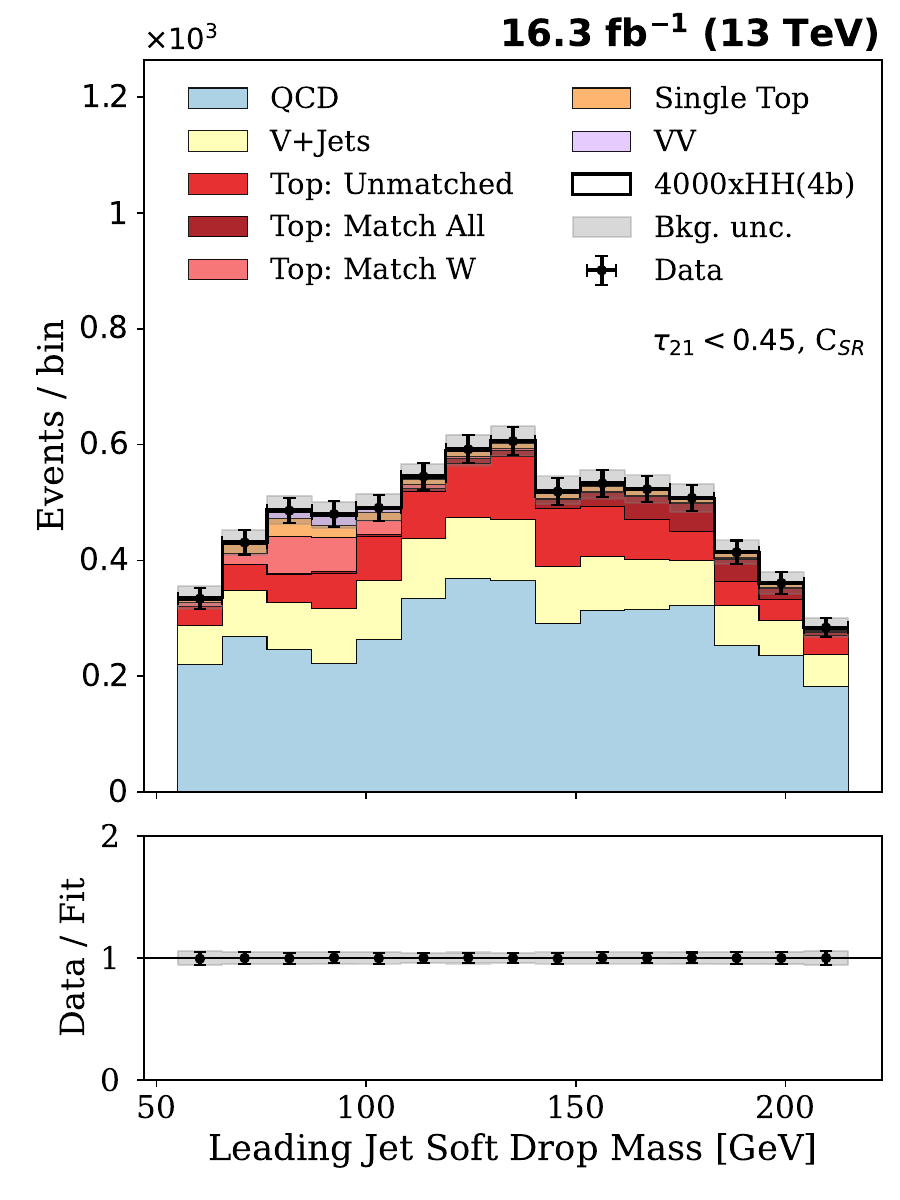}
        \includegraphics[width=.23\textwidth]{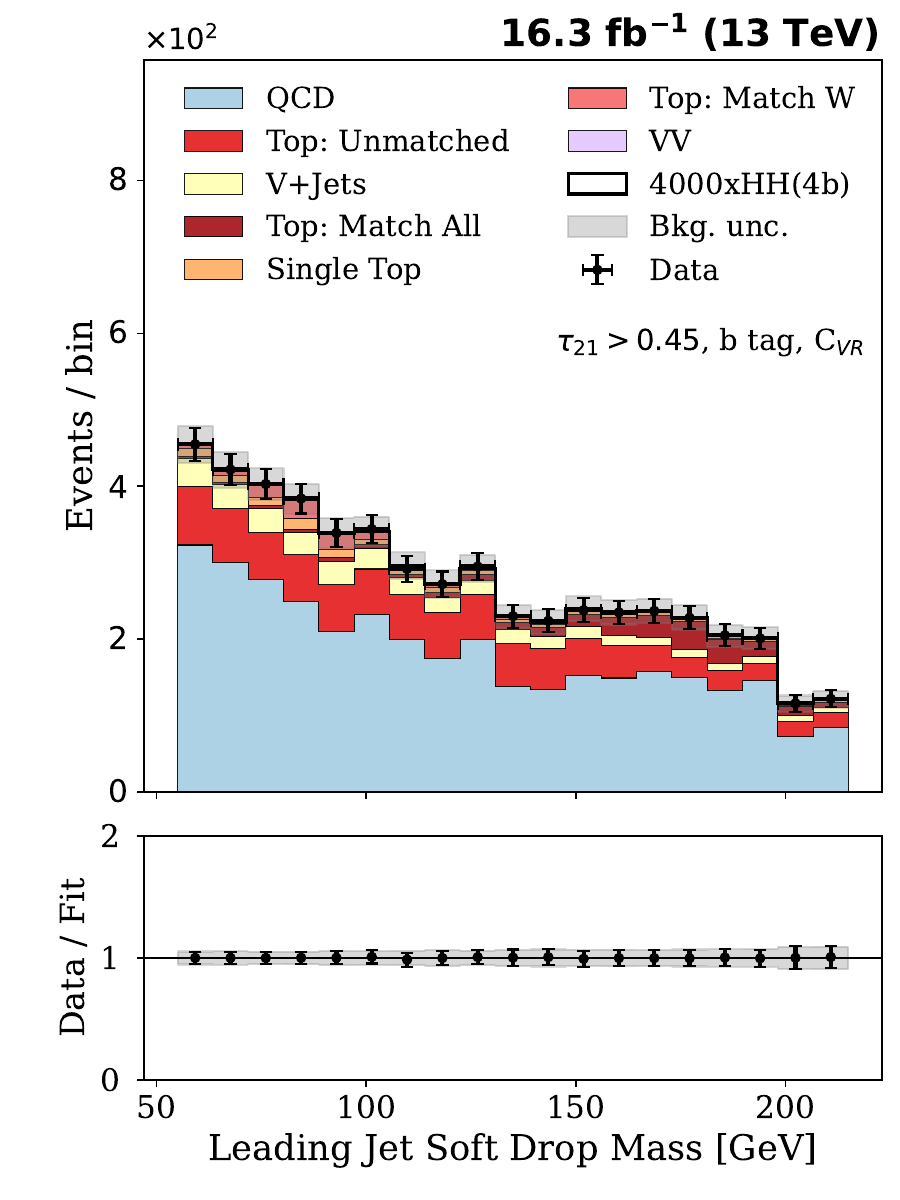}
        \includegraphics[width=.23\textwidth]{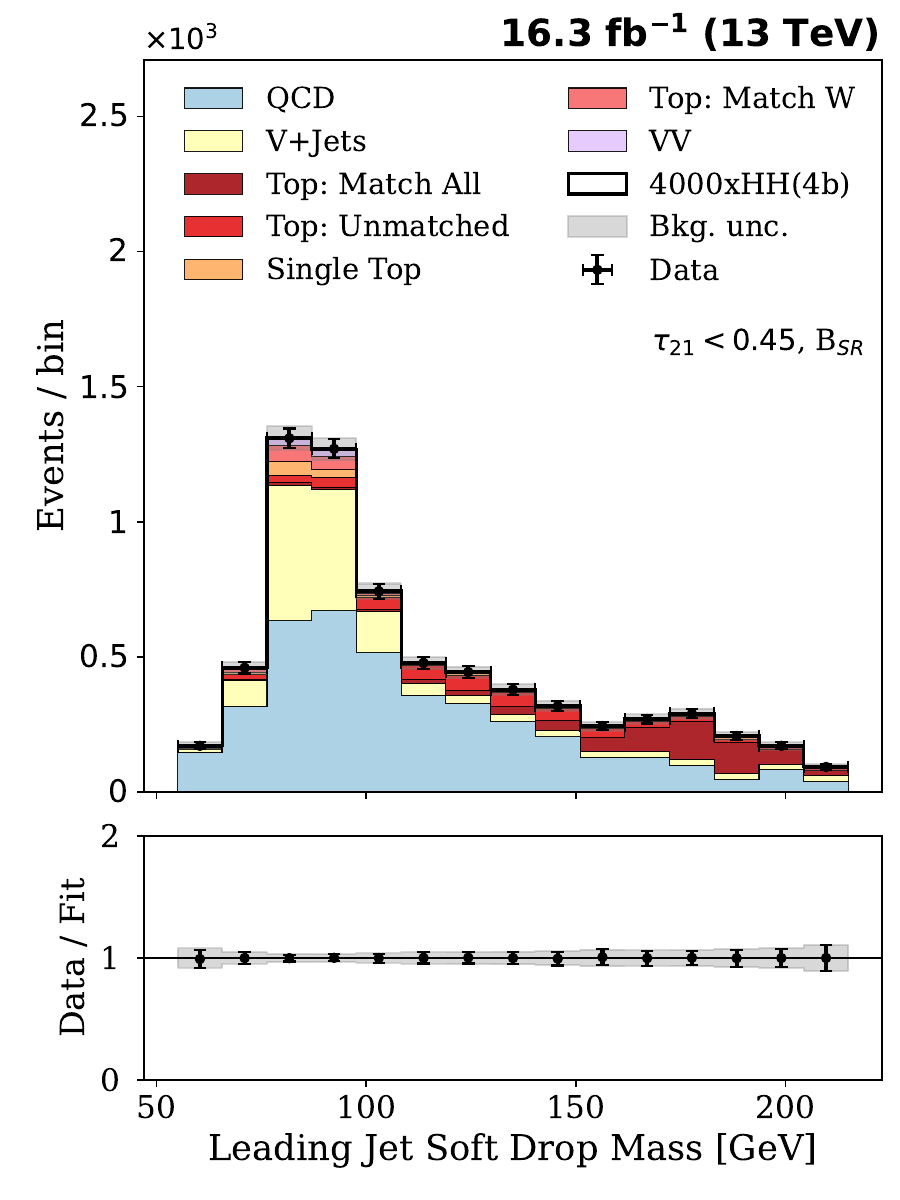}
        \includegraphics[width=.23\textwidth]{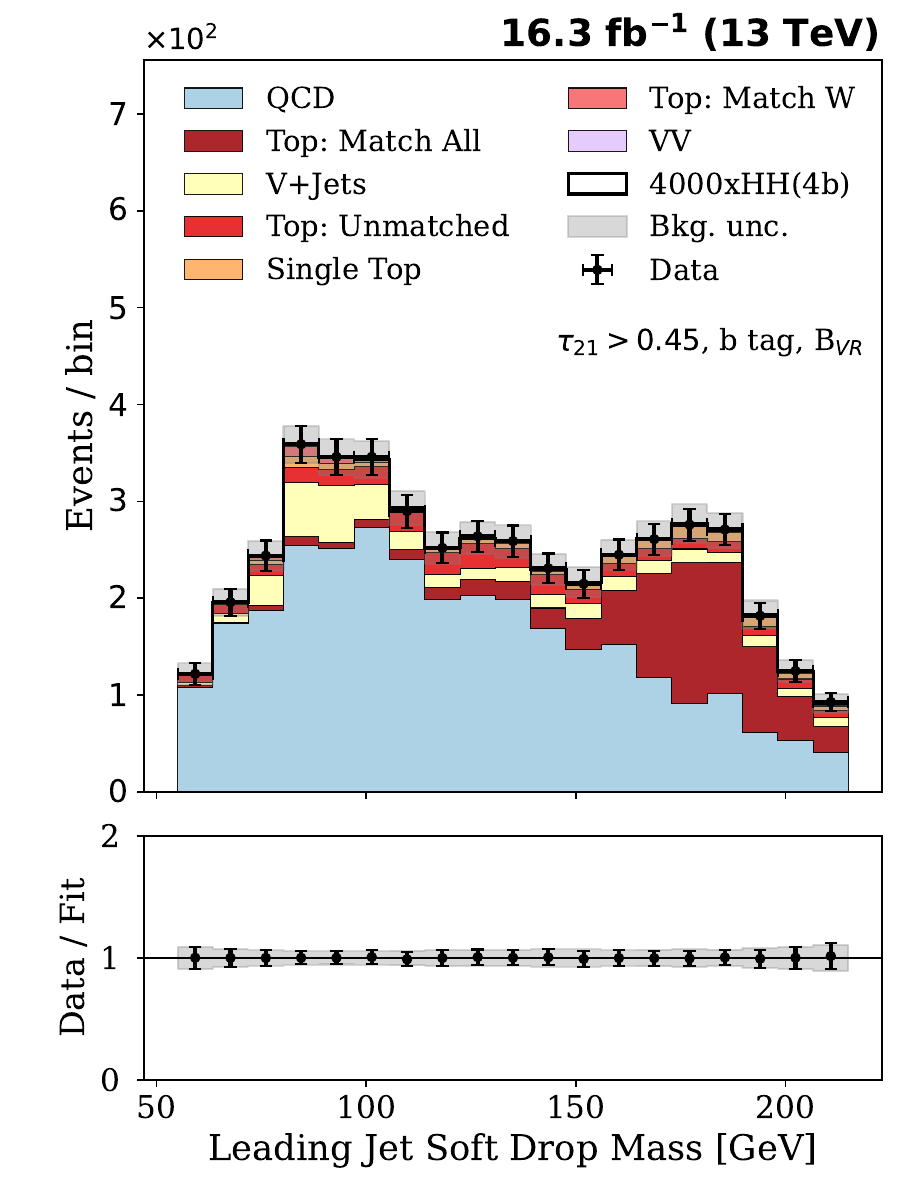}
        \includegraphics[width=.23\textwidth]{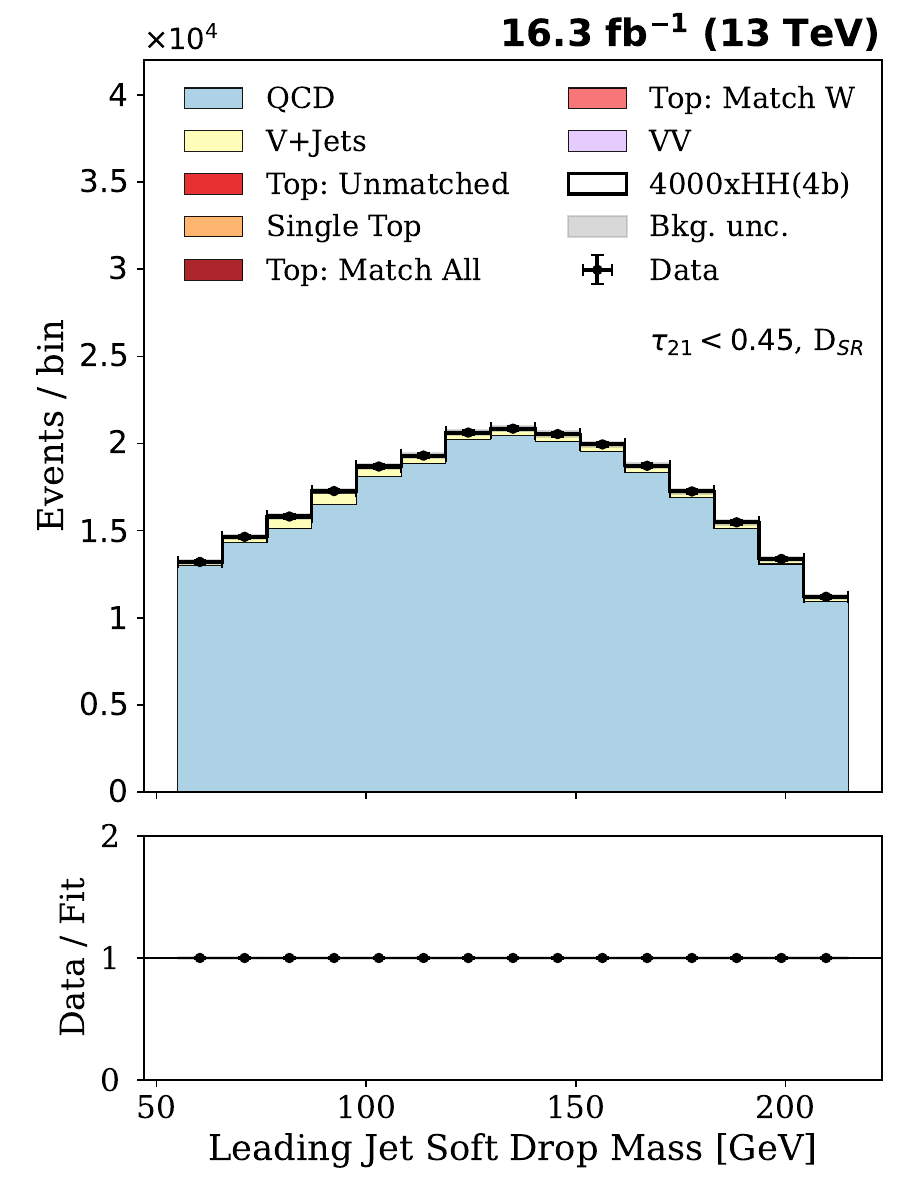}                
        \includegraphics[width=.23\textwidth]{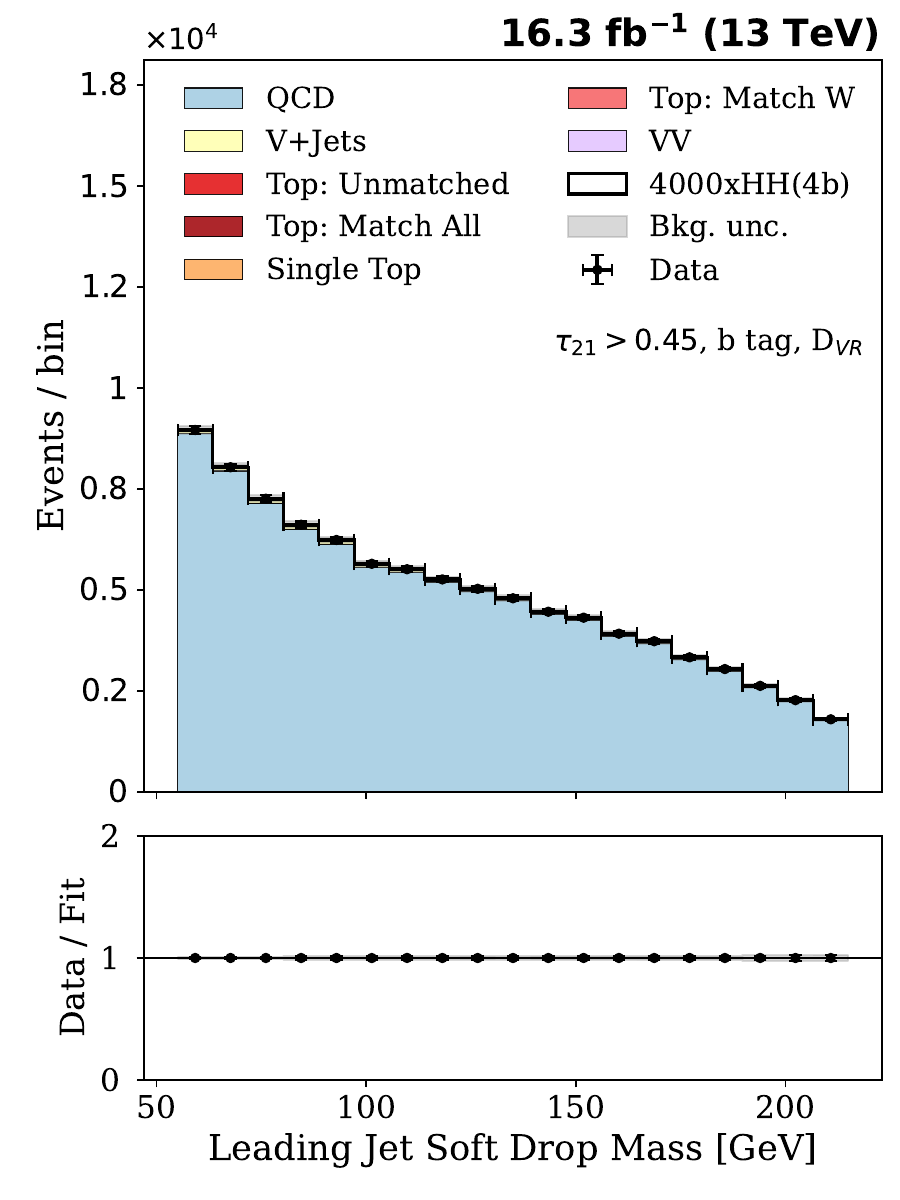}
        
    \caption{Leading jet soft drop mass where both jets are considered anomalous based on the \textsc{OmniLearned} large model score. The region where both jets have low $\tau_{21}$ values is shown at the left while the region where at least one jet fails the $\tau_{21}$ selection is shown at the right. The different regions used for the ABCD calculation are shown as rows. Shaded regions represent the total background uncertainty.}
    \label{fig:hh_baseline_split}
\end{figure}

\begin{figure*}[ht]
    \centering
        \includegraphics[width=.31\textwidth]{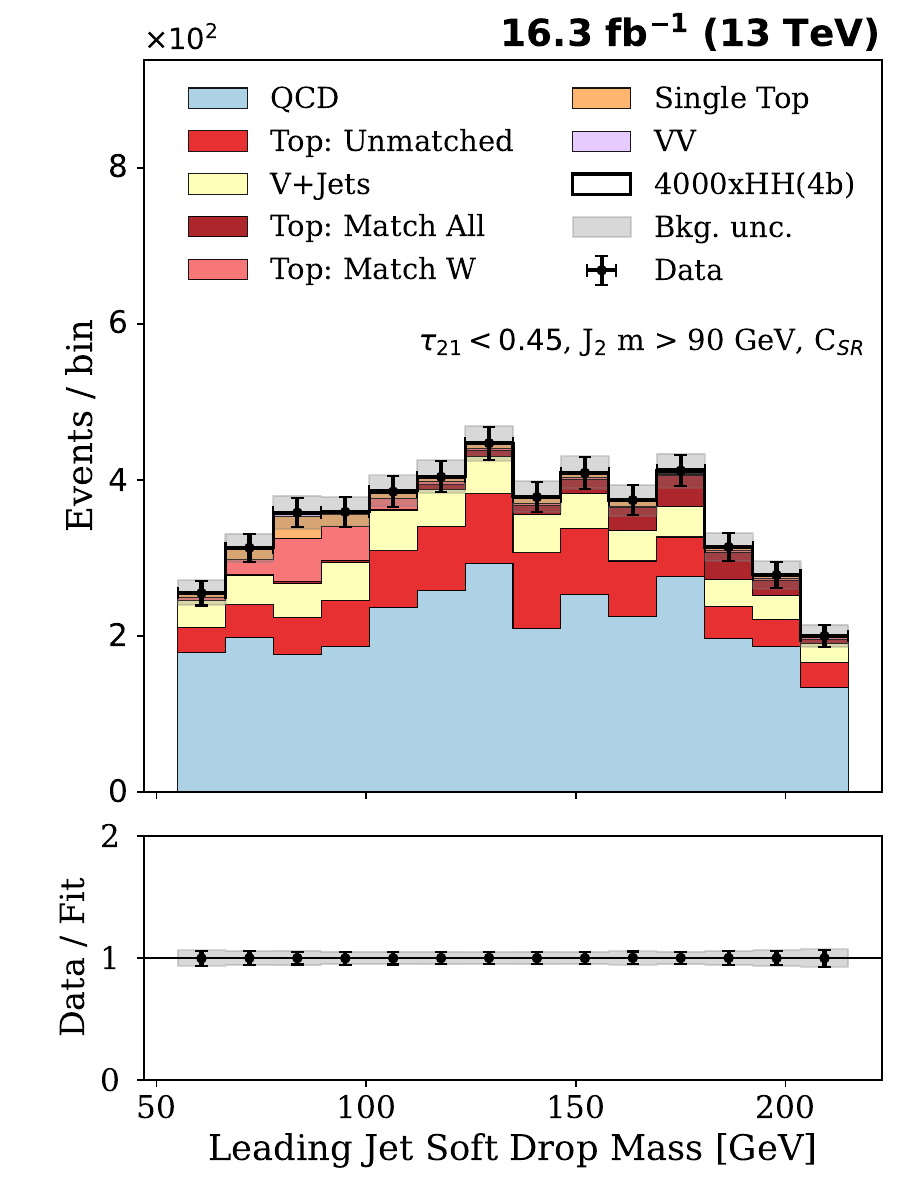}
        \includegraphics[width=.31\textwidth]{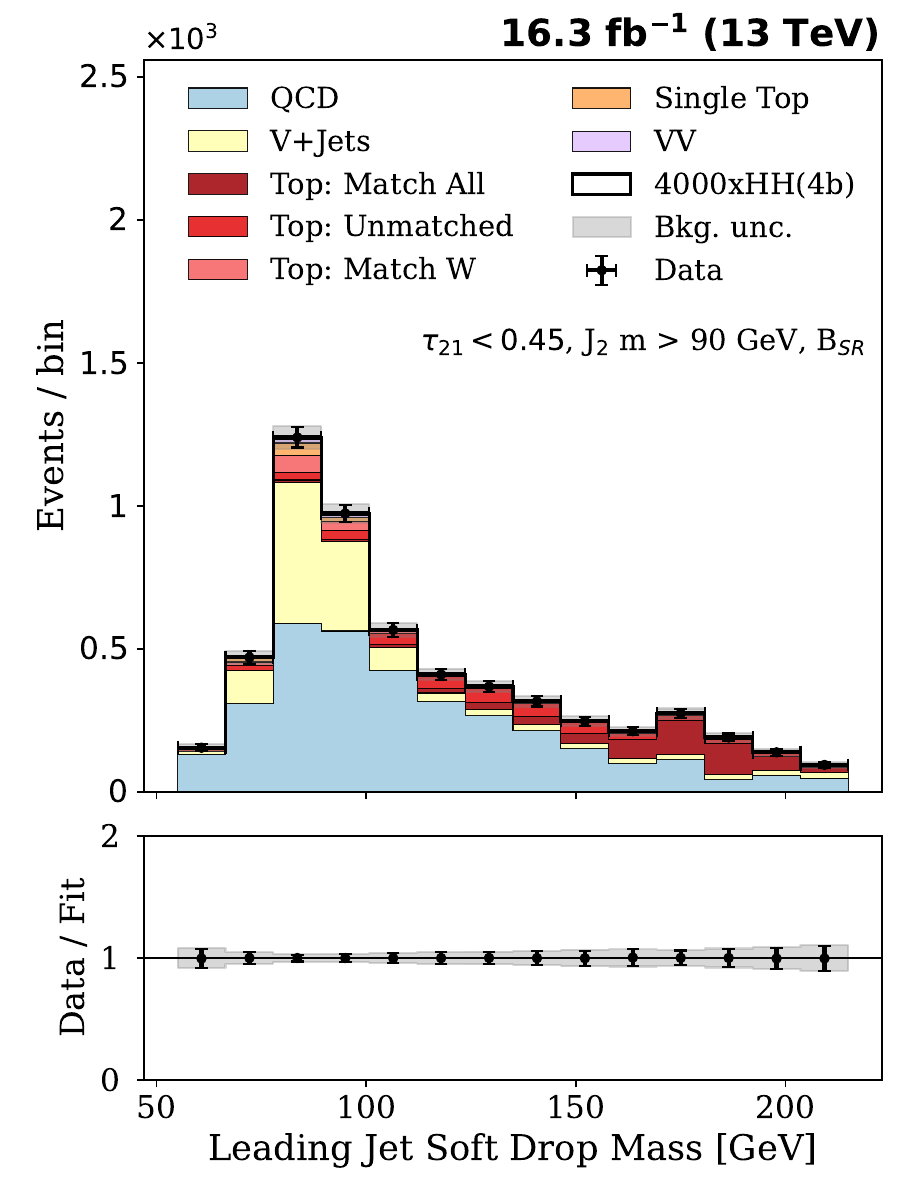}
        \includegraphics[width=.31\textwidth]{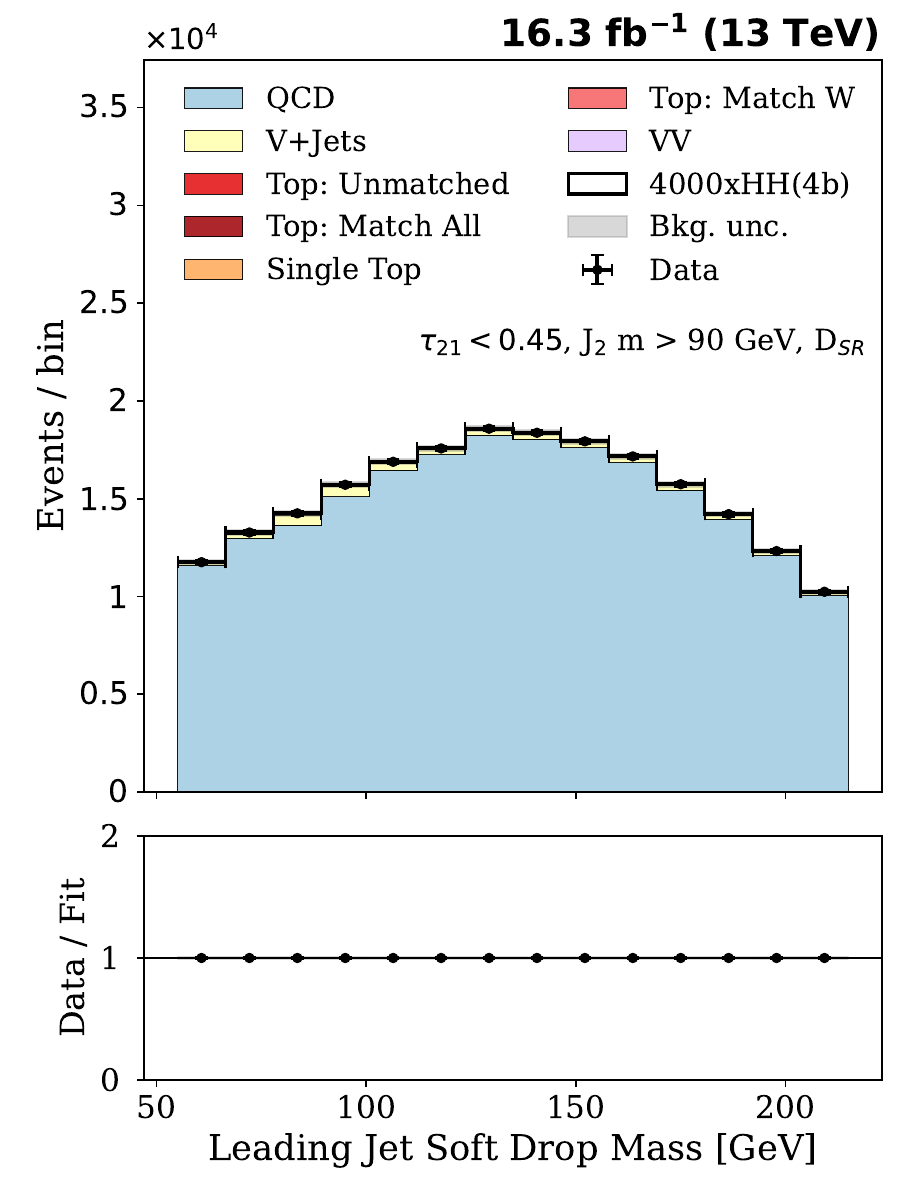}
        \includegraphics[width=.31\textwidth]{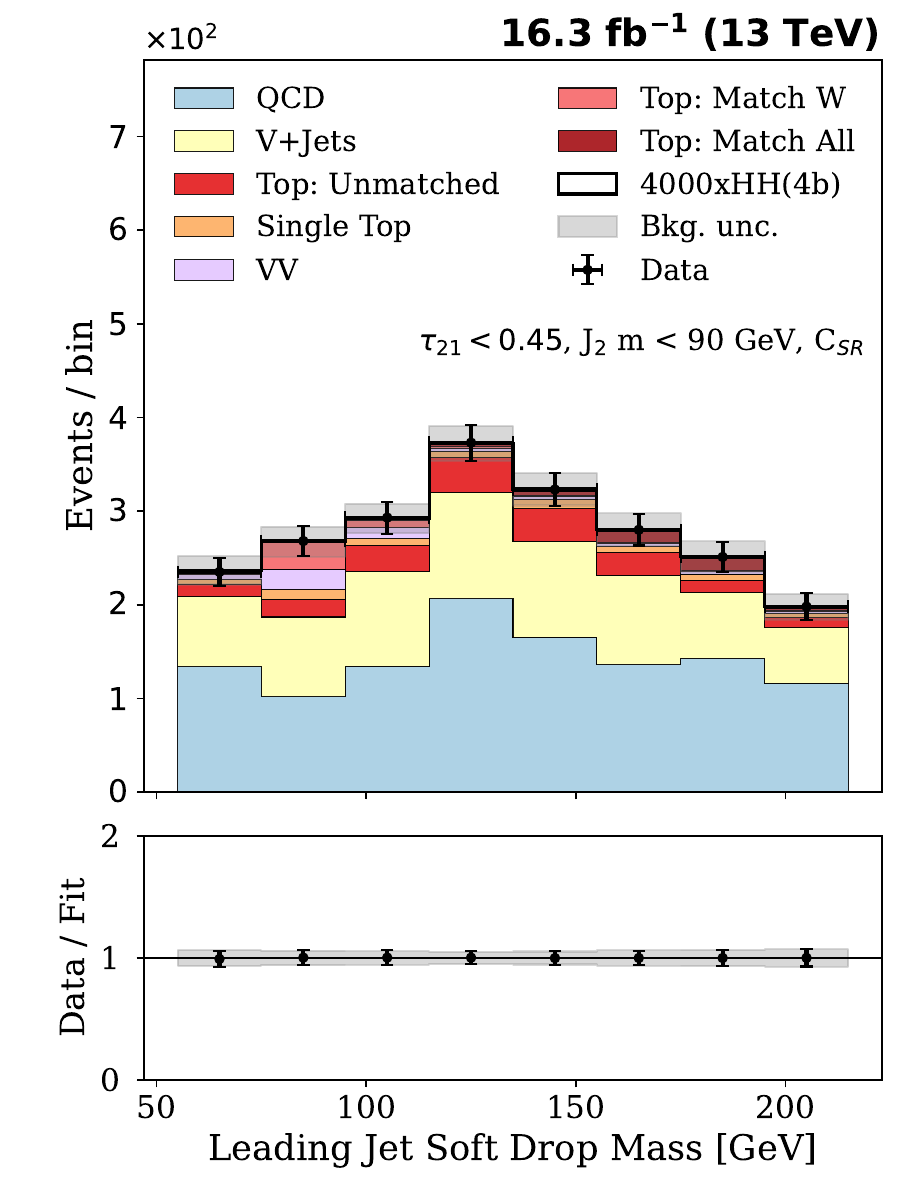}
        \includegraphics[width=.31\textwidth]{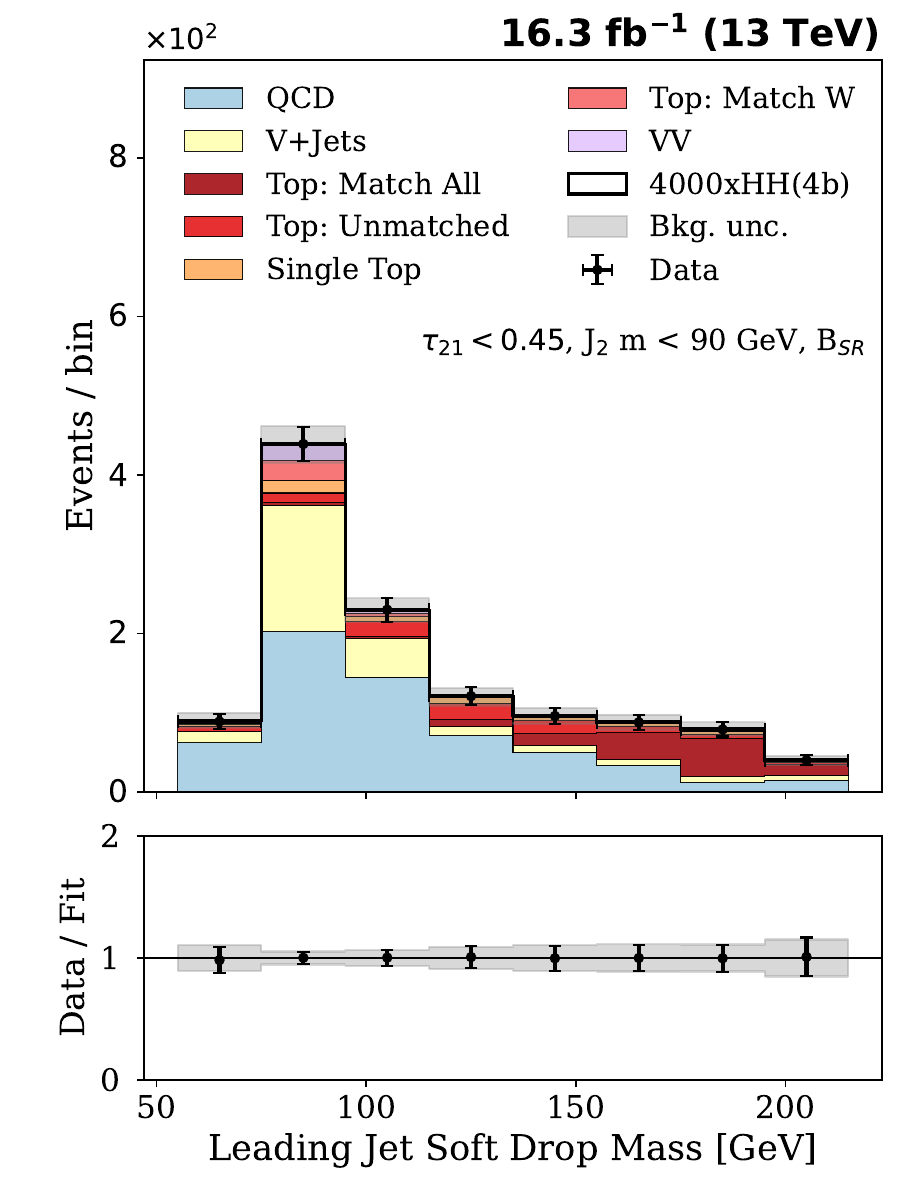}
        \includegraphics[width=.31\textwidth]{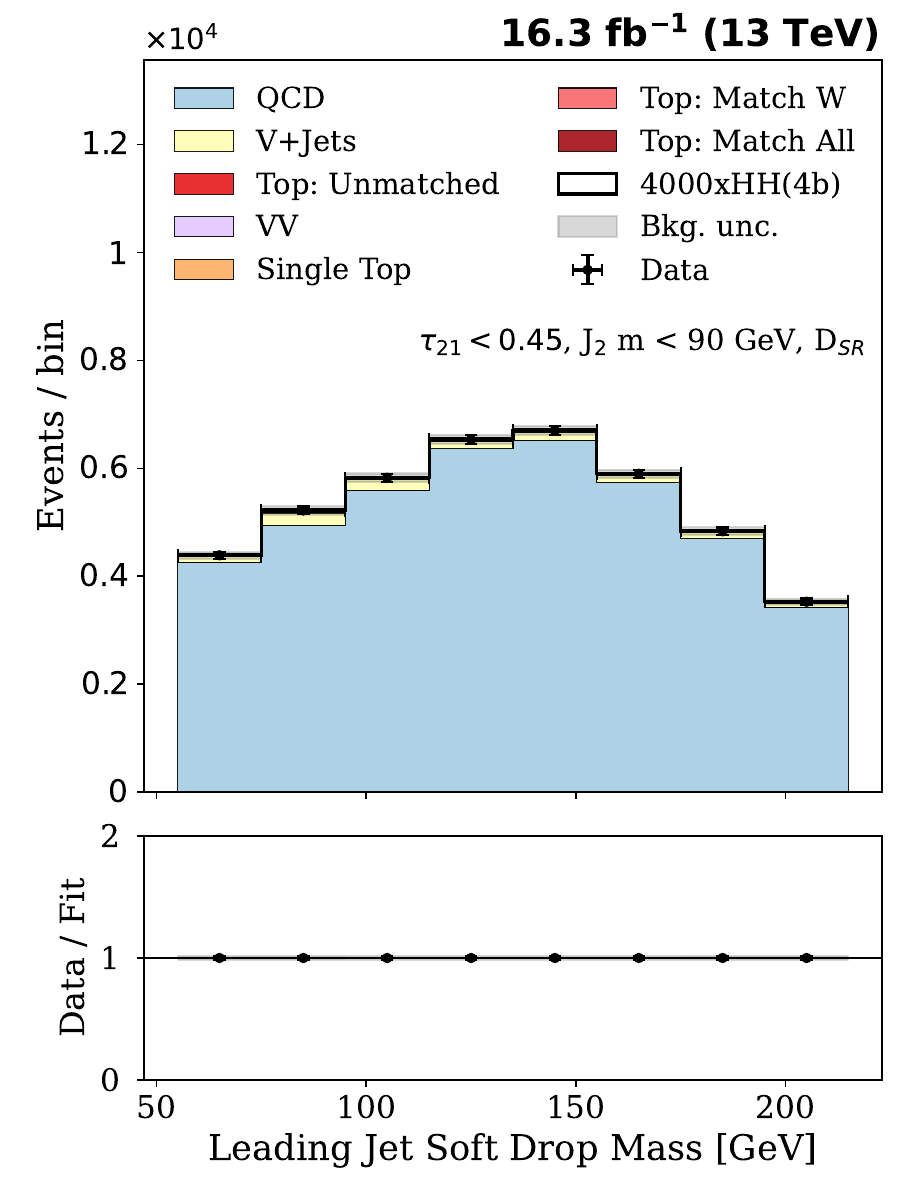}
        \includegraphics[width=.31\textwidth]{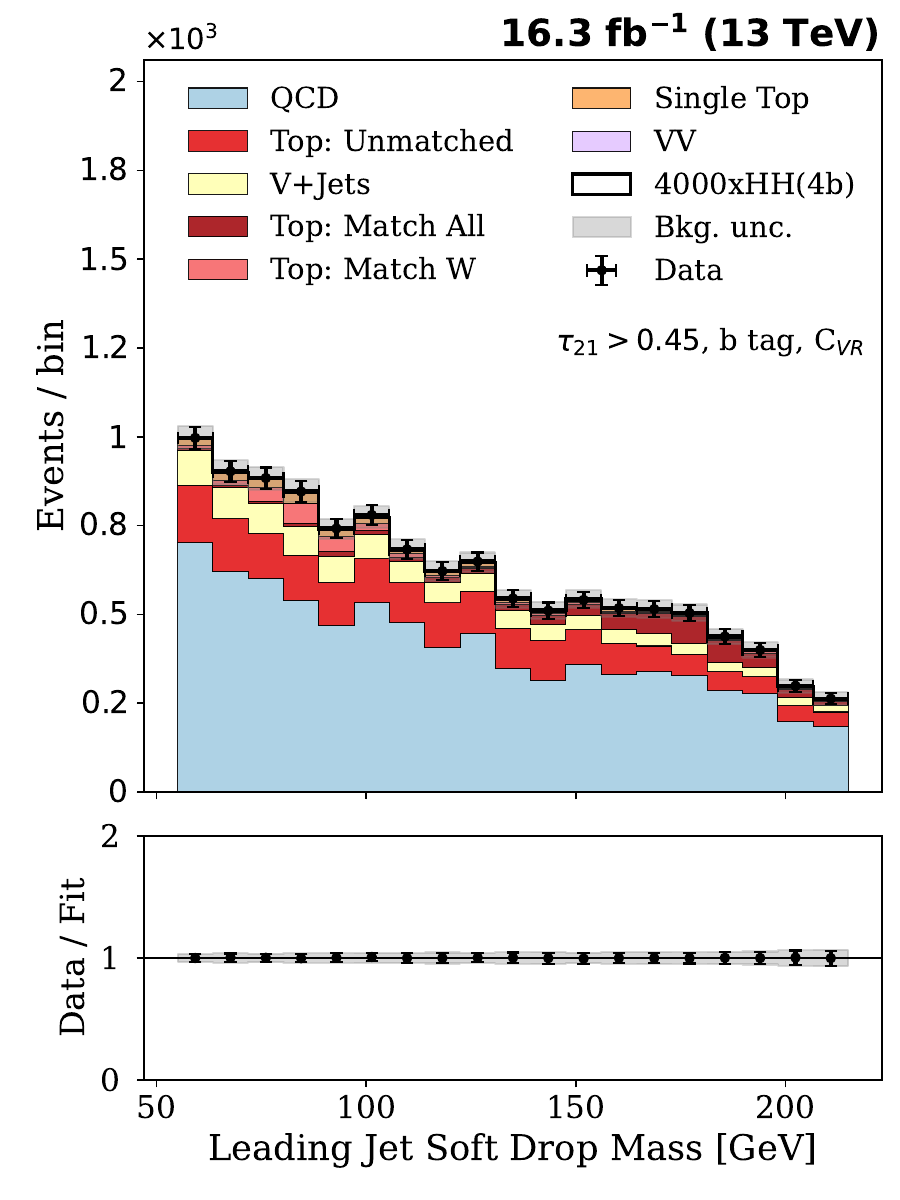}
        \includegraphics[width=.31\textwidth]{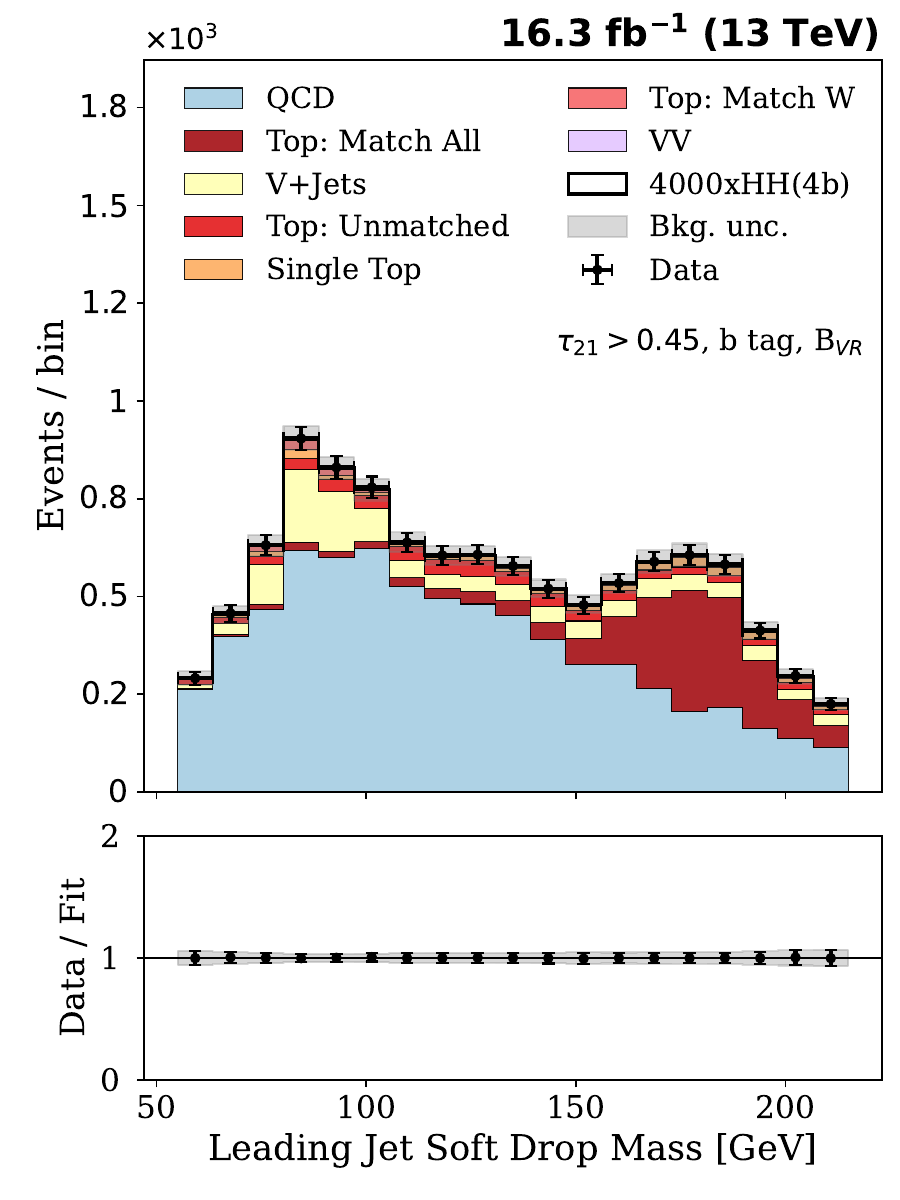}
        \includegraphics[width=.31\textwidth]{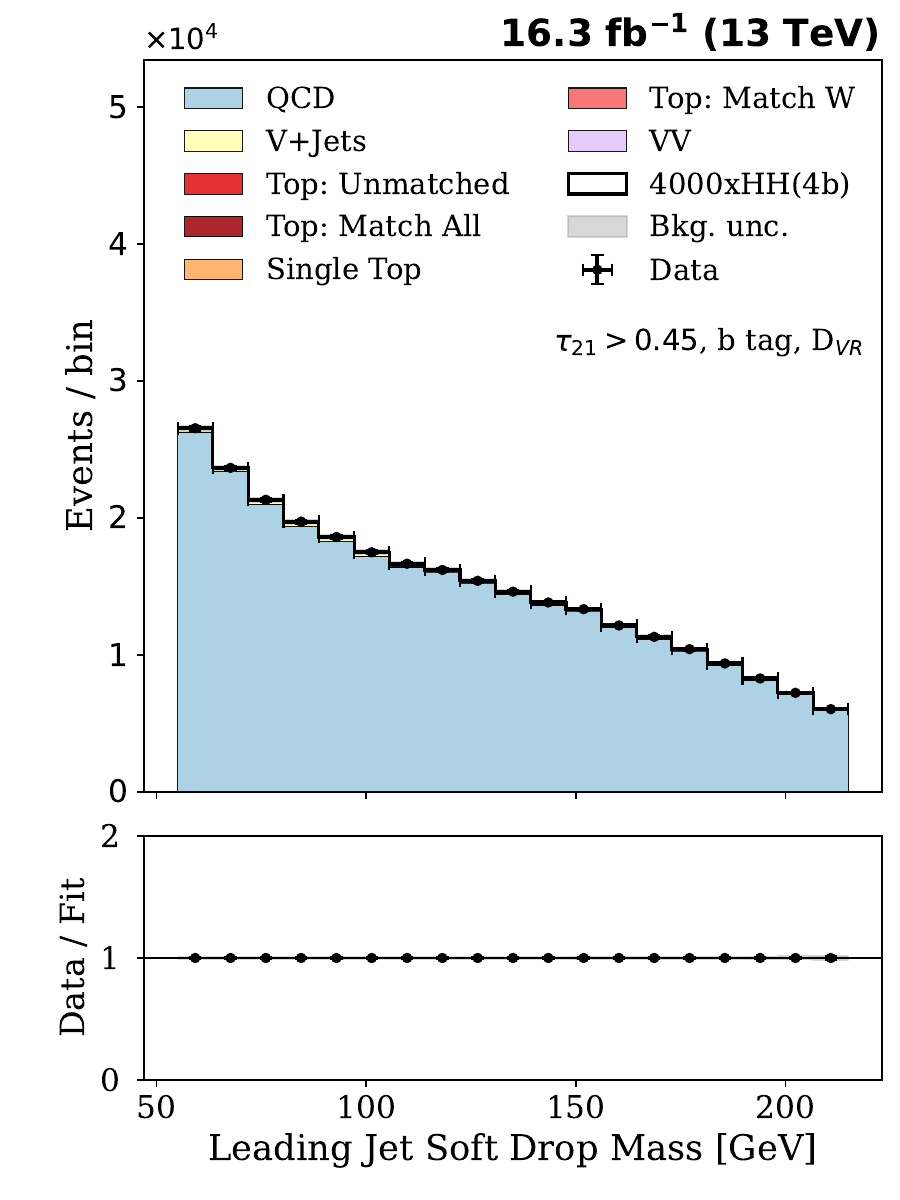}
        
    \caption{Leading jet soft drop mass where both jets are considered anomalous based on the \textsc{OmniLearned} large model score. The region where both jets have low $\tau_{21}$ values and subleading jet mass is above 90 GeV  is shown at the top while the region where the subleading jet mass is below 90 GeV is shown at the middle. The region where at least one jet fails the $\tau_{21}$ selection is shown at the bottom. The different regions used for the ABCD calculation are shown as columns. Shaded regions represent the total background uncertainty.}
    \label{fig:hh_sr12_split}
\end{figure*}

\begin{figure*}[ht]
    \centering
        \includegraphics[width=.31\textwidth]{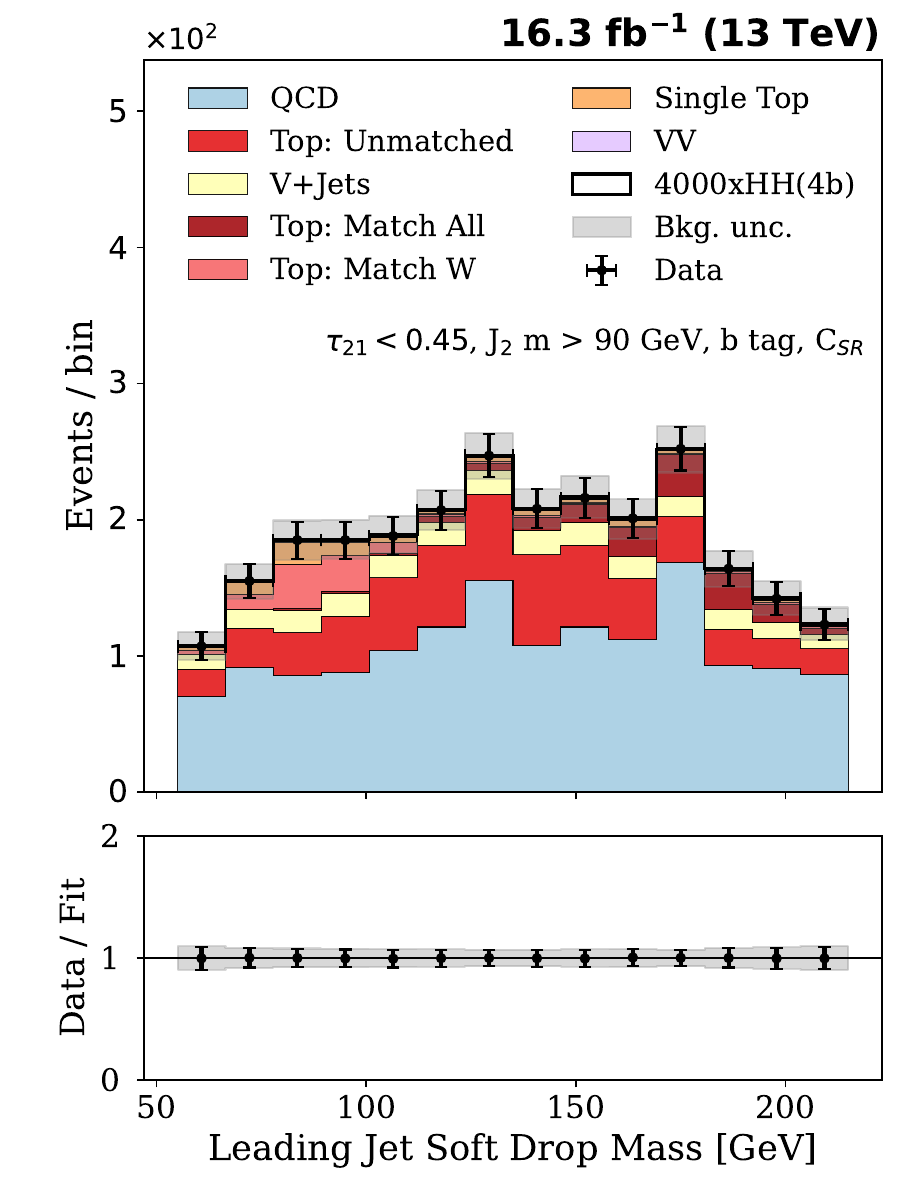}
        \includegraphics[width=.31\textwidth]{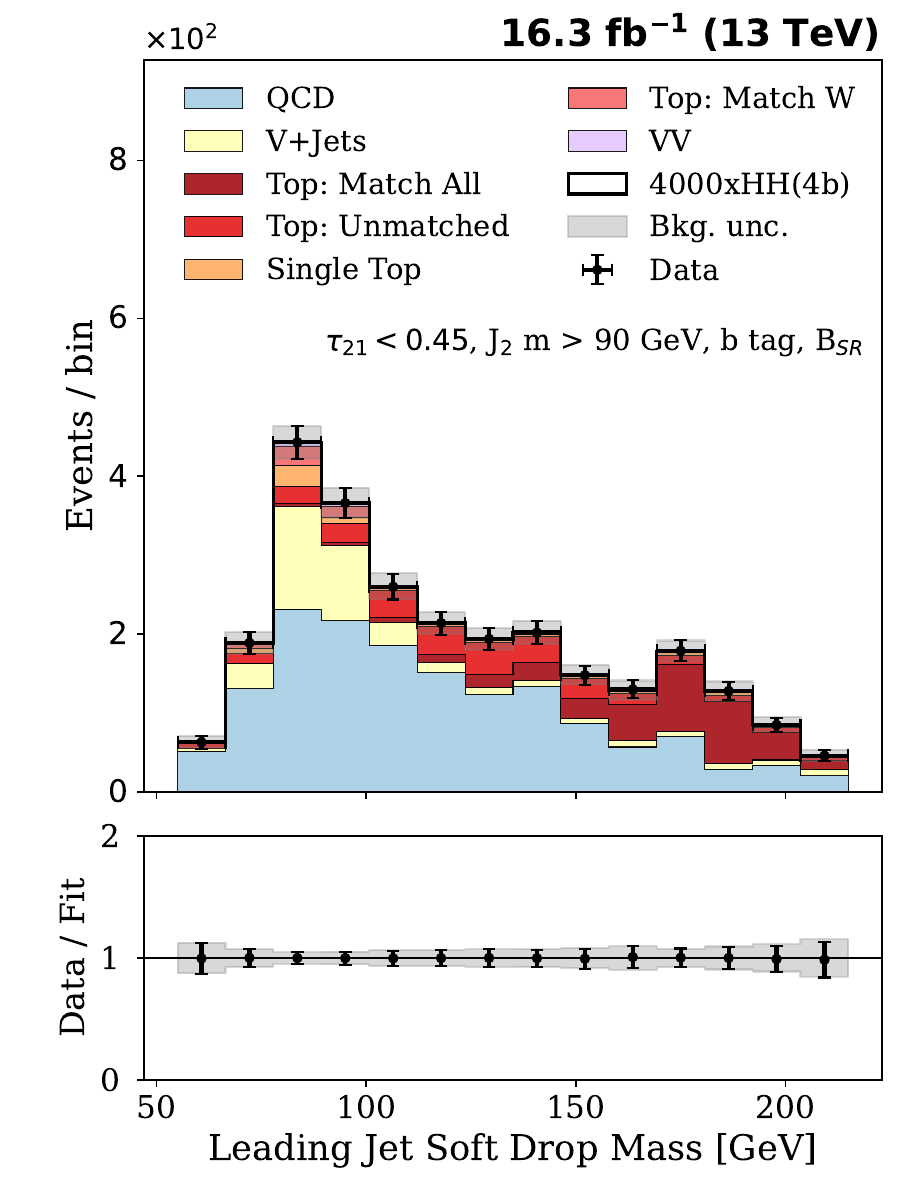}
        \includegraphics[width=.31\textwidth]{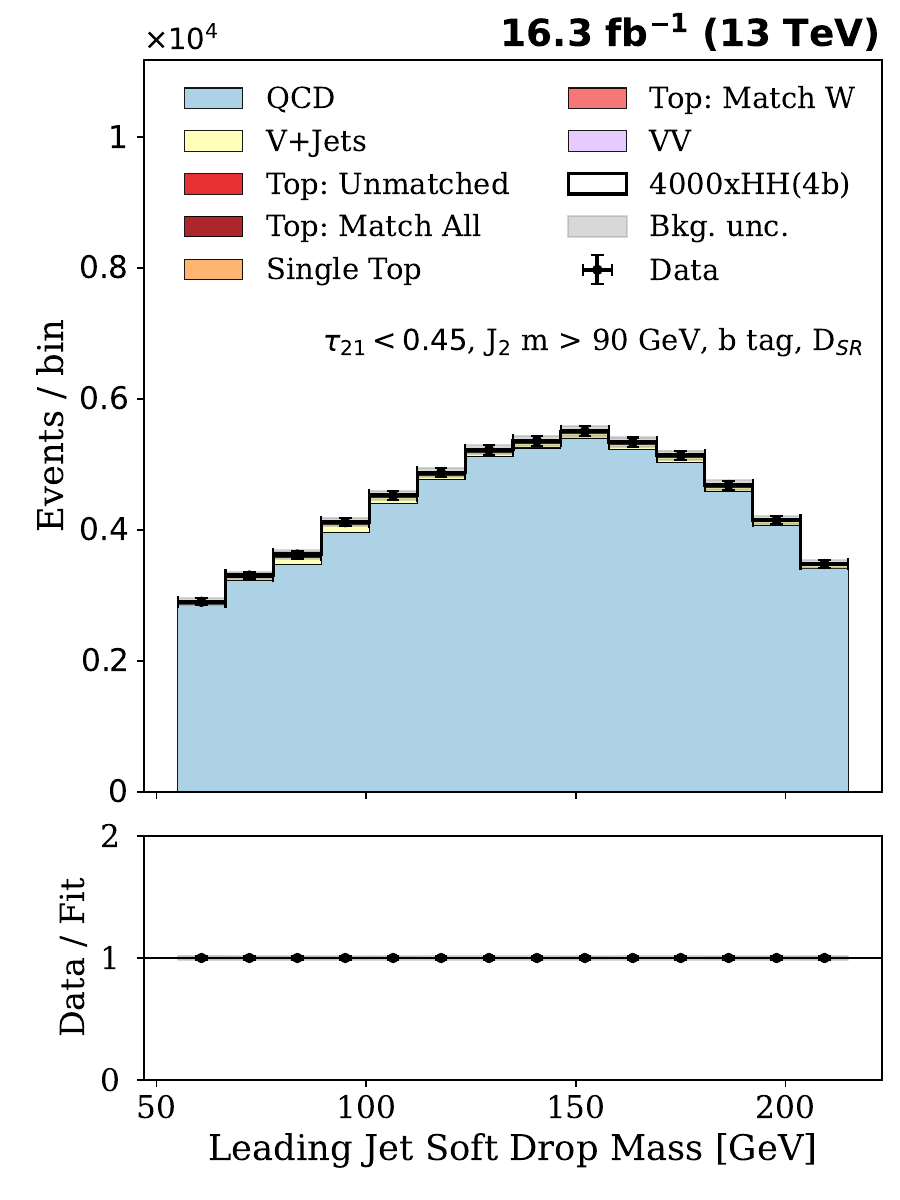}
        \includegraphics[width=.31\textwidth]{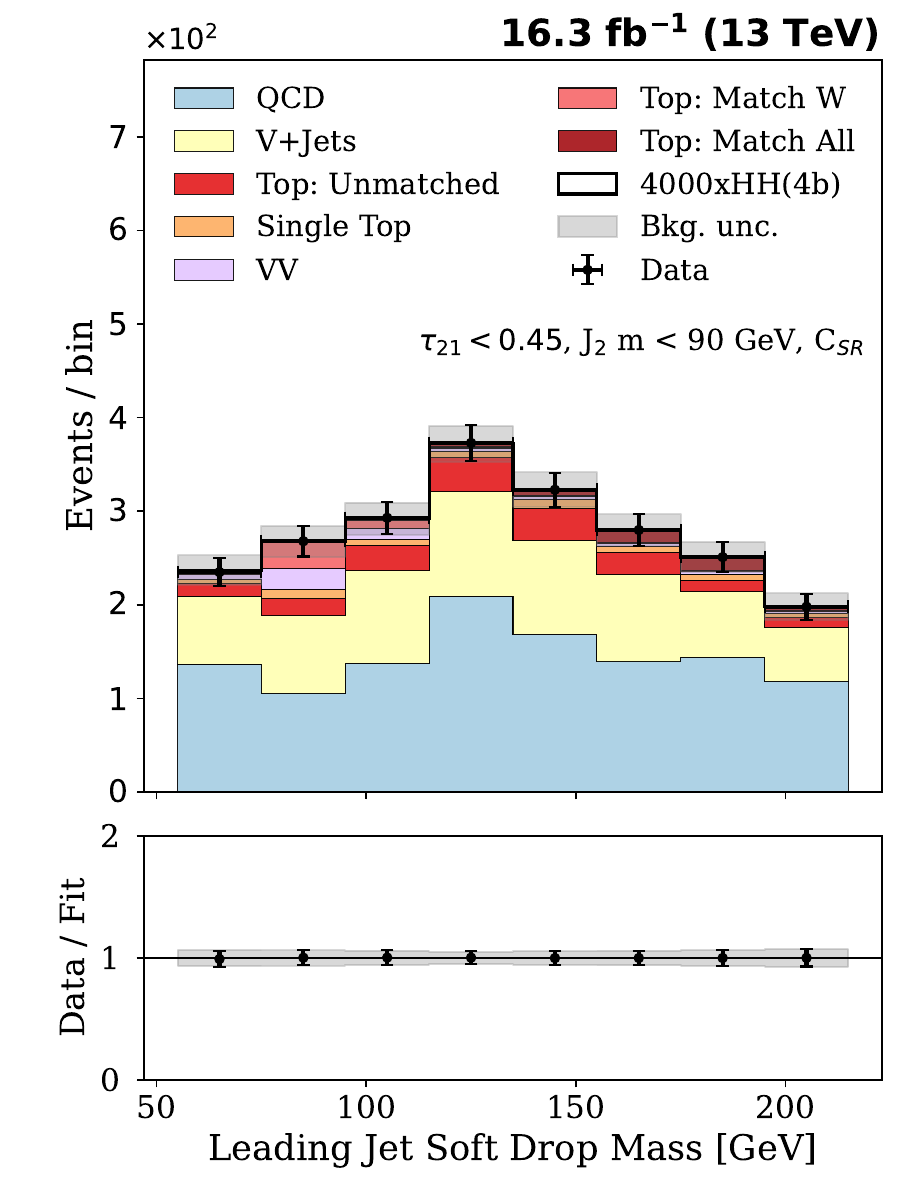}
        \includegraphics[width=.31\textwidth]{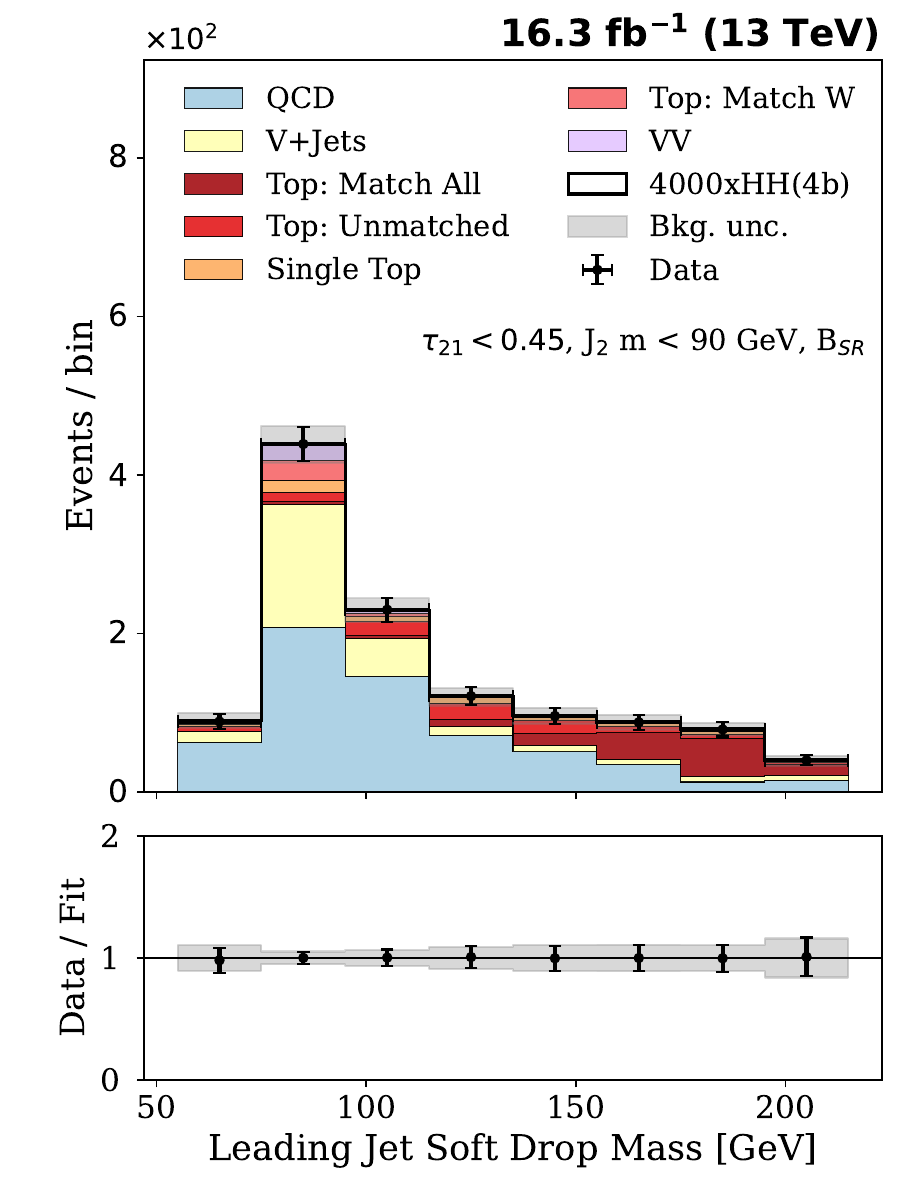}
        \includegraphics[width=.31\textwidth]{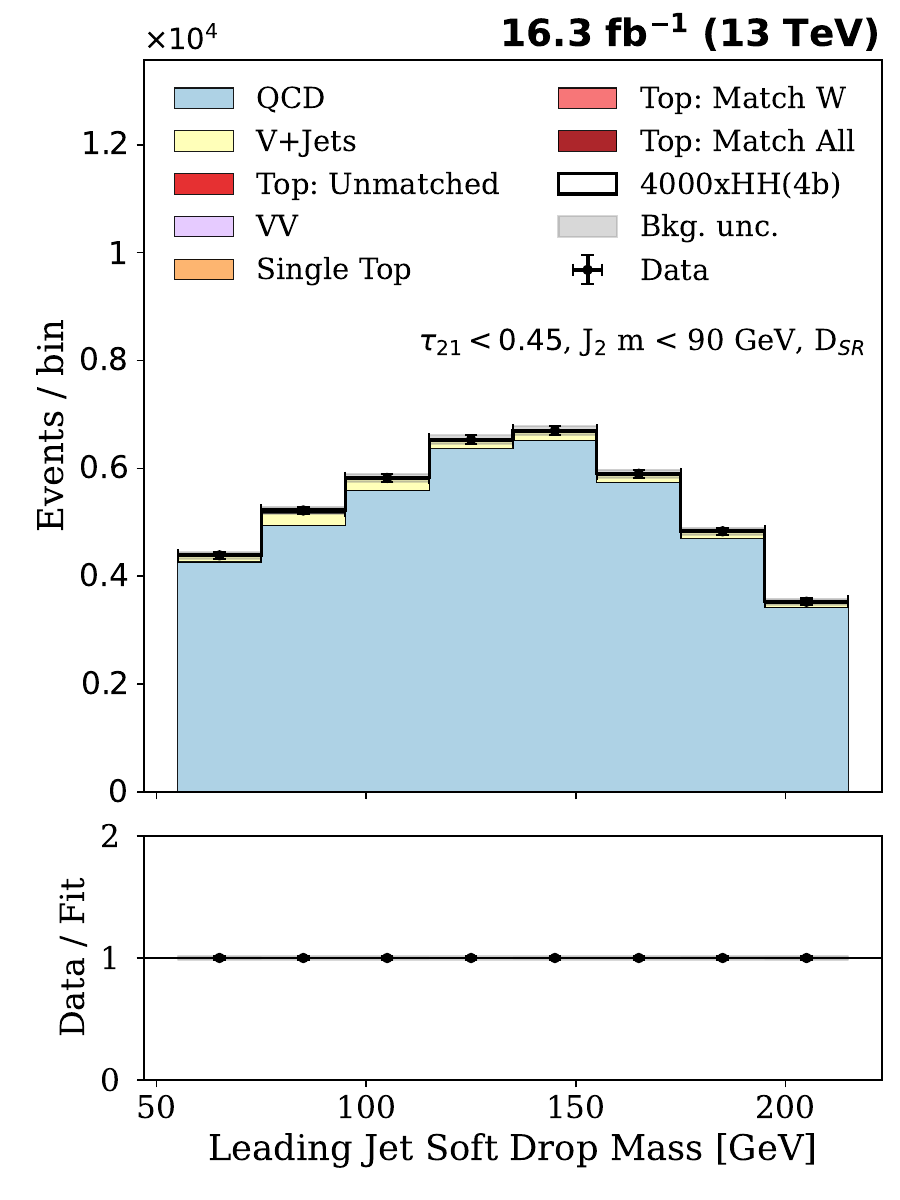}
        \includegraphics[width=.31\textwidth]{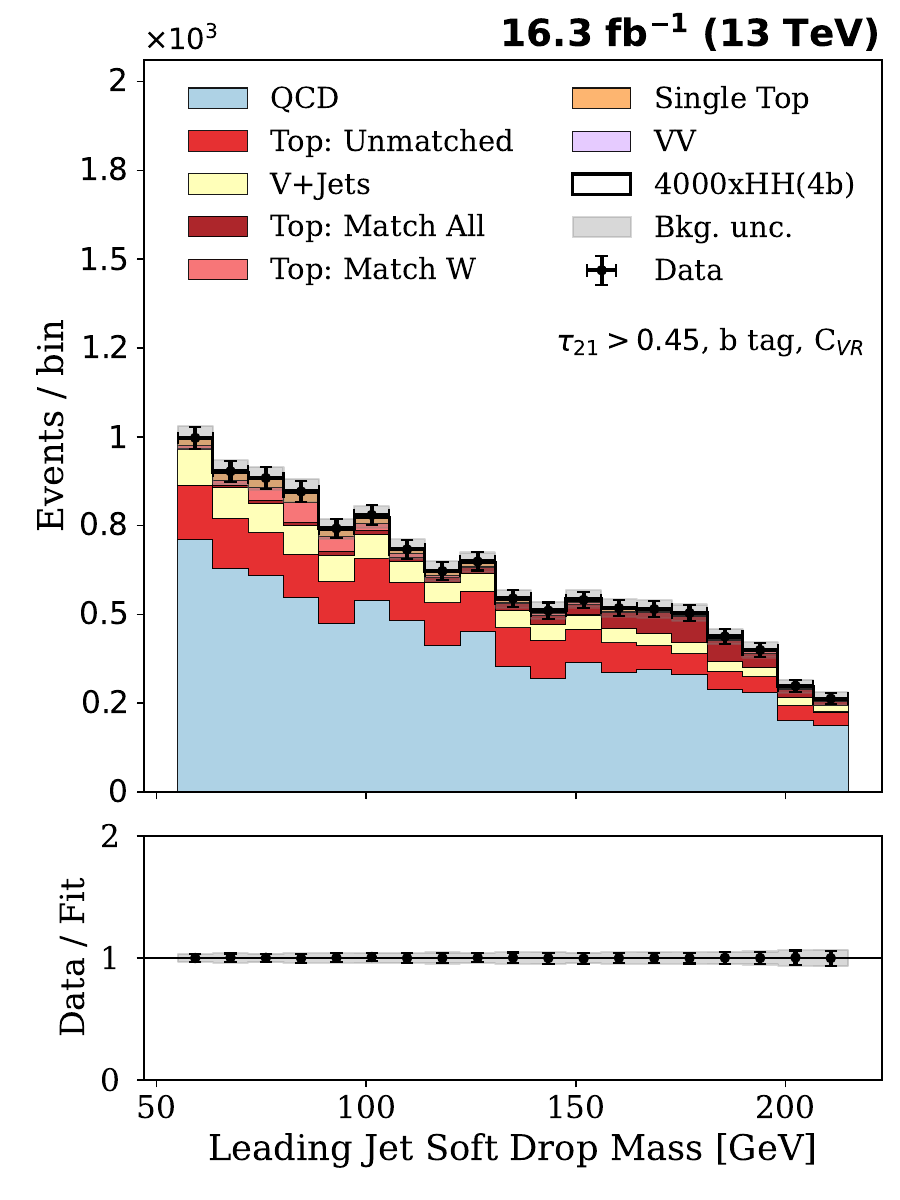}
        \includegraphics[width=.31\textwidth]{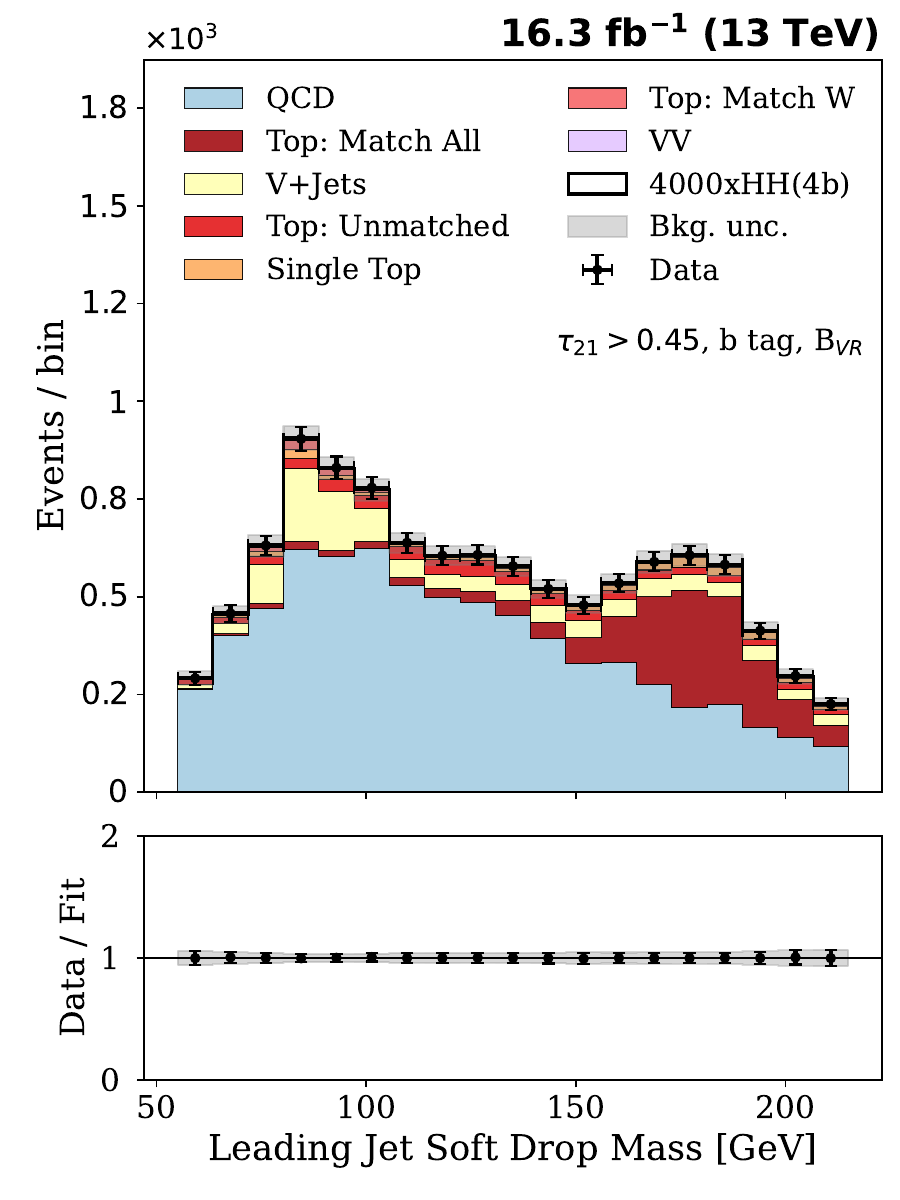}
        \includegraphics[width=.31\textwidth]{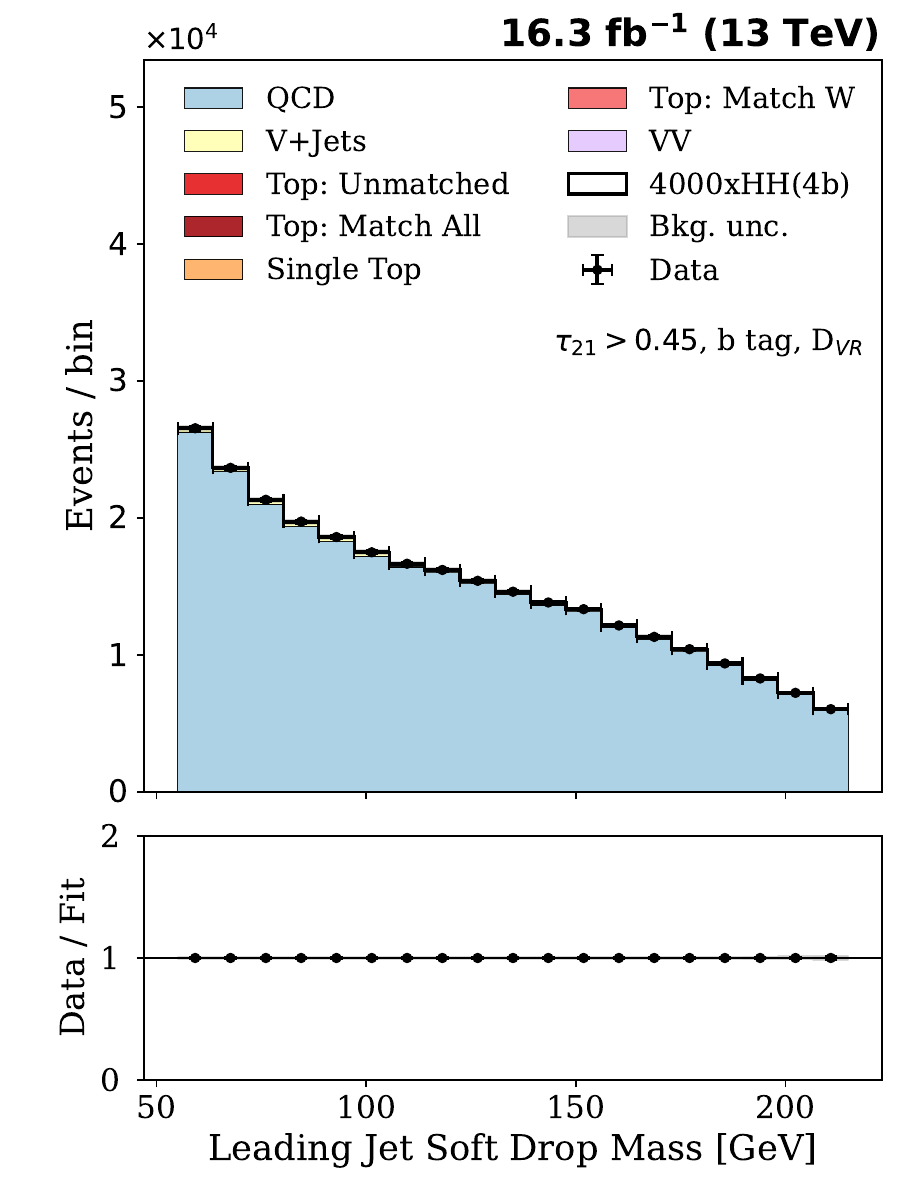}
        
    \caption{Leading jet soft drop mass where both jets are considered anomalous based on the \textsc{OmniLearned} large model score. The region where both jets have low $\tau_{21}$ values at least one jet is b-tagged, and subleading jet mass is above 90 GeV  is shown at the top while the region where the subleading jet mass is below 90 GeV is shown at the middle. The region where at least one jet fails the $\tau_{21}$ selection is shown at the bottom. The different regions used for the ABCD calculation are shown as columns. Shaded regions represent the total background uncertainty.}
    \label{fig:hh_sr32_split}
\end{figure*}

For completeness we also include the distributions in the control regions used to estimate the QCD contribution in Figs.~\ref{fig:hh_baseline_split},~\ref{fig:hh_sr12_split}, and~\ref{fig:hh_sr32_split}.

\subsection{Anomaly Score Studies}
\begin{figure}[ht]
    \centering
        \includegraphics[width=.23\textwidth]{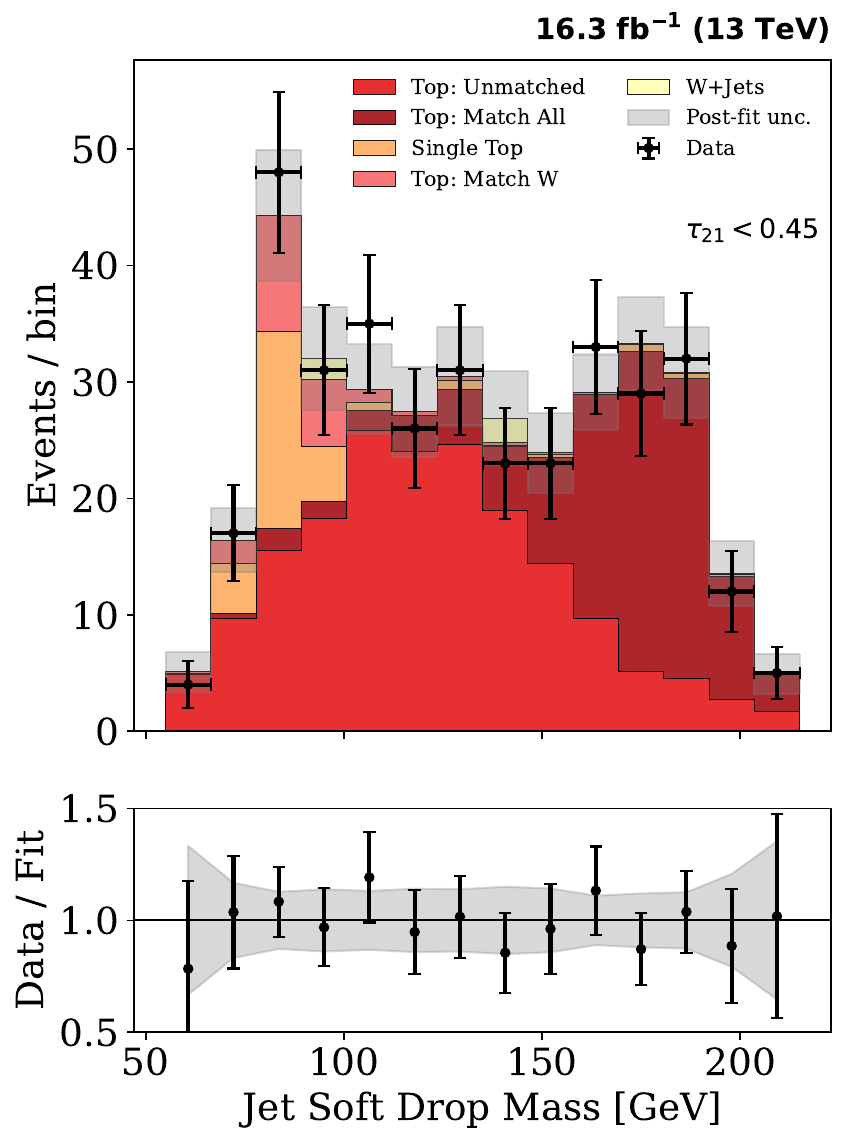}
        \includegraphics[width=.23\textwidth]{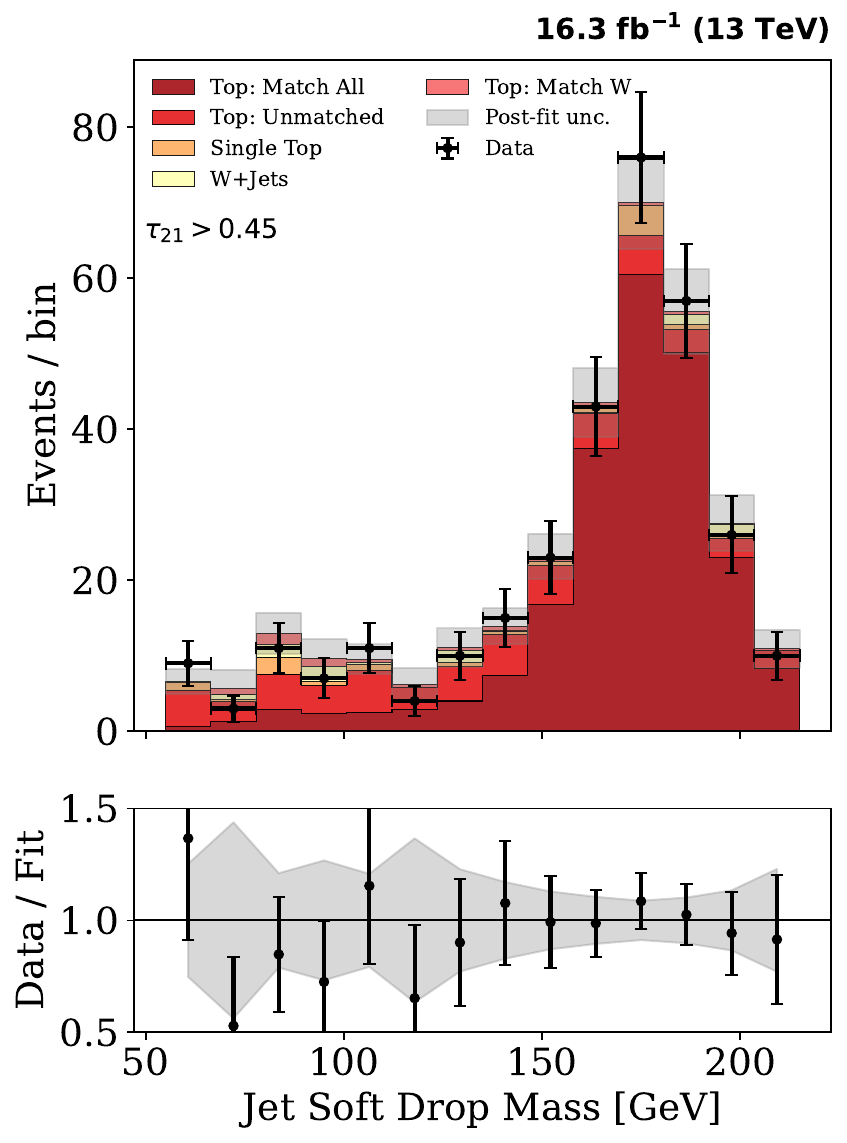}                
    \caption{Large-radius jet soft drop mass for jets passing the anomaly detection threshold used in the anomaly detection search. The region where the jet has low $\tau_{21}$ is shown on the left while region where the jet fails the $\tau_{21}$ selection is shown on the right.  Shaded regions represent the total background uncertainty.}
    \label{fig:muon_fit}
\end{figure}

The anomaly score used in this work is based on the output of the \textsc{OmniLearned} model that selects generic multi-prong jets based on signatures seen during the pretraining phase of the training. To compare events selected as anomalous in data and by simulated processes we define a control region dominated by top quark pair production where one of the top quarks decays hadronically and is reconstructed within a large radius jet while the other decays leptonically resulting in an isolated muon. In this selection, we require collision events to have a single isolated muon with $\pt > 55$ GeV, one small radius jet with radius parameter of 0.4 to pass the b-tagging threshold and the presence of a large radius jet with transverse momentum above 450 GeV, matching the original analysis requirements. We also require the distance between the muon and b-tagged jet to be below 0.8 while the distance between the large-radius jet and the isolated muon to be 0.8 or more. For events passing this criteria we apply the same anomaly detector score threshold to the large radius jet and create two categories based on the value of the $\tau_{21}$ observable. Next, we perform the same fit procedure we use in these studies to fit the soft drop mass distribution of the large-radius jet. The results of the fit are shown in Fig.~\ref{fig:muon_fit} where again we split the top quark contribution as multiple categories based on the matching to generation level objects.

We observe a good agreement between the background model and the data.

\bibliography{HEPML,other}
\bibliographystyle{apsrev4-1}

\clearpage

\end{document}